\documentclass[12pt]{article}
\usepackage[dvips]{graphicx}
\usepackage{epic}
\usepackage{eepic}
\usepackage{epsfig}

\begin{document}

\newcommand{\be}{\begin{equation}}
\newcommand{\ee}{\end{equation}}
\newcommand{\bea}{\begin{eqnarray}}
\newcommand{\eea}{\end{eqnarray}}
\newcommand{\vmat}[4]{\left(\begin{array}{cc}#1&#2\\#3&#4\end{array}\right)}

%
%
%

\title{ {\Huge\bf Bayesian Field Theory}\\ 
{\Large\bf Nonparametric Approaches 
to Density Estimation, Regression,
Classification, and Inverse Quantum Problems
}}
\author{J\"org C. Lemm\footnote{
Email: lemm@uni-muenster.de,
WWW: http://pauli.uni-muenster.de/${}^\sim$lemm/}
\vspace{0.5cm}
\\{Institut f\"ur Theoretische Physik I}
\\{Universit\"at M\"unster}
\\{Wilhelm--Klemm--Str.9}
\\{D--48149 M\"unster, Germany}
}
\date{}
\maketitle

\begin{abstract}
Bayesian field theory denotes a
nonparametric Bayesian approach
for learning functions from observational data.
Based on the principles of Bayesian statistics,
a particular Bayesian field theory
is defined by combining two models:
a likelihood model, 
providing a probabilistic description 
of the measurement process,
and a prior model,
providing the information
necessary to generalize from training to non--training data.
The particular likelihood models discussed in the paper
are those of
general density estimation,
Gaussian regression, clustering, 
classification, 
and models specific for inverse quantum problems.
Besides problem typical hard constraints,
like normalization and non--negativity for probabilities,
prior models have to implement all the specific, and often vague,
{\it a priori} knowledge available for a specific task.
Nonparametric prior models discussed in the paper are
Gaussian processes, mixtures of Gaussian processes, and
non--quadratic potentials.
Prior models are made flexible 
by including hyperparameters.
In particular, the adaption of
mean functions and covariance operators of Gaussian process components
is discussed in detail.
Even if constructed using Gaussian process building blocks,
Bayesian field theories are typically non--Gaussian
and have thus to be solved numerically.
According to increasing computational resources
the class of non--Gaussian Bayesian field theories of practical interest 
which are numerically feasible is steadily growing.
Models which turn out to be computationally too demanding
can serve as starting point to construct easier to solve 
parametric approaches, using for example variational techniques.
\end{abstract}

\tableofcontents

\section{Introduction}

The last decade has seen a rapidly growing interest 
in learning from observational data.
Increasing computational resources enabled successful applications of 
empirical learning algorithms in various areas including, for example,
time series prediction,
image reconstruction,
speech recognition,
computer tomography,
and 
inverse scattering and inverse spectral problems
for quantum mechanical systems.
Empirical learning, i.e.,
the problem of finding underlying general laws from observations,
represents a typical inverse problem
and is usually ill--posed in the sense of Hadamard
\cite{Tikhonov-1963,Tikhonov-Arsenin-1977,Vapnik-1982,Louis-1989,Kirsch-1996,Vapnik-1998}.
It is well known that a successful solution of such problems
requires additional {\it a priori} information.
It is {\it a priori} information which
controls the generalization ability of a learning system
by providing the link between available empirical ``training'' data 
and unknown outcome in future ``test'' situations.

We will focus mainly on nonparametric approaches,
formulated directly in terms of the function values of interest.
Parametric methods, on the other hand, 
impose typically implicit restrictions
which are often extremely difficult to relate 
to available {\it a priori} knowledge.
Combined with a Bayesian framework
\cite{Bayes-1763,Berger-1980,Bretthorst-1988,Loredo-1990,Robert-1994,OHagen-1994,Bernado-Smith-1994,Gelman-Carlin-Stern-Rubin-1995,Sivia-1996,Carlin-Louis-1996,West-Harrison-1997,D'Agostini-1999,Jeffrey-1999,Jaynes-in-prep},
a nonparametric approach allows a very flexible and interpretable
implementation of {\it a priori} information 
in form of stochastic processes.  
Nonparametric Bayesian methods 
can easily be adapted to different learning situations
and have therefore been applied 
to a variety of empirical learning problems,
including regression, classification, density estimation
and inverse quantum problems
\cite{Neal-1996,Williams-Rasmussen-1996,Lemm-1999b,Lemm-1999c,Lemm-1999d,Uhlig-PhD}.
Technically, they are related to 
kernel and regularization methods 
which often appear in the form of a roughness penalty approach 
\cite{Tikhonov-Arsenin-1977,Vapnik-1982,Poggio-Torre-Koch-1985,Silverman-1986,Marroquin-Mitter-Poggio-1987,Wahba-1990,Haerdle-1990,Green-Silverman-1994,Kitagawa-Gersch-1996,Vapnik-1998}.
Computationally, working with stochastic processes, 
or discretized versions thereof,
is more demanding
than, for example, fitting a small number of parameters.
This holds especially for such applications 
where one cannot take full advantage of the convenient analytical features
of Gaussian processes.
Nevertheless, it seems to be the right time 
to study nonparametric Bayesian approaches
also for non--Gaussian problems 
as they become computationally feasible now
at least for low dimensional systems
and, even if not directly solvable,
they provide a well defined basis for further approximations.

In this paper we will in particular study general density estimation problems.
Those include, as special cases,
regression, classification, and certain types of clustering.
In density estimation the functions of interest are 
the probability densities $p(y|x,h)$,
of producing output (``data'') $y$ under condition $x$
and unknown state of Nature $h$.
Considered as function of $h$, for fixed $y$, $x$, the function $p(y|x,h)$ 
is also known as {\it likelihood} function
and a Bayesian approach to density estimation
is based on a probabilistic model for likelihoods $p(y|x,h)$.
We will concentrate on situations where $y$ and $x$ are real variables,
possibly multi--dimensional.
In a nonparametric approach,
the variable $h$ represents the whole likelihood function $p(y|x,h)$.
That means, $h$ may be seen as the collection 
of the numbers $0\le p(y|x,h)\le 1$ for all $x$ and all $y$.
The dimension of $h$ is thus infinite,
if the number of values which the variables $x$ and/or $y$ can take 
is infinite.
This is the case for real $x$ and/or $y$.

A learning problem with discrete $y$ variable
is also called a {\it classification problem}.
Restricting to Gaussian probabilities $p(y|x,h)$ with fixed variance
leads to (Gaussian) {\it regression problems}.
For regression problems
the aim is to find an optimal regression function $h(x)$.
Similarly, adapting a mixture of Gaussians 
allows {\it soft clustering} of data points.
Furthermore, extracting relevant features from the 
predictive density $p(y|x,{\rm data})$
is the Bayesian analogue of {\it unsupervised learning}.
Other special density estimation problems
are, for example,
{\it inverse problems in quantum mechanics} where $h$ represents a unknown
potential to be determined from observational data
\cite{Lemm-1999b,Lemm-1999c,Lemm-1999d,Uhlig-PhD}.
Special emphasis will be put on the explicit and flexible implementation
of {\it a priori} information using, for example,
mixtures of Gaussian prior processes
with adaptive, non--zero mean functions
for the mixture components.

Let us now shortly explain what is meant by
the term ``Bayesian Field Theory'':
From a physicists point of view functions, like $h(x,y)$ = $p(y|x,h)$,
depending on continuous variables $x$ and/or $y$,
are often called a `field'.\footnote{We may also remark 
that for example statistical field theories,
which encompass quantum mechanics and quantum field theory
in their Euclidean formulation,
are technically similar to a nonparametric Bayesian approach
\cite{Zinn-Justin-1989,Itzykson-Drouffe-1989,Le_Bellac-1991}.}
Most times in this paper we will, 
as common in field theories in physics, 
not parameterize these fields 
and formulate the relevant 
probability densities or stochastic processes, 
like the prior $p(h)$ or the posterior $p(h|f)$, directly 
in terms of the field values $h(x,y)$, e.g.,
$p(h|f)$ = $p(h(x,y),x\in X,y\in Y|f)$.
(In the parametric case, discussed in Chapter \ref{variational},
we obtain a probability density $p(h|f)$ = $p(\xi|f)$
for fields $h(x,y,\xi)$ parameterized by $\xi$.)

The possibility to solve Gaussian integrals analytically
makes Gaussian processes,
or (generalized) free fields in the language of physicists,
very attractive for nonparametric learning.
Unfortunately, only the case of Gaussian regression
is completely Gaussian.
For general density estimation problems 
the likelihood terms are non--Gaussian,
and even for Gaussian priors 
additional non--Gaussian restrictions
have to be included to ensure non--negativity and normalization
of densities.
Hence, in the general case, density estimation 
corresponds to a non--Gaussian, i.e., interacting field theory.

As it is well known from physics, 
a continuum limit for non-Gaussian theories,
based on the definition of a renormalization procedure,
can be highly nontrivial to construct.
(See  \cite{Bialek-Callan-Strong-1996,Aida-1999}
for an renormalization approach to density estimation.)
We will in the following not discuss such renormalization procedures
but focus more on practical, numerical learning algorithms, 
obtained by discretizing the problem
(typically, but not necessarily in coordinate space).
This is similar, for example, to what is done in lattice field theories.

Gaussian problems live effectively in a space
with dimension not larger than the number of training data.
This is not the case for
non--Gaussian problems.
Hence, numerical implementations of learning algorithms
for non--Gaussian problems 
require to discretize the functions of interest.
This can be computationally challenging.

For low dimensional problems, however,
many non--Gaussian models
are nowadays solvable on a standard PC.
Examples include predictions of
one--dimensional time series or 
the reconstruction of two--dimensional images.
Higher dimensional problems require
additional approximations,
like projections into lower dimensional subspaces
or other variational approaches.
Indeed, it seems that a most solvable high dimensional problems
live effectively in some low dimensional subspace.

There are special situations in classification
where non--negativity and normalization constraints 
are fulfilled automatically. 
In that case, the calculations can still
be performed in a space of dimension not larger
than the number of training data.
Contrasting Gaussian models, however
the equations to be solved are then typically nonlinear.

Summarizing, we will call a nonparametric Bayesian model
to learn a function one or more continuous variables
a {\it Bayesian field theory}, 
having especially in mind non--Gaussian models.
A large variety of Bayesian field theories 
can be constructed by combining a specific likelihood models
with specific functional priors
(see Tab.\ \ref{model-scheme}).
The resulting flexibility of nonparametric Bayesian approaches
is probably their main advantage.

\begin{table}[ht]
\begin{center}
\begin{tabular}{|c|c|}
\hline\rule[-3mm]{0mm}{9mm}
likelihood model &prior model \\
\hline
\hline
\multicolumn{2}{|c|}{describes\rule[-3mm]{0mm}{9mm}} \\
\hline\rule[-3mm]{0mm}{9mm}
measurement process (Chap.~\ref{Bayesian})
& generalization behavior (Chap.~\ref{Bayesian})\\
\hline
\multicolumn{2}{|c|}{is determined by\rule[-3mm]{0mm}{9mm}} \\
\hline\rule[-3mm]{0mm}{9mm}
parameters (Chap.~\ref{Gaussian-priors}, \ref{variational})
& hyperparameters (Chap.~\ref{hyperparameters})\\
\hline
\multicolumn{2}{|c|}{Examples include\rule[-3mm]{0mm}{9mm}} \\
\hline\rule[-3mm]{0mm}{9mm}
$\!$$\!$$\!$$\!$$\!$
density estimation(Sects.~\ref{gaussL}--\ref{quad-de}, \ref{mix-de})$\!$
&hard constraints (Chap.~\ref{Bayesian})
\\
\rule[-3mm]{0mm}{9mm}
regression (Sects.~\ref{regression}, \ref{mixtures-for-regression})
&Gaussian prior factors 
(Chap.~\ref{Gaussian-priors})\\
\rule[-3mm]{0mm}{9mm}
classification (Sect.~\ref{classification})
&$\!$mixtures of Gauss. (Sects.~\ref{mix-of-G}--\ref{local-mix})$\!$
\\
\rule[-3mm]{0mm}{9mm}
inverse quantum theory (Sect.~\ref{inverse-quantum-mechanics})
&$\!$$\!$non--quadratic potentials(Sect.~\ref{other-non-gaussian})
$\!$$\!$$\!$$\!$\\
\hline
\multicolumn{2}{|c|}{Learning algorithms are treated in Chapter \ref{learning}.
\rule[-3mm]{0mm}{9mm}} \\
\hline
\end{tabular}
\end{center}
\caption{A Bayesian approach is based on the combination of two models,
a likelihood model, describing
the measurement process used to obtain the training data,
and a prior model, 
enabling generalization to non--training data.
Parameters of the prior model are commonly called hyperparameters.
In ``nonparametric'' approaches
the collection of all values of the likelihood function itself
are considered as the parameters.
A nonparametric Bayesian approach
for likelihoods depending on one or more real variables
is in this paper called a
Bayesian field theory.
}
\label{model-scheme}
\end{table}
\clearpage

The paper is organized as follows:
Chapter \ref{Bayesian}
summarizes the Bayesian framework as needed for the subsequent chapters.
Basic notations are defined,
an introduction to Bayesian decision theory is given,
and the role of {\it a priori} information is discussed together with  
the basics of a Maximum A Posteriori Approximation (MAP),
and the specific constraints for density estimation problems.
Gaussian prior processes,
being the most commonly used prior processes 
in nonparametric statistics,
are treated in Chapter \ref{Gaussian-priors}.
In combination with Gaussian prior models,
this section also introduces the likelihood models of density estimation,
(Sections \ref{gaussL}, \ref{gaussP}, \ref{other-Gaussian})
Gaussian regression and clustering 
(Section \ref{regression}),
classification (Section \ref{classification}),
and inverse quantum problems (Section \ref{inverse-quantum-mechanics}).
Notice, however, that all these likelihood models
can also be combined with 
the more elaborated 
prior models discussed in the following sections of the paper.
Parametric approaches,
useful if a numerical solution of a full nonparametric approach
is not feasible,
are the topic of Chapter \ref{variational}.
Hyperparameters, parameterizing prior processes
and making them more flexible,
are considered in Section \ref{hyperparameters}.
Two possibilities to go beyond Gaussian processes,
mixture models and non--quadratic potentials,
are presented in Section \ref{non-Gaussian}.
Chapter \ref{learning}
focuses on learning algorithms, i.e., 
on methods to solve the stationarity equations 
resulting from a Maximum A Posteriori Approximation.
In this section one can also find 
numerical solutions of Bayesian field theoretical models 
for general density estimation.

\section{Bayesian framework}
\label{Bayesian}

\subsection{Basic model and notations}
\subsubsection{Independent, dependent, and hidden variables}

Constructing theories means introducing concepts
which are not directly observable.
They should, however, explain empirical findings and thus
have to be related to observations.
Hence, it is useful and common 
to distinguish observable (visible)
from non--observable (hidden) variables.
Furthermore, it is often convenient
to separate visible variables 
into dependent variables, 
representing results of such measurements
the theory is aiming to explain,
and independent variables, 
specifying the kind of measurements performed
and not being subject of the theory.
 
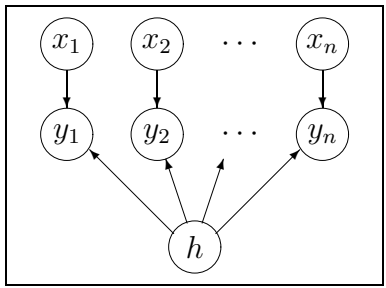
\begin{figure}
\begin{center}
\setlength{\unitlength}{1mm}
\begin{picture}(50,37)
\put(0,0){\framebox(50,37)[]{}}
\put(8,32){\circle{7}}\put(8,32){\makebox(0,0){$x_1$}}
\put(20,32){\circle{7}}\put(20,32){\makebox(0,0){$x_2$}}
\put(42,32){\circle{7}}\put(42,32){\makebox(0,0){$x_n$}}
\put(8,20){\circle{7}}\put(8,20){\makebox(0,0){$y_1$}}
\put(20,20){\circle{7}}\put(20,20){\makebox(0,0){$y_2$}}
\put(42,20){\circle{7}}\put(42,20){\makebox(0,0){$y_n$}}
\put(25,5){\circle{7}}\put(25,5){\makebox(0,0){${h}$}}
\put(31,32){\makebox(0,0){$\cdots$}}  
\put(31,20){\makebox(0,0){$\cdots$}}  
\put(8,28.5){\vector(0,-1){5}}   
\put(20,28.5){\vector(0,-1){5}}  
\put(42,28.5){\vector(0,-1){5}}  
\put(24,8.2){\vector(-1,3){2.85}}  
\put(26,8.2){\vector(1,3){2.85}}  
\put(22.2,6.8){\vector(-1,1){11.2}}  
\put(27.8,6.8){\vector(1,1){11.2}}  
\end{picture}
\end{center}
\caption{
Directed acyclic graph
for the factorization model (\protect\ref{factorization}).
}
\label{graph-model}
\end{figure}

Hence,  we will consider the following three groups of variables
\begin{itemize}
\item[1.]
observable (visible) independent variables $x$,
\item[2.]
observable (visible) dependent variables $y$,
\item[3.]
not directly observable (hidden, latent) variables ${h}$.
\end{itemize}
This characterization of variables
translates to the following factorization property, 
defining the model we will study,
\begin{equation}
p(x,y,{h}) = p(y|x,{h}) \, p(x) \, p({h}).
\label{factorization}
\end{equation}
In particular, we will be interested in scenarios
where $x$ = $(x_1,x_2,\cdots )$
and analogously $y$ = $(y_1,y_2,\cdots )$
are decomposed into independent components,
meaning that 
$p(y|x,{h})$ = $\prod_i p(y_i|x_i,{h})$
and
$p(x)$ = $\prod_i p(x_i)$ factorize.
Then,
\begin{equation}
p(x,y,{h}) = \prod_i p(y_i|x_i,{h}) \,p(x_i) \,p({h}).
\label{factorization2}
\end{equation}
Fig.\ref{graph-model} shows 
a graphical representation of the factorization model (\ref{factorization2})
as a directed acyclic graph 
\cite{Pearl-1988,Lauritzen-1996,Jensen-1996,Ripley-1996}.
The $x_i$ and/or $y_i$ itself can also be vectors.

The interpretation will be as follows:
Variables ${h}\in {H}$ represent possible 
{\it states of (the model of) Nature},
being the invisible conditions for dependent variables $y$.
The set ${H}$ defines the space of all possible states of Nature
for the model under study.
We assume that states ${h}$ are not directly observable
and all information about $p({h})$ 
comes from observed variables (data) $y$, $x$.
A given set of observed data 
results in a {\it state of knowledge} $f$ 
numerically represented by the {\it posterior density} 
$p({h}|f)$ over states of Nature.

Independent variables $x\in X$ describe the visible conditions
(measurement situation, measurement device)
under which dependent variables (measurement results) $y$ 
have been observed (measured).
According to Eq.\ (\ref{factorization})
they are independent of ${h}$, i.e., $p(x|{h})$ = $p(x)$.
The conditional density $p(y|x,{h})$ of the dependent variables $y$
is also known as {\it likelihood} of $h$ (under $y$ given $x$).
Vector--valued $y$ can be treated as a collection of one--dimensional $y$
with the vector index being part of the $x$ variable, i.e.,
$y_\alpha (x)  = y(x,\alpha) = y(\tilde x)$
with $\tilde x = (x,\alpha)$.

In the setting of {\it empirical learning} 
available knowledge is usually separated into
a finite number of {\it training data} 
$D$ = $\{(x_i,y_i)|1\le i\le n\}$
=$\{(x_D,y_D)$
and, to make the problem well defined,
additional {\it a priori} information  $D_0$.
For data $D\cup D_0$
we write $p({h}|f) = p({h}|D,D_0)$.
Hypotheses $h$ represent in this setting functions
$h(x,y)$ = $p(y|x,{h})$ 
of two (possibly multidimensional) variables $y$, $x$. 
In density estimation $y$ is a continuous variable
(the variable $x$ may be constant and thus be skipped),
while in classification problems $y$ takes only discrete values.
In regression problems on assumes $p(y|x,{h})$ to be Gaussian
with fixed variance,
so the function of interest becomes the regression function
${h} (x) = \int \,dy\, y p(y|x,{h})$.

\subsubsection{Energies, free energies, and errors}

Often it is more convenient to work with log--probabilities
$L$ = $\ln p$
than with probabilities.
Firstly, 
this ensures non--negativity 
of probabilities $p$ = $e^L \ge 0$ for arbitrary $L$.
(For $p$ = 0 the log--probability becomes $L$ = $-\infty$.)
Thus, when working with log--probabilities
one can skip the  non--negativity constraint 
which would be necessary when working with probabilities.
Secondly, 
the multiplication of probabilities for independent events,
yielding their joint probability,
becomes a sum when written in terms of $L$.
Indeed,
from 
$p(A,B)$ = $p(A \,\mbox{\sc and}\, B)$ = $p(A)p(B)$
it follows for
$L(A,B)$ 
=
$\ln P(A,B)$
that
$L(A,B)$ 
=$L(A \, \mbox{\sc and} \,B)$ 
= $L(A)L(B)$.
Especially in the limit where an infinite number of events
is combined by {\sc and}, 
this would result in
an infinite product for $p$
but yields an integral for $L$,
which is typically easier to treat.

Besides  the requirement of being non--negative,  
probabilities have to be normalized,
e.g., $\int \!dx\, p(x)$ = 1.
When dealing with a large set of elementary events 
normalization is numerically a nontrivial task.
It is then convenient 
to work as far as possible with 
unnormalized probabilities $Z(x)$
from which normalized probabilities are obtained
as $p(x)$ = $Z(x)/Z$ with partition sum $Z$ = $\sum_x Z(x)$.
Like for probabilities, it is also often advantageous
to work with the logarithm of unnormalized probabilities,
or to get positive numbers (for $p(x)<1$)
with the negative logarithm 
$E(x)$ = $-(1/\beta)\ln Z(x)$,
in physics also known as {\it energy}.
(For the role of $\beta$ see below.)
Similarly, $F$ = $-(1/\beta) \ln Z$ 
is known as {\it free energy}.

Defining the energy we have introduced a parameter $\beta$.
Varying the parameter $\beta$ 
generates an exponential family of densities
which is frequently used in practice 
by (simulated or deterministic) annealing techniques
for minimizing free energies
\cite{Kirkpatrick-Gelatt-Vecchi-1983,Mezard-Parisi-Virasoro-1987,Ripley-1987,Davis-1987,Aarts-Korts-1989,Rose-Gurewitz-Fox-1990,Yuille-1990,Gelfand-Mitter-1993,Yuille-Kosowski-1994,Yuille-Stolorz-Utans-1994}.
In physics $\beta$ is known as {\it inverse temperature}
and plays the role of a Lagrange multiplier 
in the maximum entropy approach to statistical physics.
Inverse temperature
$\beta$ can also be seen as an external field coupling to the energy.
Indeed, the free energy 
$F$
is a generating function for the cumulants of the energy,
meaning that cumulants of $E$ can be obtained by taking derivatives of $F$
with respect to $\beta$ 
\cite{Gardiner-1990,Balian-1991,Beck-Schloegl-1993,Montvay-Muenster-1994}.
For a detailled discussion 
of the relations between probability, log--probability, energy, free energy,
partition sums, generating functions, 
and also bit numbers and information see \cite{Lemm-1998}.

The posterior $p({h}|f)$, for example, can so be written
as 
\bea
p({h}|f) 
&=& e^{L({h}|f)}
 = \frac{Z({h}|f)}{Z({H}|f)}
= \frac{e^{-\beta E({h}|f)}}{Z({H}|f)}
 \nonumber\\&=&
  e^{-\beta \left( E({h}|f)-F({H}|f) \right)}
= e^{-\beta E({h}|f)+c({H}|f)}
,
\label{posterior}
\eea
with 
(posterior)
{\it log--probability}  
\be
L({h}|f) = \ln p({h}|f),
\ee
unnormalized (posterior) probabilities or {\it partition sums} 
\be
Z({h}|f)
,\qquad
Z({H}|f) = \int\! d{h} \, Z({h}|f) ,  
\ee
(posterior) {\it energy}
\be
E({h}|f) = -\frac{1}{\beta} \ln Z({h}|f)
\ee
and (posterior) {\it free energy}
\bea
F({H}|f) 
&=& -\frac{1}{\beta} \ln Z({H}|f)
\\&=&
    -\frac{1}{\beta} \ln \int\! d{h}\, e^{-\beta E({h}|f)}
,
\eea
yielding
\bea
Z({h}|f) 
&=&  
e^{-\beta E({h}|f)}
,
\\
Z({H}|f) 
&=&  
\int\! d{h} \, e^{-\beta E({h}|f)}
,
\eea
where $\int\! d{h}$ represent a (functional) integral,
for example over variables (functions)  ${h}(x,y)$ = $p(y|x,{h})$,
and 
\be
c({H}|f) =  -\ln Z({H}|f) = \beta F({H}|f)
.
\ee
Note that
we did not include the $\beta$--dependency 
of the functions $Z$, $F$, $c$ in the notation.

For the sake of clarity, 
we have chosen to use the common notation for conditional probabilities
also for energies and the other quantities derived from them.
The same conventions will also be used for other probabilities,
so we will write for example for likelihoods
\be
p(y|x,{h}) = e^{-\beta^\prime \left( E(y|x,{h}) - F(Y|x,{h}) \right) }
,
\label{beta-likelihood}
\ee
for $y\in Y$.
Inverse temperatures 
may be different for prior and likelihood.
Thus, we may choose $\beta^\prime \ne \beta$ 
in
Eq.\ (\ref{beta-likelihood})
and Eq.\ (\ref{posterior}).

In Section \ref{map} we will discuss the maximum a posteriori approximation
where an optimal $h$ is found by maximizing the posterior $p(h|f)$.
Since maximizing the posterior means minimizing
the posterior energy $E({h}|f)$ 
the latter plays the role of an {\it error functional} for $h$ to be minimized.
This is technically similar to the
minimization of an regularized error functional
as it appears in regularization theory or
empirical risk minimization,
and which is discussed in
Section \ref{empirical-risk}.

Let us have a closer look
to the integral over model states ${h}$.
The variables ${h}$ represent the parameters describing 
the data generating probabilities or likelihoods
$p(y|x,{h})$.
In this paper we will mainly be interested in
``nonparametric'' approaches where 
the $(x,y,h)$--dependent numbers $p(y|x,{h})$ 
itself are considered to be the primary degrees of freedom
which ``parameterize'' the model states ${h}$.
Then, the integral over ${h}$
is an integral over a set of real variables 
indexed by $x$, $y$,
under additional non--negativity and normalization condition.
\be
\int\!d{h} \rightarrow \int\! \left(\prod_{x,y}dp(y|x,{h})\right)
.
\ee
Mathematical difficulties arise for the case
of continuous
$x$, $y$ where 
$p({h}|f)$ represents a stochastic process.
and the integral over ${h}$ 
becomes a functional integral
over (non--negative and normalized) functions $p(y|x,{h})$.
For Gaussian processes
such a continuum limit can be defined
\cite{Doob-1953,Glimm-Jaffe-1987,Wahba-1990,Lifshits-1995,MacKay-1998}
while the construction of continuum limits
for non--Gaussian processes is highly non--trivial
(See for instance \cite{DiCastro-Jona-Lasinio-1976,Collins-1984,Itzykson-Drouffe-1989,Zinn-Justin-1989,Perskin-Schroeder-1995,Weinstein-1995,Weinstein-1996,Cardy-1996,Ryder-1996}
for perturbative approaches
or \cite{Glimm-Jaffe-1987} for a non--perturbative $\phi^4$--theory.)
In this paper we will take the numerical point of view
where all functions are considered to be finally discretized,  
so the ${h}$--integral is well--defined
(``lattice regularization'' 
\cite{Creutz-1983,Rothe-1992,Montvay-Muenster-1994}).

\subsubsection{Posterior and likelihood}

Bayesian approaches require the calculation of posterior densities.
Model states ${h}$ are commonly specified
by giving the data generating probabilities or likelihoods $p(y|x,{h})$.
Posteriors are linked to likelihoods
by Bayes' theorem 
\be
p(A|B)=\frac{p(B|A)p(A)}{p(B)}
,
\ee
which follows at once from
the definition of conditional probabilities, i.e.,
$p(A,B)$ 
= $p(A|B)p(B)$
= $p(B|A)p(A)$.
Thus, one finds
\be
p({h}|f) 
=
p({h}|D,D_0) 
= 
\frac{p(D|{h}) \, p({h}|D_0)}{p(D|D_0)}
= \frac{p(y_D|x_D,{h}) \, p({h}|D_0)}{p(y_D|x_D,D_0)}
\label{bayes-formula}
\ee
\be
 = \frac{\prod_i p(x_i, y_i|{h}) p({h}|D_0)}
        {\int \!d{h}\,  \prod_i p(x_i,y_i|{h}) p({h}|D_0)}
=
\frac{\prod_i p(y_i|x_i,{h}) p({h}|D_0)}
        {\int \!d{h}\,  \prod_i p(y_i|x_i,{h}) p({h}|D_0)}
\label{s-asymm}
,
\ee
using 
$p(y_D|x_D,D_0,{h})$ = $p(y_D|x_D,{h})$
for the training data likelihood of ${h}$ and
$p({h}|D_0,x_i)$  = $p({h}|D_0)$.
The terms of Eq.\ (\ref{bayes-formula}) 
are in a Bayesian context often referred to as
\be
{\rm posterior} = \frac{{\rm likelihood} \times {\rm prior}}{{\rm evidence}}
.
\ee
Eqs.(\ref{s-asymm}) show that
the posterior can be expressed equivalently
by the joint likelihoods $p(y_i,x_i|{h})$ or conditional
likelihoods $p(y_i|x_i,{h})$.
When working with joint likelihoods, a distinction between 
$y$ and $x$ variables is not necessary.
In that case $x$ can be included in $y$ and skipped from the notation.
If, however, $p(x)$ is already known or is not of interest
working with conditional likelihoods is preferable.
Eqs.(\ref{bayes-formula},\ref{s-asymm}) can be interpreted 
as updating (or learning) formula
used to obtain a new posterior 
from a given prior probability 
if new data $D$ arrive.

In terms of energies Eq.\ (\ref{s-asymm}) reads,
\be
p({h}|f)
=
\frac{e^{-\beta \sum_i E(y_i|x_i,{h}) - \beta E({h}|D_0) }}
{Z(Y_D|x_D,{h})\, Z({H}|D_0)}
\int\! d{h}\, \frac{Z(Y_D|x_D,{h})\, Z({H}|D_0)}
{e^{-\beta \sum_i E(y_i|x_i,{h}) - \beta E({h}|D_0) }}
,
\ee
where the same temperature $1/\beta$
has been chosen for both energies
and the normalization constants are
\bea
Z(Y_D|x_D,{h})
&=&
\prod_i \int \!dy_i\, e^{-\beta E(y_i|x_i,{h}) }
,
\\
Z({H}|D_0)
&=&
\int \!d{h}\, e^{- \beta E({h}|D_0) }
.
\eea

The predictive density we are interested in can be written as 
the ratio of two correlation functions under  $p_0 ({h})$, 
\bea
p(y|x,f)  
&=& <p(y|x,{h}) >_{{H}|f} 
\\&=&
\frac{<p(y|x,{h})  \prod_i p(y_i|x_i,{h}) >_{{H}|D_0} }
{<\prod_i p(y_i|x_i,{h})>_{{H}|D_0}  },
\label{pred2}
\\&=&
\frac{\int \!d{h}\, p(y|x,{h}) \, e^{-\beta E_{\rm comb}}}
     {\int \!d{h}\, e^{-\beta E_{\rm comb}}}
\label{pred3}
\eea
where $< \cdots >_{{H}|D_0}$ denotes the expectation 
under the prior density 
$p_0({h})$ = $p({h}|D_0)$
and the {\it combined likelihood and prior energy} $E_{\rm comb}$
collects the ${h}$--dependent energy and free energy terms
\be
E_{\rm comb} =
\sum_i E(y_i|x_i,{h}) + E({h}|D_0) - F(Y_D|x_D,{h}),
\label{Ecomb}
\ee
with
\be
F(Y_D|x_D,{h})
= -\frac{1}{\beta} \ln Z(Y_D|x_D,{h})
.
\ee
Going from Eq.\ (\ref{pred2})
to Eq.\ (\ref{pred3})
the normalization factor 
$Z({H}|D_0)$
appearing in numerator and denominator
has been canceled.

We remark that for continuous $x$ and/or $y$
the likelihood energy term $E(y_i|x_i,{h})$ describes an ideal,
non--realistic measurement because
realistic measurements cannot be arbitrarily sharp.
Considering the function $p(\cdot|\cdot,{h})$
as element of a Hilbert space
its values may be written as scalar product
$p(x|y,{h})$ = $(v_{xy},\, p(\cdot|\cdot,{h})\,)$
with a function $v_{xy}$ being also an element in that Hilbert space.
For continuous $x$ and/or $y$ this notation is only formal
as $v_{xy}$ becomes unnormalizable.
In practice a measurement of
$p(\cdot|\cdot,{h})$ corresponds to a normalizable $v_{\tilde x\tilde y}$ = 
$\int \!dy \int \!dx \,\vartheta (x,y) v_{xy}$
where the kernel $\vartheta (x,y)$ 
has to ensure normalizability.
(Choosing normalizable $v_{\tilde x\tilde y}$
as coordinates
the Hilbert space of $p(\cdot|\cdot,{h})$
is also called a reproducing kernel Hilbert space
\cite{Parzen-1963,Kimmeldorf-Wahba-1970a,Kimmeldorf-Wahba-1970b,Wahba-1990,Lifshits-1995}.)
The data terms then become
\be
p(\tilde y_i|\tilde x_i,{h}) 
=\frac{\int \!dy \int \!dx \,\vartheta_i (x,y) p(y,x|{h})}
{\int \!dy \,\vartheta_i (x,y) p(y,x|{h})}
.
\ee
The notation $p(y_i|x_i,{h})$ is understood as limit
$\vartheta (x,y)\rightarrow 
\delta (x-x_i)\delta (y-y_i)$
and means in practice 
that $\vartheta (x,y)$ is very sharply centered.
We will assume that the discretization,
finally necessary to do numerical calculations,
will implement such an averaging.

\subsubsection{Predictive density}

Within a Bayesian approach predictions 
about (e.g., future) events are based 
on the {\it predictive probability density},
being the expectation of probability for $y$
for given (test) situation $x$, training data $D$ 
and prior data $D_0$
\be
p(y|x,f) 
=
p(y|x,D,D_0) 
= \int \!d{h} \, p({h}|f) \, p(y|x,{h})
= 
\,\,< p(y|x,{h})>_{{H}|f}
.
\label{predictive1}
\ee
Here 
$< \cdots >_{{H}|f}$ 
denotes the expectation under the posterior 
$p({h}|f)$ = $p({h}|D,D_0)$, the state of knowledge $f$
depending on prior and training data.
Successful applications of Bayesian approaches
rely strongly on an adequate choice of the model space ${H}$
and model likelihoods $p(y|x,{h})$.

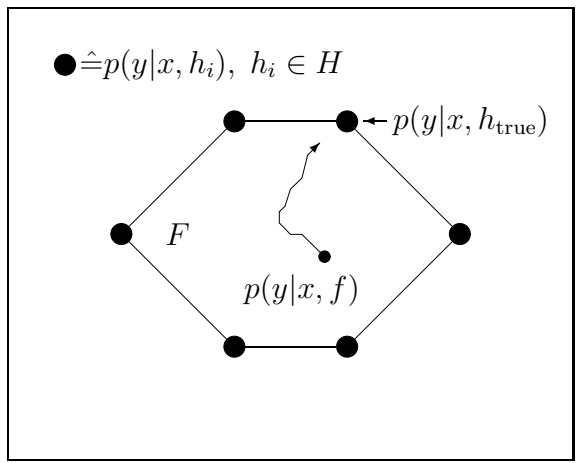
\begin{figure}
\begin{center}
\setlength{\unitlength}{1.5mm}
\begin{picture}(50,40)
\put(0,0){\framebox(50,40)[]{}}
\put(5,35){\circle*{1.9}}
\put(14,35)
{\makebox(8,0){$\hat = p(y|x,h_i),\;h_i\in {H}$}}
\put(20,30){\circle*{1.9}}
\put(30,30){\circle*{1.9}}
\put(21,30){\line(1,0){8}}  
\put(41,30){\makebox(0,0){$p(y|x,h_{\rm true})$}}
\put(33.5,30.){\vector(-1,0){2}}   
\put(10,20){\circle*{1.9}}
\put(40,20){\circle*{1.9}}
\put(10.7,20.7){\line(1,1){8.65}}  
\put(39.3,20.7){\line(-1,1){8.65}}  
\put(10.7,19.3){\line(1,-1){8.65}}  
\put(39.3,19.3){\line(-1,-1){8.65}}  
\put(20,10){\circle*{1.9}}
\put(30,10){\circle*{1.9}}
\put(21,10){\line(1,0){8}}  
\put(28,18){\circle*{1.0}}
\put(26,15){\makebox(0,0){$p(y|x,f)$}}
\put(15,20){\makebox(0,0){${F}$}}
\drawline[2](27.85,18.15)(26.,20.0)
(25.,20.)(24.,21.)
(24.,22.)(24.5,22.5)
(25.,24.)(26,25)
(26.5,27.)(27.5,28.)
\put(27.5,28.){\vector(1,1){0.1}}   
\end{picture}
\end{center}
\caption{The predictive density $p(y|x,f)$
for a state of knowledge $f$ = $f(D,D_0)$
is in the convex hull spanned
by the possible states of Nature $h_i$
characterized by the likelihoods $p(y|x,h_i)$.
During learning the actual predictive density $p(y|x,f)$
tends to move stochastically towards the extremal point
$p(y|x,h_{\rm true})$
representing the ``true'' state of Nature.
}
\label{learning-scheme}
\end{figure}

Note that 
$p(y|x,f)$ 
is in the convex cone spanned by the possible states of Nature
${h}\in {H}$,
and typically not equal 
to one  of these $p(y|x,{h})$.
The situation is illustrated in Fig.\ \ref{learning-scheme}.
During learning the predictive density $p(y|x,f)$
tends to approach the true $p(y|x,h)$.
Because the training data are random variables,
this approach is stochastic.
(There exists an extensive literature analyzing 
the stochastic properties of learning and generalization 
from a statistical mechanics perspective
\cite{Gardner-1987,Gardner-1988,Gardner-Derrida-1988,Watkin-Rau-Biehl-1993,Wolpert-1995,Opper-Kinzel-1996}).

\subsubsection{Mutual information and learning}
\label{mutual}

The aim of learning is to generalize
the information obtained from training data to non--training situations.
For such a {\it generalization} to be possible,
there must exist a, at least partially known, 
relation between the likelihoods $p(y_i|x_i,h)$
for training and for non--training data.
This relation is typically provided by {\it a priori} knowledge.

One possibility to quantify 
the relation between two random variables $y_1$ and $y_2$, 
representing for example training and non--training data,
is to calculate their {\it mutual information},
defined as
\be
M(Y_1,Y_2)
= 
\sum_{y_1\in Y_1,y_2\in Y_2} p(y_1,y_2)\ln \frac{p(y_1,y_2)}{p(y_1) p(y_2)}
\label{mutual-def1}
.
\ee
It is also instructive to express the mutual information
in terms of (average) information content or entropy, 
which, for a probability function $p(y)$, is defined as
\be
H(Y) = -\ln \sum_{y\in Y} p(y)\ln p(y)
.
\ee
We find
\be
M(Y_1,Y_2)
= H(Y_1)+H(Y_2)-H(Y_1,Y_2)
,
\ee
meaning that the mutual information 
is the sum of the two individual entropies
diminished by the entropy  common to both variables.

To have a compact notation
for a family of predictive densities
$p(y_i|x_i,f)$ 
we choose a vector
$x$ = $(x_1,x_2,\cdots)$
consisting of all possible values $x_i$
and corresponding vector
$y$ = $(y_1,y_2,\cdots)$,
so we can write
\be
p(y|x,f)
=
p(y_1,y_2,\cdots |x_1,x_2,\cdots ,f).
\ee
We now would like to characterize a state of knowledge $f$
corresponding to predictive density $p(y|x,f)$
by its mutual information.
Thus, we generalize
the definition (\ref{mutual-def1}) from two random variables
to a random vector $y$ with components $y_i$,
given vector $x$ with components $x_i$
and obtain the conditional mutual information
\be
M(Y|x,f) = 
\int \left(\prod_i \,dy_i\right)
p(y|x,f) \ln \frac{p(y|x,f)}{\prod_j p(y_j|x_j,f)}
,
\label{cond-mutual}
\ee
or
\be
M(Y|x,f) = 
\left(\int\!dy_i\, H(Y_i|x,f) - H(Y|x,f)\right)
,
\ee
in terms of conditional entropies
\be
H(Y|x,f) = - 
\int \!dy\, p(y|x,f) \ln  p(y|x,f)
.
\ee
In case not a fixed vector $x$ is given, 
like for example $x$ = $(x_1,x_2,\cdots)$, 
but a density $p(x)$,
it is useful to average
the conditional mutual information and 
conditional entropy 
by including the integral $\int\!dx \,p(x)$
in the above formulae.

It is clear from Eq.\ (\ref{cond-mutual})
that predictive densities
which factorize 
\be
p(y|x,f)
=
\prod_i p(y_i|x_i,f)
,
\ee
have a mutual information of zero.
Hence,
such {\it factorial states}
do not allow any generalization from training to non--training data.
A special example are the possible states of Nature or
pure states $h$,
which factorize according to the definition of our model
\be
p(y|x,h)
=
\prod_i p(y_i|x_i,h)
.
\label{pure-factorizing}
\ee
Thus,  pure states do not allow any further generalization.
This is consistent with the fact that 
pure states represent the natural endpoints of any learning process.

It is interesting to see, however,
that there are also other states for which the predictive
density factorizes.
Indeed, from Eq.\ (\ref{pure-factorizing}) it follows that
any (prior or posterior) probability $p(h)$
which factorizes leads to a factorial state,
\be
p(h) 
= \prod_i p(h(x_i))
\Rightarrow
p(y|x,f)
=
\prod_i p(y_i|x_i,f)
.
\ee
This means generalization, i.e., (non--local) learning, 
is impossible when starting from a {\it factorial prior}.

A factorial prior provides a very clear reference
for analyzing the role of a--priori information in learning.
In particular, with respect to a prior
factorial in local variables $x_i$, 
learning may be decomposed into two steps, one increasing, 
the other lowering mutual information:
\begin{itemize}
\item[1.] Starting from a factorial prior,
new {\it non--local data} $D_0$
(typically called {\it a priori} information)
produce a new non--factorial state
with non--zero mutual information.
\item[2.]
{\it Local data} $D$
(typically called training data)
stochastically
reduce the mutual information. 
Hence, learning with local data corresponds to a
{\it stochastic decay of mutual information}.
\end{itemize}

Pure states, 
i.e., the extremal points in the space of possible predictive densities,
do not have to be deterministic.
Improving measurement devices,
stochastic pure states 
may be further decomposed into finer components $g$, so that
\be
p(y_i|x_i,h) = \int\! dg\, p(g) \, p(y_i|x_i,g)
.
\ee
Imposing a non--factorial prior $p(g)$ on the new, finer hypotheses $g$
enables again non--local learning with local data,
leading asymptotically 
to one of the new pure states $p(y_i|x_i,g)$.

Let us exemplify
the stochastic decay of mutual information 
by a simple numerical example.
Because the mutual information requires
the integration over all $y_i$ variables
we choose a problem with only two of them,
$y_a$ and $y_b$ corresponding to 
two $x$ values $x_a$ and $x_b$.
We consider a model with four
states of Nature $h_l$, $1\le l\le 4$,
with Gaussian likelihood
$p(y|x,h)$ =
$(\sqrt{2 \pi} \sigma)^{-1} 
\exp{\left(-(y- h_i(x))^2/(2\sigma^2)\right)}$
and local means $h_l(x_j)$ = $\pm 1$.

Selecting a ``true'' state of Nature $h$,
we sample 50 data points $D_i$ = $(x_i,y_i)$
from the corresponding Gaussian likelihood
using $p(x_a)$ = $p(x_b)$ = $0.5$.
Then, starting from a given, factorial or non--factorial, prior
$p(h|D_0)$
we sequentially update the predictive density,
\be
p(y|x,f(D_{i+1},\cdots,D_{0})) = 
\sum_{l=1}^4 p(y|x,h_l)\, p(h_l|D_{i+1},\cdots,D_{0})
,
\ee
by calculating the posterior
\be
p(h_l|D_{i+1},\cdots,D_{0}) 
= \frac{p(y_{i+1}|x_{i+1},h_l) \, p(h_j|D_{i}\cdots,D_{0})}
       {p(y_{i+1}|x_{i+1},D_{i}\cdots,D_{0})}
.
\label{sequential-update}
\ee
It is easily seen from Eq.\ (\ref{sequential-update})
that factorial states remain factorial under local data.

\begin{figure}
\begin{center}
\epsfig{file=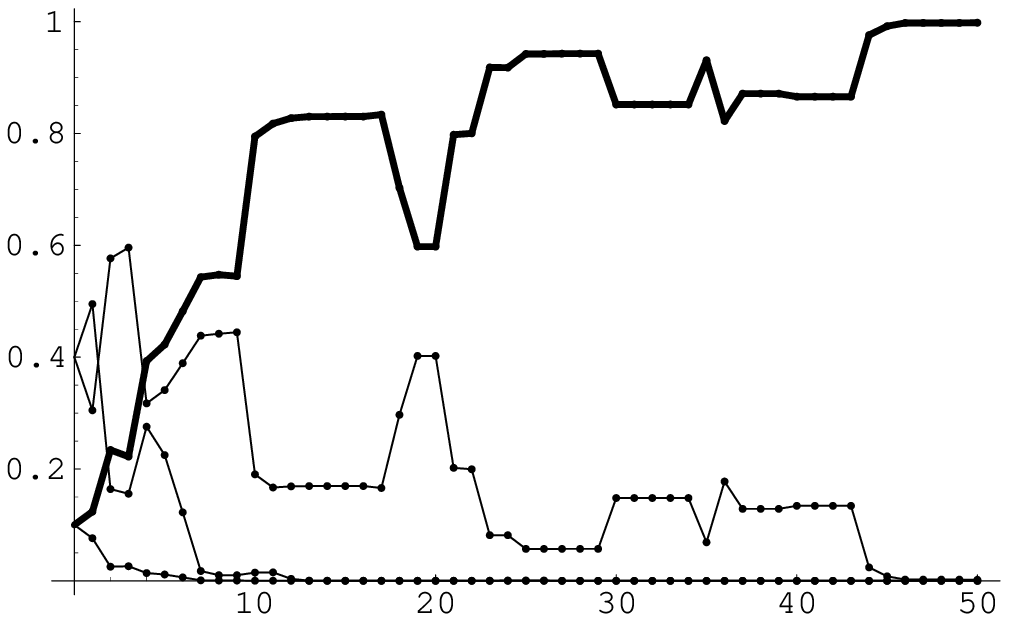, width=65mm}
\epsfig{file=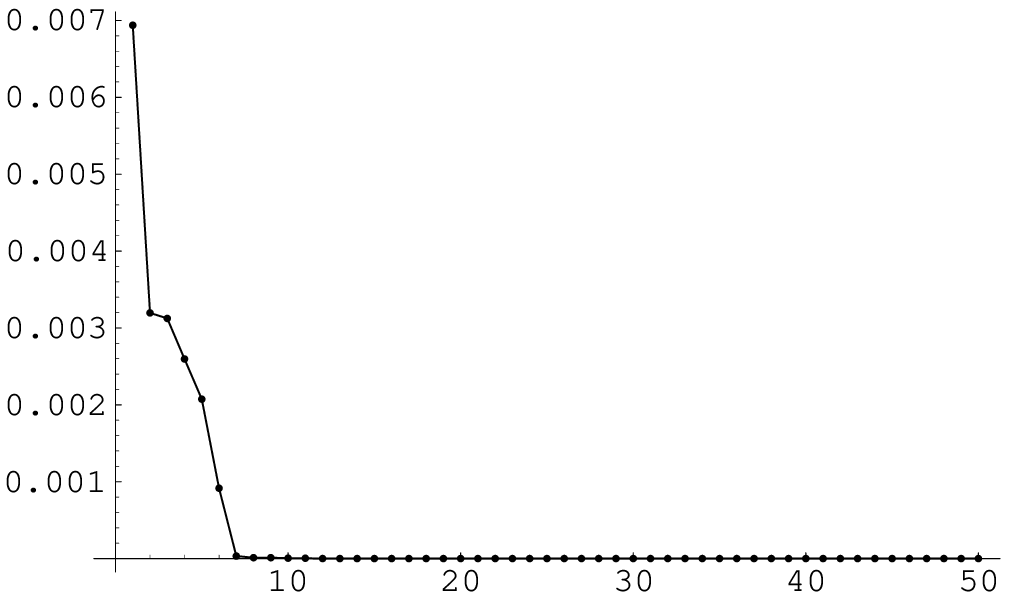, width=65mm}
\epsfig{file=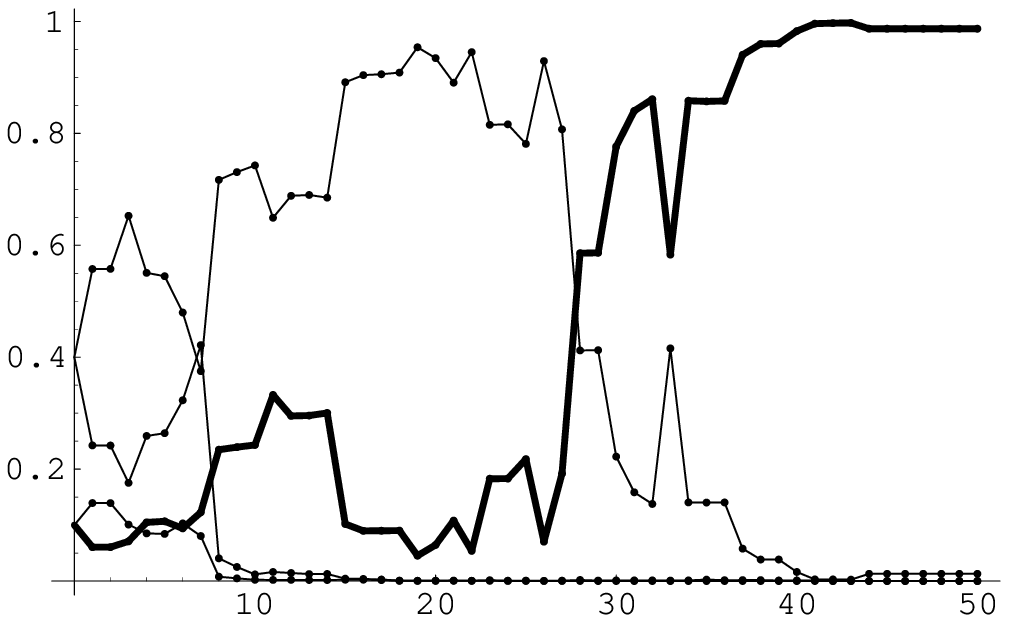, width=65mm}
\epsfig{file=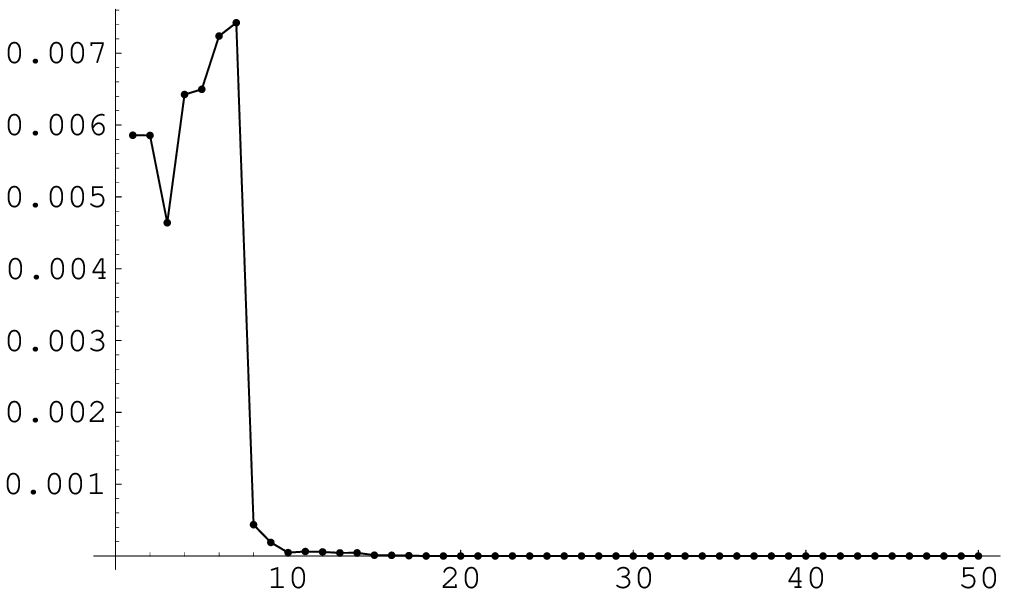, width=65mm}
\epsfig{file=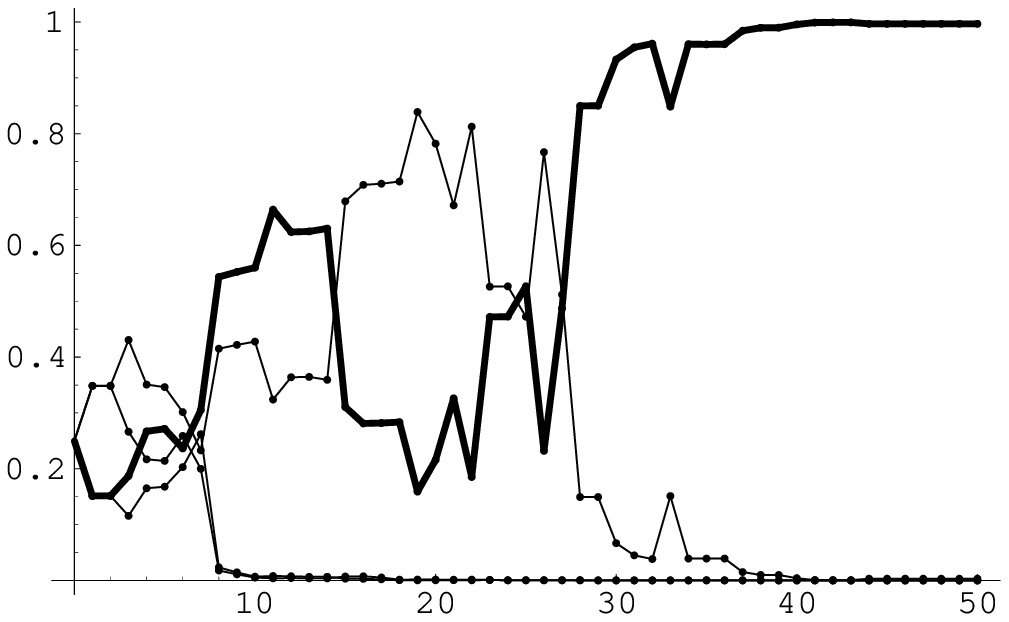, width=65mm}
\epsfig{file=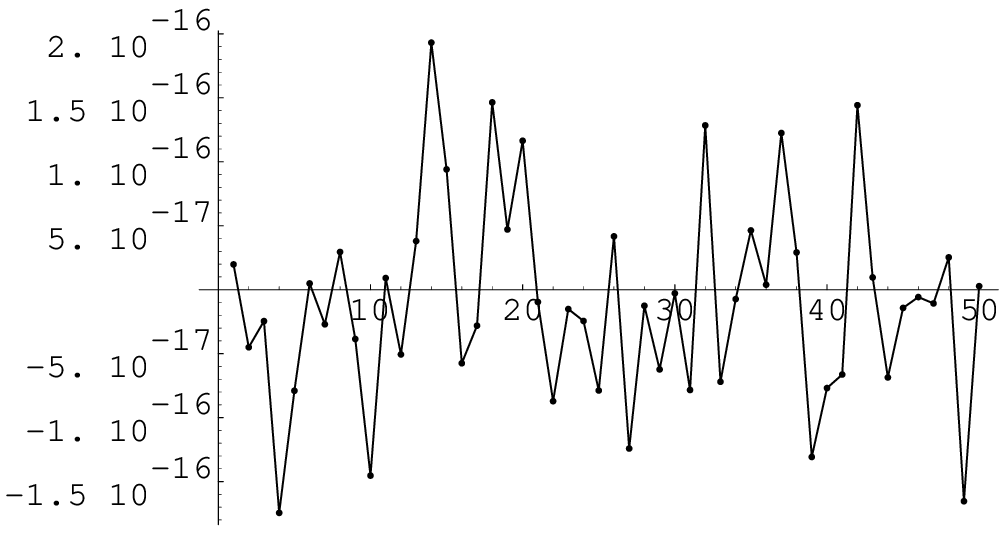, width=65mm}
\end{center}
\setlength{\unitlength}{1mm}
\begin{picture}(50,10)
\put(40,145){\makebox(0,0){posterior}}
\put(107,145){\makebox(0,0){mutual information}}
\put(14,134){\makebox(0,0){(a)}}
\put(130,134){\makebox(0,0){(b)}}
\put(14,92){\makebox(0,0){(c)}}
\put(130,92){\makebox(0,0){(d)}}
\put(14,50){\makebox(0,0){(e)}}
\put(130,50){\makebox(0,0){(f)}}
\end{picture}
\vspace{-1.5cm}
\caption{
The decay of mutual information during learning:
Model with 4 possible states $h_l$
representing Gaussian likelihoods $p(y_i|x_i,h_l)$
with means $\pm 1$ for two different $x_i$ values.
Shown are posterior probabilities 
$p(h_l|f)$ ($a$, $c$, $e$, on the left hand side, 
the posterior of the true $h_l$ is shown by a thick line)
and mutual information $M(y)$
($b$, $d$, $f$, on the right hand side)
during learning 50 training data.
($a$, $b$): The mutual information decays during learning
and becomes quickly practically zero.
($c$, $d$): For ``unlucky'' training data
the wrong hypothesis $h_i$ can dominate at the beginning.
Nevertheless, the mutual information decays and
the correct hypothesis has finally to be found through
``local'' learning.
($e$, $f$): Starting with a factorial prior
the mutual information is and remains zero, 
up to artificial numerical fluctuations.
For ($e$, $f$) the same random data
have been used as for ($c$, $d$). 
}
\label{mutual-pic}
\end{figure}

Fig.\ \ref{mutual-pic} shows that indeed the mutual information decays rapidly.
Depending on the training data,
still the wrong hypothesis $h_l$ may survive
the decay of mutual information.
Having arrived at a factorial state, 
further learning has to be local.
That means,  data points for $x_i$
can then only influence the predictive density for the corresponding $y_i$
and do not allow generalization to the other $y_j$ with $j\ne i$.

For a factorial prior
$p(h_l)$ = $p(h_l(x_a))p(h_l(x_b))$
learning is thus local from the very beginning.
Only very small numerical random fluctuations of the mutual information
occur, quickly eliminated by learning.
Thus, the predictive density
moves through a sequence of factorial states.

\subsection{Bayesian decision theory}

\subsubsection{Loss and risk}

In {\it Bayesian decision theory}
a set $A$ of possible actions $a$ is considered,
together with a function $l(x,y,a)$
describing the {\it loss} $l$ suffered
in situation $x$ if $y$ appears and action $a$ is selected
\cite{Berger-1980,LeCam-1986,Pearl-1988,Robert-1994}.
The loss averaged over {\it test data} $x$, $y$,  
and possible states of Nature ${h}$
is known as {\it expected risk},
\bea
r(a,f) 
&=& \int \!dx\,dy\, p(x)\,p(y|x,f)\,l(x,y,a).
\\&=&
  < l(x,y,a) >_{X,Y|f}
\\&=&
< r(a,{h}) >_{{H}|f}
\eea
where 
$< \cdots >_{X,Y|f}$ 
denotes the expectation under the joint predictive density
$p(x,y|f)$ = $p(x) p(y|x,f)$
and
\be
r(a,{h}) = \int \!dx\,dy\, p(x)\,p(y|x,{h})\,l(x,y,a).
\ee
The aim is to find an optimal action $a^*$
\be
a^* ={\rm argmin}_{a\in A} r(a,f)
.
\ee

\subsubsection{Loss functions for approximation}
\label{some-loss-functions}

\vspace{0.3cm}\noindent{\bf Log--loss:}
A typical loss function for {\it density estimation problems}
is the {\it log--loss}
\be
l(x,y,a) = -b_1(x)\ln p(y|x,a)+b_2(x,y)
\label{log-loss}
\ee
with some $a$--independent $b_1(x)>0$, $b_2(x,y)$ and
actions $a$ describing probability densities
\be
\int \!dy\, p(y|x,a) = 1, \,\,\forall x\in X, \forall a\in A
.
\ee
Choosing 
$b_2(x,y)$ = $p(y|x,f)$
and $b_1(x)$ = $1$
gives
\bea
r(a,f) &=&
\int\! dx\, dy\, p(x)p(y|x,f) \ln \frac{p(y|x,f)}{p(y|x,a)}
\\&=&
< \ln \frac{p(y|x,f)}{p(y|x,a)} >_{X,Y|f}
\\&=&
<{\rm KL}( \,{p(y|x,f)},\, {p(y|x,a)}\, )>_{X}
,
\eea
which shows that minimizing log--loss is equivalent to minimizing
the ($x$--averaged) {\it Kullback--\-Leibler entropy} 
${\rm KL}( \, {p(y|x,f)},\, {p(y|x,a)}\, )$\cite{Kullback-Leibler-1951,Kullback-1951,Beck-Schloegl-1993,Deco-Obradovic-1996,Ebeling-Freund-Schweitzer-1998}.

While the paper will concentrate on log--loss 
we will also give a short summary of loss functions
for {\it regression problems}.
(See for example \cite{Berger-1980,Robert-1994} for details.) 
Regression problems are special density estimation problems
where the considered possible actions are restricted to 
$y$--independent functions $a(x)$.

\vspace{0.3cm}\noindent{\bf Squared--error loss:}
The most common loss function for regression problems
(see Sections \ref{regression}, \ref{exact})
is the squared--error loss. It reads
for one--dimensional $y$
\be
l(x,y,a) = b_1(x) \left( y-a(x) \right)^2 +b_2(x,y)
,
\label{squared-loss}
\ee
with arbitrary $b_1(x)>0$ and  $b_2(x,y)$.
In that case the optimal function $a(x)$ is 
the {\it regression function} of the posterior
which is the mean of the predictive density
\be
a^*(x)
= \int \!dy \, y \, p(y|x,f)
= \,\, <y>_{Y|x,f}
.
\ee
This can be easily seen by writing
\bea
\left( y-a(x) \right)^2
&=& 
\left( y\;-\!<y>_{Y|x,f}+<y>_{Y|x,f}-\;a(x) \right)^2
\\&=& 
\left( y\;-\!<y>_{Y|x,f} \right)^2 
+\left( a(x)\;- <y>_{Y|x,f} \right)^2
\nonumber\\&&- 2 \left( y\,-<y>_{Y|x,f} \right) 
 \left( a(x)\;- <y>_{Y|x,f} \right)^2
\label{proofsq}
,
\eea
where the first term in (\ref{proofsq}) is independent of $a$
and the last term vanishes after integration over $y$
according to the definition of $<y>_{Y|x,f}$.
Hence,
\be
r(a,f)=\int\!dx\,b_1(x) p(x)\left( a(x)\,-\!<y>_{Y|x,f} \right)^2+{\rm const.}
\ee
This is minimized by 
$a(x) = <y>_{Y|x,f}$.
Notice that for Gaussian $p(y|x,a)$ with fixed variance
log--loss and squared-error loss are equivalent.
For multi--dimensional $y$
one--dimensional loss functions like Eq.\ (\ref{squared-loss})
can be used
when the component index of $y$ is considered part of the $x$--variables.
Alternatively, loss functions depending explicitly on multidimensional $y$
can be defined.
For instance, a general quadratic loss function would be 
\be
l(x,y,a)=
\sum_{k,k^\prime }(y_k-a_k){\bf K}(k,k^\prime)(y_{k^\prime }
-a_{k^\prime }(x)).
\ee
with symmetric, positive definite kernel ${\bf K}(k,k^\prime )$.

\vspace{0.3cm}\noindent{\bf Absolute loss:}
For absolute loss
\be
l(x,y,a) = b_1(x) |y-a(x)| +b_2(x,y)
,
\ee
with arbitrary $b_1(x)>0$ and  $b_2(x,y)$.
The risk becomes
\bea
r(a,f)&=&
\int \!dx\, b_1(x) p(x) \int_{-\infty}^{a(x)}\!dy \left (a(x)-y\right) p(y|x,f)
\nonumber\\&&+
\int \!dx\, b_1(x)p(x)\int_{a(x)}^\infty\!dy \left (y-a(x)\right) p(y|x,f)
+{\rm const.}
\\&=&
2\int \!dx\, b_1(x) p(x) \int_{m(x)}^{a(x)}\!dy \left (a(x)-y\right) p(y|x,f)
+{\rm const.}^\prime
,
\eea
where the integrals have been rewritten as 
$\int_{-\infty}^{a(x)}$ = $\int_{-\infty}^{m(x)}$ +
$\int_{m(x)}^{a(x)}$
and
$\int_{a(x)}^\infty$ = $\int_{a(x)}^{m(x)}$ + $\int_{m(x)}^{\infty}$
introducing a median function $m(x)$ 
which satisfies
\be
\int_{-\infty}^{m(x)} \!dy\, p(y|x,f) = \frac{1}{2},\, \forall x\in X
,
\ee
so that
\be
a(x) \left( \int_{-\infty}^{m(x)} \!dy\, p(y|x,f) 
-\int_{m(x)}^\infty \!dy\, p(y|x,f)
\right) 
= 0,\, \forall x\in X
.
\ee
Thus the risk is minimized by any {\it median function} $m(x)$.

\vspace{0.3cm}\noindent{\bf $\delta$--loss and $0$--$1$ loss :}
Another possible loss function, 
typical for classification tasks 
(see Section \ref{classification}),
like for example image segmentation
\cite{Marroquin-Mitter-Poggio-1987},
is
the $\delta$--loss for continuous $y$
or $0$--$1$--loss for discrete $y$
\be
l(x,y,a) = - b_1(x) \delta \left( y-a(x) \right) +b_2(x,y),
\ee
with arbitrary $b_1(x)>0$ and  $b_2(x,y)$.
Here $\delta$ denotes
the Dirac $\delta$--functional
for continuous $y$ 
and the Kronecker $\delta$ for discrete $y$.
Then, 
\be
r(a,f) = \int\!dx\, b_1(x) p(x) \, p(\,y\!=\!a(x)\,|x,f)
+{\rm const.}
,
\ee
so the optimal $a$
corresponds to any {\it mode function}
of the predictive density.
For Gaussians mode and median are unique, 
and coincide with the mean.

\subsubsection{General loss functions and unsupervised learning}

Choosing actions $a$ in specific situations
often requires the use of specific loss functions.
Such loss functions may
for example contain additional terms
measuring {\it costs} of choosing action $a$
not related to approximation of the predictive density.
Such costs can quantify aspects like
the simplicity, implementability,
production costs, sparsity,
or understandability of action $a$.

Furthermore, instead of approximating a whole density
it often suffices to extract some of its features.
like identifying clusters of similar $y$--values, 
finding independent components for multidimensional $y$,
or mapping to an approximating density with lower dimensional $x$.
This kind of {\it exploratory data analysis}
is the Bayesian analogue to {\it unsupervised learning methods}.
Such methods are on one hand often utilized as a preprocessing step
but are, on the other hand,
also important to choose actions for situations
where specific loss functions can be defined.

From a Bayesian point of view general loss functions 
require in general
an explicit two--step procedure \cite{Lemm-1996}:
1. Calculate (an approximation of) the predictive density,
and 2. Minimize the expectation of the loss function
under that (approximated) predictive density.
(Empirical risk minimization, on the other hand,
minimizes the empirical average
of the (possibly regularized) loss function,
see Section \ref{empirical-risk}.)
(For a related example see for instance \cite{Lemm-Beiu-Taylor-1995}.)

For a Bayesian version of cluster analysis, for example, partitioning
a predictive density obtained from empirical data
into several clusters, 
a possible loss function is
\be
l(x,y,a) = (y-a(x,y) )^2
,
\ee
with action $a(x,y)$ being a mapping of $y$ for given $x$
to a finite number of cluster centers (prototypes). 
Another example of a clustering method based
on the predictive density is
Fukunaga's valley seeking procedure \cite{Fukunaga-1990}.

For multidimensional $x$ 
a space of actions $a( {\bf P}_x x,y)$ can be chosen 
depending only on a (possibly adaptable) lower dimensional 
projection of $x$.

For multidimensional $y$ with components $y_i$
it is often useful to identify independent components.
One may look, say, for a linear mapping 
$\tilde y$ = ${\bf M} y$
minimizing the 
correlations between different components of 
the `source' variables $\tilde y$
by minimizing the loss function
\be
l(y,y^\prime,{\bf M}) 
=
\sum_{i\ne j} \tilde y_i\, \tilde y_j^\prime
,
\ee
with respect to ${\bf M}$
under the joint predictive density
for $y$ and $y^\prime$ given
$x,x^\prime,D,D_0$.
This includes a Bayesian version
of blind source separation
(e.g. applied to the so called cocktail party problem
\cite{Bell-Sejnowski-1995, Amari-Cichocki-Yang-1996}), 
analogous to 
the treatment of Molgedey and Schuster \cite{Molgedey-Schuster-1994}.
Interesting projections of multidimensional $y$ 
can for example be found
by projection pursuit techniques
\cite{Friedman-Tukey-1974,Huber-1985,Jones-Sibson-1987,Silverman-1986}.

\subsection{Maximum A Posteriori Approximation}
\label{map}

In most applications
the (usually very high or even formally infinite dimensional) 
${h}$--integral over model states 
in Eq.\ (\ref{pred3})
cannot be performed exactly.
The two most common methods used to calculate
the ${h}$ integral approximately are
{\it Monte Carlo integration}
\cite{Metropolis-Rosenbluth-Rosenbluth-Teller-Teller-1953,Hammersley-Handscomb-1964,Hastings-1970,Ripley-1977,Berger-1980,Geman-Geman-1984,Ripley-1987,Binder-Heermann-1988,Tierney-1994,Winkler-1995,Gelman-Carlin-Stern-Rubin-1995,Neal-1996,O-Ruanaidh-Fitzgerald-1996,Rodriguez-1997,Neal-1997}
and {\it saddle point approximation}
\cite{Berger-1980,De-Bruijn-1981,Bleistein-Handelsman-1986,Negele-Orland-1988,Berger-Wolpert-1988,Zinn-Justin-1989,Robert-1994,Gelman-Carlin-Stern-Rubin-1995,Girosi-Jones-Poggio-1995,Lemm-1996}.
The latter approach will be studied in the following.

For that purpose, we expand $E_{\rm comb}$ of Eq.~(\ref{Ecomb})
with respect to ${h}$ around some ${h}^{*}$
\bea
E_{\rm comb} ({h}) 
&=& 
e^{(\Delta {h},\nabla) }E({h})\Big|_{{h}={h}^{*}}
\label{Eexpansion}
\\&=&
E_{\rm comb} ({h}^{*}) 
+ ( \Delta {h} , \nabla ({h}^{*}) )
+ \frac{1}{2}( \Delta {h}, {\bf H}({h}^{*}) \Delta {h}) 
+ \cdots
\nonumber
\eea
with 
$\Delta {h}$
= $( {h}-{h}^{*})$,
gradient $\nabla$ (not acting on $\Delta {h}$),
Hessian ${\bf H}$,
and round brackets $( \cdots , \cdots)$ denoting scalar products.
In case $p(y|x,{h})$ is parameterized independently for
every $x$, $y$ the states ${h}$ represent a parameter set
indexed by $x$ and $y$, hence
\be
\nabla ({h}^{*}) 
= \frac{\delta E_{\rm comb}({h})}
       {\delta {h}(x,y)}\Bigg|_{{h}={h}^{*}}
= \frac{\delta E_{\rm comb}(p(y^\prime |x^\prime ,{h}))}
       {\delta p(y|x,{h})}\Bigg|_{{h}={h}^{*}}
,
\ee
\be
{\bf H}({h}^{*}) 
= \frac{\delta^2 E_{\rm comb}({h})}
       {\delta {h}(x,y)\delta {h}(x^\prime,y^\prime)}
\Bigg|_{{h}={h}^{*}}
= \frac{\delta^2 E_{\rm comb}(p(y^{\prime\prime}|x^{\prime\prime},{h}))}
{\delta p(y|x,{h})\delta p(y^\prime|x^\prime,{h})}
\Bigg|_{{h}={h}^{*}}
,
\ee
are functional derivatives
\cite{Hilbert-Courant-1989,Jeggle-1979,Blanchard-Bruening-1982,Choquet-Bruhat-DeWitt-Morette-Dillard-Bleick-1982}
(or partial derivatives for discrete $x$, $y$)
and  for example
\be
(\Delta {h}, \nabla ({h}^{*}) )
= \int\!dx\,dy\, 
\left( {h}(x,y)-{h}^{*}(x,y) \right)  \nabla ({h}^{*})(x,y)
.
\ee

Choosing ${h}^{*}$ to be the location of a local minimum
of $E_{\rm comp}({h})$
the linear term in (\ref{Eexpansion})
vanishes. The second order term
includes the Hessian
and corresponds to a Gaussian integral over ${h}$
which could be solved analytically
\be
\int\!d{h}\, e^{-\beta (\Delta {h}, {\bf H}\Delta {h})}
=
\pi^\frac{d}{2} \beta^{-\frac{d}{2}} 
(\det {\bf H})^{-\frac{1}{2}} 
\label{gauss}
,\ee
for a $d$--dimensional ${h}$--integral.
However, 
using the same approximation for
the ${h}$--integrals in numerator and denominator of Eq.\ (\ref{pred3}),
expanding then also $p(y|x,{h})$ around ${h}^{*}$,
and restricting to the first (${h}$--independent) term
$p(y|x,{h}^{*})$ of that expansion,
the factor (\ref{gauss}) cancels, even for infinite $d$.
(The result is the zero order term
of an expansion of the predictive density in powers of $1/\beta$.
Higher order contributions can be calculated by using Wick's theorem
\cite{De-Bruijn-1981,Bleistein-Handelsman-1986,Negele-Orland-1988,Zinn-Justin-1989,Kaku-1993,Montvay-Muenster-1994,Lemm-1996}.)
The final approximative result for the predictive density
(\ref{predictive1})
is very simple and intuitive
\be
p(y|x,f) \approx p(y|x,{h}^{*})
,
\ee
with 
\be
 {h}^{*} 
\!= {\rm argmin}_{{h}\in {H}} E_{comb}
\!= {\rm argmax}_{{h}\in {H}} \,p({h}|f)
\!= {\rm argmax}_{{h}\in {H}} \,p(y_D|x_D,{h})p(h|D_0)
.
\label{minimum}
\ee
The saddle point (or Laplace) approximation
is therefore also called {\it Maximum A Posteriori Approximation} (MAP).
Notice that the same $h^*$ also maximizes 
the integrand of the evidence of the data $y_D$
\be
p(y_D|x_D,D_0)
= 
\int dh\, p(y_D|x_D,h) p(h|D_0)
\label{yd-evidence}
.
\ee
This is due to the assumption that
$p(y|x,h)$ is slowly varying at the stationary point
and has not to be included in the saddle point approximation
for the predictive density.
For (functional) differentiable $E_{comb}$
Eq.\ (\ref{minimum}) yields the stationarity equation, 
\be
\frac{\delta E_{\rm comb}({h})}{\delta {h}(x,y)} = 0
.
\ee
The functional $E_{\rm comb}$ including
training and prior data (regularization, stabilizer) terms
is also known as (regularized) {\it error functional} for ${h}$.

In practice a saddle point approximation may be expected
useful if the posterior is peaked enough around a single maximum,
or more general, if the posterior is well approximated
by a Gaussian centered at the maximum.
For asymptotical results one would have to require
\be
-\frac{1}{\beta} \sum_i L(y_i|x_i,{h})
,
\ee
to become $\beta$--independent 
for $\beta\rightarrow \infty$ with some $\beta$ being the same
for the prior and data term.
(See \cite{Cressie-1993,Yakowitz-Szidarovsky-1985}).
If for example $\frac{1}{n} \sum_i L(y_i|x_i,{h})$
converges for large number $n$ of training data
the low temperature limit $1/\beta\rightarrow 0$ can be interpreted
as large data limit $n\rightarrow\infty$,
\be
n E_{\rm comb} =
n \left( -\frac{1}{n} \sum_i L(y_i|x_i,{h}) 
+ \frac{1}{n} E({h}|D_0) \right)
\label{n-asym}
.
\ee
Notice, however, the factor $1/n$ in front of the prior energy.
For Gaussian $p(y|x,{h})$ temperature $1/\beta$
corresponds to variance $\sigma^2$
\be
\frac{1}{\sigma^2} 
E_{\rm comb} =
\frac{1}{\sigma^2}  
\left( \frac{1}{2} \sum_i (y_i-{h}(x_i))^2
+ \sigma^2
E({h}|D_0) \right)
.
\ee
For Gaussian prior this would require simultaneous scaling
of data and prior variance.

We should also remark that for continuous $x$,$y$
the stationary solution ${h}^{*}$
needs not to be a typical representative of the process
$p({h}|f)$.
A common example is a
Gaussian stochastic process $p({h}|f)$ 
with prior energy $E({h}|D_0)$
related to some smoothness measure of ${h}$
expressed by derivatives of $p(y|x,{h})$.
Then, even if the stationary ${h}^{*}$ is smooth, 
this needs not to be the case for a typical ${h}$ 
sampled according to $p({h}|f)$.
For Brownian motion, for instance,
a typical sample path is not even differentiable (but continuous) 
while the stationary path is smooth.
Thus, for continuous variables
only expressions like  $\int \!d{h}\,e^{-\beta E({h})}$ 
can be given an exact meaning as a Gaussian measure, 
defined by a given covariance with existing normalization factor, 
but not the expressions $d{h}$ and $E({h})$ alone
\cite{Doob-1953,Gardiner-1990,Wahba-1990,van-Kampen-1992,Green-Silverman-1994,Lifshits-1995}.

Interestingly, the stationary ${h}^{*}$
yielding maximal posterior $p({h}|f)$
is not only useful to obtain an approximation
for the predictive density
$p(y|x,f)$
but is also the optimal solution $a^*$
for a Bayesian decision problem with log--loss 
and $a\in A={H}$. 
\vspace{0.2cm}

\noindent 
{\it Indeed, for a Bayesian decision problem with log--loss (\ref{log-loss})
\be
{\rm argmin}_{a\in {H}} r(a,{h}) = {h}
,
\label{th1}
\ee
and analogously,}
\be
{\rm argmin}_{a\in F} r(a,f) = f
.
\label{th2}
\ee
\vspace{0.2cm}

\noindent
This is proved as follows:
Jensen's inequality states that
\be
\int\!dy\, p(y)g(q(y))
\ge
g( \int\!dy\, p(y) q(y) )
,
\ee
for any convex function $g$ and probability $p(y)\ge 0$ with
$\int\!dy\, p(y) =1$.
Thus, because the logarithm is concave
\be
-\int\!dy\, p(y|x,{h}) \ln\frac{p(y|x,a)}{p(y|x,{h})} 
\ge 
-\ln \int\!dy\, p(y|x,{h}) \frac{p(y|x,a)}{p(y|x,{h})} 
=0
\ee
\be
\Rightarrow
-\int\!dy\, p(y|x,{h}) \ln p(y|x,a) 
\ge
-\int\!dy\, p(y|x,{h}) \ln p(y|x,{h})
,
\ee
with equality for $a={h}$.
Hence
\bea
r(a,{h}) 
  &=& -\int\!dx \int\!dy\, p(x) p(y|x,{h})
    \left( b_1(x) \ln p(y|x,a) +b_2(x,y) \right)
\\&=& 
-\int\!dx\, p(x)b_1(x) \int\!dy\, p(y|x,{h}) \ln p(y|x,a) + {\rm const.}
\label{before}
\\&\ge&
-\int\!dx\, p(x)b_1(x) \int\!dy\, p(y|x,{h}) \ln p(y|x,{h}) + {\rm const.}
\label{after}
\\&=& r({h},{h})
,
\eea
with equality for $a={h}$.
For $a\in F$ replace ${h}\in {H}$ by $f\in F$.
This proves Eqs.~(\ref{th1}) and (\ref{th2}).

\subsection{Normalization, non--negativity, and specific priors}

Density estimation problems are characterized by 
their normalization and non--negativity condition for 
$p(y|x,{h})$.
Thus, the prior density $p({h}|D_0)$
can only be non--zero for such ${h}$
for which $p(y|x,{h})$ is positive and normalized over $y$ 
for all $x$.
(Similarly, when solving for a distribution function, 
i.e., the integral of a density,
the non--negativity constraint is replaced by monotonicity and
the normalization constraint 
by requiring the distribution function to be 1
on the right boundary.)
While the non--negativity constraint
is local with respect to
$x$ and $y$, the normalization constraint 
is nonlocal with respect to $y$.
Thus, implementing a normalization constraint
leads to nonlocal and in general non--Gaussian priors.

For classification problems, having discrete $y$ values (classes), 
the normalization constraint
requires simply to sum over the different classes 
and a Gaussian prior structure with respect to the $x$--dependency
is not altered \cite{Williams-Barber-1998}.
For general density estimation problems, however, i.e.,
for continuous $y$, the loss of the Gaussian 
structure with respect to $y$ is more severe,
because non--Gaussian functional integrals can in general not
be performed analytically.
On the other hand,
solving the learning problem 
numerically by discretizing the $y$ and $x$ variables,
the normalization term is 
typically not a severe complication.

To be specific, 
consider a Maximum A Posteriori Approximation,
minimizing
\be
\beta E_{\rm comb} 
=-\sum_i L(y_i|x_i,{h}) + \beta E({h}|D_0)
,
\ee
where the likelihood free energy $F(Y_D|x_D,{h})$ 
is included,
but not the prior free energy $F({H}|D_0)$ 
which, being ${h}$--independent, 
is irrelevant for minimization with respect to $h$.
The prior energy $\beta E({h}|D_0)$ has to implement 
the non--negativity and normalization conditions
\bea
Z_X(x,{h})=
\int \!dy_i\, p(y_i|x_i,{h}) = 1, 
&&
\forall x_i \in X_i, \forall {h}\in {H}\\
p(y_i|x_i,{h}) \ge 0,
&&
\forall y_i \in Y_i,  
\forall x_i \in X_i, 
\forall {h}\in {H}
.
\eea
It is useful to isolate the normalization condition
and non--negativity constraint defining
the class of density estimation problems
from the rest of the problem specific priors.
Introducing the {\it specific prior information} $\tilde D_0$
so that
$D_0$ = $\{ \tilde D_0, {\rm normalized,positive} \}$,
we have
\be
p({h}|\tilde D_0,{\rm norm.,pos.})
=
\frac{p({\rm norm.,pos.}|{h})p({h}|\tilde D_0)}{p({\rm norm.,pos.}|\tilde D_0)}
,
\label{specific-prior}
\ee
with deterministic, $\tilde D_0$--independent
\be
p({\rm norm.,pos.}|{h}) 
= p({\rm norm.,pos.}|{h}, \tilde D_0) 
\ee
\be
=p({\rm norm.}|{h}) p({\rm pos.}|{h}) 
= \delta (Z_X - 1) \prod_{xy}\Theta \Big(p(y|x,h)\Big)
,
\label{delta-step}
\ee
and step function $\Theta$.
( The density $p({\rm norm.}|{h})$ 
is normalized over all possible normalizations of $p(y|x,h)$, 
i.e., over all possible values of $Z_X$, and
$p({\rm pos.}|{h})$ over all possible sign combinations.)
The ${h}$--independent denominator $p({\rm norm.,pos.}|\tilde D_0)$
can be skipped for error minimization with respect to ${h}$.
We define the {\it specific prior} as
\be
{p({h}|\tilde D_0)} \propto
e^{-E({h}|\tilde D_0)}
.
\label{spez-prior-energy}
\ee

In Eq.\ (\ref{spez-prior-energy})
the specific prior appears
as posterior of a ${h}$--generating process
determined by the parameters $\tilde D_0$.
We will call therefore Eq.\ (\ref{spez-prior-energy})
the {\it posterior form} of the specific prior.
Alternatively, 
a specific prior can also be in {\it likelihood form}
\be
p(\tilde D_0,h|{\rm norm.,pos.}) = p(\tilde D_0|h) 
\,p(h|{\rm norm.,pos.})
.
\label{specific-prior-likeli}
\ee
As the likelihood $p(\tilde D_0|h)$
is conditioned on ${h}$ this means
that the normalization 
$Z$ =
$\int \!d\tilde D_0 \,e^{-E(\tilde D_0|{h})}$
remains in general ${h}$--dependent and must be included
when minimizing with respect to ${h}$.
However,
Gaussian specific priors with ${h}$--independent covariances
have the special property
that according to Eq.\ (\ref{gauss}) 
likelihood and posterior interpretation coincide.
Indeed, 
representing Gaussian
specific prior data $\tilde D_0$ by a mean function 
$t_{\tilde D_0}$ and covariance ${{\bf K}}^{-1}$
(analogous to standard training data in the case of Gaussian regression,
see also Section \ref{Non--zero-means})
one finds 
due to the fact that the normalization 
of a Gaussian is independent of the mean
(for uniform (meta) prior $p({h})$)
\bea
p({h}|\tilde D_0) 
&= &
\frac{e^{-\frac{1}{2}( {h}-t_{\tilde D_0}, {{\bf K}} ({h}-t_{\tilde D_0}))}}
{\int\!d{h}\,e^{-\frac{1}{2}( {h}-t_{\tilde D_0},{{\bf K}}({h}-t_{\tilde D_0}))}}
\\
=p(t_{\tilde D_0}|{h},{{\bf K}}) 
&=& 
\frac{e^{-\frac{1}{2}( {h}-t_{\tilde D_0}, {{\bf K}} ({h}-t_{\tilde D_0}))}}
{\int\!dt\,e^{-\frac{1}{2}( {h}-t, {{\bf K}} ({h}-t))}}
.
\eea
Thus, for Gaussian 
$p(t_{\tilde D_0}|{h},{{\bf K}})$
with ${h}$--independent
normalization the specific prior energy in likelihood form
becomes analogous to Eq.\ (\ref{spez-prior-energy})
\be
{p(t_{\tilde D_0}|{h},{{\bf K}})} \propto
e^{-E(t_{\tilde D_0}|{h},{{\bf K}})}
,
\ee
and specific prior energies can be interpreted
both ways.

Similarly,
the complete likelihood factorizes 
\be
p(\tilde D_0,{\rm norm.,pos.}|{h})
=
p({\rm norm.,pos.}|{h})
\, p(\tilde D_0 | {h})
.
\ee

According to Eq.\ (\ref{delta-step})
non--negativity and normalization conditions are implemented by step
and $\delta$--functions.
The non--negativity constraint is only active 
when there are locations with $p(y|x,h)$ = $0$.
In all other cases the gradient has no component 
pointing into forbidden regions.
Due to the combined effect 
of data, where $p(y|x,h)$ has to be larger than zero by definition,
and smoothness terms 
the non--negativity condition for $p(y|x,{h})$
is usually (but not always) fulfilled.
Hence, if strict positivity is checked for the final solution,
then it is not necessary to include 
extra non--negativity terms in the error (see Section \ref{error-e-p}).
For the sake of simplicity we will therefore not include non--negativity
terms explicitly in the following.
In case a non--negativity constraint has to be included
this can be done using Lagrange multipliers, or
alternatively, by writing the step functions in
$p({\rm pos.}|h) \propto \prod_{x,y} \Theta (p(y|x,{h}))$ 
\be
\Theta(x-a) 
= \int_a^\infty \!d\xi \int_{-\infty}^{\infty} d\eta e^{i\eta(\xi-x)}
,
\ee
and solving the $\xi$--integral in saddle point approximation
(See for example \cite{Gardner-1987,Gardner-1988,Gardner-Derrida-1988}).

Including the normalization condition
in the prior $p_0({h}|D_0)$ in form of a $\delta$--functional
results in a posterior probability
\begin{equation}
p({h}|f) \!= 
e^{ \sum_i L_i(y_i|x_i,{h}) - E({h}|\tilde D_0) 
+ \tilde c({H}|\tilde D_0)}
\prod_{x\in X} \delta \left( \int\!dy\,e^{L(y|x,{h})} -1 \right)
\end{equation}
with constant
$\tilde c({H}|\tilde D_0)$ 
= $-\ln \tilde Z({h}|\tilde D_0)$ related to the normalization
of the specific prior $e^{-E({h}|\tilde D_0)}$.
Writing the $\delta$--functional 
in its Fourier representation
\begin{equation}
\delta (x)
= \frac{1}{2 \pi} \int_{-\infty}^{\infty} \!dk\, e^{i k x }
=  \frac{1}{2 \pi i} \int_{-i\infty}^{i\infty} \!dk\, e^{- k x },
\label{fourierdelta}
\end{equation}
i.e.,  
\begin{equation}
\delta ( \int \! dy \, e^{L(y|x,{h})}-1)
= \frac{1}{2 \pi i} 
\int_{-i \infty}^{i \infty} \!d\Lambda_X (x)\, 
e^{\Lambda_X (x)  
\left( 1- \int \! dy \, e^{L(y|x,{h})} \right) }
,
\end{equation}
and performing a saddle point approximation 
with respect to
$\Lambda_X (x)$ (which is exact in this case)
yields
\begin{equation}
P ({h}|f) = 
e^{ \sum_i L_i(y_i|x_i,{h}) - E({h}|\tilde D_0)
+ \tilde c({H}|\tilde D_0)
+\int \!\!dx\, \Lambda_X (x) \left( 1-\int\!\! dy e^{L(y|x,{h})} \right)}.
\end{equation}
This is equivalent to the Lagrange multiplier approach.
Here the stationary 
$\Lambda_X (x)$ is the Lagrange multiplier vector (or function) 
to be determined by the normalization conditions
for $p(y|x,{h})=e^{L(y|x,{h})}$.
Besides the Lagrange multiplier terms 
it is numerically sometimes useful to add additional 
terms to the log--posterior which vanish for normalized $p(y|x,{h})$.

\subsection{Empirical risk minimization}
\label{empirical-risk}

In the previous sections
the error functionals we will try to minimize in the following
have been given a Bayesian interpretation 
in terms of the log--posterior density.
There is, however, an alternative justification
of error functionals using the Frequentist approach of
{\it empirical risk minimization}
\cite{Vapnik-1982,Vapnik-1995,Vapnik-1998}.

Common to both approaches is the aim to minimize
the {\it expected risk} for action $a$
\be
r(a,f) = \int\!dx\,dy\,p(x,y|f(D,D^0)) \, l(x,y,a)
.
\ee
The expected risk, however, cannot be calculated without 
knowledge of the true $p(x,y|f)$.
In contrast to the Bayesian approach of modeling $p(x,y|f)$
the Frequentist approach approximates the expected risk
by the {\it empirical risk}
\be
E(a) = \hat r(a,f) = \sum_i l(x_i,y_i,a)
,
\ee
i.e., by replacing the unknown true probability by
an observable empirical probability.
Here it is essential for obtaining asymptotic convergence results
to assume that training data are sampled
according to the true $p(x,y|f)$
\cite{Vapnik-1982,Dudley-1984,Pollard-1984,LeCam-1986,Vapnik-1998}.
Notice that in contrast in a Bayesian approach the density $p(x_i)$ 
for training data $D$
does according to Eq.\ (\ref{s-asymm}) not enter the formalism
because $D$ enters as conditional variable.  
For a detailed discussion of the relation
between quadratic error functionals
and Gaussian processes 
see for example 
\cite{Parzen-1962a,Parzen-1963,Parzen-1970,Kimmeldorf-Wahba-1970a,Kimmeldorf-Wahba-1970b,Marroquin-Mitter-Poggio-1987,Wahba-1990,Lifshits-1995}.

From that Frequentist point of view one is 
not restricted to logarithmic data terms 
as they arise from the posterior--related Bayesian interpretation.
However, like in the Bayesian approach,
training data terms are not enough to 
make the minimization problem well defined.
Indeed this is a typical inverse problem 
\cite{Vapnik-1982,Kirsch-1996,Vapnik-1998}
which can,
according to the classical regularization approach
\cite{Tikhonov-1963,Tikhonov-Arsenin-1977,Morozov-1984},
be treated
by including additional {\it regularization (stabilizer) terms}
in the loss function $l$.
Those regularization terms, which correspond
to the prior terms in a Bayesian approach,
are thus from the point of view of empirical risk minimization 
a technical tool
to make the minimization problem well defined.

The {\it empirical generalization error}
for a test or validation data set independent 
from the training data $D$, on the other hand, 
is measured using only the data terms of the error functional 
without regularization terms.
In empirical risk minimization 
this empirical generalization error 
is used, for example,
to determine adaptive (hyper--)parameters
of regularization terms.
A typical example is a factor multiplying the regularization terms 
controlling the trade--off between data and regularization terms.
Common techniques 
using the empirical generalization error
to determine such parameters
are {\it cross--validation} or {\it bootstrap} like techniques
\cite{Mosteller-Wallace-1963,Allen-1974,Wahba-Wold-1975,Stone-1974,Stone-1977,Golup-Heath-Wahba-1979,Craven-Wahba-1979,Wahba-1990,Efron-Tibshirani-1993}.
From a strict Bayesian point of view
those parameters would have to be integrated out
after defining an appropriate prior
\cite{Berger-1980,MacKay-1992c,MacKay-1994d,Bishop-1995b}.

\subsection{Interpretations of Occam's razor}
\label{Occam}

The principle to prefer simple models over complex models
and to find an optimal trade--off between fitting data 
and model complexity
is often referred to as Occam's razor (William of Occam, 1285--1349).
Regularization terms, 
penalizing for example non--smooth (``complex'') functions,
can be seen as an implementation of Occam's razor. 

The related phenomena appearing in practical learning is called
over--fitting \cite{Vapnik-1982,Hertz-Krogh-Palmer-1991,Bishop-1995b}.
Indeed, when studying the generalization behavior
of trained models on a test set different from the training set,
it is often found
that there is an optimal model complexity.
Complex models can due to their higher flexibility
achieve better performance on the training data than simpler models.
On a test set independent from the training set, however,
they can perform poorer than simpler models.

Notice, however, that the Bayesian interpretation
of regularization terms as ({\it a priori}) information about Nature
and the Frequentist interpretation as additional 
cost terms in the loss function are {\it not} equivalent.
Complexity priors reflects the case
where Nature is known to be simple
while complexity costs
express the wish for simple models
without the assumption of a simple Nature.
Thus, while the practical procedure of minimizing an error functional
with regularization terms
appears to be identical for empirical risk minimization
and a Bayesian Maximum A Posteriori Approximation,
the underlying interpretation for this procedure is different.
In particular, because the Theorem in Section \ref{map}
holds only for log--loss,
the case of loss functions differing from log--loss requires
from a Bayesian point of view
to distinguish explicitly between model states ${h}$
and actions $a$.
Even in saddle point approximation,
this would result in a two step procedure,
where in a first step the hypothesis ${h}^{*}$,
with maximal posterior probability is determined,
while the second step minimizes the risk for action $a\in A$
under that hypothesis ${h}^{*}$  \cite{Lemm-1996}.

\subsection{{\it A priori} information and {\it a posteriori} control}

Learning is based on data, which includes
training data as well as {\it a priori} data.
It is prior knowledge which,
besides specifying the space of local hypothesis,
enables generalization
by providing the necessary link 
between measured training data 
and not yet measured or non--training data.
The strength of this connection
may be quantified by the mutual information
of training and non--training data,
as we did in Section \ref{mutual}.

Often, the role of {\it a priori} information seems to be underestimated.
There are theorems, for example, proving that 
asymptotically learning results become independent 
of {\it a priori} information
if the number of training data goes to infinity.
This, however,is correct only if the space of
hypotheses $h$ is already sufficiently restricted
and if {\it a priori} information
means knowledge in addition to that restriction.

In particular, let us assume that the
number of potential test situations $x$, 
is larger than the number of training data 
one is able to collect.
As the number of actual training data has to be finite,
this is always the case if $x$ can take an infinite number of values,
for example if $x$ is a continuous variable.
The following arguments, however, are not restricted
to situations were one considers
an infinite number of test situation,
we just assume that their number 
is too large to be completely included in the training data.

If there are $x$ values for which
no training data are available, 
then learning for such $x$
must refer to the mutual information
of such test data and the available training data.
Otherwise, training would be useless for these test situations.
This also means, 
that the generalization to non--training situations
can be arbitrarily modified by varying 
{\it a priori} information.

To make this point very clear,
consider the rather trivial situation
of learning a deterministic function $h(x)$
for a $x$ variable which can take only two values $x_1$ and $x_2$,
from which only one can be measured.
Thus, having measured for example $h(x_1)$ = 5,
then ``learning'' $h(x_2)$ is not possible
without linking it to $h(x_1)$.
Such prior knowledge may have the form of a 
``smoothness'' constraint,
say $|h(x_1)-h(x_2)|\le 2$
which would allow a learning algorithm to ``generalize''
from the training data and obtain $3\le h(x_2)\le 7$.
Obviously, arbitrary results can be obtained for $h(x_2)$
by changing the prior knowledge.
This exemplifies
that generalization can be considered
as a mere reformulation
of available information, i.e., of training data and prior knowledge.
Except for such a rearrangement of knowledge,
a learning algorithm does not add any new information to the problem.
(For a discussion of the related ``no--free-lunch'' theorems see
\cite{Wolpert-1996a,Wolpert-1996b}.)

Being extremely simple, this example nevertheless shows
a severe problem.
If the result of learning can be arbitrary modified by 
{\it a priori} information, then it is critical
which prior knowledge is implemented in the learning algorithm.
This means, that prior knowledge 
needs an empirical foundation, just like standard training
data have to be measured empirically.
Otherwise, the result of learning cannot expected to be
of any use.

Indeed, the problem of appropriate {\it a priori} information
is just the old induction problem, i.e., the problem
of learning general laws from a finite number of observations,
as already been discussed by the ancient Greek philosophers.
Clearly, this is not a purely academic problem,
but is extremely important for every system which depends on a successful
control of its environment.
Modern applications of learning algorithms,
like speech recognition
or image understanding,
rely essentially on correct {\it a priori} information.
This holds especially for situations where
only few training data are available,
for example, because sampling is very costly.

Empirical measurement of {\it a priori} information, however, 
seems to be impossible.
The reason is that we must link every possible
test situation to the training data. 
We are not able to do this in practice if, as we assumed, 
the number of potential test situations
is larger than the number of measurements one is able to perform.

Take as example again
a deterministic learning problem like the one discussed above.
Then measuring {\it a priori} information
might for example be done by measuring (e.g., bounds on)
all differences $h(x_1)-h(x_i)$.
Thus, even if we take the deterministic structure of the problem for granted,
the number of such differences is equal to the number
of potential non--training situations $x_i$ we included in our model.
Thus, measuring {\it a priori} information does not require
fewer measurements 
than measuring directly all potential non--training data.
We are interested in situations where this is impossible.

Going to a probabilistic setting
the problem remains the same. 
For example, even if we assume Gaussian hypotheses
with fixed variance, 
measuring a complete mean function
$h(x)$, say for continuous $x$, 
is clearly impossible in practice.
The same holds thus 
for a Gaussian process prior on $h$.
Even this very specific prior
requires the determination
of a covariance and a mean function
(see Chapter \ref{Gaussian-priors}).

As in general empirical measurement of {\it a priori} information
seems to be impossible,
one might thus just try to guess some prior.
One may think, for example, of some ``natural''
priors.
Indeed, the term ``{\it a priori}'' goes back to Kant
\cite{Kant--1911}
who assumed certain knowledge to be necessarily be given ``{\it a priori}''
without reference to empirical verification.
This means that we are either only able
to produce correct prior assumptions,
for example because incorrect prior assumptions are ``unthinkable'',
or that one must typically be lucky to implement the right {\it a priori} information.
But looking at the huge number of different prior assumptions 
which are usually possible (or ``thinkable''),
there seems no reason why one should be lucky.
The question thus remains, how can prior assumptions
get empirically verified.

Also, one can ask whether there are ``natural'' priors 
in practical learning tasks.
In Gaussian regression one might maybe consider
a ``natural'' prior to be a Gaussian process with constant mean function
and smoothness--related covariance.
This may leave a single regularization parameter
to be determined for example by cross--validation.
Formally, one can always even use a zero mean function for the prior process
by subtracting a base line or reference function.
Thus does, however, not solve the problem of finding a correct prior,
as now that reference function has to be known
to relate the results of learning to empirical measurements.
In principle {\it any} function could be chosen as reference function.
Such a reference function
would for example enter a smoothness prior.
Hence, there is no ``natural'' constant function
and from an abstract point of view
no prior is more ``natural'' than any other.

Formulating a general law refers 
implicitly (and sometimes explicitly)
to a ``ceteris paribus'' condition,
i.e., the constraint that
all relevant variables, not explicitly mentioned in the law, 
are held constant.
But again, verifying a ``ceteris paribus'' condition 
is part of an empirical measurement of {\it a priori} information
and by no means trivial.

Trying to be cautious 
and use only weak or ``uninformative'' priors
does also not solve the principal problem. 
One may hope that such priors 
(which may be for example an improper constant prior 
for a one--dimensional real variable)
 do not introduce a completely wrong bias, so that
the result of learning is essentially determined by the training data.
But, besides the problem to define what exactly  
an uninformative prior has to be,
such priors are in practice only useful
if the set of possible hypothesis is already sufficiently restricted,
so ``the data can speak for themselves''
\cite{Gelman-Carlin-Stern-Rubin-1995}.
Hence, the problem remains
to find that priors which impose the necessary 
restrictions, so that uninformative priors can be used.

Hence, 
as measuring {\it a priori} information seems impossible
and finding correct {\it a priori} information by pure luck
seems very unlikely, 
it looks like also successful learning is impossible.
It is a simple fact, however,
that learning can be successful.
That means there must be a way
to control {\it a priori} information empirically.

Indeed, the problem of measuring  {\it a priori} information
may be artificial,
arising from the introduction of a large number of 
{\it potential} test situations
and correspondingly
a large number of hidden variables $h$ 
(representing what we call ``Nature'')
which are not all observable.

In practice, the number
of {\it actual} test situations is also always finite,
just like the number of training data has to be.
This means,
that not {\it all} potential test data but only the actual
test data must be linked to the training data.
Thus, in practice it is only a finite number of relations
which must be under control to allow successful generalization.
(See also Vapnik's distinction between
 induction and transduction problems.
\cite{Vapnik-1998}:
In induction problems one tries to infer a whole function,
in transduction problems one is only interested in predictions
for a few specific test situations.)

This, however, opens a possibility to
control {\it a priori} information empirically.
Because
we do not know which test situation will occur,
such an empirical control cannot take place
at the time of training.
This means {\it a priori} information has to be implemented
at the time of measuring the test data.
In other words,
{\it a priori} information 
has to be implemented by the measurement process
\cite{Lemm-1996,Lemm-1998b}.

Again, a simple example may clarify this point.
Consider the prior information,
that a function $h$ is bounded, i.e., 
$a\le h(x)\le b$, $\forall x$.
A direct measurement of this
prior assumption is practically not possible,
as it would require to check every value $h(x)$. 
An implementation
within the  measurement process is however trivial.
One just has to use a measurement device
which is only able to to produce
output in the range between $a$ and $b$.
This is a very realistic assumption and valid
for all real measurement devices.
Values smaller than $a$ and larger than $b$
have to be filtered out or actively projected into that range.
In case we nevertheless find a value out of that range
we either have to adjust the bounds
or we exchange the ``malfunctioning''
measurement device with a proper one.
Note, that this range filter is only needed
at the finite number of actual measurements.
That means,
{\it a priori} information can be implemented by
{\it a posteriori} control at the time of testing.

\begin{figure}
\epsfig{file=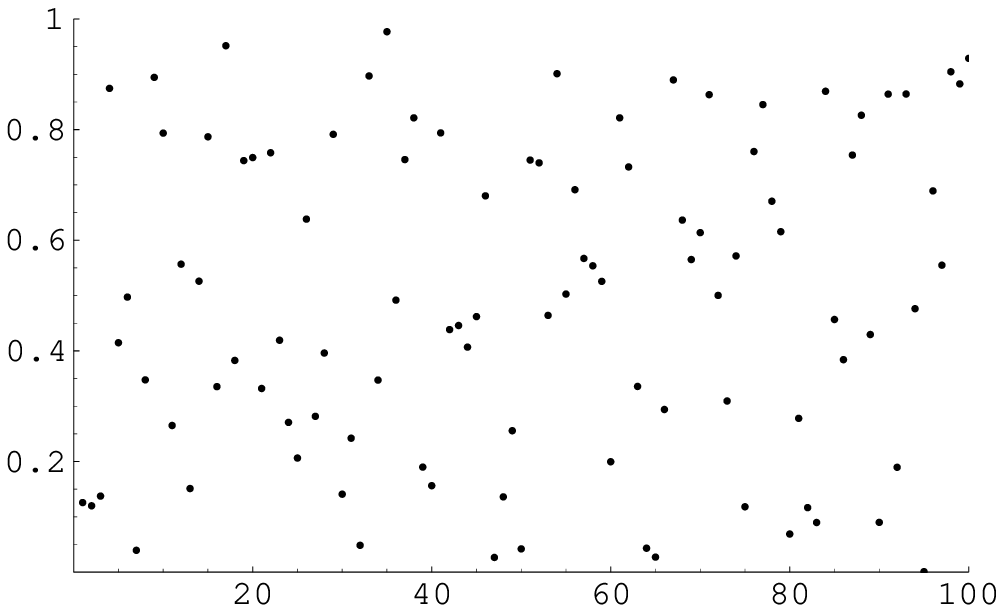, width=60mm}
\epsfig{file=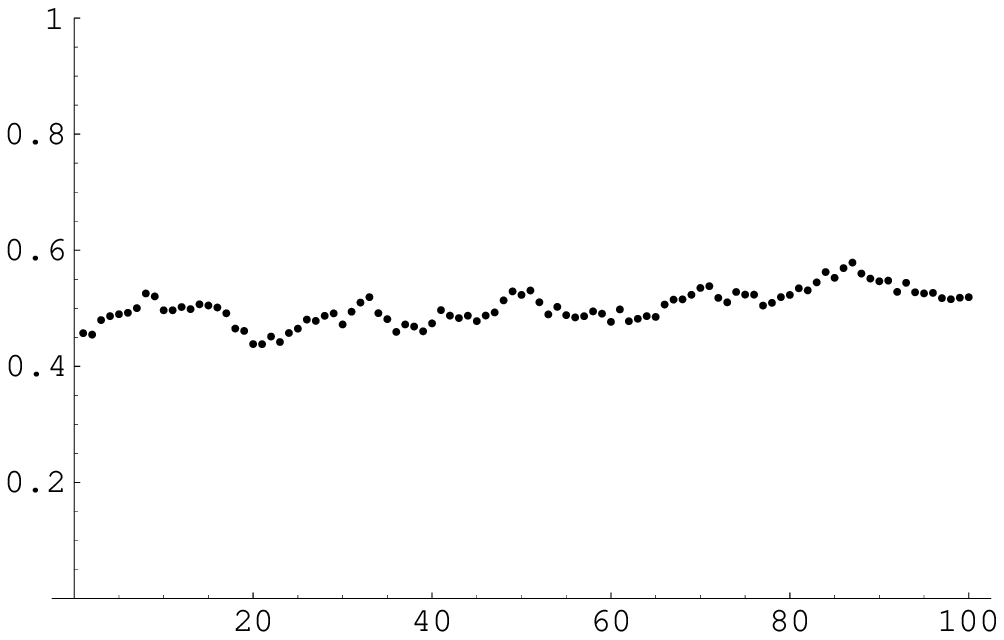, width=60mm}
\caption{The l.h.s. shows a bounded random function 
which does not allow
generalization from training to non--training data.
Using a measurement device with input averaging (r.h.s.)
or input noise the function becomes learnable.
}
\label{input-random}
\end{figure}

A realistic measurement device
does not only produce bounded output
but shows also always {\it input noise} or {\it input averaging}.
A device with input noise has noise in the $x$ variable.
That means if one intends to measure at $x_i$
the device measures instead at $x_i+\Delta$
with $\Delta$ being a random variable.
A typical example is translational noise,
with $\Delta$ being a, possibly multidimensional, Gaussian 
random variable with mean zero.
Similarly, a device with
input averaging returns a weighted average
of results for different $x$ values
instead of a sharp result.
Bounded devices with translational input noise, for example,
will always measure smooth functions
\cite{Leen-1995,Bishop-1995a,Lemm-1996}.
(See Fig.\ \ref{input-random}.)
This may be an explanation for the success
of smoothness priors.

The last example shows, that 
to obtain adequate {\it a priori} information 
it can be helpful in practice
to analyze the measurement process
for which learning is intended.
The term ``measurement process'' 
does here not only refer to a specific device, e.g., a box on the table,
but to the collection
of all processes which lead to a measurement result.

We may remark
that measuring a measurement process
is as difficult or impossible as a 
direct measurement of {\it a priori} information.
What has to be ensured is the validity of the necessary restrictions
during a finite number of actual measurements.
This is nothing else than the implementation of a probabilistic rule
producing $y$ given the test situation and the training data.
In other words, what has to be implemented
is the predictive density $p(y|x,D)$.
This predictive density indeed only depends 
on the actual test situation and the finite number of training data.
(Still, the probability density for a real $y$
cannot strictly be empirically verified or controlled.
We may take it here, for example,
as an approximate statement about frequencies.)
This shows the tautological character of learning,
where measuring {\it a priori} information means 
controlling directly the corresponding predictive density.

The {\it a posteriori} interpretation of {\it a priori} information
can be related to a  constructivistic point of view.
The main idea of constructivism can be characterized
by a sentence  of Vico (1710):
{\it Verum ipsum factum}
--- the truth is the same as the made
\cite{Vico-1710}.
(For an introduction to constructivism see
\cite{Watzlawick-1984} and references therein,
for constructive mathematics
see \cite{Bishop-Bridges-1985}.)

\section{Gaussian prior factors}
\label{Gaussian-priors}
\subsection{Gaussian prior factor for log--probabilities}
\label{gaussL}
\subsubsection{Lagrange multipliers: Error functional $E_L$}
\label{error-e-l}

In this chapter 
we look at density estimation problems
with Gaussian prior factors.
We begin with a discussion of functional priors
which are Gaussian in probabilities 
or in log--probabilities,
and continue with general Gaussian prior factors.
Two section are devoted to
the discussion of 
covariances and means of Gaussian prior factors,
as their adequate choice is essential for 
practical applications.
After exploring some
relations of Bayesian field theory and empirical risk minimization,
the last three sections introduce 
the specific likelihood models of 
regression, classification,
inverse quantum theory.

We begin a discussion of Gaussian prior factors in $L$.
As Gaussian prior factors correspond 
to quadratic error (or energy) terms,
consider an error functional with a quadratic regularizer in $L$
\be
(L,{{\bf K}}L)=||L||^2_{{\bf K}} =
\frac{1}{2}\int\!dx\,dy\,dx^\prime dy^\prime  
L(x,y) {{\bf K}}(x,y;x^\prime,y^\prime ) L(x^\prime ,y^\prime)
,
\label{quaderr}
\ee
writing for the sake of simplicity
from now on $L(x,y)$ for the log--probability $L(y|x,{h})$ = $\ln p(y|x,{h})$. 
The operator ${{\bf K}}$ 
is assumed symmetric and positive
semi--definite and positive definite on some subspace.
(We will understand positive semi--definite to include symmetry
in the following.)
For positive (semi) definite ${{\bf K}}$
the scalar product 
defines a (semi) norm 
by 
\be
||L||_{{\bf K}} = \sqrt{(L,{{\bf K}}L)}
, 
\ee
and a corresponding distance by
$||L-L^\prime||_{{\bf K}}$. 
The quadratic error term (\ref{quaderr})
corresponds to a Gaussian factor of the prior density 
which have been called the specific prior  
$p({h}|\tilde D_0)$ = $p(L|\tilde D_0)$
for $L$.
In particular, we will consider here the posterior density 
\begin{equation}
p({h}|f) \!= 
e^{ \sum_i L_i(x_i,y_i) 
- \!\frac{1}{2}\int\! dx dy dx^\prime dy^\prime  
  L(x,y) {{\bf K}}(x,y;x^\prime,y^\prime ) L(x^\prime ,y^\prime)
+ \int \!\!dx\,  \Lambda_X (x)  \left( 1 - \int\!dy\, e^{L(x,y)} \right) 
+ \tilde c,
}
\label{L-functional}
,
\end{equation}
where prefactors like $\beta$
are understood to be included in ${{\bf K}}$.
The constant $\tilde c$ referring to the specific prior is determined 
by the determinant of ${{\bf K}}$ according to Eq.\ (\ref{gauss}).
Notice however that not only the likelihood $\sum_i L_i$ 
but also the complete prior 
is usually {\it not} Gaussian 
due to the presence of the normalization conditions.
(An exception is Gaussian regression, see Section \ref{regression}.)
The posterior (\ref{L-functional}) corresponds to an error functional 
\be
E_L = \beta E_{\rm comb} = -(L,N) + \frac{1}{2} (L,{{\bf K}} L)
+(e^L-\delta(y),\Lambda_X)
\label{L-functional1}
,
\ee
with {\it likelihood  vector (or function)} 
\be
L(x,y) = L(y|x,{h})
,
\ee
{\it data vector (function)} 
\be
N(x,y) = \sum_i^n \delta(x-x_i)\delta(y-y_i) 
,
\label{nmat}
\ee
{\it Lagrange multiplier vector (function)} 
\be
\Lambda_X(x,y) = \Lambda_X(x) 
,
\ee
{\it probability vector (function)}
\be
e^L(x,y) = e^{L(x,y)} = P(x,y) = p(y|x,{h})
,
\ee
and 
\be
\delta(y)(x,y) =\delta(y)
.
\ee
According to Eq.\ (\ref{nmat}) $N/n$ = $P_{\rm emp}$
is an {\it empirical density function} 
for the joint probability $p(x,y|{h})$.

We end this subsection by defining some notations. 
Functions of vectors (functions) and matrices (operators),
different from multiplication, 
will be understood element-wise
like for example $(e^L)(x,y)$ = $e^{L(x,y)}$.
Only multiplication of matrices (operators) 
will be interpreted as matrix product.
Element-wise multiplication has then to be written 
with the help of diagonal matrices.
For that purpose we introduce diagonal matrices 
made from vectors (functions) 
and denoted by the corresponding bold letters.
For instance, 
\bea
{\bf I} (x,y ; x^\prime,y^\prime) &=& \delta (x-x^\prime) \delta (y-y^\prime),
\\
{\bf L} (x,y;x^\prime,y^\prime )
       &=&\delta (x-x^\prime)\delta (y-y^\prime) L(x,y),
\\
{\bf P} (x,y;x^\prime,y^\prime )
&=& {\bf e^L} (x,y;x^\prime,y^\prime )
\\
       &=&\delta (x-x^\prime)\delta (y-y^\prime) P(x,y),
\\
{\bf N} (x,y ; x^\prime,y^\prime) 
&=&
\delta (x-x^\prime) \delta (y-y^\prime) \, N(x,y),
\\
{\bf \Lambda}_X (x,y;x^\prime,y^\prime )
&=&
\delta (x-x^\prime)\delta (y-y^\prime) \Lambda_X (x)
,
\eea
or
\be
L = {\bf L} I,\quad
P = {\bf P} I,\quad
e^L = {\bf e^L} I,\quad
N = {\bf N} I,\quad
\Lambda_X = {\bf \Lambda}_X I
,
\ee
where
\be
I(x,y) = 1
.
\ee
Being diagonal all these matrices commute with each other.
{\it Element-wise multiplication} can now be expressed as
\bea
({{\bf K}} {\bf L}) (x^\prime,y^\prime,x,y)
&=& \int \!dx^{\prime\prime}dy^{\prime\prime}
   {{\bf K}}(x^\prime,y^\prime,x^{\prime\prime},y^{\prime\prime}) 
   {\bf L}(x^{\prime\prime},y^{\prime\prime},x,y)  
\nonumber\\&=& \int \!dx^{\prime\prime}dy^{\prime\prime}
   {{\bf K}}(x^\prime,y^\prime,x^{\prime\prime},y^{\prime\prime}) 
L(x,y) \delta(x-x^{\prime\prime})\delta(y-y^{\prime\prime})  
\nonumber\\&=&
{{\bf K}}(x^\prime,y^\prime,x,y) L(x,y)
.
\eea
In general this is not equal to
$ L(x^\prime,y^\prime ) {{\bf K}}(x^\prime,y^\prime,x,y)$.
In contrast, the {\it matrix product} ${{\bf K}} L$ with vector $L$  
\be
({{\bf K}} L) (x^\prime,y^\prime) 
= \int \!dx\,dy\,{{\bf K}}(x^\prime,y^\prime,x,y) L(x,y)
,
\ee
does not depend on $x$, $y$ anymore,
while the {\it tensor product} or  outer product,
\be
({{\bf K}} \otimes L)(x^{\prime\prime},y^{\prime\prime},x,y,x^\prime,y^\prime) 
= {{\bf K}}(x^{\prime\prime},y^{\prime\prime},x^{\prime},y^{\prime})
L(x,y)
,
\ee
depends on additional $x^{\prime\prime}$, $y^{\prime\prime}$.

Taking the variational derivative of (\ref{L-functional}) 
with respect to $L(x,y)$ using 
\begin{equation}
\frac{ \delta L(x^\prime,y^\prime) }{\delta L(x,y )}
= \delta (x-x^\prime ) \delta (y-y^\prime )
\end{equation}
and setting the gradient equal to zero 
yields the stationarity equation
\begin{equation}
0 =  N - {{\bf K}} L - {\bf e^L} \Lambda_X
.
\label{stat-eq-bra-ket}
\end{equation}
Alternatively, we can write
${\bf e^L} \Lambda_X$
= ${\bf \Lambda}_X e^L$  
= ${\bf P} \Lambda_X$.

The Lagrange multiplier function $\Lambda_X$ is determined 
by the normalization condition 
\begin{equation}
Z_X(x) = \int \!dy \, e^{L(x,y)} = 1, \quad \forall x\in X
,
\label{normalization1}
\end{equation}
which can also be written
\begin{equation}
Z_X 
= {\bf I}_X P
= {\bf I}_X e^L
= I
\quad \mbox{\rm or} \quad
{\bf Z}_X = {\bf I},
\end{equation}
in terms of normalization vector,
\be
Z_X(x,y) 
= Z_X(x)
,
\ee
normalization matrix,
\be
{\bf Z_X} (x,y ; x^\prime,y^\prime) 
=\delta (x-x^\prime) \delta (y-y^\prime) \, Z_X(x)
,
\ee
and 
identity on $X$, 
\be
{\bf I}_X (x,y;x^\prime,y^\prime) = \delta(x-x^\prime)
.
\ee
Multiplication of a vector with ${\bf I}_X$ corresponds to $y$--integration.
Being a non--diagonal matrix
${\bf I}_X$ does in general not commute
with diagonal matrices like ${\bf L}$ or ${\bf P}$.
Note also that despite
${\bf I}_X e^L$ 
= ${\bf I}_X {\bf e^L} I$
= ${\bf I}I$
= $I$
in general
${\bf I}_X {\bf P}$
=
${\bf I}_X {\bf e^L}\ne {\bf I}$
= ${\bf Z}_X$.
According to the fact that 
${\bf I}_X$ and ${\bf \Lambda}_X$ commute, i.e.,
\be
{\bf I}_X  {\bf \Lambda}_X ={\bf \Lambda}_X {\bf I}_X  
\Leftrightarrow
[{\bf \Lambda}_X , {\bf I}_X] 
= {\bf \Lambda}_X {\bf I}_X - {\bf I}_X {\bf \Lambda}_X 
= 0
\label{gam-i-com}
,
\ee
(introducing the commutator $[A,B]$ = $AB-BA$),
and that the same holds for the diagonal matrices
\be
[{\bf \Lambda}_X , {\bf e^L}  ] =
[{\bf \Lambda}_X , {\bf P}  ] = 0
,
\ee
it follows from the  normalization condition
${\bf I}_X P$ = $I$
that
\be
{\bf I}_X {\bf P} \Lambda_X
= {\bf I}_X {\bf \Lambda_X} P 
= {\bf \Lambda_X} {\bf I}_X P 
= {\bf \Lambda_X} I 
= \Lambda_X  
,
\label{lambda-norm}
\ee
i.e.,
\be
0=
({\bf I} - {\bf I}_X {\bf e^L}) \Lambda_X
=({\bf I} - {\bf I}_X {\bf P}) \Lambda_X
.
\label{lambda-norm2}
\ee
For $\Lambda_X(x) \ne 0$
Eqs.(\ref{lambda-norm},\ref{lambda-norm2})
are equivalent to
the normalization (\ref{normalization1}).
If there exist directions at the stationary point $L^*$
in which the normalization of $P$ changes, 
i.e., the normalization constraint is active,
a $\Lambda_X(x) \ne 0$ restricts the gradient 
to the normalized subspace 
(Kuhn--Tucker conditions
\cite{Fletcher-1987,Bertsekas-1995,Horst-Pardalos-Thoai-1995,Polak-1997}).
This will clearly be the case for the unrestricted 
variations of $p(y,x)$ which we are considering here.
Combining
$\Lambda_X$ = ${\bf I}_X {\bf P} \Lambda_X$ 
for $\Lambda_X(x) \ne 0$
with the stationarity equation (\ref{stat-eq-bra-ket}) 
the Lagrange multiplier function is obtained  
\begin{equation}
\Lambda_X 
= {\bf I}_X \left( N - {{\bf K}} L \right) 
= N_X -  ({\bf I}_X{{\bf K}} L)
.
\label{lambda}
\end{equation}
Here we introduced the vector 
\be
N_X = {\bf I}_X N
,
\ee
with components
\be
N_X (x,y)  
= N_X (x)  
= \sum_i \delta (x - x_i)
= n_x
,
\ee
giving the number of data available for $x$.
Thus, Eq.\ (\ref{lambda}) reads in components
\be
\Lambda_X (x)
= 
\sum_i \delta (x - x_i)
- \int \! dy^{\prime\prime}\, dx^\prime dy^\prime \, 
{{\bf K}} (x,y^{\prime\prime};x^\prime,y^\prime ) L(x^\prime, y^\prime ).
\ee
Inserting now this equation
for $\Lambda_X$ 
into the stationarity equation 
(\ref{stat-eq-bra-ket}) yields
\begin{equation}
0 =   N  - {{\bf K}} L  -  {\bf e^L} ( N_X -  {\bf I}_X {{\bf K}} L) 
= \left( {\bf I}-{\bf e^L} {\bf I}_X \right) \left(N- {{\bf K}} L \right) 
\label{LwithLambda}
.
\end{equation}
Eq.\ (\ref{LwithLambda})
possesses, besides normalized solutions we are looking for,
also possibly unnormalized solutions fulfilling  $N={{\bf K}}L$
for which Eq.\ (\ref{lambda}) yields $\Lambda_X = 0$.
That happens because we used Eq.\ (\ref{lambda-norm})
which is also fulfilled for $\Lambda_X(x) = 0$.
Such a $\Lambda_X(x)=0$ 
does not play the role of a Lagrange multiplier.
For parameterizations of $L$ where the normalization constraint
is not necessarily active at a stationary point
$\Lambda_X(x)=0$ can be possible for a normalized solution $L^*$. 
In that case normalization has to be checked.

It is instructive to define
\begin{equation}
 T_L  =   N -  \Lambda_X e^L ,
\label{T-L}
\end{equation}
so the stationarity equation (\ref{stat-eq-bra-ket}) 
acquires the form 
\begin{equation}
{{\bf K}}  L  =  T_L,
\label{OL=T}
\end{equation}
which reads in components
\begin{equation}
\int \! dx^\prime dy^\prime \, {{\bf K}} (x,y;x^\prime,y^\prime )  
L(x^\prime,y^\prime ) 
=  \sum_i \delta (x-x_i) \delta (y-y_i)  -  {\bf \Lambda}_X (x) \, e^{L(x,y)}
,
\end{equation}
which is in general a non--linear equation because $T_L$ depends on $L$.
For existing (and not too ill--conditioned) ${{\bf K}}^{-1}$
the form (\ref{OL=T})
suggest however an iterative 
solution of the stationarity equation
according to 
\be
L^{i+1} = {{\bf K}}^{-1} T_L(L^i)
,
\ee
for discretized $L$,
starting from an initial guess $L^0$.
Here the Lagrange multiplier $\Lambda_X$ has to 
be adapted so it fulfills condition (\ref{lambda})
at the end of iteration. 
Iteration procedures will be discussed in detail
in Section \ref{learning}.

\subsubsection{Normalization by parameterization: Error functional $E_g$}

Referring to the discussion in Section \ref{map}
we show that Eq.\ (\ref{LwithLambda}) can
alternatively be obtained by ensuring normalization, 
instead of using Lagrange multipliers, 
explicitly by the parameterization
\begin{equation}
L(x,y)  = g(x,y) - \ln \int \!dy^\prime \, e^{g(x,y^\prime )},
\quad
L = g - \ln Z_X
,
\end{equation}
and considering the functional
\begin{equation}
E_{g} = 
-\Big(N , \, g -\ln Z_X\, \Big)
+\frac{1}{2}\Big( \,g -\ln Z_X \,, \,{{\bf K}}\,(g -\ln Z_X)\,\Big)
.
\end{equation}
The stationary equation for $g(x,y)$   
obtained by setting the functional derivative $\delta E_{g}/\delta g$
to zero yields again Eq.\ (\ref{LwithLambda}).
We check this, using
\begin{equation}
\frac{\delta \ln Z_X (x^\prime)}{\delta g(x,y)}
= \delta (x-x^\prime )  e^{L(x,y)} 
,
\quad
\frac{\delta \ln Z_X}{\delta g}
=   {\bf I}_X {\bf e^L} 
= \left( 
    {\bf e^L} {\bf I}_X 
  \right)^T ,
\end{equation}
and
\begin{equation}
\frac{\delta L (x^\prime,y^\prime)}{\delta g(x,y)}
= \delta (x-x^\prime ) \delta (y-y^\prime )
- \delta (x-x^\prime ) e^{L(x,y)},
\label{dldg}
\quad
\frac{\delta L}{\delta g}
= {\bf I} - {\bf I}_X {\bf e^L} 
,
\end{equation}
where $\frac{\delta L}{\delta g}$ denotes a matrix,
and the superscript ${}^T$ the transpose of a matrix.
We also note
that despite ${\bf I}_X =  {\bf I}_X^T$
\be
{\bf I}_X {\bf e^L} \ne {\bf e^L} {\bf I}_X 
= ({\bf I}_X {\bf e^L})^T
,
\ee 
is not symmetric
because ${\bf e^L}$ depends on $y$ and 
does not commute with the non--diagonal 
${\bf I}_X$.
Hence, we obtain 
the stationarity equation
of functional $E_g$ written in terms of $L(g)$
again Eq.\ (\ref{LwithLambda})
\begin{equation}
0 
= -\left( \frac{\delta L}{\delta g} \right)^T 
          \frac{\delta E_g}{\delta L}
= G_L - {\bf e^L} \Lambda_X
= \left( {\bf I} - {\bf e^L} {\bf I}_X \right)  \left(N- {{\bf K}} L \right)
.
\end{equation}
Here $G_L = N - {{\bf K}} L = -\delta E_g/\delta L$
is the $L$--gradient of $-E_g$.
Referring to the discussion following Eq.\ (\ref{LwithLambda})
we note, however, that solving for $g$ instead for $L$
no unnormalized solutions
fulfilling $N={{\bf K}}L$ are possible.

In case $\ln Z_X$ is in the zero space of ${{\bf K} }$
the functional $E_g$ corresponds to 
a Gaussian prior in $g$ alone.
Alternatively, we may also directly
consider a Gaussian prior in $g$
\begin{equation}
\tilde E_{g} = 
-\Big(N , \, g -\ln Z_X\, \Big)
+\frac{1}{2}\Big( \,g\,, \,{{\bf K}}\,g\,\Big)
,
\end{equation}
with stationarity equation
\begin{equation}
0=
N -{\bf K} g - {\bf e^L}N_X 
.
\label{stat-eq-data-sp}
\end{equation}
Notice, that expressing the density estimation problem in terms of $g$,
nonlocal normalization terms have not disappeared but are 
part of the likelihood term.
As it is typical for density estimation problems,
the solution $g$ can be calculated in 
$X$--data space, i.e., in the space 
defined by the $x_i$ of the training data.
This still allows to use a Gaussian prior structure 
with respect to the $x$--dependency
which is especially useful for classification problems
\cite{Williams-Barber-1998}.

\subsubsection{The Hessians ${\bf H}_L$, ${\bf H}_g$}
\label{Hessians-L}

The Hessian ${\bf H}_L$ of $-E_L$
is defined as the matrix or operator of second derivatives
\begin{equation}
{\bf H}_L (L) (x,y; x^\prime y^\prime)
=
\frac{\delta^2 (-E_L)}
{\delta L (x,y)\delta L(x^\prime ,y^\prime )}\Bigg|_{L}.
\end{equation}
For functional (\ref{L-functional1})
and fixed $\Lambda_X$ we find the Hessian by taking the derivative
of the gradient in (\ref{stat-eq-bra-ket}) with respect to $L$ again.
This gives
\begin{equation}
{\bf H}_L(L) (x,y; x^\prime y^\prime )
=-{{\bf K}}(x,y;x^\prime y^\prime ) 
\label{Hessian-L}
-\delta (x-x^\prime )\delta (y-y^\prime ) \Lambda_X (x )e^ {L(x,y)}
\end{equation}
or
\begin{equation}
{\bf H}_L
= 
-{{\bf K}} 
-  {\bf \Lambda}_X {\bf e^L}.
\label{H_L}
\end{equation}
The addition of the diagonal matrix 
${\bf \Lambda}_X {\bf e^L}$ = ${\bf e^L} {\bf \Lambda}_X$ 
can result in a negative definite
${\bf H}$ even if ${{\bf K}}$ has zero modes.
like in the case where
${{\bf K}}$ is a differential operator
with periodic boundary conditions.
Note, however, that 
${\bf \Lambda}_X {\bf e^L}$
is diagonal and therefore symmetric,
but not necessarily positive definite, because $\Lambda_X(x)$ can be negative
for some $x$. Depending on the sign of $\Lambda_X (x)$
the normalization condition $Z_X(x)=1$ for  that $x$ can be replaced
by the inequality $Z_X(x)\le 1$ or $Z_X(x)\ge 1$.
Including the $L$--dependence of $\Lambda_X$ and with
\begin{equation}
\frac{\delta e^{L(x^\prime,y^\prime)} }{\delta g(x,y)}
= \delta (x-x^\prime) \delta (y-y^\prime) e^{{L}(x,y)}
-
\delta (x-x^\prime) e^{L(x,y)} e^{L(x^\prime,y^\prime )}
,
\end{equation}
i.e.,
\begin{equation}
\frac{\delta e^{L} }{\delta g}
= \left( {\bf I} - {\bf e^L}\,  {\bf I}_X \right) {\bf e^L}
= {\bf e^L} - {\bf e^L}\,  {\bf I}_X {\bf e^L},
\end{equation}
we find, written in terms of $L$,
\[
{\bf H}_g (L)(x,y;x^\prime,y^\prime ) 
= \frac{\delta^2 (-E_g)}
   {\delta g (x,y)\delta g(x^\prime ,y^\prime )}\Bigg|_{L}
\]
\[
= \!\! \int \!\! dx^{\prime\prime} dy^{\prime\prime} \!
\left(
  \frac{\delta^2 (-E_g)}
   {\delta L (x,y)\delta L(x^{\prime\prime} ,y^{\prime\prime} )}
  \frac{\delta L(x^{\prime\prime},y^{\prime\prime} )}
   {\delta g (x^\prime,y^\prime) }
+ \frac{\delta(-E_g)}
   {\delta L(x^{\prime\prime} ,y^{\prime\prime} )}
  \frac{\delta^2 L(x^{\prime\prime},y^{\prime\prime} )}
   {\delta g (x,y)\delta g (x^\prime,y^\prime) }
\right)\!\Bigg|_{L}
\]
\bea
&=&
-{{\bf K}}(x,y;x^\prime,y^\prime ) 
- e^{L(x^\prime, y^\prime)}e^{L(x, y)} 
  \int \!dy^{\prime\prime} dy^{\prime\prime\prime} 
  {{\bf K}}( x^\prime,y^{\prime\prime} ; x,y^{\prime\prime\prime} ) 
\nonumber\\&&
+ e^{L(x^\prime, y^\prime)} 
  \int \!dy^{\prime\prime} {{\bf K}}(x^\prime,y^{\prime\prime};x,y) 
+e^{L(x,y)} 
  \int \!dy^{\prime\prime} {{\bf K}}(x^\prime,y^\prime;x,y^{\prime\prime})
\nonumber\\&&
-\delta (x-x^\prime) \delta (y-y^\prime) e^{L(x,y)} 
\left(  N_X(x ) -\int\!dy^{\prime \prime} ({{\bf K}} L)(x,y^{\prime\prime})   \right)
\nonumber\\&&
+
\delta(x- x^\prime)
e^{L(x,y)} e^{L(x^\prime,y^\prime)} 
\left( N_X(x)-\int \!dy^{\prime \prime} ({{\bf K}} L)(x,y^{\prime\prime})\right).
\eea
The last term, diagonal in $X$, has dyadic structure in $Y$,
and therefore for fixed $x$ at most one non--zero eigenvalue.
In matrix notation the Hessian becomes
\bea
{\bf H}_g 
&=& 
- \left( {\bf I} - {\bf e^L} {\bf I}_X \right)
{{\bf K}} \left( {\bf I} - {\bf I}_X {\bf e^L} \right)
- \left( {\bf I} - {\bf e^L} {\bf I}_X \right) 
{\bf \Lambda}_X {\bf e^L} 
\nonumber\\&=& 
- \left( {\bf I} - {\bf P} {\bf I}_X \right)
\left[
 {{\bf K}} \left( {\bf I} - {\bf I}_X {\bf P} \right)
+ {\bf \Lambda}_X {\bf P} 
\right]
,
\label{L-Hessian}
\eea
the second line written in terms of the probability matrix.
The expression is symmetric under
$x\leftrightarrow x^\prime$,$y\leftrightarrow y^\prime$,
as it must be for a Hessian and as can be verified using the symmetry of 
${{\bf K}} = {{\bf K}}^T$ and the fact that
${\bf \Lambda}_X$ and ${\bf I}_X$ commute, i.e.,
$[{\bf \Lambda}_X , {\bf I}_X] = 0$.
Because functional $E_g$ is invariant under a shift transformation,
$g(x,y) \rightarrow g^\prime(x,y) + c(x)$, the Hessian has 
a space of zero modes with the dimension of $X$.
Indeed, 
any $y$--independent function
(which can have finite 
$L^1$--norm only in finite $Y$--spaces)
is a left eigenvector of
$\left( {\bf I} - {\bf e^L} {\bf I}_X \right)$ with eigenvalue zero.
The zero mode can be removed by 
projecting out the zero modes and using 
where necessary instead of the inverse 
a pseudo inverse of ${\bf H}$, 
for example obtained by singular value decomposition,
or by including additional conditions on $g$
like for example boundary conditions.

\subsection{Gaussian prior factor for probabilities}
\label{gaussP}
\subsubsection{Lagrange multipliers: Error functional $E_P$}
\label{error-e-p}

We write $P(x,y)=p(y|x,{h})$ for 
the probability of $y$ conditioned on $x$ and ${h}$.
We consider now a regularizing term which is quadratic in $P$ instead of $L$.
This corresponds to a factor 
within the posterior probability (the specific prior) 
which is Gaussian with respect to $P$.
\begin{equation}
p(\!{h}|f) \!=\!\!
e^{
\sum_i \!\ln P_i(x_i,y_i) 
- \!\frac{1}{2}\!\!\int\!\! dx dy dx^\prime dy^\prime  P(x,y) 
{{\bf K}}(x,y;x^\prime,y^\prime ) P(x^\prime ,y^\prime)
+ \!\int \!\!dx\,  \Lambda_X (x)  \left(  1 - \int\!dy\,P(x,y) \right) 
+ \tilde c,
}
\end{equation}
or written in terms of $L=\ln P$ for comparison,
\begin{equation}
p(\!{h}|f) \!
=\!\!
e^{
\sum_i \!L_i(x_i,y_i) 
- \!\frac{1}{2}\!\!\int\!\! dx dy dx^\prime dy^\prime e^{L(x,y)} 
{{\bf K}}(x,y;x^\prime,y^\prime ) e^{L(x^\prime ,y^\prime)}
+ \!\int \!\!dx\,  \Lambda_X (x)  \left( 1 - \int\!dy\,e^{L(x,y)} \right) 
+ \tilde c.
}
\end{equation}
Hence, the error functional is
\begin{equation}
E_P=\beta E_{\rm comb}=-(\ln P,N) + \frac{1}{2} (P,{{\bf K}}\,P)  
+ (\,P-\delta(y)\,, \Lambda_X ).
\label{P-functional}
\end{equation}
In particular, the choice ${{\bf K}}$ = $\frac{\lambda}{2}{\bf I}$, i.e.,
\be
\frac{\lambda}{2}(P,\,P) = \frac{\lambda}{2}||P||^2,
\label{quadP}
\ee
can be interpreted as a smoothness prior with respect to the
distribution function of $P$ (see Section \ref{other-Gaussian}).

In functional (\ref{P-functional}) we have only implemented
the normalization condition for $P$ by a Lagrange multiplier
and not the non--negativity constraint.
This is sufficient if $P(x,y)>0$ (i.e., $P(x,y)$ not equal zero)
at the stationary point because then 
$P(x,y)>0$ holds also in some neighborhood 
and there are no components of the gradient pointing
into regions with negative probabilities.
In that case the non--negativity
constraint is not active at the stationarity point.
A typical smoothness constraint, for example, together
with positive probability at data points
result in positive probabilities everywhere
where not set to zero explicitly by boundary conditions.
If, however, the stationary point has locations
with $P(x,y)$ = $0$ at non--boundary points, then
the component of the gradient 
pointing in the region with negative probabilities
has to be projected out by introducing 
Lagrange parameters for each $P(x,y)$.
This may happen, for example, if the regularizer
rewards oscillatory behavior.

The stationarity equation for $E_P$ is
\begin{equation}
0 =  {\bf P}^{-1} N -{{\bf K}} P -  \Lambda_X  
,
\label{stat-eq-P}
\end{equation}
with the diagonal matrix 
${\bf P} (x^\prime,y^\prime;x,y)$ =
$\delta (x-x^\prime)\delta (y-y^\prime) P(x,y)$,
or multiplied by ${\bf P}$
\begin{equation}
0 =  N - {\bf P}{{\bf K}} P -  {\bf P} \Lambda_X  
.
\label{stat-eq-PP}
\end{equation}
Probabilities $P(x,y)$
are unequal zero at observed data points $(x_i,y_i)$
so ${\bf P}^{-1} N$ is well defined.

Combining the normalization condition Eq.\ (\ref{lambda-norm})
for $\Lambda_X(x)\ne 0$
with Eq.\ (\ref{stat-eq-P}) or (\ref{stat-eq-PP})
the Lagrange multiplier function $\Lambda_X$
is found as
\begin{equation}
\Lambda_X
= {\bf I}_X \left( N - {\bf P}{{\bf K}} P \right) 
= N_X - {\bf I}_X  {\bf P}{{\bf K}} P,
\label{lambdaP}
\end{equation}
where
\[
{\bf I}_X {\bf P}{{\bf K}}P (x,y)
= \int \!dy^\prime dx^{\prime\prime} dy^{\prime\prime} 
\, P(x,y^\prime )
{{\bf K}}(x,y^{\prime}; x^{\prime\prime} ,y^{\prime\prime})
P(x^{\prime\prime},y^{\prime\prime}).
\]
Eliminating $\Lambda_X$ in Eq.\ (\ref{stat-eq-P})
by using Eq.\ (\ref{lambdaP})
gives finally
\be
0 = ({\bf I} - {\bf I}_X {\bf P})
({\bf P}^{-1}N - {{\bf K}} P)
,
\ee
or for Eq.\ (\ref{stat-eq-PP})
\be
0 = ({\bf I} - {\bf P}{\bf I}_X)
(N - {\bf P}{{\bf K}} P)
.
\ee
For similar reasons as has been discussed
for Eq.\ (\ref{LwithLambda})
unnormalized solutions fulfilling $N-{\bf P}{{\bf K}} P$
are possible.
Defining 
\begin{equation}
T_P =  {\bf P}^{-1} N -  \Lambda_X
    =  {\bf P}^{-1} N- N_X - {\bf I}_X {\bf P}{{\bf K}} P  ,
\end{equation}
the stationarity equation can be written 
analogously to Eq.\ (\ref{OL=T}) as
\begin{equation}
{{\bf K}} P = T_P,
\label{OP=T}
\end{equation}
with $T_P = T_P(P)$,
suggesting for existing 
${{\bf K}}^{-1}$ an iteration
\be
P^{i+1} = {{\bf K}}^{-1} T_P(P^i)
,
\ee
starting from some initial guess $P^0$.

\subsubsection{Normalization by parameterization: Error functional $E_z$}

Again, normalization can also be ensured by parameterization of $P$
and solving for unnormalized probabilities $z$, i.e.,
\begin{equation}
P(x,y)  = \frac{z(x,y)}{\int \!dy\, z(x,y)},
\quad
P = \frac{z}{Z_X}
.
\end{equation}
The corresponding functional reads
\begin{equation}
E_z = - \left(N , \ln \frac{z}{Z_X}\right)
+ \frac{1}{2}\left( \frac{z}{Z_X}, {{\bf K}}\,\frac{z}{Z_X}\right).
\label{E_z}
\end{equation}
We have
\begin{equation}
\frac{\delta z}{\delta z} =  {\bf I},
\quad
\frac{\delta Z_X}{\delta z} = {\bf I}_X,
\quad
\frac{\delta \ln z}{\delta z} 
=  {\bf z}^{-1} = ({\bf Z}_X {\bf P})^{-1},
\quad
\frac{\delta \ln Z_X}{\delta z} 
= {\bf Z}_X^{-1} \, {\bf I}_X ,
\end{equation}
with diagonal matrix ${\bf z}$
built analogous to ${\bf P}$ and ${\bf Z}_X$,
and 
\begin{equation}
\frac{\delta P }{\delta z} 
=\frac{\delta (z/Z_X) }{\delta z} 
= {\bf Z}_X^{-1} \left( {\bf I} -  {\bf P} {\bf I}_X \right),
\quad
\frac{\delta \ln P }{\delta z} 
= {\bf Z}_X^{-1} \left( {\bf P}^{-1} - {\bf I}_X \right),
\end{equation}
\begin{equation}
\frac{\delta Z_X^{-1} }{\delta z} 
= - {\bf Z}_X^{-2} {\bf I}_X,
\quad
\frac{\delta P^{-1} }{\delta z} 
= - {\bf P}^{-2} \, {\bf Z}_X^{-1} \left( {\bf I}  - {\bf P} {\bf I}_X  \right).
\end{equation}
The diagonal matrices $[{\bf Z}_X , {\bf P}  ] = 0$ commute,
as well as $[{\bf Z}_X , {\bf I}_X] = 0$,
but $[ {\bf P}  , {\bf I}_X] \ne 0$.
Setting the gradient to zero and using 
\be
\left( {\bf I} - {\bf P} {\bf I}_X  \right)^T
= \left( {\bf I} - {\bf I}_X {\bf P} \right)
,
\ee
we find
\[
0 
= -\left( \frac{\delta P}{\delta z} \right)^T  
\frac{\delta E_z}{\delta P} 
\]
\[
={\bf Z}_X^{-1} \left[
\left( {\bf P}^{-1} - {\bf I}_X \right) N
- \left({\bf I} - {\bf I}_X {\bf P} \right) {{\bf K}} P
\right]
\]
\[
= {\bf Z}_X^{-1} 
\left( {\bf I}  - {\bf I}_X {\bf P} \right)
\left( {\bf P}^{-1} N  - {{\bf K}} P \right)
\]
\be
= {\bf Z}_X^{-1} 
\left( {\bf I}  - {\bf I}_X {\bf P} \right) G_P
= {\bf Z}_X^{-1} \left( G_P - \Lambda_X \right)
= \left( {\bf G}_P - {\bf \Lambda}_X \right) {Z}_X^{-1} ,
\label{grad_z}
\ee
with $P$--gradient 
$G_P = {\bf P}^{-1} N - {{\bf K}} P$ = $-\delta E_z/\delta P$
of $-E_z$
and ${\bf G}_P$ the corresponding diagonal matrix.
Multiplied by ${\bf Z}_X$ this gives the stationarity equation (\ref{OP=T}).

\subsubsection{The Hessians ${\bf H}_P$, ${\bf H}_z$}
\label{Hessians-P}

We now calculate the Hessian of the functional $-E_P$.
For fixed $\Lambda_X$ one finds the Hessian 
by differentiating again the gradient (\ref{stat-eq-P}) 
of $-E_P$
\begin{equation}
{\bf H}_P(P) (x,y; x^\prime y^\prime )
=  
-{{\bf K}}(x^\prime y^\prime ;x,y) 
-\delta (x-x^\prime )\delta (y-y^\prime ) 
\sum_i \frac{\delta (x-x_i )\delta (y-y_i)}{P^2(x,y)},
\label{Hessian-P}
\end{equation}
i.e., 
\begin{equation}
{\bf H}_P 
= -{{\bf K}} - {\bf P}^{-2} {\bf N}.
\label{H_P}
\end{equation}
Here the diagonal matrix 
${\bf P}^{-2} {\bf N}$
is non--zero only at data points.

Including the dependence of $\Lambda_X$ on $P$ 
one obtains for the Hessian of $-E_z$ in (\ref{E_z}) 
by calculating the derivative of the gradient in (\ref{grad_z})
\[
{\bf H}_z (x,y;x^\prime,y\prime) =
-\frac{1}{Z_X(x)} \Big[
{{\bf K}} (x,y;x^\prime,y\prime) 
\]\[
- \int\! dy^{\prime\prime} 
  \Big(
  p(x,y^{\prime\prime}){{\bf K}}(x,y^{\prime\prime};x^\prime,y^\prime) 
+  {{\bf K}}(x,y;x^\prime,y^{\prime\prime}) p(x^\prime,y^{\prime\prime}) 
\Big)
\]
\[
+ \int\! dy^{\prime\prime} dy^{\prime\prime\prime}
  p(x,y^{\prime\prime}) 
  {{\bf K}}(x,y^{\prime\prime};x^\prime,y^{\prime\prime\prime}) 
  p(x^\prime,y^{\prime\prime\prime}) 
\]
\[
+ \delta (x-x^\prime) \delta (y-y^\prime)
 \sum_i \frac{ \delta (x-x_i)\delta (y-y_i) }{ p^2(x,y) }
- \delta (x-x^\prime)  \sum_i \delta (x-x_i)
\]
\[
-\delta (x-x^\prime) 
  \int\! dx^{\prime\prime} dy^{\prime\prime} 
\Big( 
 {{\bf K}}(x,y;x^{\prime\prime},y^{\prime\prime}) 
  p(x^{\prime\prime},y^{\prime\prime}) 
+  p(x^{\prime\prime},y^{\prime\prime}) 
  {{\bf K}}(x^{\prime\prime},y^{\prime\prime};x^\prime,y^\prime) 
\Big)
\]
\be
+ 2\, \delta (x-x^\prime) 
  \int\! dy^{\prime\prime} dx^{\prime\prime\prime} dy^{\prime\prime\prime} 
  p(x,y^{\prime\prime}) 
  {{\bf K}}(x,y^{\prime\prime};x^{\prime\prime\prime},y^{\prime\prime\prime}) 
  p(x^{\prime\prime\prime},y^{\prime\prime\prime}) 
\Big] \frac{1}{ Z_X (x^\prime) }, 
\ee
i.e.,
\bea
{\bf H}_z &=&
 {\bf Z}_X^{-1} 
 \left( {\bf I}-{\bf I}_X {\bf P} \right) 
 \left(-{{\bf K}}-{\bf P}^{-2} {\bf N} \right)
 \left( {\bf I}- {\bf P} {\bf I}_X \right)
{\bf Z}_X^{-1}
\nonumber\\
&&- {\bf Z}_X^{-1} \left( {\bf I}_X
  \left( {\bf G}_P -   {\bf \Lambda}_X \right)
+ 
  \left( {\bf G}_P -   {\bf \Lambda}_X \right)
{\bf I}_X \right) {\bf Z}_X^{-1},
\label{hessz}
\\&=&
- {\bf Z}_X^{-1} \Big[
\left( {\bf I}-{\bf I}_X {\bf P} \right) 
 {{\bf K}}
 \left( {\bf I} - {\bf P} {\bf I}_X \right)
+{\bf P}^{-2} {\bf N} 
\nonumber\\&&
- {\bf I}_X {\bf P}^{-1} {\bf N}
- {\bf N} {\bf P}^{-1} {\bf I}_X 
+{\bf I}_X {\bf N} {\bf I}_X
\nonumber
\\
&&+ {\bf I}_X  {\bf G}_P + {\bf G}_P  {\bf I}_X 
- 2\,   {\bf I}_X  {\bf \Lambda}_X 
\Big] {\bf Z}_X^{-1}.
\eea
Here we used
$[{\bf \Lambda}_X,{\bf I}_X]$ = 0.
It follows from the normalization $\int \! dy \, p(x,y) = 1$
that any $y$--independent function 
is right eigenvector of $\left( {\bf I}-{\bf I}_X {\bf P} \right)$
with zero eigenvalue.
Because $\Lambda_X$ = ${\bf I}_X {\bf P} G_P$
this factor or its transpose is 
also contained in the second line of
Eq.\ (\ref{hessz}),  
which means that ${\bf H}_z$ has a zero mode.
Indeed, functional $E_z$ is invariant under multiplication
of $z$ with a $y$--independent factor. The zero modes can be projected out
or removed by including additional conditions, e.g.\ by 
fixing one value of $z$ for every $x$.

\subsection{General Gaussian prior factors}
\label{other-Gaussian}
\subsubsection{The general case}
\label{general-Gaussian}

In the previous sections we studied
priors consisting of a factor (the specific prior) which was Gaussian 
with respect to $P$ or $L= \ln P$
and additional normalization (and non--negativity) conditions.
In this section we consider the situation where the 
probability $p(y|x,h)$
is expressed in terms of a function $\phi(x,y)$.
That means, we assume 
a, possibly non--linear, operator
$P$ = $P(\phi)$ which maps the function $\phi$ 
to a probability.
We can then formulate a learning problem in terms 
of the function $\phi$, meaning that
$\phi$ now represents the hidden variables
or unknown state of Nature $h$.\footnote{
Besides $\phi$  also the hyperparameters 
discussed in Chapter \ref{hyperparameters}
belong to the  hidden variables $h$.}
Consider the case of a specific prior
which is Gaussian in $\phi$,
i.e., which has a log--probability quadratic in $\phi$
\begin{equation}
-\frac{1}{2} (\,\phi\,, \,{{\bf K}}\,\phi\, )
.
\end{equation}
This means we are lead to error functionals of the form
\begin{equation}
E_\phi =
-(\,\ln P(\phi)\,,\, N \,)
+\frac{1}{2} (\,\phi\,,\,{{\bf K}}\,\phi \,)
+(\,P(\phi)\,,\,\Lambda_X\,),
\label{phifunctional}
\end{equation}
where we have skipped the $\phi$--independent part of the $\Lambda_X$--terms.
In general cases also  
the non--negativity constraint has to be implemented.

To express
the functional derivative of functional (\ref{phifunctional})
with respect to $\phi$ we define
besides the diagonal matrix ${\bf P}$ = ${\bf P}(\phi)$ 
the Jacobian, i.e., the matrix of derivatives
${\bf P}^\prime$ 
= ${\bf P}^\prime(\phi)$
with matrix elements 
\begin{equation}
{\bf P}^\prime (x,y;x^\prime,y^\prime;\phi)=
\frac{\delta P (x^\prime,y^\prime;\phi)}{\delta \phi (x,y)}.
\end{equation}
The matrix ${\bf P}^\prime$ is diagonal
for point--wise transformations, i.e., for
$P(x,y;\phi)$ = $P(\,\phi(x,y)\,)$.
In such cases we use $P^\prime$ to denote the vector of diagonal elements
of ${\bf P}^\prime$.
An example is the previously discussed transformation $L=\ln P$
for which ${\bf P}^\prime$ = ${\bf P}$.
The stationarity equation for functional (\ref{phifunctional}) becomes
\begin{equation}
0 =
{\bf P}^\prime (\phi) {\bf P}^{-1}(\phi) N  
- {{\bf K}}\phi
-{\bf P}^\prime (\phi) \Lambda_X
,
\label{phistat1}
\end{equation}
and for existing
${\bf P} {{\bf P}^\prime}^{-1}$
=$({\bf P}^\prime {\bf P}^{-1})^{-1}$
(for nonexisting inverse 
see Section \ref{general-variational}),
\begin{equation}
0 =
N  - {\bf P} {{\bf P}^\prime}^{-1}{{\bf K}}\, \phi
- {\bf P} \Lambda_X 
.
\label{phistat2}
\end{equation}
From Eq.\ (\ref{phistat2})
the Lagrange multiplier function can be found by integration,
using the normalization condition 
${\bf I}_X P $ = $I$,
in the form 
${\bf I}_X {\bf P} \Lambda_X$ = $\Lambda_X$
for $\Lambda_X(x) \ne 0$.
Thus, multiplying Eq.\ (\ref{phistat2}) by ${\bf I}_X$
yields
\begin{equation}
\Lambda_X = 
{\bf I}_X \left( N - {\bf P} {{\bf P}^\prime}^{-1} {{\bf K}}\, \phi \right)
=
N_X - {\bf I}_X 
{\bf P} {{\bf P}^\prime}^{-1} {{\bf K}}\, \phi .
\label{philag}
\end{equation}
$\Lambda_X$ is now eliminated by
inserting Eq.\ (\ref{philag}) into Eq.\ (\ref{phistat2}) 
\begin{equation}
0 =
\left( {\bf I} - {\bf P}{\bf I}_X\right)
\left( N - {\bf P} {{\bf P}^\prime}^{-1}{{\bf K}}\, \phi \right).
\label{lambdain}
\end{equation}
A simple iteration procedure, 
provided ${{\bf K}}^{-1}$ exists,
is suggested by writing
Eq.\ (\ref{phistat1}) in the form
\be
{{\bf K}} \phi = T_\phi 
,\quad
\phi^{i+1} = 
{{\bf K}}^{-1} T_\phi (\phi^i) 
,
\label{Ophi=Tphi}
\ee
with 
\be
T_\phi (\phi)
= {\bf P}^\prime {\bf P}^{-1}N  
-{\bf P}^\prime \Lambda_X 
.
\ee

Table \ref{var-tab}
lists constraints to be implemented explicitly
for some choices of $\phi$.

\begin{table}
\begin{center}
\begin{tabular}{|c|c|c|c|}
\hline
$\phi$  & $P(\phi)$ & \multicolumn{2}{c|}{constraints} \\
\hline
\hline\rule[-3mm]{0mm}{9mm}
$P(x,y)$  & $P=P$ & norm & non--negativity \\
\hline\rule[-3mm]{0mm}{9mm}
$z(x,y)$  & $P=z/\int\!z\,dy$ & --- & non--negativity \\
\hline\rule[-3mm]{0mm}{9mm}
$L(x,y)=\ln P$  & $P=e^L$ & norm & --- \\
\hline\rule[-3mm]{0mm}{9mm}
$g(x,y)$  & $P={e^g}/{\int\!e^g\,dy}$ & --- & --- \\
\hline\rule[-3mm]{0mm}{9mm}
$\Phi=\int^y dy^\prime \,P$  & $P={d\Phi}/{dy}$ & boundary & monotony \\
\hline
\end{tabular}
\end{center}
\caption{Constraints for specific choices of $\phi$}
\label{var-tab}
\end{table}

\subsubsection{Example: Square root of $P$}
\label{Square-root}

We already discussed the cases
$\phi = \ln P$ with $P^\prime = P = e^L$, $P/P^\prime = 1$
and $\phi = P$ with $P^\prime = 1$, $P/P^\prime = P$.
The choice $\phi = \sqrt{P}$
yields the common $L_2$--normalization condition over $y$
\begin{equation}
1 =\int \!dy\, \phi^2(x,y), \quad \forall x\in X,
\end{equation}
which is quadratic in $\phi$,
and $P=\phi^2$, $P^\prime = 2 \phi$, $P/P^\prime = \phi/2$.
For real $\phi$ 
the non--negativity condition $P\ge 0$ is automatically 
satisfied \cite{Good-Gaskins-1971,Silverman-1986}.

For $\phi$ = $\sqrt{P}$ 
and a negative Laplacian inverse covariance ${{\bf K}}$ = $-\Delta$,
one can relate the corresponding Gaussian prior 
to the {\it Fisher information}
\cite{Cox-Hinkley-1974,Silverman-1986,Schervish-1995}.
Consider, for example, a problem with fixed $x$ 
(so $x$ can be skipped from the notation and one can write $P(y)$)
and a $d_y$--dimensional $y$.
Then one has, assuming the necessary differentiability conditions
and vanishing boundary terms, 
\begin{equation}
(\,\phi \,,\, {{\bf K}} \, \phi \,)
=-(\,\phi \,,\, \Delta \, \phi \,)
= \int \!dy\,
\sum_k^{d_y} \left| \frac{\partial \phi}{\partial y_k}\right|^2 
\end{equation}
\begin{equation}
=  \sum_k^{d_y} \int \frac{dy }{ 4 P(y) }
   \left( \frac{ \partial  P(y) }{ \partial y_k } \right)^2 
= \frac{1}{4} \sum_k^{d_y} I^F_k (0)
,
\end{equation}
where $I^F_k (0)$
is the Fisher information,
defined as 
\begin{equation}
I^F_k (y_0) 
= \int \! dy 
     \frac{ 
  \left| \frac{\partial P(y - y^0)}{\partial y^0}\right|^2}{ P(y-y^0) }
= \int \! dy 
     \left|\frac{ \partial \ln P(y - y^0)  }{\partial y^0_k}\right|^2
     P(y-y^0_k),
\end{equation}
for the family $P(\cdot-y^0)$ with location parameter 
vector $y^0$.

A connection to quantum mechanics can be found 
considering the training data free case
\begin{equation}
E_{\phi} = \frac{1}{2}(\,\phi,\,\,{{\bf K}}\,\phi\,) + (\Lambda_X,\,\phi)
,
\label{ritz}
\end{equation}
has the homogeneous stationarity equation
\begin{equation}
{{\bf K}} \, \phi = -2 \Phi \Lambda_X
.
\end{equation}
For $x$--independent $\Lambda_X$
this is an eigenvalue equation.
Examples include the quantum mechanical
Schr\"odinger equation
where ${{\bf K}}$ corresponds to the system Hamiltonian          
and
\begin{equation}
-2 \Lambda_X
=
\frac{( \phi ,\, {{\bf K}} \, \phi )}{(\phi ,\,\phi)},
\label{Ritz}
\end{equation}
to its ground state energy.
In quantum mechanics Eq. (\ref{Ritz}) is the basis 
for variational methods (see Section \ref{variational})
to obtain approximate solutions for ground state energies
\cite{Eisenberg-Greiner-1972,Ring-Schuck-1980,Blaizot-Ripka-1986}.

Similarly, one can take
$\phi = \sqrt{-(L-L_{\rm max})}$
for $L$ bounded from above by $L_{\rm max}$
with the normalization
\begin{equation}
1 =\int \!dy\, e^{-\phi^2(x,y)+L_{\rm max}}, \quad \forall x\in X,
\end{equation}
and  $P = e^{-\phi^2 + L_{\rm max}}$,
$P^\prime = - 2 \phi P$,
$P/P^\prime$ = $-1/(2\phi )$.

\subsubsection{Example: Distribution functions}

Instead in terms of the probability density function,
one can formulate the prior in terms of its integral,
the distribution function. The density $P$ is then recovered
from the distribution function $\phi$ by differentiation, 
\begin{equation}
P(\phi) = \prod_k^{d_y} \frac{\partial \phi}{\partial y_k} 
         = \prod_k^{d_y}  \nabla_{y_k} \phi
         = \bigotimes_k^{d_y}  {\bf R}_k^{-1} \phi.
         = {\bf R}^{-1} \phi,
\label{denstrans}
\end{equation} 
resulting in a non--diagonal ${\bf P}^\prime$.
The inverse of the derivative operator ${\bf R}^{-1}$ is
the integration operator ${\bf R}$ = $\bigotimes_k^{d_y} {\bf R}_k P$
with matrix elements 
\be
{\bf R} (x,y;x^\prime,y^\prime)
= \delta (x-x^\prime) \theta (y-y^\prime),
\ee 
i.e.,
\be
{\bf R}_k (x,y;x^\prime,y^\prime)
= \delta (x-x^\prime) \prod_{l\ne k}\delta (y_l-y_l^\prime) 
\theta (y_k-y^\prime_k).
\ee
Thus, (\ref{denstrans}) 
corresponds to the transformation of ($x$--conditioned) 
density functions $P$ in ($x$--conditioned) 
distribution functions 
$\phi$ = ${\bf R} P$, i.e.,
$\phi (x,y)$ = $\int_{-\infty}^y P(x,y^\prime ) dy^\prime$.
Because ${\bf R^T} {{\bf K}} {\bf R}$ is (semi)--positive definite
if ${{\bf K}}$ is, 
a specific prior which is Gaussian in the distribution function $\phi $ 
is also Gaussian in the density $P$.
${\bf P}^\prime$ becomes
\begin{equation}
{\bf P}^\prime (x,y;x^\prime,y^\prime )
= \frac{\delta 
  \left(\prod_k^{d_y}\nabla_{y_{k^\prime}}\phi (x^\prime,y^\prime \right)}
 {\delta \phi (x,y)}
= \delta (x-x^\prime ) \prod_k^{d_y}\delta^\prime (y_k-y_k^\prime ).
\end{equation}
Here the derivative of the $\delta$--function is defined
by formal partial integration
\begin{equation}
\int_{-\infty}^\infty \!dy^\prime\, f(y^\prime) \delta^\prime (y-y^\prime )
=
f(y^\prime ) \delta(y^\prime -y)|_{-\infty}^\infty 
- f^\prime (y)
.
\end{equation}
Fixing $\phi(x,-\infty) = 0$
the variational derivative $\delta/( \delta \phi (x, -\infty) )$ 
is not needed.
The normalization condition for $P$
becomes 
for the distribution function $\phi$ = ${\bf R}P$
the boundary condition $\phi(x,\infty) = 1$, $\forall x\in X$.
The non--negativity condition for $P$ corresponds
to the monotonicity condition
$\phi(x,y) \ge \phi(x,y^\prime)$,
$\forall y\ge y^\prime$,
$\forall x\in X$
and to $\phi(x,-\infty) \ge 0$, $\forall x\in X$.

\subsection{Covariances and invariances}
\subsubsection{Approximate invariance}

Prior terms can often be related to
the assumption of 
approximate invariances
or approximate symmetries.
A Laplacian smoothness functional, for example,
measures the deviation from translational symmetry 
under infinitesimal translations.

Consider for example a linear mapping  
\be
\phi \rightarrow 
\tilde \phi = {\bf S} \phi 
,
\ee
given by the operator ${\bf S}$.
To compare $\phi$ with $\tilde \phi$
we define a (semi--)distance defined by choosing a
positive (semi--)definite ${\bf K}_S$,
and use as error measure
\be
\frac{1}{2}\Big( (\phi-{\bf S}\phi),\, {\bf K}_S(\phi-{\bf S}\phi) \Big) 
=\frac{1}{2}\Big( \phi,\,{\bf K} \phi\Big)
.
\ee
Here
\be
{\bf K} 
=
({\bf I}-{\bf S})^T {\bf K}_S({\bf I}-{\bf S})
\label{symK}
\ee
is positive semi--definite 
if ${\bf K}_S$ is.
Conversely, every positive semi--definite {\bf K}
can be written 
${\bf K}$ = ${\bf W}^T {\bf W}$
and is thus of form (\ref{symK}) with
${\bf S}$ = ${\bf I}-{\bf W}$ and ${\bf K}_S = {\bf I}$.
Including terms of the form of (\ref{symK})
in the error functional forces $\phi$
to be similar to $\tilde\phi$.

A special case are mappings 
leaving the norm invariant
\be
(\phi,\, \phi) 
= ({\bf S} \phi, {\bf S} \phi) 
= (\phi, \,{\bf S}^T {\bf S} \phi) 
.
\label{norm-inv}
\ee
For real $\phi$ and $\tilde \phi$
i.e., $({\bf S}\phi)$ 
= $({\bf S}\phi)^*$,
this requires
${\bf S}^T = {\bf S}^{-1}$
and ${\bf S}^* = {\bf S}$.
Thus, in that case
${\bf S}$ has to be an orthogonal matrix
$\in O(N)$ and can be written
\be
{\bf S}(\theta) 
= e^{\bf A}
= e^{\sum_i \theta_i  {\bf A}_i}
=
\sum_{k=0}^\infty \frac{1}{k!}\left(\sum_i \theta_i{\bf A}_i\right)^k
,
\label{reell-lie}
\ee
with antisymmetric
${\bf A} = - {\bf A}^T$
and real parameters $\theta_i$.
Selecting a set of 
(generators) ${\bf A}_i$
the matrices obtained be varying the parameters $\theta_i$
form a Lie group.
Up to first order the expansion of the exponential function reads
${\bf S} \approx  1+\sum_i\theta_i {\bf A}_i$.
Thus, we can define an error measure with respect to an
infinitesimal transformation by
\begin{equation}
\frac{1}{2}
\sum_i  \left(\frac{\phi - 
      (1 + \theta_i {\bf A}_i) \phi}{\theta_i},\,{\bf K}_S
      {\frac{\phi - (1 + \theta_i {\bf A}_i)\phi}{\theta_i}} \right) 
=
\frac{1}{2}(\phi,\, \sum_i {\bf A}_i^T  {\bf K}_S {\bf A}_i \phi )
\label{Lie-error}
.
\end{equation}

\subsubsection{Approximate symmetries}

Next we come to the special case of symmetries, i.e.,
invariance under under coordinate transformations.
Symmetry transformations ${\bf S}$ change the arguments of a function $\phi$.
For example for the translation of a function
$\phi(x)\rightarrow\tilde \phi(x)={\bf S} \phi(x) = \phi(x-c)$.
Therefore it is useful to see how 
${\bf S}$ acts on the arguments of a function.
Denoting the 
(possibly improper) eigenvectors
of the coordinate operator ${\bf x}$ with
eigenvalue $x$ by $(\cdot,\,x)$ = $|x)$, i.e.,
${\bf x} |x) = x |x)$,
function values can be expressed as scalar products, e.g.
$\phi(x)$ = $(x,\, \phi)$ for a function in $x$, 
or, in two variables, 
$\phi(x,y)$ = $(x\otimes y,\, \phi)$.
(Note that in this `eigenvalue' notation,
frequently used by physicists,
for example $2|x)\ne|2x)$.)
Thus, we see that the action of 
${\bf S}$ on some function $h(x)$
is equivalent to the action of 
${\bf S}^T$ ( = ${\bf S}^{-1}$ if orthogonal)
on $|x)$
\be
{\bf S} \phi(x)=(x,{\bf S} \phi)=({\bf S}^T x, \phi)
,
\ee
or for
$\phi(x,y)$
\be
{\bf S} \phi(x,y)
=\left( {\bf S}^T (x \otimes y), \, \phi \right)
.
\ee
Assuming ${\bf S }$ = ${\bf S}_x{\bf S}_y$
we may also split the action of ${\bf S}$,
\be
{\bf S} \phi(x,y)
=\left( ({\bf S}_x^T x) \otimes y, \, {\bf S}_y \phi \right)
.
\ee
This is convenient for example
for vector fields in physics
where $x$ and $\phi(\cdot,y)$  
form three dimensional vectors
with $y$ representing a linear combination of component labels of $\phi$.

Notice that, for a general operator ${\bf S}$, 
the transformed argument ${\bf S}|x)$
does not have to be an eigenvector 
of the coordinate operator ${\bf x}$ again.
In the general case ${\bf S}$ can map a specific $|x)$
to arbitrary vectors being linear combinations
of all $|x^\prime)$, i.e.,
${\bf S}|x)$ = $\int \!dx^\prime\, S(x,x^\prime) |x^\prime)$.
A general orthogonal ${\bf S}$
maps an orthonormal basis 
to another orthonormal basis.
Coordinate transformations, however, are represented by operators ${\bf S}$,
which map coordinate eigenvectors $|x)$ to other 
coordinate eigenvectors
$|\sigma (x))$. 
Hence, such coordinate transformations
${\bf S}$ just changes 
the argument $x$ of a function $\phi$ into $\sigma(x)$, i.e.,
\be
{\bf S} \phi(x) = \phi(\sigma (x))
,
\ee
with $\sigma (x)$ a permutation
or a one--to--one coordinate transformation.
Thus, even for an arbitrary nonlinear coordinate transformation
$\sigma$ the corresponding operator ${\bf S}$
in the space of $\phi$ is linear. 
(This is one of the reasons why linear functional analysis
is so useful.)

A special case are linear coordinate transformations
for which we can write
$\phi(x)\rightarrow\tilde \phi(x) = {\bf S} \phi (x) = \phi(Sx)$, 
with $S$ (in contrast to ${\bf S}$)
acting in the space of $x$.
An example of such $S$ are coordinate rotations 
which preserve the norm in $x$--space,
analogously to Eq.\ (\ref{norm-inv}) for $\phi$,
and form a Lie group 
$S(\theta)=e^{\sum_i\theta_i A_i}$ acting on coordinates,
analogously to Eq.\ (\ref{reell-lie}).

\subsubsection{Example: Infinitesimal translations}

A Laplacian smoothness prior, for example, 
can be related to 
an approximate symmetry 
under infinitesimal translations.
Consider the group of 
$d$--dimensional translations which is generated
by the gradient operator $\nabla$.
This can be verified by recalling the multidimensional Taylor formula 
for expansion of $\phi$ at $x$ 
\begin{equation}
{\bf S}(\theta) \phi(x) 
= e^{ \sum_i \theta_i \nabla_i } \phi(x)
= \sum_{k=0}^\infty 
\frac{\left(\sum_i \theta_i \nabla_i\right)^{k}}{k!} \phi(x)
= \phi(x+\theta).
\end{equation}
Up to first order 
${\bf S} \approx  1+\sum_i\theta_i \Delta_i$.
Hence, for infinitesimal translations,
the error measure of Eq.\ (\ref{Lie-error}) becomes
\begin{equation}
\frac{1}{2}\sum_i  \left(\frac{\phi - 
      (1 + \theta_i {\Delta}_i) \phi}{\theta_i},\,
      {\frac{\phi - (1 + \theta_i {\Delta}_i)\phi}{\theta_i}} \right) 
\!=\! \frac{1}{2}(\phi,\, \sum_i \nabla_i^T \nabla_i \phi )
\!=\!-\frac{1}{2}(\phi,\, \Delta  \phi )
.
\end{equation}
assuming vanishing boundary terms
and choosing ${\bf K}_S$ = ${\bf I}$.
This is the classical Laplacian smoothness term.

\subsubsection{Example: Approximate periodicity}

As another example, lets us discuss the implementation of 
approximate periodicity.
To measure the deviation from exact periodicity
let us define the difference operators
\bea
\nabla^{R}_\theta \phi(x)
&=&
\phi(x)-\phi(x+\theta),
\\
\nabla^{L}_\theta \phi(x)
&=&
\phi(x-\theta)-\phi(x).
\eea
For periodic boundary conditions
$(\nabla^{L}_\theta)^T$
=
$-\nabla^{R}_\theta$,
where $(\nabla^{L}_\theta)^T$ denotes the transpose of $\nabla^{L}_\theta$.
Hence, the operator,
\be
\Delta_\theta 
= \nabla^L_\theta\nabla^R_\theta
= -(\nabla^R_\theta)^T\nabla^R_\theta
,
\ee
defined similarly to the Laplacian,
is positive definite,
and a possible error term, 
enforcing approximate periodicity with period $\theta$,
is
\be
\frac{1}{2}(\nabla_R (\theta)\phi,\;\nabla_R (\theta) \phi)
=-\frac{1}{2}(\phi,\;\Delta_\theta \phi)
=\frac{1}{2}\int \!dx\; |\phi(x)-\phi(x+\theta)|^2
.
\label{periodic-error}
\ee
As every periodic function with $\phi(x)=\phi(x+\theta)$
is in the null space of $\Delta_\theta$
typically another error term has to be added 
to get a unique solution of the stationarity equation.
Choosing, for example, a Laplacian smoothness term,
yields
\be
-\frac{1}{2}
(\phi,\;\left(\Delta+\lambda \Delta_\theta\right) \phi)
.
\ee
In case $\theta$ is not known, it can be treated
as hyperparameter as discussed in Section \ref{hyperparameters}.

Alternatively to an implementation by choosing a
semi--positive definite operator ${\bf K}$ 
with symmetric functions in its null space,
approximate symmetries can be
implemented by giving explicitly a symmetric reference function 
$t(x)$.
For example,
$
\frac{1}{2} \big(\phi -t,\; {\bf K} (\phi-t)\, \big)
$
with 
$t(x)$ = $t(x+\theta)$.
This possibility will be discussed in the next section.

\subsection{Non--zero means}
\label{Non--zero-means}

A prior energy term $(1/2)(\phi,\, {{\bf K}}\,\phi)$
measures the squared ${{\bf K}}$--distance 
of $\phi$ to the zero function $t\equiv 0$.
Choosing a zero mean function for the prior process
is calculationally convenient for Gaussian priors, 
but by no means mandatory.
In particular, 
a function $\phi$ is
in practice often measured relative to some non--trivial base line.
Without further {\it a priori} information
that base line can in principle be an arbitrary function.
Choosing a zero mean function that base line does not enter 
the formulae 
and remains hidden in the realization of the measurement process.
On the the other hand,
including explicitly a non--zero mean function $t$, playing the role of 
a function ${\it template}$
(or reference, target, prototype, base line)
and being technically relatively straightforward,
can be a very powerful tool.
It allows, for example, to 
parameterize $t(\theta$) by introducing hyperparameters
(see Section \ref{hyperparameters})
and to specify explicitly different maxima of multimodal functional priors
(see Section \ref{non-Gaussian}.
\cite{Lemm-1996,Lemm-1998,Lemm-1998a,Lemm-1998b,Lemm-1998c}).
All this cannot be done by referring to a single baseline.

Hence, in this section we consider error terms of the form
\begin{equation}
\frac{1}{2} \Big(\phi - t,\, {{\bf K}}\,(\phi - t)\,\Big)
.
\label{templates}
\end{equation}
Mean or template functions $t$ allow an easy and straightforward implementation
of prior information in form of examples for $\phi$.
They are the continuous analogue of standard training data.
The fact that template functions $t$ 
are most times chosen equal to zero, and thus do not appear explicitly 
in the error functional,
should not obscure the fact that they are of key importance
for any generalization.
There are many situations where
it can be very valuable to include non--zero prior means explicitly.
Template functions for $\phi$ can for example result
from learning done in the past for the same or for similar tasks. 
In particular, 
consider for example
$\tilde \phi(x)$ 
to be the output
of an empirical learning system 
(neural net, decision tree, nearest neighbor methods, $\ldots$) 
being the result of learning 
the same or a similar task.
Such a $\tilde \phi(x)$ would be a natural candidate
for a template function $t(x)$.
Thus, we see that template functions 
could be used for example 
to allow {\it transfer} of knowledge between similar tasks
or to {\it include the results of earlier learning} on the same task
in case the original data are lost but the output of another
learning system is still available.

Including non--zero template functions
generalizes functional $E_\phi$ of Eq.\ (\ref{phifunctional})
to
\bea
E_\phi &=&
-(\ln P(\phi),\,N)
+\frac{1}{2} \Big(\phi-t,\,{{\bf K}}\,(\phi-t)\Big)
+ (P(\phi),\, \Lambda_X )
\label{templatefunctional}
\\&=&
-(\ln P(\phi),\,N)
+\frac{1}{2} (\phi,\,{{\bf K}}\,\phi)
-(J,\,\phi)
\!+\!(P(\phi),\, \Lambda_X )
\!+\!{\rm const}
.
\eea
In the language of physics
$J$ = ${{\bf K}} t$ represents an {\it external field}
coupling to $\phi (x,y)$, 
similar, for example, to a magnetic field. 
A non--zero field leads to a non--zero expectation of $\phi$
in the no--data case.
The $\phi$--independent constant
stands for the term $\frac{1}{2} (t,\,{{\bf K}}\,t)$,
or $\frac{1}{2} (J,\,{{\bf K}}^{-1}\,J)$ for invertible ${{\bf K}}$,
and can be skipped from the error/energy functional $E_\phi$.

The stationarity equation for an $E_\phi$ with non--zero 
template $t$
contains an inhomogeneous term 
${{\bf K}}t$ = $J$ 
\begin{equation}
0 =
{\bf P}^\prime (\phi) {\bf P}^{-1}(\phi) N  
- {\bf P}^\prime (\phi) \Lambda_X
- {{\bf K}}\left( \phi - t\right)
,
\label{stat-eq-nzm}
\end{equation}
with, for invertible ${\bf P} {{\bf P}^\prime}^{-1}$
and $\Lambda_X\ne 0$,
\be
\Lambda_X = 
{\bf I}_X \left( N - {\bf P} {{\bf P}^\prime}^{-1} {{\bf K}}\, (\phi-t) \right)
.
\label{lambdatemp}
\ee
Notice that functional (\ref{templatefunctional})
can be rewritten as a functional with zero template $t\equiv 0$ 
in terms of $\widetilde \phi$ = $\phi - t$.
That is the reason why we have not included non--zero templates
in the previous sections.
For general non--additive combinations of 
squared distances of the form (\ref{templates})
non--zero templates cannot be removed from the functional 
as we will see in Section \ref{non-Gaussian}.
Additive combinations of squared error terms,
on the other hand, can again be written as one 
squared error term,
using a generalized `bias--variance'--decomposition
\begin{equation}
\frac{1}{2}
\sum_{j=1}^N \Big( \phi - t_j,\, {{\bf K}}_j\,(\phi - t_j)\Big)
=
\frac{1}{2}\Big(\phi - t,\, {{\bf K}}\, (\phi - t)\Big)
+ E_{\rm min}
\label{ANDing}
\end{equation}
with {\it template average}
\begin{equation}
t = {{\bf K}}^{-1} \sum_{j=1}^N {{\bf K}}_j t_j,
\end{equation}
assuming the existence of the inverse of
the operator
\begin{equation}
{{\bf K}} = \sum_{j=1}^N {{\bf K}}_j.
\end{equation}
and {\it minimal energy/error}
\begin{equation}
E_{\rm min} = \frac{N}{2} V(t_1,\cdots t_N)
= \frac{1}{2}
\sum_{j=1}^N (t_j,\, {{\bf K}}_j\,t_j)
- (t,\, {{\bf K}}\,t),
\end{equation}
which up to a factor $N/2$ represents a generalized 
template variance $V$.
We end with the remark that adding error terms 
corresponds
in its probabilistic Bayesian interpretation
to ANDing independent events.
For example, if we wish to implement that $\phi$
is likely to be smooth AND mirror symmetric,
we may add two squared error terms, one related to smoothness
and another to mirror symmetry.
According to (\ref{ANDing}) the result will be 
a single squared error term of form (\ref{templates}).

Summarizing, we have seen
that there are many potentially useful applications 
of non--zero template functions.
Technically, however,
non--zero template functions can be removed from the formalism by a simple
substitution $\phi^\prime = \phi-t$
if the error functional consists of
an additive combination of quadratic prior terms.
As most regularized error functionals used in practice have additive
prior terms this is probably the reason that they 
are formulated for $t\equiv 0$,
meaning that non--zero templates functions (base lines)
have to be treated by including a preprocessing step switching  
from $\phi$ to $\phi^\prime$.
We will see in Section \ref{non-Gaussian}
that for general error functionals templates cannot be removed
by a simple substitution and do enter the error functionals
explicitly.

\subsection{Quadratic density estimation and empirical risk minimization}
\label{quad-de}

Interpreting an energy or error functional $E$ probabilistically,
i.e., assuming $-\beta E +c$ 
to be the logarithm of a posterior probability under study, 
the form of the training data term has to be $-\sum_i \ln P_i$. 
Technically, however, it would be easier to replace
that data term by one which is quadratic in the probability $P$ 
of interest. 

Indeed, we have mentioned in Section \ref{empirical-risk}
that such functionals can be justified
within the framework of empirical risk minimization.
From that Frequentist point of view an error functional $E(P)$,
is not derived from a log--posterior,
but represents an empirical risk 
$\hat r(P,f) = \sum_i l(x_i,y_i,P)$,
approximating an
expected risk $r(P,f)$ for action $a$ = $P$.
This is possible under the assumption that
training data are sampled according to the true $p(x,y|f)$.
In that interpretation
one is therefore not restricted to
a log--loss for training data
but may as well choose for training data a quadratic loss like 
\begin{equation}
\frac{1}{2}
\Big( P-P_{\rm emp},\, {{\bf K}_D}\,(P-P_{\rm emp})\Big)
,
\label{quad-data-loss}
\end{equation}
choosing a reference density $P{\rm emp}$
and a
real symmetric positive \mbox{(semi--)}/-definite ${{\bf K}_D}$.

Approximating  a joint probability $p(x,y|h)$
the reference density $P_{\rm emp}$
would have to be the joint empirical density 
\be
P_{\rm emp}^{\rm joint} (x,y) = 
\frac{1}{n} \sum_i^n \delta (x-x_i) \delta(y-y_i)
,
\label{jointPemp}
\ee
i.e.,
$P_{\rm emp}^{\rm joint}$ = $N/n$, 
as obtained from the training data. 
Approximating  conditional probabilities $p(y|x,h)$
the reference $P_{\rm emp}$ has to be chosen
as conditional empirical density,
\be
P_{\rm emp}(x,y) 
=
\frac{\sum_i \delta(x-x_i)\delta(y-y_i)}
     {\sum_i \delta(x-x_i) }
= \frac{N(x,y)}{n_x}
,
\label{cond-P-emp1}
\ee
or, 
defining the diagonal matrix 
${\bf N}_X(x,x^\prime,y,y^\prime)$ 
= $\delta(x-x^\prime)\delta(y-y^\prime) N_X(x)$
= $\delta(x-x^\prime)\delta(y-y^\prime) \sum_i \delta(x-x_i)$
\be
P_{\rm emp} = {\bf N}_X^{-1}N
.
\label{cond-P-emp2}
\ee
This, however, is only a valid expression if
$N_X(x)\ne0$, meaning that for all $x$
at least one measured value has to be available.
For $x$ variables with a large number of possible values, 
this cannot be assumed.
For continuous $x$ variables it is even impossible.

Hence, approximating conditional empirical densities  
either non--data $x$--values must be excluded
from the integration in (\ref{quad-data-loss})
by using an operator ${\bf K}_D$ containing the projector
$\sum_{x^\prime\in x_D} \delta(x-x^\prime)$,
or $P_{\rm emp}$ 
must be defined also for such non--data $x$--values. 
For existing 
$V_X$ = ${\bf I}_X 1$ = $\int \! dy\, 1$,
a possible extension $\tilde P_{\rm emp}$ of $P_{\rm emp}$ 
would be to assume a uniform density for non--data $x$ values,
yielding
\be
\tilde P_{\rm emp}(x,y) 
=
\left\{
{
\frac{\sum_i \delta(x-x_i)\delta(y-y_i)}
     {\sum_i \delta(x-x_i) }
\quad {\rm for} \quad 
\sum_i \delta(x-x_i) \ne 0
,
\atop 
\;\qquad\frac{1}
     {\int \! dy\, 1}
\;\quad\qquad {\rm for} \quad 
\sum_i \delta(x-x_i) = 0
.
}
\right.
\label{tilde-P-emp}
\ee
This introduces a bias towards uniform probabilities,
but has the advantage
to give a empirical density for all $x$
and to fulfill the conditional normalization requirements.

Instead of a quadratic term in $P$,
one might consider a quadratic term in the log--probability $L$. 
The log--probability, however,
is minus infinity at all non--data points $(x,y)\not\in D$.
To work with a finite expression, one can choose 
small $\epsilon (y)$ and approximate $P_{\rm emp}$
by 
\be
P^\epsilon_{\rm emp} (x,y) = 
\frac{\epsilon(y) + \sum_i \delta(x-x_i)\delta(y-y_i)}
     {\int\!dy \, \epsilon (y) + \sum_i \delta(x-x_i)}
,
\label{Pepsilon}
\ee
provided
$\int\!dy \, \epsilon (y)$
exists.
For $\epsilon (y)\ne 0$
also $P^\epsilon_{\rm emp} (x,y)\ne 0$, $\forall x$
and $L^\epsilon_{\rm emp}$ = $\ln P^\epsilon_{\rm emp}>-\infty$
exists.

A quadratic data term in $P$ 
results in an error functional
\begin{equation}
\tilde E_P =
\frac{1}{2} \Big(P-P_{\rm emp},\, {{\bf K}_D}\,(P-P_{\rm emp})\Big) 
+ \frac{1}{2} (P,\, {{\bf K}}\,P)
+(P,\, \Lambda_X ),  
\label{P-quadratic}
\end{equation}
skipping the constant part of the $\Lambda_X$--terms.
In (\ref{P-quadratic})
the empirical density 
$P_{\rm emp}$
may be replaced by $\tilde P_{\rm emp}$ of (\ref{tilde-P-emp}).

Positive (semi--)definite operators ${{\bf K}_D}$
have a square root and can be written
in the form ${\bf R}^T{\bf R}$.
One possibility, 
skipping for the sake of simplicity $x$ in the following,
is to choose
as square root ${\bf R}$
the integration operator, i.e.,
${\bf R}$ = $\bigotimes_k {\bf R}_k$ and
${\bf R} (y,y^\prime)$ = $\theta (y-y^\prime )$.
Thus, $\phi ={\bf R} P$
transforms the density function $P$
in the distribution function $\phi$,
and we have $P = P(\phi) = {\bf R}^{-1} \phi$.
Here the inverse ${\bf R}^{-1}$ is the differentiation operator
$\prod_k \nabla_{y_k}$ (with appropriate boundary condition)
and $\left({\bf R}^T\right)^{-1}{\bf R}^{-1}$ = $-\prod_k \Delta_k$
is the product of one--dimensional Laplacians
$\Delta_k = \partial^2 /\partial y_k^2$.
Adding for example a regularizing term (\ref{quadP})
$\frac{\lambda}{2}(P,\,P)$ 
gives 
\begin{equation}
\tilde E_P =
\frac{1}{2} \Big(\,P-P_{\rm emp}\,,\, {\bf R}^T{\bf R}\,(P-P_{\rm emp})\,\Big) 
+\frac{\lambda}{2} (P,\,P)
\label{P-quad-dens}
\end{equation}
\begin{equation}
=
\frac{1}{2} \left(
\Big(\phi-\phi_{\rm emp},\, \phi-\phi_{\rm emp} \,\Big)   
- \lambda \Big(\phi,\,\prod_k \Delta_k \,\phi\,\Big)
\right)
\label{P-quad-dis}
\end{equation}
\begin{equation}
=
\frac{1}{2 m^2} \Big(\phi,\, (- \prod_k \Delta_k + m^2 {\bf I})\phi\Big)
- (\phi\,,\phi_{\rm emp}) 
+ \frac{1}{2} (\phi_{\rm emp},\, \phi_{\rm emp}).
\label{P-quad-dis2}
\end{equation}
with $m^2=\lambda^{-1}$.
Here the empirical distribution function 
$\phi_{\rm emp}$ = ${\bf R} P_{\rm emp}$ 
is
given by $\phi_{\rm emp} (y)$ = $\frac{1}{n}\sum_i \theta (y-y_i)$
(or, including the $x$ variable,
$\phi_{\rm emp} (x,y)$ 
= $\frac{1}{N_X(x)}\sum_{x^\prime\in x_D} \delta(x-x^\prime)\theta (y-y_i)$
for $N_X(x)\ne 0$ 
which could be extended to a linear 
$\tilde \phi$ =
${\bf R} \tilde P_{\rm emp}$ for $N_X(x)$ = $0$).
The stationarity equation yields
\begin{equation}
\phi =
m^2 \left( - \prod_k \Delta_k + m^2 {\bf I} \right)^{-1} \phi_{\rm emp}. 
\label{parzen}
\end{equation}
For $d_y$ = $1$ (or $\phi = \prod_k \phi$)
the operator becomes
$\left( - \Delta + m^2{\bf I}\right)^{-1}$
which has the structure of 
a free massive propagator for a scalar field with mass
$m^2$ and is calculated below.
As already mentioned the normalization and non--negativity condition for $P$
appear for $\phi$ as boundary
and monotonicity conditions.
For non--constant $P$ the monotonicity condition 
has not to be implemented by Lagrange multipliers
as the gradient at the stationary point has no components
pointing into the forbidden area.
(But the conditions nevertheless have to be checked.)
Kernel methods of density estimation,
like the use of Parzen windows,
can be founded on such quadratic regularization functionals 
\cite{Vapnik-1982}.
Indeed, the one--dimensional Eq.\ (\ref{parzen}) 
is equivalent to the use of Parzen´s kernel
in density estimation
\cite{Parzen-1962b,Nadaraya-1965}.

\subsection{Regression}
\label{regression}

\subsubsection{Gaussian regression}

An important special case of density estimation 
leading to quadratic data terms
is regression for independent training data 
with Gaussian likelihoods 
\be
p(y_i|x_i,{h})
= \frac{1}{\sqrt{2\pi}\sigma}e^{-\frac{(y_i-{h}(x_i))^2}{2 \sigma^2}}
,
\ee
with fixed, but possibly $x_i$--dependent, variance $\sigma^2$.
In that case $P(x,y)$ = $p(y_i|x_i,{h})$
is specified by $\phi$ = $h$ and
the logarithmic term $\sum_i \ln P_i$
becomes quadratic in the regression function ${h}(x_i)$,
i.e., of the form (\ref{templates}).
In an interpretation as empirical risk minimization
quadratic error terms corresponds to the choice of 
a squared error loss function $l(x,y,a)$ = $(y-a(x))^2$ for action
$a(x)$.
Similarly, the
technical analogon of Bayesian priors
are additional (regularizing) cost terms.

We have remarked in Section \ref{map}
that for continuous $x$ measurement
of ${h}(x)$ has to be understood as
measurement of a
${h} (\tilde x)$ =
$\int \!dx\, \vartheta (x) {h} (x)$ 
for sharply peaked $\vartheta (x)$.
We assume here that the discretization
of ${h}$ used in numerical calculations takes care of that averaging.
Divergent quantities like $\delta$--functionals,
used here for convenience, will then not be present.

We now combine Gaussian data terms and a Gaussian (specific) prior 
with prior operator 
${{\bf K}}_0 (x,x^\prime)$ 
and define for training data $x_i$, $y_i$ the operator
\be
{{\bf K}}_i(x,x^\prime ) = 
\delta(x-x_i)\delta(x-x^\prime)  
,
\label{Oidiag}
\ee
and training data templates $t=y_i$.
We also allow a general  prior template $t_0$ 
but remark that it is often chosen identically zero.
According to (\ref{ANDing}) the resulting functional can be written 
in the following forms, useful for different purposes,
\bea
E_{h} &=&
 \frac{1}{2}\sum_{i=1}^n ( {h} (x_i) -y_i)^2
 +\frac{1}{2}(\,{h}-t_0,\, {{\bf K}}_0 \, ({h}-t_0) \,)_X
\label{regression-functional}
\\&=&
 \frac{1}{2}\sum_{i=1}^n (\,{h} -t_i,\, {{\bf K}}_i ({h}-t_i)\,)_X
 +\frac{1}{2}(\,{h}-t_0,\, {{\bf K}}_0 \, ({h}-t_0) \,)_X
\label{Eregression}
\\&=&
 \frac{1}{2}({h} -t_D,\, {{\bf K}}_D ({h}-t_D))_X
\!+\!\frac{1}{2}({h}-t_0,\, {{\bf K}}_0 ({h}-t_0))_X
  \!+\!E_{{\rm min}}^D
\label{Edataregression}
\\&=&
\frac{1}{2}(\,{h} - t,\, {{\bf K}} \, ({h}-t)\,)_X + E_{\rm min}
\label{Ereg2}
,
\eea
with
\be
{{\bf K}}_D = \sum_{i=1}^n {{\bf K}}_i
,\quad
t_D = {{\bf K}}_D^{-1} \sum_{i=1}^n {{\bf K}}_i t_i
,
\ee
\begin{equation}
{{\bf K}} = \sum_{i=0}^n {{\bf K}}_i
,\quad
t = {{\bf K}}^{-1} \sum_{i=0}^n {{\bf K}}_i t_i,
\label{t-def}
\end{equation}
and 
${h}$--independent minimal errors,
\bea
E_{{\rm min}}^D 
 &=& 
 \frac{1}{2}
 \left( 
      \sum_{i=1}^n \left( t_i,\; {\bf K}_i t_i\right)_X 
     +\left( t_D,\; {\bf K}_D t_D\right)_X 
 \right)
,\\ 
E_{\rm min} 
&=& 
 \frac{1}{2}
 \left( 
      \sum_{i=0}^n \left( t_i,\; {\bf K}_i t_i\right)_X 
     +\left( t,\; {\bf K} t\right)_X
 \right)
,
\eea
being proportional to the ``generalized variances''
$V_D$ = $ 2 E_{{\rm min}}^D/n$
and
$V$ = $2E_{{\rm min}}/(n+1)$.
The scalar product $(\cdot,\cdot)_X$ 
stands for $x$--integration only,
for the sake of simplicity however, we will skip the subscript $X$ 
in the following.
The data operator ${{\bf K}}_D$ 
\be
{{\bf K}}_D (x,x^\prime ) 
= \sum_{i=1}^n \delta (x-x_i)  
\delta (x-x^\prime)   
= n_x\, \delta (x-x^\prime)
,
\ee
contains for discrete $x$ on its diagonal
the number of measurements at $x$,
\be
n_x 
=N_X(x)
=\sum_{i=1}^n \delta (x-x_i)
,
\ee
which is zero for $x$ not in the training data.
As already mentioned for continuous $x$ a integration 
around a neighborhood of $x_i$ is required.
${{\bf K}}_D^{-1}$ is a short hand notation for
the inverse within the space of training data
\be
{{\bf K}}_D^{-1}  = 
\left({\bf I}_D {{\bf K}}_D{\bf I}_D\right)^{-1} = 
\delta (x-x^\prime)/n_x
,
\ee
${\bf I}_D$ denoting 
the projector into the space of training data 
\be
{\bf I}_D = \delta ( x-x^\prime )
             \sum_{i=1}^{\tilde n} \delta (x-x_i)
.
\label{dataprojector}
\ee
Notice that the sum is not over all $n$ training points $x_i$
but only over the $\tilde n\le n$ different $x_i$.
(Again for continuous $x$ an integration around $x_i$ is required
to ensure ${\bf I}_D^2$ = ${\bf I}_D$).
Hence, the data template
$t_D$ becomes the mean of $y$--values measured at $x$
\be
t_D (x) = \frac{1}{n_x}\sum_{j=1 \atop x_j=x}^{n_x} y (x_j)
,
\ee  
and $t_D (x)$ = $0$ for $n_x$ = $0$.
Normalization of $P(x,y)$ is not influenced by a change in ${h} (x)$
so the Lagrange multiplier terms have been skipped.

The stationarity equation is most easily obtained from (\ref{Ereg2}),
\be
0={{\bf K}} ({h}-t)
.
\ee
It is linear and has on a space where ${{\bf K}}^{-1}$ exists
the unique solution
\be
{h}=t
.
\label{phi=t}
\ee
We remark that ${{\bf K}}$ can be invertible
(and usually is so the learning problem is well defined) 
even if ${{\bf K}}_0$ is not invertible.
The inverse ${{\bf K}}^{-1}$, necessary to calculate $t$,
is training data dependent
and represents the covariance operator/matrix
of a Gaussian posterior process.
In many practical cases, however, the prior covariance
${{\bf K}}_0^{-1}$ 
(or in case of a null space a pseudo inverse of ${{\bf K}}_0$)
is directly given or can be calculated.
Then an inversion of a finite dimensional matrix
in data space is sufficient to find the minimum of the energy $E_{h}$
\cite{Wahba-1990,Girosi-Jones-Poggio-1995}.

{\bf Invertible ${\bf K}_0$}:
Let us assume first deal with the case of an invertible ${\bf K}_0$.
It is the best to begin the stationarity equation 
as obtained from (\ref{Eregression}) or (\ref{Edataregression})
\bea
0 
&=&
\sum_{i=1}^n {{\bf K}}_i ({h}-t_i) + {{\bf K}}_0 ({h} - t_0)
\\&=& 
{{\bf K}}_D ({h}-t_D) + {{\bf K}}_0 ({h} - t_0)
.
\label{datastat}
\eea
For existing ${{\bf K}}_0^{-1}$
\be
{h} = t_0 + {{\bf K}}_0^{-1} {{\bf K}}_D (t_D-{h} )
,
\ee
one can introduce
\be
a = {{\bf K}}_D (t_D -{h})
,
\label{b1}
\ee
to obtain
\be
{h} = t_0 + {{\bf K}}_0^{-1} a 
.
\label{b2}
\ee
Inserting Eq.\ (\ref{b2}) into Eq.\ (\ref{b1}) one finds
an equation for $a$
\be
\left(I + {{\bf K}}_D {{\bf K}}_0^{-1}\right) a 
=
{{\bf K}}_D (t_D - t_0)
.
\label{b3}
\ee
Multiplying Eq.\ (\ref{b3}) 
from the left by the projector ${\bf I}_D$
and  using
\be
{{\bf K}}_D {\bf I}_D = {\bf I}_D {{\bf K}}_D 
,\quad 
a = {\bf I}_D a
,\quad 
t_D = {\bf I}_D t_D
,
\ee
one obtains an equation in data space
\be
\left(I_D + {{\bf K}}_D {{\bf K}}_{0,DD}^{-1}\right) a 
=
{{\bf K}}_D (t_D - t_{0,D})
,
\label{dataspace}
\ee
where
\be
{{\bf K}}_{0,DD}^{-1} 
= ({{\bf K}}_0^{-1})_{DD}
=  {\bf I}_D {{\bf K}}_0^{-1} {\bf I}_D 
\ne ({{\bf K}}_{0,DD})^{-1}
,\quad
t_{0,D} = {\bf I}_D t_0
.
\ee
Thus, 
\be
a = {{\bf C}_{DD}}\; b
,
\label{dataspaceeq}
\ee
where 
\be
{\bf C}_{DD} = 
\left({\bf I}_D + {{\bf K}}_D {{\bf K}}_{0,DD}^{-1}\right)^{-1}  
,
\ee
and
\be
b ={{\bf K}}_D (t_D - t_0)
.
\ee
In components Eq.\ (\ref{dataspaceeq}) reads,
\be
\sum_l \left( \delta_{kl} + n_{x_k} {{\bf K}}_{0}^{-1}(x_k,x_l)\right) a(x_l)
=
n_{x_k} \left( t_D(x_k) -t_0(x_k) \right)
.
\ee
Having calculated $a$ the solution ${h}$
is given by Eq.\ (\ref{b2})
\be
{h} 
= t_0 + {{\bf K}}_0^{-1} {{\bf C}_{DD}} b
= t_0 + {{\bf K}}_0^{-1} 
\left({\bf K}_D^{-1} + {{\bf K}}_{0,DD}^{-1} \right)^{-1}  (t_D - t_0)
\label{regression-solution}
.
\ee
Eq.\ (\ref{regression-solution})
can also be obtained directly from
Eq.\ (\ref{phi=t}) and the definitions (\ref{t-def}), 
without introducing the auxiliary variable $a$,
using the decomposition
${\bf K}_0 t_0$ = $-{\bf K}_D t_0$ + $({\bf K}_0 + {\bf K}_D )t_0$
and
\be
{\bf K}^{-1}{\bf K}_D 
={\bf K}_0^{-1} \left({\bf I} + {\bf K}_D {\bf K}_0^{-1}\right)^{-1}{\bf K}_D 
={\bf K}_0^{-1} \sum_{m=0}^\infty
\left(-{\bf K}_D {\bf K}_0^{-1}\right)^m {\bf K}_D 
\ee
\be
={\bf K}_0^{-1} \sum_{m=0}^\infty
\left(-{\bf K}_D {\bf I}_D {\bf K}_0^{-1} {\bf I}_D \right)^m {\bf K}_D 
={\bf K}_0^{-1} \left({\bf I}_D 
  + {\bf K}_D {\bf K}_{0,DD}^{-1}\right)^{-1}{\bf K}_D 
.
\ee

${{\bf K}}_0^{-1} {{\bf C}_{DD}}$ is also known as equivalent kernel
due to its relation to kernel smoothing techniques
\cite{Silverman-1984,Hastie-Tibshirani-1990,Haerdle-1990,Girosi-Jones-Poggio-1995}.

Interestingly,
Eq.\ (\ref{b2}) still holds
for non--quadratic data terms 
of the form $g_D(h)$ with any differentiable function 
fulfilling 
$g(h)$ = $g(h_D)$, 
where $h_D$ = ${\bf I}_D h$
is the restriction of $h$ to data space.
Hence, also
the function of functional derivatives
with respect to $h(x)$
is restricted to data space, i.e., 
$g^\prime (h_D)$ = $g^\prime_D(h_D)$ 
with $g^\prime_D$ = ${\bf I}_D g^\prime$
and $g^\prime (h,x)=\delta g(h)/\delta h(x)$.
For example,
$g(h)$ = $\sum_{i=1}^nV({h} (x_i) -y_i)$
with $V$ a differentiable function.
The finite dimensional vector $a$
is then found by solving a nonlinear equation instead of a linear one
\cite{Girosi-1991,Girosi-Poggio-Caprile-1991}.

Furthermore, one can study vector fields, i.e., the case where,
besides possibly $x$,
also $y$, and thus ${h}(x)$, is a vector for given $x$.
(Considering the variable indicating the vector components of $y$
as part of the $x$--variable,
this is a  situation where a fixed number of one--dimensional $y$,
corresponding to a subspace of $X$ with fixed dimension,
is always measured simultaneously.)
In that case the diagonal ${{\bf K}}_i$ of Eq.\ (\ref{Oidiag})
can be replaced by a version with non--zero off--diagonal elements 
${{\bf K}}_{\alpha,\alpha^\prime}$
between the vector components $\alpha$ of $y$. 
This corresponds
to a multi--dimensional Gaussian data generating probability
\be
p(y_i|x_i,{h})
= \frac{\det{{{\bf K}}_i}^\frac{1}{2}}{(2\pi)^\frac{k}{2}}
e^{-\frac{1}{2} \sum_{\alpha,\alpha^\prime}
(y_{i,\alpha}-{h}_\alpha (x_i))\,{\bf K}_{i,\alpha,\alpha^\prime}(x_i) 
(y_{i,\alpha^\prime}-{h}_{\alpha^\prime}(x_i))}
,
\ee
for $k$--dimensional vector $y_i$ with components $y_{i,\alpha}$.

{\bf Non-invertible ${\bf K}_0$}:
For non--invertible ${{\bf K}}_0$
one can solve for ${h}$
using the Moore--Penrose inverse 
${{\bf K}}_0^\#$.
Let us first recall some basic facts
\cite{Fredholm-1903,Moore-1920,Ben-Israel-Greville-1974,Koecher-1985}. 
A pseudo inverse of (a possibly non--square) ${\bf A}$ 
is defined by the conditions
\be
{\bf A}^\# {\bf A} {\bf A}^\# = {\bf A}
,\quad
{\bf A} {\bf A}^\# {\bf A} = {\bf A}^\#
,
\ee
and becomes for real ${\bf A}$ 
the unique Moore--Penrose inverse  ${\bf A}^\#$
if 
\be
({\bf A} {\bf A}^\#)^T = {\bf A} {\bf A}^\# 
,\quad
({\bf A}^\# {\bf A})^T = {\bf A}^\# {\bf A}
.
\ee
A linear equation
\be
{\bf A} x = b
\ee 
is solvable if
\be
{\bf A}{\bf A}^\# b = b
.
\label{solvcond}
\ee 
In that case the solution is
\be
x 
= {\bf A}^\#b + x^0
= {\bf A}^\#b + y-{\bf A}^\#{\bf A}y
,
\label{MPsol}
\ee
where $x^0 = y-{\bf A}^\#{\bf A}y$
is solution of the homogeneous equation ${\bf A}x^0 = 0$ 
and vector $y$ is arbitrary.
Hence, $x^0$ can be expanded in an orthonormalized basis $\psi_l$
of the null space of ${\bf A}$
\be
x^0 = \sum_l c_l \psi_l
.
\ee
For an ${\bf A}$ which can be diagonalized, i.e., 
${\bf A}$ = ${\bf M}^{-1} {\bf D} {\bf M}$
with diagonal ${\bf D}$,
the Moore--Penrose inverse is
${\bf A}^\#$ = ${\bf M}^{-1} {\bf D}^\# {\bf M}$.
Therefore
\be
{\bf A}{\bf A}^\# 
= {\bf A}^\#{\bf A} 
= {\bf I}_1 = {\bf I} - {\bf I}_0
.
\label{MPproj}
\ee
where
${\bf I}_0$ = $\sum_l \psi_l \psi_l^T$
is the projector into the zero space of ${\bf A}$
and 
${\bf I}_1$ = ${\bf I} - {\bf I}_0$
= ${\bf M}^{-1} {\bf D} {\bf D}^\# {\bf M}$.
Thus, the solvability condition Eq.\ (\ref{solvcond})
becomes
\be
{\bf I}_0 b = 0
,
\ee
or in terms of $\psi_l$
\be
(\, \psi_l , \, b) = 0,\; \forall l
,
\ee
meaning that the inhomogeneity $b$ must have no components
within the zero space of ${\bf A}$.

Now we apply this to Eq.\ (\ref{datastat})
where ${{\bf K}}_0$ is
diagonalizable because positive semi definite.
(In this case ${\bf M}$ is an orthogonal matrix
and the entries of $D$ are real and larger or equal to zero.)
Hence, one obtains under the condition
\be
{\bf I}_0 \left( {{\bf K}}_0 t_0 + {{\bf K}}_D (t_D -{h}) \right)  = 0
,
\label{MPOc}
\ee
for Eq.\ (\ref{MPsol})
\be
{h} = 
{{\bf K}}_0^\# \left(
{{\bf K}}_0 t_0 + {{\bf K}}_D (t_D -{h})
\right)
+{h}^0
,
\label{MPOs}
\ee
where 
${{\bf K}}_0 {h}^0$ = $0$
so that ${h}^0 = \sum_l c_l \psi_l$
can be expanded in an orthonormalized basis $\psi_l$
of the null space of ${{\bf K}}_0$, assumed here 
to be of finite dimension.
To find an equation in data space
define the vector
\be
a = {{\bf K}}_D (t_D - {h})
,
\label{defa}
\ee
to get from Eqs.(\ref{MPOc}) and (\ref{MPOs})
\bea
0 &=& (\, \psi_l,\, {{\bf K}}_0 t_0) + (\, \psi_l,\, a),\; \forall l
\\
{h} &=& 
{{\bf K}}_0^\# \left(
{{\bf K}}_0 t_0 + a
\right)
+\sum_l c_l \psi_l
.
\label{MPaa}
\eea
These equations have to be solved for $a$ and the coefficients $c_l$.
Inserting Eq.\ (\ref{MPaa}) into the 
definition (\ref{defa}) gives 
\be
({\bf I} + {{\bf K}}_D{{\bf K}}_0^\#) a =
{{\bf K}}_D t_D -
{{\bf K}}_D {\bf I}_1 t_0 - {{\bf K}}_D \sum_l c_l \psi_l
,
\label{MPa}
\ee
using 
${{\bf K}}_0^\#{{\bf K}}_0 $ = ${\bf I}_1$ 
according to Eq.\ (\ref{MPproj}).
Using $a$ = ${\bf I}_Da $
the solvability condition (\ref{MPOc}) becomes
\be
\sum_{i=1}^{\tilde n} \psi_l(x_i) a
= -(\, \psi_l,\, {{\bf K}}_0 t_0\,), \forall l
,
\ee
the sum going over different $x_i$ only.
Eq.\ (\ref{MPa}) for $a$ and $c_l$
reads in data space,
similar to Eq.\ (\ref{dataspace}),
\be
a = \tilde {{\bf C}} \tilde b
,
\ee
where 
$\tilde {{\bf C}}^{-1}$ = ${\bf I} + {{\bf K}}_D{{\bf K}}_0^\#$
has been assumed invertible
and $\tilde b$
is given by the right hand side of Eq.\ (\ref{MPa}). 
Inserting into Eq.\ (\ref{MPaa})
the solution finally can be written
\be
{h} = 
{\bf I}_1 t_0 
+ {{\bf K}}_0^\# \tilde {{\bf C}} \tilde b 
+\sum_l c_l \psi_l
.
\ee

Again, general non--quadratic data terms $g(h_D)$ can be allowed.
In that case
$\delta g(h_D)/\delta h(x)$
=
$g^\prime(h_D,x)$ 
=
$({\bf I}_D g^\prime)(h_D,x)$ 
and Eq.\ (\ref{defa})
becomes the nonlinear equation
\be
a = g^\prime (h_D) = 
g^\prime \!\left(
{\bf I}_D\left({{\bf K}}_0^\# \left(
{{\bf K}}_0 t_0 + {{\bf K}}_D (t_D -{h})
\right)+{h}^0\right)
\right)
.
\ee
The solution(s) $a$ of that equation have then to be inserted
in Eq.\ (\ref{MPaa}).

\subsubsection{Exact predictive density}
\label{exact}

For Gaussian regression the predictive density
under training data $D$ and prior $D_0$
can be found analytically
without resorting to a saddle point approximation.
The predictive density is defined as the $h$-integral
\bea
p(y|x,D,D_0) 
&=& \int\!d{h} \, p(y|x,{h}) p({h}|D,D_0)
\nonumber\\
&=& \frac{\int\!d {h} \, p(y|x,{h}) p(y_D|x_D,{h}) p({h}|D_0)}
       {\int\!d {h} \, p(y_D|x_D,{h}) p({h}|D_0)}
\nonumber\\
&=& \frac{p(y,y_D|x,x_D,D_0)}
       {p(y_D|x_D,D_0)}
.
\eea
Denoting 
training data values $y_i$ by $t_{i}$
sampled with covariance ${\bf K}_{i}$
concentrated on $x_i$ 
and analogously 
test data values $y$ = $y_{n+1}$ by $t_{n+1}$
sampled with (co--)variance ${\bf K}_{n+1}$,
we have for $1\le i\le n+1$
\be
p(y_i|x_i,{h}) = 
\det ({\bf K}_i/2\pi )^{\frac{1}{2}}
e^{-\frac{1}{2} 
             \Big( {h}-t_{i},\, {\bf K}_{i} ({h}-t_{i}) \Big) }
,
\ee
and
\be
p({h}|D_0) = 
\det ({\bf K_0}/2\pi)^{\frac{1}{2}}
e^{-\frac{1}{2} 
             \Big( {h}-t_0,\, {\bf K}_0 ({h}-t_0) \Big) }
,
\ee
hence
\be
p(y|x,D,D_0) = 
\frac{\int\!d{h} \, e^{-\frac{1}{2} \sum_{i=0}^{n+1}
           \Big( {h}-t_i,\, {\bf K}_i ({h}-t_i) \Big) 
         +\frac{1}{2} \sum_{i=0}^{n+1} \ln \det_i ({\bf K}_i/2\pi)}}
{\int\!d{h} \,e^{-\frac{1}{2} \sum_{i=0}^{n}
           \Big( {h}-t_i,\, {\bf K}_i ({h}-t_i) \Big) 
         +\frac{1}{2} \sum_{i=0}^{n} \ln \det_i ({\bf K}_i/2\pi)}}
.
\ee
Here we have this time written explicitly
$\det_i({\bf K}_i/2\pi)$ for a determinant calculated
in that space where ${\bf K}_i$ is invertible.
This is useful because for example in general
$\det_i {\bf K}_i \det {\bf K}_0 \ne \det_i {\bf K}_i {\bf K}_0$. 
Using the generalized `bias--variance'--decomposition (\ref{ANDing})
yields
\be
p(y|x,D,D_0) = 
\frac{\int\!d{h} \, e^{-\frac{1}{2} 
           \Big( {h}-t_+,\, {\bf K}_+ ({h}-t_+) \Big) + \frac{n}{2} V_+
         +\frac{1}{2} \sum_{i=0}^{n+1} \ln \det_i ({\bf K}_i/2\pi)}}
{\int\!d{h} \,e^{-\frac{1}{2} 
           \Big( {h}-t,\, {\bf K} ({h}-t) \Big) + \frac{n}{2} V 
         +\frac{1}{2} \sum_{i=0}^{n} \ln \det_i ({\bf K}_i/2\pi)}}
,
\ee
with
\bea
t         &=& {\bf K}^{-1} \sum_{i=0}^{n} {\bf K}_i t_i
,\qquad {\bf K}  = \sum_{i=0}^{n} {\bf K}_i,\\
t_+       &=& {\bf K}_+^{-1} \sum_{i=0}^{n+1} {\bf K}_i t_i 
,\qquad {\bf K}_+ = \sum_{i=0}^{n+1} {\bf K}_i,\\
V         &=& \frac{1}{n} \sum_{i=0}^{n} 
              \Big( t_i, \, {\bf K}_i t_i \Big)
             -\Big( t, \, \frac{{\bf K}}{n} \,t \Big),\\
V_+       &=& \frac{1}{n} \sum_{i=0}^{n+1} 
              \Big( t_i,\, {\bf K}_i t_i \Big)
             -\Big( t_+,\, \frac{{\bf K}_+}{n}\, t_+ \Big)
.
\eea
Now the ${h}$--integration can be performed
\be
p(y|x,D,D_0) =
\frac{e^{-\frac{n}{2} V_+
         +\frac{1}{2} \sum_{i=0}^{n+1} \ln \det_i ({\bf K}_i/2\pi)
         -\frac{1}{2} \ln \det ({\bf K}_+/2\pi)}}
{e^{-\frac{n}{2} V 
         +\frac{1}{2} \sum_{i=0}^{n} \ln \det_i ({\bf K}_i/2\pi)
         -\frac{1}{2} \ln \det ({\bf K}/2\pi)}}
\ee
Canceling common factors,
writing again $y$ for $t_{n+1}$,
${\bf K}_x$ for  ${\bf K}_{n+1}$,
$\det_x$ for $\det_{n+1}$,
and
using 
${\bf K}_+ t_+$ = ${\bf K}t + {\bf K}_{x}y$, 
this becomes
\be
p(y|x,D,D_0) =
e^{-\frac{1}{2} 
(y,{\bf K}_y \,y) + (y,{\bf K}_y \,t)
+ \frac{1}{2}\big(t,({\bf K}{\bf K}_+^{-1}{\bf K}-{\bf K})\,t\big)
         +\frac{1}{2} \ln\det_{x}({\bf K}_{x}{\bf K}_+^{-1} {\bf K}/2\pi)}
.
\ee
Here we introduced
${\bf K}_y$ = ${\bf K}_y^T$ =
${\bf K}_{x}-{\bf K}_{x}{\bf K}_+^{-1} {\bf K}_{x}$
and used that
\be
\det {\bf K}^{-1}{\bf K}_+
=\det ( {\bf I}-{\bf K}^{-1}{\bf K}_{x})
=\det\!{}_{x} {\bf K}^{-1}{\bf K}_+
\ee
can be calculated in the space of test data $x$.
This follows from 
${\bf K}$ = ${\bf K}_+ - {\bf K}_{x}$ 
and the equality 
\be
\det\vmat{1-A}{0}{B}{1} = \det(1-A)
\ee
with 
$A$ = ${\bf I}_x {\bf K}^{-1} {\bf K}_{x}$, 
$B$ = $({\bf I} - {\bf I}_x) {\bf K}^{-1} {\bf K}_{x}$,
and ${\bf I}_{x}$ denoting the projector
into the space of test data $x$.
Finally
\be
{\bf K}_y
={\bf K}_{x}-{\bf K}_{x}{\bf K}_+^{-1} {\bf K}_{x}
= {\bf K}_{x}{\bf K}_+^{-1} {\bf K}
= ({\bf K}-{\bf K}{\bf K}_+^{-1}{\bf K})
,
\ee
yields the correct normalization of the predictive density 
\be
p(y|x,D,D_0) =
e^{-\frac{1}{2} \Big( y - \bar y,\, {\bf K}_y (y-\bar y)\Big)
+\frac{1}{2} \ln \det\!{}_{x} ({\bf K}_y/2\pi)
}
,
\label{predictive-dens}
\ee
with mean and covariance
\bea
\bar y &=& t
= {\bf K}^{-1} \sum_{i=0}^n {\bf K}_i t_i,
\label{pmean1}
\\
{\bf K}_y^{-1} &=& 
\left({\bf K}_{x}
-{\bf K}_{x}{\bf K}_+^{-1} {\bf K}_{x}\right)^{-1}
= {\bf K}_{x}^{-1} + {\bf I}_{x}{\bf K}^{-1}{\bf I}_{x}
\label{pcov1}
.
\eea
It is useful to express
the posterior covariance ${\bf K}^{-1}$ by
the prior covariance ${\bf K}_0^{-1}$.
According to
\be
\vmat{1+A}{B}{0}{1}^{-1}
=
\vmat{(1+A)^{-1}}{-(1+A)^{-1}B}{0}{1},
\ee
with 
$A$ = ${\bf K}_D {\bf K}_{0,DD}^{-1}$, 
$B$ = ${\bf K}_D {\bf K}_{0,D\bar D}^{-1}$,
and
${\bf K}_{0,DD}^{-1}$ = ${\bf I}_{D}{\bf K}_0^{-1}{\bf I}_{D}$,
${\bf K}_{0,D\bar D}^{-1}$ = ${\bf I}_{D}{\bf K}_0^{-1}{\bf I}_{\bar D}$,
${\bf I}_{\bar D}$ = ${\bf I}-{\bf I}_{D}$
we find
\bea
{\bf K}^{-1}
&=&{\bf K}_0^{-1} \left( {\bf I}+{\bf K}_D {\bf K}_0^{-1}\right)^{-1}
\label{aux1}
\\
&=&{\bf K}_0^{-1} \! \left(
 \left( {\bf I}_D+{\bf K}_D {\bf K}_{0,DD}^{-1}\right)^{-1}
\!\!
 -\left( {\bf I}_D+{\bf K}_D {\bf K}_{0,DD}^{-1}\right)^{-1}
  {\bf K}_D {\bf K}_{0,D\bar D}^{-1} + {\bf I}_{\bar D}\right)
\nonumber
.
\eea
Notice that while
${\bf K}_{D}^{-1}$ = $({\bf I}_{D}{\bf K}_D {\bf I}_{D})^{-1}$
in general
${\bf K}_{0,DD}^{-1}$ 
= ${\bf I}_{D}{\bf K}_0^{-1}{\bf I}_{D}
\ne ({\bf I}_{D}{\bf K}_0{\bf I}_{D})^{-1}$.
This means for example that
${\bf K}_{0}^{-1}$ has to be 
known to find ${\bf K}_{0,DD}^{-1}$ and it is not enough to invert
${\bf I}_{D}{\bf K}_0{\bf I}_{D}$
= ${\bf K}_{0,DD}\ne ({\bf K}_{0,DD}^{-1})^{-1}$.
In data space 
$\left( {\bf I}_D+{\bf K}_D {\bf K}_{0,DD}^{-1}\right)^{-1}$
=
 $\left( {\bf K}_D^{-1}+{\bf K}_{0,DD}^{-1}\right)^{-1}{\bf K}_D^{-1}$
,
so Eq.\ (\ref{aux1}) can be manipulated to give
\be
{\bf K}^{-1}
=
{\bf K}_0^{-1} \left( {\bf I}- {\bf I}_D
       \left({\bf K}_D^{-1}+{\bf K}_{0,DD}^{-1}\right)^{-1}
       {\bf I}_D{\bf K}_0^{-1}
\right)
.
\ee
This allows now to express
the predictive mean (\ref{pmean1}) and covariance (\ref{pcov1}) 
by the prior covariance
\bea
\bar y &=& 
t_0 + {{\bf K}}_0^{-1} 
\left({\bf K}_D^{-1} + {{\bf K}}_{0,DD}^{-1} \right)^{-1}  (t_D - t_0)
,
\label{pmean2}
\\
{\bf K}_y^{-1} &=& 
{\bf K}_{x}+
{\bf K}_{0,xx}^{-1}
-{\bf K}_{0,xD}^{-1}
\left({\bf K}_D^{-1}+{\bf K}_{0,DD}^{-1}\right)^{-1} 
{\bf K}_{0,Dx}^{-1}
.
\label{pcov2}
\eea
Thus, for given prior covariance
${\bf K}_0^{-1}$ both, 
$\bar y$ and ${\bf K}_y^{-1}$,
can be calculated
by inverting the $\tilde n\times\tilde n$ matrix
$\widetilde {\bf K}$ 
= 
$\left({\bf K}_{0,DD}^{-1} + {\bf K}_D^{-1}\right)^{-1}$.

Comparison of Eqs.(\ref{pmean2},\ref{pcov2}) 
with the maximum posterior solution ${h}^*$ of
Eq.\ (\ref{regression-solution})
now shows that for Gaussian regression
the exact predictive density
$p(y|x,D,D_0)$ and its
maximum posterior approximation 
$p(y|x,{h}^*)$ have the same
mean 
\be
t = \int \!dy\, y\, p(y|x,D,D_0) = \int \!dy\, y\, p(y|x,{h}^*)
.
\ee
The variances, however, differ 
by the term ${\bf I}_x {\bf K}^{-1}{\bf I}_x$.

According to the results of Section \ref{some-loss-functions}
the mean of the predictive density
is the optimal choice under squared--error loss (\ref{squared-loss}).
For Gaussian regression, therefore
the optimal regression function
$a^*(x)$ is the same for squared--error loss
in exact and in maximum posterior treatment 
and thus also for log--loss 
(for Gaussian $p(y|x,a)$ with fixed variance)
\be
a^*_{\rm MPA,log}
=a^*_{\rm exact,log}
= a^*_{\rm MPA,sq.}
= a^*_{\rm exact,sq.}
={h}^* = t.
\ee
In case the space of possible $p(y|x,a)$ 
is not restricted to Gaussian densities with fixed variance,
the variance of the optimal density under log--loss
$p(y|x,a^*_{\rm exact,log})$
= $p(y|x,D,D_0)$ differs by ${\bf I}_x {\bf K}^{-1}{\bf I}_x$
from its maximum posterior approximation
$p(y|x,a^*_{\rm MPA,log})$
= $p(y|x,{h}^*)$.

\subsubsection{Gaussian mixture regression (cluster regression)}

Generalizing Gaussian regression
the likelihoods may be modeled by a mixture of $m$ Gaussians
\be
p(y|x,{h}) 
=
\frac{\sum_k^m p(k)\, e^{-\frac{\beta}{2} (y-h_k(x))^2}}
{\int \!dy\,\sum_k^m p(k)\, e^{-\frac{\beta}{2} (y-h_k(x))^2}}
,
\ee
where the normalization factor is found as
$\sum_k p(k) \left(\frac{\beta}{2\pi}\right)^{\frac{m}{2}}$.
Hence, $h$ is here specified by mixing coefficients $p(k)$
and a vector of regression functions
$h_k(x)$ 
specifying the $x$--dependent location of the $k$th cluster centroid
of the mixture model.
A simple prior for $h_k(x)$ 
is a smoothness prior diagonal in the cluster components.
As any density $p(y|x,h)$ can be approximated arbitrarily
well by a mixture with large enough $m$ 
such cluster regression models 
allows to interpolate between Gaussian regression
and more flexible density estimation.

The posterior density becomes for independent data
\be
p(h|D,D_0)
=
\frac{p(h|D_0)}{p(y_D|x_D,D_0)}
\prod_i^n \frac{\sum_k^m p(k)\, e^{-\frac{\beta}{2} (y_i-h_k(x_i))^2}}
{\sum_k^m p(k)\, \left(\frac{\beta}{2\pi}\right)^{\frac{m}{2}}}
.
\ee
Maximizing that posterior is 
--- for fixed $x$, uniform $p(k)$ and $p(h|D_0)$ ---
equivalent to the clustering approach
of Rose, Gurewitz, and Fox 
for squared distance costs \cite{Rose-Gurewitz-Fox-1990}.

\subsubsection{Support vector machines and regression}

\begin{figure}[tb]
\vspace{-6cm}
\begin{center}
\epsfig{file=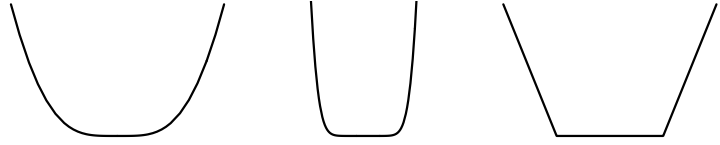, width=100mm}
\end{center}
\vspace{-6.5cm}
\caption{Three 
robust error functions which are 
insensitive to small errors.
Left: Logarithm of mixture with two Gaussians with equal variance 
and different means.
Middle: Logarithm of mixture with 11 Gaussians with equal variance 
and different means.
Right: $\epsilon$--insensitive error.}
\label{robust-pic}
\end{figure}

Expanding the regression function $h(x)$ in a basis of
eigenfunctions $\Psi_k$ of ${\bf K}_0$
\be
 K_0 = \sum_k \lambda_k \Psi_k \Psi_k^T 
,\quad
h(x) = \sum_k n_k \Psi_k (x)
\ee
yields for functional (\ref{regression-functional})
\be
E_h = \sum_i \left(\sum_k n_k \Psi_k (x_i)-y_i\right)^2
+\sum_k \lambda_k |n_k |^2 
.
\ee
Under the assumption of output noise for training data
the data terms may for example be replaced by the logarithm
of a mixture of Gaussians.
Such mixture functions with varying mean can develop flat regions
where the error is insensitive (robust) to changes of $h$.
Analogously, Gaussians with varying mean can be added
to obtain errors which are flat compared to Gaussians for large 
absolute errors.
Similarly to such Gaussian mixtures 
the mean--square error data term
$(y_i-h(x_i))^2$ may be replaced
by an $\epsilon$--insensitive error $|y_i-h(x_i)|_\epsilon$,
which is zero for absolute errors smaller $\epsilon$ and linear 
for larger absolute errors (see Fig.\ref{robust-pic}).
This results in a quadratic programming problem
and is equivalent to Vapnik's support vector machine
\cite{Vapnik-1995,Girosi-1997,Vapnik-1998,Smola-Schoelkopf-1998,Smola-Schoelkopf-Mueller-1998,Dietrich-Opper-Sompolinsky-1999}.
For a more detailed discussion 
of the relation between support vector machines
and Gaussian processes see 
\cite{Wahba-1997,Schoelkopf-Burges-Smola-1998}.

\subsection{Classification}
\label{classification}

In classification 
(or pattern recognition) tasks 
the independent visible variable 
$y$ takes discrete values (group, cluster or pattern labels)
\cite{Berger-1980,Fukunaga-1990,Bishop-1995b,Devroye-Gyorfi-Lugosi-1996}.
We write
$y$ = $k$ and 
$p(y|x,h)$ = $P_k(x,h)$,
i.e., $\sum_k P_k(x,h)$ = $1$.
Having received 
classification data $D$ = $\{(x_i,k_i)|1\le i\le n\}$
the density estimation error functional 
for a prior on function $\phi$
(with components $\phi_k$
and $P$ = $P(\phi)$)
reads
\be
E_{\rm cl.}
=
\sum_i^n \ln P_{k_i}(x_i;\phi) 
+\frac{1}{2}\Big(\phi-t,\, {\bf K}\,(\phi-t)  \Big)
+(P(\phi), \Lambda_X)
.
\ee
In classification the scalar product corresponds to an integral
over $x$ and a summation over $k$, e.g.,
\be
\Big(\phi-t,\, {\bf K}\,(\phi-t)  \Big)
=
\sum_{k,k^\prime} \int\! dx\,dx^\prime
(\phi_k(x)-t_k(x)) {\bf K}_{k,k^\prime}(x,x^\prime)
(\phi_{k^\prime}(x^\prime)-t_{k^\prime}(x^\prime))
,
\ee
and
$(P,\Lambda_X)$ = $\int\!dx\,\Lambda_X(x)\sum_k P_k(x)$.

For zero--one loss $l(x,k,a)$ = $\delta_{k,a(x)}$
--- a typical loss function for classification problems ---
the optimal decision (or {\it Bayes classifier}) is given by the mode
of the predictive density
(see Section \ref{some-loss-functions}), i.e.,
\be
a(x) = {\rm argmax}_k \, p(k|x,D,D_0)
.
\ee
In saddle point approximation
$p(k|x,D,D_0)\approx p(k|x,\phi^*)$
where $\phi^*$ minimizing $E_{\rm cl.}(\phi)$
can be found by solving the stationarity equation
(\ref{stat-eq-nzm}).

For the choice $\phi_k=P_k$ 
non--negativity and normalization 
must be ensured.
For $\phi=L$ with $P=e^L$
non--negativity is automatically fulfilled
but the Lagrange multiplier must 
be included to ensure normalization.

Normalization is guaranteed by using
unnormalized probabilities
$\phi_k=z_k$, $P=z_k/\sum_l z_l$
(for which non--negativity has to be checked)
or 
shifted log--likelihoods
$\phi_k=g_k$ with
$g_k = L_k +\ln \sum_l e^{L_l}$, i.e.,
$P_k$ = $e^{g_k }/\sum_l e^{g_l}$.
In that case the nonlocal normalization terms are part of the likelihood
and no Lagrange multiplier has to be used 
\cite{Williams-Barber-1998}.
The resulting equation can be solved in the space defined by the $X$--data
(see Eq.\ (\ref{stat-eq-data-sp})).
The restriction of $\phi_k$ = $g_k$ 
to linear functions $\phi_k(x) = w_k x +b_k$
yields log--linear models \cite{McCullagh--Nelder-1989}.
Recently  a mean field theory for 
Gaussian Process classification has been developed
\cite{Opper-1999,Opper-Winther-1999}.

Table \ref{type-table} lists
some special cases of density estimation.
The last line of the table, referring to
inverse quantum mechanics,
will be discussed in the next section.

\begin{table}
\begin{center}
\begin{tabular}{|c|c|}
\hline\rule[-3mm]{0mm}{9mm}
likelihood $p(y|x,h)$&problem type\\
\hline
\hline\rule[-3mm]{0mm}{9mm}
of general form  & density estimation \\ 
\hline\rule[-3mm]{0mm}{9mm}
discrete $y$  & classification \\ 
\hline\rule[-3mm]{0mm}{9mm}
Gaussian with fixed variance  & regression \\ 
\hline\rule[-3mm]{0mm}{9mm}
mixture of Gaussians  & clustering \\ 
\hline\rule[-3mm]{0mm}{9mm}
quantum mechanical likelihood& inverse quantum mechanics \\ 
\hline
\end{tabular}
\end{center}
\caption{Special cases of density estimation}
\label{type-table}
\end{table}

\subsection{Inverse quantum mechanics}
\label{inverse-quantum-mechanics}

Up to now we have formulated the learning problem
in terms of a function $\phi$
having a simple, e.g., pointwise, relation to $P$.
Nonlocalities in the relation between $\phi$ and $P$ 
was only due to the
normalization condition,
or, working with the distribution function,
due to an integration.
{\it Inverse problems for quantum mechanical systems}
provide examples
of more complicated, nonlocal
relations between likelihoods 
$p(y|x,h)$ = $p(y|x,\phi)$
and the hidden variables $\phi$ 
the theory is formulated in.
To show the flexibility of Bayesian Field Theory 
we will give in the following a short introduction
to its application to inverse quantum mechanics.
A more detailed discussion of inverse quantum problems 
including numerical applications
can be found in
\cite{Lemm-1998,Lemm-1999b,Lemm-1999c,Lemm-1999d,Uhlig-PhD}.
 
The state of a quantum mechanical systems can be 
completely described by giving its density operator
$\rho$.
The density operator of a specific system
depends on its preparation
and its Hamiltonian,
governing the time evolution of the system.
The inverse problem of quantum mechanics
consists in the
reconstruction of $\rho$ from observational data.
Typically, one studies systems with identical preparation
but differing Hamiltonians.
Consider for example Hamiltonians of the form
${\bf H} =  {\bf T} + {\bf V}$,
consisting of a kinetic energy part ${\bf T}$ 
and a potential ${\bf V}$.
Assuming the kinetic energy to be fixed,
the inverse problem 
is that of reconstructing the potential ${\bf V}$
from measurements.
A local potential 
${\bf V}(y,y^\prime )$ = $V(y)\delta (y-y^\prime )$
is specified by a function $V(y)$. 
Thus, for reconstructing a local potential
it is the function $V(y)$
which determines the likelihood
$p(y|x,h)$ = $p(y|{\bf X},\rho)$ = $p(y|{\bf X},V)$ = $P(\phi)$ 
and it is natural to formulate the prior in terms of 
the function $\phi$ = $V$.
The possibilities of implementing prior information for $V$
are similar to those
we discuss in this paper for general density estimation problems.
It is the likelihood model where inverse quantum mechanics differs
from general density estimation.

Measuring quantum systems
the variable $x$ corresponds 
to a hermitian operator ${\bf X}$.
The possible outcomes $y$ of measurements are given by
the eigenvalues of ${\bf X}$,
i.e.,
\be
{\bf X} |y> = y |y>
,
\ee
where $|y>$, with dual $<y|$,
denotes the eigenfunction with eigenvalue $y$.
(For the sake of simplicity we assume
nondegenerate eigenvalues,
the generalization to the degenerate case being 
straightforward.)
Defining the projector 
\be
\Pi_{{\bf X},y} = |y\!><\!y|
\ee
the likelihood model of quantum mechanics
is given by
\be
p(y|x,\rho) = {\rm Tr} (\Pi_{{\bf X},y} \rho)
.
\label{qm-likelihood}
\ee

In the simplest case, where the system
is in a pure state, say the ground state $\varphi_0$
of ${\bf H}$
fulfilling 
\be
{\bf H}|\varphi_0> = E_0|\varphi_0>
,
\label{H-eigen}
\ee
the density operator is
\be
\rho = \rho^2 = |\varphi_0\!><\!\varphi_0|
, 
\ee
and the likelihood (\ref{qm-likelihood})
becomes 
\be
p(y|x,h) 
=p(y|{\bf X},\rho) 
= {\rm Tr} (|\varphi_0\!><\!\varphi_0|y><y|)
= |\varphi_0(y)|^2
.
\ee
Other common choices for $\rho$ are shown in
Table \ref{qm-table}.

In contrast to ideal measurements on classical systems,
quantum measurements change the state of the system.
Thus, in case one is interested in repeated measurements 
for the same $\rho$, 
that density operator has to be prepared
before each measurement.
For a stationary state at finite temperature, for example, 
this can be achieved by waiting until the system is again
in thermal equilibrium.

\begin{table}
\begin{center}
\begin{tabular}{|c|c|}
\hline\rule[-3mm]{0mm}{9mm}
&$\rho$\\
\hline
\hline\rule[-3mm]{0mm}{9mm}
general pure state & $|\psi><\psi|$ \\
\hline\rule[-3mm]{0mm}{9mm}
stationary pure state & $|\varphi_i({\bf H}) ><\varphi_i({\bf H})|$ \\
\hline\rule[-3mm]{0mm}{9mm}
ground state & $|\varphi_0({\bf H})|><\varphi_0({\bf H})|$ \\
\hline\rule[-3mm]{0mm}{9mm}
time--dependent pure state &
$
     |{\bf U}(t,t_0)\psi(t_0)><{\bf U}(t,t_0)\psi(t_0)|
$\\
\hline\rule[-3mm]{0mm}{9mm}
scattering &
$
\lim_{t\rightarrow\infty\atop t_0\rightarrow-\infty}
     |{\bf U}(t,t_0)\psi(t_0)><{\bf U}(t,t_0)\psi(t_0)|
$
\\
\hline\rule[-3mm]{0mm}{9mm}
general mixture state & $\sum_k p(k) 
\;|\psi_k><\psi_k|$ \\
\hline\rule[-3mm]{0mm}{9mm}
stationary mixture state & $\sum_i p(i|{\bf H}) 
\;|\varphi_i ({\bf H}) ><\varphi_i ({\bf H}) |$ \\
\hline\rule[-3mm]{0mm}{9mm}
canonical ensemble & $
({\rm Tr}\, e^{-\beta {\bf H}})^{-1}
e^{-\beta {\bf H}} 
$\\ 
\hline
\end{tabular}
\end{center}
\caption{
The most common examples of density operators
for quantum systems.
In this table 
$\psi$ denotes an arbitrary pure state,
$\varphi_i$ represents an eigenstate
of Hamiltonian $H$.
The unitary time evolution operator 
for a time--independent Hamiltonian ${\bf H}$ 
is given by
${\bf U}$ = $e^{-i(t-t_0) {\bf H}}$.
In scattering one imposes typically
additional specific 
boundary conditions on the initial and final states.
}
\label{qm-table}
\end{table}

For a Maximum A Posteriori Approximation
the functional derivative of
the likelihood is needed.
Thus, for reconstructing a local potential
we have to calculate
\be
\delta_{V(y)} p(y_i|{\bf X},V)
.
\ee
To be specific, let us assume we measure
particle coordinates, meaning we have chosen ${\bf X}$
to be the coordinate operator.
For a system prepared 
in the ground state of its Hamiltonian ${\bf H}$,
we then have to find,
\be
\delta_{V(y)}
 |\varphi_0 (y_i)|^2
\label{qm-likelihood-stationary}.
\ee
For that purpose, we take the functional derivative
of Eq.\ (\ref{H-eigen}), which yields
\be
({\bf H}-E_0)|\delta_{V(y)} \varphi_0\!>
=
(\delta_{V(y)} {\bf H} -\delta_{V(y)} E_0)|\varphi_0\!>
.
\ee
Projecting from the left by $<\!\varphi_0|$,
using again Eq.\ (\ref{H-eigen}) 
and the fact that for a local potential
$\delta_{V(y)} {\bf H}(y^\prime,y^{\prime\prime})$ 
= $\delta (y-y^\prime)\delta(y^\prime-y^{\prime\prime})$,
shows that
\be
\delta_{V(y)} E_0
=
<\!\varphi_0|\delta_{V(y)} {\bf H}|\varphi_0\!>
= |\varphi_0(y)|^2
.
\ee
Choosing
$<\!\varphi_0|\delta_{V(y)} \varphi_0\!>$ = 0
and inserting
a complete basis of eigenfunctions $|\varphi_j\!>$
of ${\bf H}$,
we end up with
\be
\delta_{V(y)} \varphi_0 (y_i)
=
\sum_{j\ne 0} \frac{1}{E_0-E_j}
\varphi_j (y_i) \varphi_j^* (y) \varphi_0 (y)
. 
\ee
From this the functional derivative of the 
quantum mechanical log--likelihood (\ref{qm-likelihood-stationary})
corresponding to data point $y_i$
can be obtained easily,
\be
\delta_{V(y)} \ln p(y_i|{\bf X},V)
=
2 {\rm Re} \left( \varphi_0(y_i)^{-1} \delta_{V(y)} \varphi_0 (y_i)\right) 
.
\label{derivative-qm-likelihood}
\ee
The MAP equations
for inverse quantum mechanics are obtained
by including the functional derivatives of the
prior term for $V$.
In particular, for a Gaussian prior with mean $V_0$
and inverse covariance ${\bf K}_V$,
acting in the space of potential functions $V(y)$,
its negative logarithm, i.e., its prior error functional,
reads
\be
\frac{1}{2} \Big(V-V_0,\; {\bf K}_V \,(V-V_0)\Big)
+\ln Z_V
,
\ee
with $Z_V$ being the $V$--independent constant
normalizing the prior over $V$.
Collecting likelihood and prior terms,
the stationarity equation finally becomes
\be
0=
\sum_i \delta_{V(y)} \ln p(y_i|{\bf X},V)
-
{\bf K}_V \,(V-V_0)
.
\ee

The Bayesian approach to inverse quantum problems
is quite flexible and can be used for many different learning
scenarios and quantum systems.
By adapting Eq.\ (\ref{derivative-qm-likelihood}),
it can deal with measurements of different observables,
for example, coordinates, momenta, energies,
and with other density operators, 
describing, for example, 
time--dependent states
or systems at finite temperature \cite{Lemm-1999b}.

The treatment of bound state or scattering problems 
for quantum many--body systems
requires additional approximations.
Common are, for example, mean field methods,
for bound state problems 
\cite{Eisenberg-Greiner-1972,Ring-Schuck-1980,Blaizot-Ripka-1986}
as well as for scattering theory
\cite{Goeke-Cusson-Gruemmer-Reinhard-Reinhardt-1983,Blaizot-Ripka-1986,Lemm-1990,Lemm-1994,Lemm-1995a,Lemm-1995b,Uhlig-Lemm-Weiguny-1998}
Referring to such mean field methods
inverse quantum problems can also be treated
for many--body systems \cite{Lemm-1999c}.

\section[Parameterizing likelihoods: Variational methods]
        {Parameterizing likelihoods: Variational\\ methods}
\label{variational}

\subsection{General parameterizations}

\label{general-variational}

Approximate
solutions of the error minimization problem are obtained
by restricting the search (trial) space for $h(x,y)$ = $\phi (x,y)$
(or $h(x)$ in regression).
Functions $\phi$ which are in the considered search space are 
called {\it trial functions}.
Solving a minimization problem in some restricted trial space
is also called a {\it variational approach}
\cite{Hilbert-Courant-1989,Jeggle-1979,Blanchard-Bruening-1982,Choquet-Bruhat-DeWitt-Morette-Dillard-Bleick-1982,Blaizot-Ripka-1986}.
Clearly, minimal values obtained by
minimization within a trial space can only be larger or equal
than the true minimal value,
and from two variational approximations
that with smaller error is the better one.

Alternatively, 
using parameterized functions $\phi$ can also implement the prior
where $\phi$ is known to have that specific parameterized form.
(In cases where $\phi$ is only known
to be approximately of a specific parameterized form,
this should ideally be implemented using a prior with a parameterized template
and the parameters be treated as hyperparameters
as in Section \ref{hyperparameters}.)
The following discussion holds for both interpretations.

Any parameterization $\phi$ = $\phi (\{\xi_l\})$ 
together with a range of allowed values
for the parameter vector $\xi$
defines a possible trial space.
Hence we consider the error functional
\begin{equation}
E_{\phi (\xi)} =
-(\,\ln P (\xi),\, N\,)
+\frac{1}{2} (\,\phi (\xi),\, {{\bf K}}\,\phi (\xi)\,)
+ (\,P( \xi ),\,\Lambda_X\,),
\label{error-xi}
\end{equation}
for $\phi$ depending on parameters $\xi$ 
and $p(\xi )$ = $p(\,\phi (\xi)\,)$.
In the special case of Gaussian regression
this reads
\begin{equation}
E_{h (\xi)} =
\frac{1}{2} (\,h (\xi)-t_D,\, {{\bf K}_D}\,h (\xi)-t_D\,)
+\frac{1}{2} (\,h (\xi),\, {{\bf K}}\,h (\xi)\,)
.
\label{error-xi-h}
\end{equation}
Defining the matrix
\be
\Phi^\prime (l;x,y) 
= \frac{\partial \phi (x,y)}{\partial \xi_l}
\end{equation} 
the stationarity equation for the functional (\ref{error-xi})
becomes
\begin{equation}
0 =
\Phi^\prime {\bf P}^\prime {\bf P}^{-1} N  
- \Phi^\prime {{\bf K}}\phi 
-\Phi^\prime  {\bf P}^\prime \Lambda_X   
.
\label{varlag}
\end{equation}
Similarly, 
a parameterized functional $E_\phi$ with non--zero template $t$ 
as in (\ref{templatefunctional}) would give
\begin{equation}
0 =
\Phi^\prime {\bf P}^\prime  {\bf P}^{-1} N  
- \Phi^\prime {{\bf K}}\left( \phi  - t\right)
- \Phi^\prime  {\bf P}^\prime  \Lambda_X
.
\label{parametricSE}
\end{equation}
To have a convenient notation when solving for $\Lambda_X$
we introduce
\be
{\bf P}^\prime_\xi 
= \Phi^\prime (\xi) {\bf P}^\prime(\phi),
\ee 
i.e.,
\be
{\bf P}^\prime_\xi (l;x,y) 
= \frac{\partial P(x,y)}{\partial \xi_l}
=\int\!dx^\prime dy^\prime\,
\frac{\partial \phi (x^\prime,y^\prime)}{\partial \xi_l}
\frac{\delta P(x,y)}{\delta \phi (x^\prime ,y^\prime )}
,
\end{equation} 
and
\be
G_{\phi(\xi)}
={\bf P}^\prime_\xi {\bf P}^{-1} N  
- \Phi^\prime {{\bf K}}\phi 
,
\ee
to obtain for Eq.\ (\ref{varlag})
\begin{equation}
{\bf P}^\prime_\xi \Lambda_X  
= G_{\phi(\xi)}
.
\label{varlag2}
\end{equation}
For a parameterization $\xi$ 
restricting the space of possible $P$ 
the matrix
${\bf P}^\prime_\xi$ is not square
and cannot be inverted.
Thus, let 
$({\bf P}^\prime_\xi)^{\#}$
be the Moore--Penrose inverse of 
${\bf P}^\prime_\xi$, i.e.,
\be
({\bf P}^\prime_\xi)^{\#} {\bf P}^\prime_\xi 
({\bf P}^\prime_\xi)^{\#}
=
{\bf P}^\prime_\xi 
,\quad
{\bf P}^\prime_\xi 
({\bf P}^\prime_\xi)^{\#}
{\bf P}^\prime_\xi 
=
({\bf P}^\prime_\xi)^{\#}
,
\ee
and symmetric
$({\bf P}^\prime_\xi)^{\#} {\bf P}^\prime_\xi$ 
and ${\bf P}^\prime_\xi ({\bf P}^\prime_\xi)^{\#}$.
A solution for $\Lambda_X$ exists
if
\be
{\bf P}^\prime_\xi ({\bf P}^\prime_\xi)^{\#}
G_{\phi(\xi)}
=G_{\phi(\xi)}
.
\label{lamcond}
\ee
In that case the solution can be written
\be
\Lambda_X = 
({\bf P}^\prime_\xi)^{\#} G_{\phi(\xi)}
+ V_\Lambda - ({\bf P}^\prime_\xi)^{\#} {\bf P}^\prime_\xi V_\Lambda
,
\label{pseudo-sol}
\ee
with arbitrary vector $V_\Lambda$
and 
\be
\Lambda_X^0 = V_\Lambda 
               - ({\bf P}^\prime_\xi)^{\#}{\bf P}^\prime_\xi V_\Lambda
\ee
from the right null space of
${\bf P}^\prime_\xi$,
representing a solution of 
\be
{\bf P}_\xi^\prime  \Lambda_X^0 = 0
.
\ee
Inserting for $\Lambda_X(x) \ne 0$
Eq.\ (\ref{pseudo-sol})
into the normalization condition
$\Lambda_X$ = ${\bf I}_X {\bf P} \Lambda_X$
gives
\be
\Lambda_X = 
{\bf I}_X {\bf P} \left(
({\bf P}^\prime_\xi)^{\#} G_{\phi(\xi)}
+ V_\Lambda - ({\bf P}^\prime_\xi)^{\#} {\bf P}^\prime_\xi  V_\Lambda
\right)
.
\ee
Substituting back in Eq.\ (\ref{varlag})
$\Lambda_X$ is eliminated yielding as stationarity equation
\be
0 =
\left(
{\bf I} - {\bf P}^\prime_\xi {\bf I}_X {\bf P} ({\bf P}^\prime_\xi)^{\#}
\right) G_{\phi(\xi)}
- {\bf P}^\prime_\xi {\bf I}_X {\bf P} 
\left(
V_\Lambda - ({\bf P}^\prime_\xi)^{\#} {\bf P}^\prime_\xi  V_\Lambda
\right)
,
\label{xistat}
\ee
where $G_{\phi(\xi)}$ has to fulfill Eq.\ (\ref{lamcond}).
Eq.\ (\ref{xistat}) may be written in a form
similar to Eq.\ (\ref{Ophi=Tphi})
\be
{{\bf K}}_{\phi(\xi)}( \xi) = T_{\phi(\xi)} 
\label{Oxi=Txi}
\ee
with 
\be
T_{\phi(\xi)} (\xi)
= {\bf P}_\xi^\prime {\bf P}^{-1}N  
-{\bf P}_\xi^\prime \Lambda_X 
,
\ee
but with
\be
{{\bf K}}_{\phi(\xi)} (\xi)
= \Phi^\prime {{\bf K}} \Phi(\xi)
,
\ee
being in general a nonlinear operator.

\subsection{Gaussian priors for parameters}
\label{gauss-param}

Up to now we assumed the prior to be given for a function
$\phi (\xi)(x,y)$ depending on $x$ and $y$.
Instead of a prior in a function $\phi (\xi)(x,y)$ 
also a prior in another not $(x,y)$--dependent function
of the parameters $\psi(\xi)$ can be given.
A Gaussian prior in 
$\psi (\xi) = W_{\psi} \xi$ being a linear function of $\xi$,
results in a prior which is also Gaussian in 
the parameters $\xi$, giving a regularization term
\be
\frac{1}{2} (\,\xi,\, W_{\psi}^T {{\bf K}}_\psi W_{\psi}\,\xi\,)
=
\frac{1}{2} (\,\xi,\, {{\bf K}}_\xi\,\xi\,),
\ee
where ${{\bf K}}_\xi$ = $W_{\psi}^T {{\bf K}}_\psi W_{\psi}$
is not an operator in a space of functions $\phi (x,y)$
but a matrix in the space of parameters  $\xi$.
The results of Section \ref{general-variational}
apply to this case provided the following replacement is made
\be
\Phi^\prime {{\bf K}} \phi 
\rightarrow {{\bf K}}_\xi \xi
.
\ee
Similarly,
a nonlinear $\psi$ requires the replacement
\be
\Phi^\prime {{\bf K}} \phi 
\rightarrow 
{\Psi}^\prime {{\bf K}}_\psi \psi
,
\label{replace}
\ee
where
\be
\Psi^\prime (k,l)
= 
\frac{\partial \psi_l (\xi) }{\partial \xi_k}
.
\ee
Thus, in the general case where a Gaussian (specific) prior in 
$\phi(\xi)$ and $\psi(\xi)$ is given, 
\bea
E_{\phi (\xi),\psi(\xi)} &=&
-(\,\ln P (\xi),\, N\,)+ (\,P( \xi ),\,\Lambda_X\,)
\nonumber\\
&&+\frac{1}{2} (\,\phi (\xi),\, {{\bf K}}\,\phi (\xi)\,)
+\frac{1}{2} (\,\psi (\xi),\, {{\bf K}}_\psi\,\psi (\xi)\,)
,
\label{error-phi-psi-xi}
\eea
or, including also non--zero template functions (means)
$t$, $t_\psi$
for $\phi$ and $\psi$
as discussed in Section \ref{Non--zero-means},
\bea
E_{\phi (\xi),\psi(\xi)} &=&
-(\,\ln P (\xi),\, N\,)+ (\,P( \xi ),\,\Lambda_X\,)
\nonumber\\&&
+\frac{1}{2} (\,\phi (\xi)-t,\, {{\bf K}}\,(\phi (\xi)-t)\,)
\nonumber\\&&
+\frac{1}{2} (\,\psi (\xi)-t_\psi,\, {{\bf K}}_\psi\,(\psi (\xi)-t_\psi)\,)
.
\label{error-phi-psi-xi-templ}
\eea
The $\phi$ and $\psi$--terms of the energy
can be interpreted as corresponding to
a probability 
$p(\xi|t,{{\bf K}},t_\psi,{{\bf K}}_\psi)$,
($\ne p(\xi|t,{{\bf K}})$
$p(\xi|t_\psi,{{\bf K}}_\psi)$),
or, for example, 
to $p(t_\psi|\xi,{{\bf K}}_\psi)$ $p(\xi|t,{{\bf K}})$
with one of the two terms term
corresponding to a Gaussian likelihood
with $\xi$--independent normalization.

The stationarity equation becomes
\bea
 0 &=&  
{\bf P}_\xi^\prime  {\bf P}^{-1} N  
-{\Phi}^\prime {{\bf K}} (\phi-t)
-{\Psi}^\prime {{\bf K}}_\psi (\psi-t_\psi)
   -{\bf P}_\xi^\prime \Lambda_X
\\&=&
G_{\phi,\psi}
-{\bf P}_\xi^\prime \Lambda_X  
,
\label{genstat}
\eea
which defines $G_{\phi,\psi}$,
and for $\Lambda_X\ne0$   
\be
\Lambda_X
=
  {\bf I}_X {\bf P} \left(
  ({\bf P}^\prime_\xi)^{\#} G_{\phi,\psi} +\Lambda_X^0 
  \right)
,
\ee
for 
${\bf P}_\xi^\prime  \Lambda_X^0 = 0$.

\begin{table}
\begin{tabular}{|c|c|c|c|}
\hline
Variable & Error & Stationarity equation & $\Lambda_X$  
\\
\hline
\hline\rule[-3mm]{0mm}{9mm}
$\!L(x,y)$ & 
$E_L$ &
  $ {{\bf K}} L  = N - {\bf e^L} \Lambda_X$
&
  ${\bf I}_X \left( N - {{\bf K}} L \right)$
\\
\hline\rule[-3mm]{0mm}{9mm}
$\!P(x,y)$ & 
$E_P$ &
  ${{\bf K}} P =
   {\bf P}^{-1} N 
    -\Lambda_X$
&
  ${\bf I}_X (N - {\bf P}{{\bf K}} P)$
\\
\hline\rule[-3mm]{0mm}{9mm}
$\!\phi=\sqrt{P}$ & 
$E_{\sqrt{P}}$ &
  ${{\bf K}} \phi =
   2{\Phi}^{-1} N 
    -2\Phi  \Lambda_X$
&
  ${\bf I}_X (N - \frac{1}{2}\Phi {{\bf K}} \phi)$
\\
\hline\rule[-3mm]{0mm}{9mm}
$\!\phi (x,y)$ &
$E_{\phi}$ &
  ${{\bf K}}\phi =
   {\bf P}^\prime  {\bf P}^{-1} N  
   -{\bf P}^\prime \Lambda_X$
&
  ${\bf I}_X \left( N - {\bf P} {{\bf P}^\prime}^{-1} {{\bf K}}\, \phi \right)$
\\
\hline\rule[-3mm]{0mm}{9mm}
$\!\xi$ & 
$E_{\phi(\xi)}$ &
  $\Phi^\prime {{\bf K}}\phi 
  =
    {\bf P}_\xi^\prime {\bf P}^{-1} N  
   -{\bf P}_\xi^\prime \Lambda_X$   
&
  ${\bf I}_X {\bf P} \left(
  ({\bf P}^\prime_\xi)^{\#} G_{\phi(\xi)} +\Lambda^0_X 
  \right)\!$
\\
\hline\rule[0mm]{0mm}{6mm}
$\!\xi$ & 
$E_{\phi(\xi)\psi(\xi)}$ &
  $\Phi^\prime {{\bf K}}(\phi\!-\!t) 
  + \Psi^\prime {{\bf K}}_\psi (\psi\!-\!t_\psi)$ 
&
  ${\bf I}_X {\bf P} \left(
  ({\bf P}^\prime_\xi)^{\#} G_{\phi,\psi}+\Lambda^0_X 
  \right)\!$
\\
\rule[-3mm]{0mm}{0mm}
&& 
   $={\bf P}_\xi^\prime {\bf P}^{-1} N  
   -{\bf P}_\xi^\prime \Lambda_X$   
&
\\
\hline
\end{tabular}
\caption{Summary of stationarity equations.
For notations, conditions and comments see
Sections 
\ref{error-e-l},
\ref{error-e-p},
\ref{Square-root},
\ref{general-Gaussian},
\ref{general-variational}
and \ref{gauss-param}.
}
\end{table}

\subsection{Linear trial spaces}

Solving a density estimation problem numerically,
the function $\phi$ has to be discretized.
This is done by expanding $\phi$ in a basis
$B_l$ (not necessarily orthonormal)
and,
choosing some $l_{\rm max}$, 
truncating the sum to terms with $l\le l_{\rm max}$,
\begin{equation}
\phi = \sum_{l=1}^\infty c_l B_l
\rightarrow
\phi = \sum_{l=1}^{l_{\rm max}}  c_l B_l
.
\label{discrete}
\end{equation}
This, also called Ritz's  method, corresponds 
to a finite linear trial space
and is equivalent
to solving a projected stationarity equation.
Using a {\it discretization} (\ref{discrete})
the functional (\ref{phifunctional})
becomes
\begin{equation}
E_{\rm Ritz} =
-(\,\ln P(\phi),\,N\,)
+\frac{1}{2}\sum_{kl} c_k c_l (\,B_k,\,{{\bf K}}\,B_l\,)
+ (\,P(\phi),\,\Lambda_X\,).
\end{equation}
Solving for the coefficients $c_l$, $l \le  l_{\rm max}$ 
to minimize the error results 
according to Eq.[\ref{varlag}) and
\begin{equation}
\Phi^\prime (l;x,y) 
= B_l(x,y),
\end{equation} 
in
\begin{equation}
0 =
(\, B_l ,\, {\bf P}^\prime {\bf P}^{-1}\,N\,)  
- \sum_k c_k (\,B_l,\, {{\bf K}}\,B_k\,)
- (\,B_l ,\, {\bf P}^\prime \, \Lambda_X\,)
, \forall l\le l_{\rm max},
\end{equation}
corresponding to the $l_{\rm max}$--dimensional equation 
\begin{equation}
{{\bf K}}_B c =
N_B (c) 
- \Lambda_B (c),
\label{phistatlin}
\end{equation}
with 
\bea
c(l) &=& c_l,
\\
{{\bf K}}_B (l,k) &=& (\,B_l,\, {{\bf K}}\,B_k\,),  
\\
N_B (c)(l) 
&= & (\,B_l,\,{\bf P}^\prime(\phi(c))\,{\bf P}^{-1}(\phi(c))\,N\,), 
\\
\Lambda_B (c)(l) 
&=&  (\, B_l,\, {\bf P}^\prime(\phi (c)) \,\Lambda_X\,). 
\eea
Thus, for an orthonormal basis $B_l$
Eq.\ (\ref{phistatlin}) corresponds
to Eq.\ (\ref{phistat1}) projected into the trial space
by $\sum_l B_l^T\,B_l$.

The so called {\it  linear models} are obtained by the 
(very restrictive) choice
\be
\phi(z) = \sum_{l=0}^{1} c_l B_l = c_0 + \sum_l c_l z_l
\ee
with $z=(x,y)$ and 
$B_0$ = 1 and $B_l$ = $z_l$.
Interactions, i.e., terms proportional to
products of $z$--components like $c_{mn}z_mz_n$ can be included.
Including all possible interaction would correspond to a 
multidimensional Taylor expansion
of the function $\phi$(z).

If the functions $B_l(z)$ are also parameterized
this leads to mixture models for $\phi$.
(See Section \ref{mixture-models}.)

\subsection{Mixture models}
\label{mixture-models}

The function $\phi(z)$ can be approximated by a mixture model,
i.e., by a linear combination of components functions
\be
\phi(z) = \sum c_l B_l(\xi_l,z),
\ee
with parameter vectors $\xi_l$ 
and constants $c_l$ (which could also be included into the vector $\xi_l$)
to be adapted.
The functions
$B_l(\xi_l,z)$
are often chosen to depend on one--dimensional
combinations of the vectors $\xi_l$ and $z$.
For example they may depend on some distance
$||\xi_l - z||$ (`local or distance approaches')
or the projection of $z$ in $\xi_l$--direction, i.e.,
$\sum_k \xi_{l,k} z_k$ (`projection approaches').
(For projection approaches see also Sections 
\ref{additive-modell}, \ref{proj-pursuit} and \ref{neural}).

A typical example are Radial Basis Functions (RBF)
using Gaussian $B_l(\xi_l,z)$
for which centers 
(and possibly covariances and also number of components)
can be adjusted.
Other local  methods include 
$k$--nearest neighbors methods ($k$NN)
and learning vector quantizations (LVQ)
and its variants.
(For a comparison see \cite{StatLog-1994}.)

\subsection{Additive models}
\label{additive-modell}

Trial functions $\phi$ may be chosen as sum 
of simpler functions $\phi_l$ each depending only
on part of the $x$ and $y$ variables.
More precisely, we consider functions $\phi_l$
depending on projections $z_l$ = ${\bf I}_l^{(z)} z$
of the vector $z$ = $(x,y)$
of all $x$ and $y$ components.
${\bf I}_l^{(z)}$ denotes an projector in the vector space of 
$z$ (and not in the space of functions $\Phi(x,y)$).
Hence, $\phi$ becomes of the form
\begin{equation}
\phi (z) = \sum_l \phi_l (z_l)
,
\label{additive}
\end{equation}
so only one--dimensional functions
$\phi_l$ have to be determined.
Restricting the functions
$\phi_l$ to a
parameterized function space
yields 
a ``parameterized additive model''
\begin{equation}
\phi (z) = \sum_l \phi_l (\xi, z_l),
\label{additive3}
\end{equation}
which has to be solved for the parameters $\xi$.
The model can also be generalized 
to a model ``additive in parameters $\xi_l$''
\begin{equation}
\phi (z) = \sum_l \phi_l (\xi_l,x,y),
\label{additive2}
\end{equation}
where the functions $\phi_l (\xi_l,x,y)$
are not restricted to one--dimensional functions
depending only on projections $z_l$ on the coordinate axes.
If the parameters $\xi_l$ determine the component functions 
$\phi_l$ completely, this yields just the mixture models
of Section \ref{mixture-models}.
Another example is projection pursuit,
discussed in Section \ref{proj-pursuit}),
where a parameter vector $\xi_l$
corresponds to a projections $\xi_l \cdot z$.
In that case even for given $\xi_l$ still a one--dimensional function 
$\phi_l(\xi_l \cdot z)$ has to be determined.

An ansatz like (\ref{additive}) is made more flexible
by including also interactions 
\begin{equation}
\phi (x,y) = 
\sum_l \phi_l (z_l)
+\sum_{kl} \phi_{kl} (z_k, z_l)
+\sum_{klm} \phi_{klm} (z_k, z_l, z_m) +
\cdots .
\end{equation}
The functions $\phi_{kl\cdots} (z_k, z_l,\cdots)$
can be chosen to depend on
product terms
like
$z_{l,i} z_{k,j}$, 
or $z_{l,i} z_{k,j} z_{m,n}$,
where $z_{l,i}$ denotes one--dimensional 
sub-variables of $z_l$.

In additive models in the narrower sense 
\cite{Stone-1985,Hastie-Tibshirani-1986,Hastie-Tibshirani-1987,Hastie-Tibshirani-1990}
$z_l$ is a subset of $x$, $y$ components, i.e., 
$z_l \subseteq 
\{ x_i | 1\le i\le d_x \}$ $\cup$ $\{ y_j | 1\le j\le d_y \}$,
$d_x$ denoting the dimension of $x$,
$d_y$ the dimension of $y$.
In regression, for example, one takes usually the one--element subsets 
$z_l$ = $\{x_l\}$ for $1\le l \le d_x$.

In more general schemes the projections of $z$ 
do not have to be restricted to projections on the coordinates axes.
In particular, the projections can be optimized too.
For example, one--dimensional projections
${\bf I}_l^{(z)} z $ = $w\cdot z$ with $z,w\in X\times Y$
(where $\cdot$ denotes a scalar product 
in the space of $z$ variables)
are used by {\it ridge approximation} schemes. 
They include for regression problems 
one--layer (and similarly multilayer)
feedforward neural networks
(see Section \ref{neural})
projection pursuit regression 
(see Section \ref{proj-pursuit})
and hinge functions \cite{Breiman-1993}.
For a detailed discussion of the regression case see
\cite{Girosi-Jones-Poggio-1995}.

The stationarity equation for $E_\phi$
becomes 
for the ansatz (\ref{additive})
\be
0 = {\bf P}^\prime_l {\bf P}^{-1} N 
-{{\bf K}} \phi 
-{\bf P}^\prime_l \Lambda_X
,
\label{add1-stat}
\ee
with 
\be
{\bf P}^\prime_l (z_l,z^\prime) 
= \frac{\delta P(z^\prime)}{\delta \phi_l(z_l)}
.
\label{add1-der}
\ee
Considering a density $P$ being also decomposed into
components $P_l$ determined
by the components $\phi_l$ 
\be
P(z) = \sum_l P_l(\phi_l(z_l)),
\ee
the derivative (\ref{add1-der}) becomes
\be
{\bf P}^\prime_l (z_l,z_k^\prime) 
= \frac{\delta P_l(z_l^\prime)}{\delta \phi_l(z_l)}
,
\ee
so that specifying an additive prior 
\be
\frac{1}{2} \sum_{kl}(\,\phi_k -t_k,\, {{\bf K}_{kl}}\,(\phi_l-t_l)\,)
,
\ee
the stationary conditions are coupled equations 
for the component functions $\phi_l$
which, because ${\bf P}$ is diagonal,
only contain integrations over $z_l$--variables
\be
0=
\frac{\delta P_l}{\delta \phi_l}
{\bf P}^{-1} N 
-\sum_k {\bf K}_{lk} (\phi_k-t_k)
-\frac{\delta P_l}{\delta \phi_l} \Lambda_X.
\ee

For the parameterized approach (\ref{additive3}) 
one finds 
\be
0 = \Phi_l^\prime {\bf P}^\prime_l {\bf P}^{-1} N 
-\Phi_l^\prime {{\bf K}} \phi 
-\Phi_l^\prime {\bf P}^\prime_l \Lambda_X
,
\ee
with
\be
\Phi_l^\prime (k,z_l) 
= \frac{\partial \phi_l(z_l)}{\partial \xi_k}
.
\ee
For the ansatz (\ref{additive2})
$\Phi_l^\prime (k,z)$ would be restricted to a subset of $\xi_k$.

\subsection{Product ansatz}

A product ansatz has the form
\begin{equation}
\phi (z) = \prod_l \phi_l (z_l)
,
\label{product}
\end{equation}
where 
$z_l$ = ${\bf I}_l^{(z)}z$
represents projections 
of the vector $z$ consisting of all $x$ and $y$ components. 
The ansatz can be made more flexible by using sum of products
\begin{equation}
\phi (z) = \sum_k \prod_l \phi_{k,l} (z_l).
\end{equation}
The restriction of the trial space to product functions
corresponds to the Hartree approximation in physics.
(In a Hartree--Fock approximation
the product functions are antisymmetrized under coordinate exchange.)

For additive ${{\bf K}}$ 
= $\sum_l {{\bf K}}_l$ 
with ${{\bf K}}_l$ acting only on $\phi_l$, i.e.,
${{\bf K}}_l$ = ${{\bf K}}_l \otimes 
\left( \bigotimes_{l^\prime \ne l} {\bf I}_{l^\prime} \right)$, 
with ${\bf I}_l$ the projector into the space of functions 
$\phi_l$ = ${\bf I}_l \phi_l$,
the quadratic regularization term becomes,
assuming ${\bf I}_l$ ${\bf I}_{l^\prime}$ = $\delta_{l,l^\prime}$,
\begin{equation}
(\,\phi,\, {{\bf K}}\,\phi\,)
= \sum_l (\,\phi_l,\, {{\bf K}}_l \, \phi_l\,)
\prod_{l^\prime \ne l} (\, \phi_{l^\prime},\, \phi_{l^\prime}\,)
.
\end{equation}
For ${{\bf K}}$ = $\bigotimes_{l} {{\bf K}}_{l}$
with a product structure with respect to $\phi_l$
\begin{equation}
(\,\phi,\, {{\bf K}}\,\phi\,)
= 
\prod_{l} (\,\phi_l ,\, {{\bf K}}_l \, \phi_l\,). 
\end{equation}
In both cases the prior term factorizes into lower dimensional
contributions.

\subsection{Decision trees}

Decision trees \cite{Breiman-Friedman-Olshen-Stone-1993}
implement functions which are
piecewise constant on rectangular areas
parallel to the coordinate axes $z_l$.
Such an approach can be written in tree structure
with nodes only performing comparisons
of the form $x<a$ or $x>a$ which allows a very effective
hardware implementation.
Such a piecewise constant approach can be written in the form
\be
\phi(z) = \sum_l c_{l} \prod_k \Theta(z_{\nu(l,k)}-a_{lk})
\ee
with step function $\Theta$
and $z_{\nu(l,k)}$ indicating the component of $z$
which is compared with the reference value $a_{lk}$.
While there are effective constructive methods to build trees
the use of gradient--based 
minimization or maximization methods would require,
for example, to replace the step function by a sigmoid.
In particular,
decision trees correspond to neural networks
at zero temperature, where sigmoids become step functions,
and which are restricted to
weights vectors in coordinate directions
(see Section \ref{neural}).

An overview over different variants
of decision trees together with a comparison with rule--based systems,
neural networks (see Section \ref{neural})
techniques from applied statistics like linear discriminants,
projection pursuit
(see Section \ref{proj-pursuit})
and local methods like for example $k$-nearest neighbors methods ($k$NN),
Radial Basis Functions (RBF),
or learning vector quantization (LVQ)
is given in \cite{StatLog-1994}.

\subsection{Projection pursuit}
\label{proj-pursuit}

Projection pursuit models 
\cite{Friedman-Stuetzle-1981,Huber-1985,Donoho-Johnstone-1989}
are a generalization of additive models
(\ref{additive})
(and a special case of models (\ref{additive2}) additive in parameters)
where the projections of $z$ = $(x,y)$ are also adapted
\begin{equation}
\phi (z) = \xi_0+\sum_l \phi_l (\xi_{0,l}+\xi_l \cdot z)
.
\label{proj-pursuit-Eq}
\end{equation}
For such a model one has to determine 
one--dimensional `ridge' functions $\phi_l$ 
together with
projections defined by vectors $\xi_l$
and constants $\xi_0$, $\xi_{0,l}$.
Adaptive projections may also be used for product approaches
\begin{equation}
\phi (z) = \prod_l \phi_l (\xi_{0,l}+\xi_l \cdot z)
.
\end{equation}
Similarly, $\phi$ may be decomposed into functions depending
on distances to adapted reference points (centers).
That gives models of the form
\begin{equation}
\phi (z) = \prod_l \phi_l (||\xi_l -z||),
\end{equation}
which require to adapt
parameter vectors (centers) $\xi_l$ 
and distance functions $\phi_l$.
For high dimensional spaces
the number of centers necessary 
to cover a high dimensional space
with fixed density grows exponentially.
Furthermore, as the volume of a 
high dimensional sphere tends to be concentrated near its surface,
the tails become more important in higher dimensions.
Thus, typically, projection methods are better suited for
high dimensional spaces than distance methods
\cite{Silverman-1986}.

\subsection{Neural networks}
\label{neural}

While in projection pursuit--like techniques
the one--dimensional `ridge' functions $\phi_l$
are adapted optimally, neural networks use ridge functions
of a fixed sigmoidal form.
The resulting lower flexibility following from fixing the ridge function
is then compensated by iterating this parameterization.
This leads to multilayer neural networks.

Multilayer neural networks have been become a popular 
tool for regression and classification problems
\cite{Rumelhart-PDP-1986,Lapedes-Farber-1988,Minski-Papert-1990,Hertz-Krogh-Palmer-1991,Mueller-Reinhardt-1991,Watkin-Rau-Biehl-1993,Bishop-1995b,Ripley-1996,Ballard-1997}.
One-layer neural networks, also known as perceptrons,
correspond to the parameterization
\begin{equation}
\phi (z) = \sigma \left(\sum_l w_l z_l-b\right)
=\sigma (v)
\label{single-layer}
,
\end{equation}
with a sigmoidal function $\sigma$,
parameters $\xi$ = $w$, projection
$v = \sum_l w_l z_l-b$ and $z_l$ single components of the variables $x$, $y$, 
i.e.,
$z_{l}$ = $x_l$ for $1\le l\le d_x$
and 
$z_{l}$ = $y_l$ for $d_x+1\le l\le d_x+d_y$.
(For neural networks with Lorentzians instead of sigmoids
see \cite{Giraud-Lapedes-Liu-Lemm-1995}.)

Typical choices for the sigmoid are
$\sigma (v)$ = $\tanh (\beta v)$
or $\sigma (v)$ = $1/(1+e^{-2\beta v})$.
The parameter $\beta$, often called inverse temperature,
controls the sharpness of the step of the sigmoid.
In particular,
the sigmoid functions become a sharp step
in the limit $\beta\rightarrow\infty$, i.e.,
at zero temperature.
In principle the sigmoidal function $\sigma$ may depend on further parameters
which then 
--- similar to projection pursuit discussed in  Section \ref{proj-pursuit} ---
would also have to be included in the optimization process.
The threshold or bias $b$ can
be treated as weight if an additional input component is included
clamped to the value $1$.

A linear combination of perceptrons
\begin{equation}
\phi (x,y) = b+\sum_l W_l \sigma \left(\sum_k w_{lk} z_k-b_k\right)
\label{linear-NN}
,
\end{equation}
has the form of a projection pursuit approach (\ref{proj-pursuit-Eq})
but with fixed $\phi_l(v)$ = $W_l \sigma(v)$.

In multi--layer networks the parameterization (\ref{single-layer}) is cascaded,
\begin{equation}
z_{k,i} 
= \sigma \left(\sum_{l=1}^{m_{i-1}} w_{kl,i} z_{l,i-1}-b_{k,i})\right)
= \sigma (v_{k,i}),
\end{equation}
with 
$z_{k,i}$ representing
the output of the $k$th node (neuron) in layer $i$ 
and
\be
v_{k,i} = \sum_{l=1}^{m_{i-1}} w_{kl,i} z_{l,i-1} -b_{k,i}
,
\ee
being the input for that node.
This yields, skipping the bias terms for simplicity
\be
\phi(z,w) =
\sigma\left(\sum_{l_{n-1}}^{m_{n-1}} w_{l_{n-1},n}
\sigma\left(\sum_{l_{n-2}}^{m_{n-2}} w_{l_{n-1}l_{n-2},n-1}
\cdots
\sigma\left(\sum_{l_{0}}^{m_{0}} w_{l_{1}l_{0},1} z_{l_{0},0}
\right)
\cdots
\right)
\right)
,
\label{multi-layer-neural-net}
\ee
beginning with an input layer  
with $m_0$ = $d_x+d_y$ nodes
(plus possibly nodes to implement the bias)
$z_{l,0}$ = $z_l$
and going over intermediate layers with $m_i$ nodes
$z_{l,i}$, $0<i<n$, $1\le l\le m_i$ 
to a single node output layer $z_{n}$ = $\phi (x,y)$.

Commonly neural nets are used in regression and classification
to parameterize a function $\phi (x,y)$ = $h(x)$ in functionals 
\be
E=\sum_i (y_i-{h}(x_i,w))^2
\label{quad}
,
\ee
quadratic in ${h}$
and without further regularization terms.
In that case, regularization has to be assured by
using either 
1.\ a neural network architecture which is restrictive enough,
2.\ by using early stopping like training procedures
so the full flexibility of the network structure cannot 
completely develop and destroy generalization,
where in both cases the optimal architecture or algorithm
can be determined
for example by cross--validation or bootstrap techniques
\cite{Mosteller-Wallace-1963,Allen-1974,Wahba-Wold-1975,Stone-1974,Stone-1977,Golup-Heath-Wahba-1979,Craven-Wahba-1979,Wahba-1990,Efron-Tibshirani-1993},
or 
3.\ by averaging over ensembles of networks \cite{Neal-1996}.
In all these cases regularization is implicit in the parameterization
of the network.
Alternatively, explicit regularization or prior terms 
can be added to the functional. 
For regression or classification 
this is for example done in {\it learning by hints} 
\cite{Abu--Mostafa-1990,Abu--Mostafa-1993a,Abu--Mostafa-1993b}
or {\it curvature--driven smoothing} with feedforward networks
\cite{Bishop-1993}.

One may also remark
that from a Frequentist point of view the quadratic functional
is not interpreted as posterior but as 
squared--error loss $\sum_i (y_i-a(x_i,w))^2$
for actions $a(x) = a(x,w)$.
According to Section \ref{some-loss-functions}
minimization of error functional (\ref{quad}) 
for data $\{(x_i,y_i)|1\le i\le n\}$ sampled under the true density $p(x,y|f)$
yields therefore
an empirical estimate for the regression function
$\int \!dy \,y\, p(y|x,f)$.

We consider here neural nets as parameterizations for density estimation
with prior (and normalization) terms explicitly included 
in the functional $E_\phi$.
In particular, the stationarity equation for functional (\ref{error-xi})
becomes
\be
0 = \Phi_w^\prime {\bf P}^\prime {\bf P}^{-1} N 
-\Phi_w^\prime {{\bf K}} \phi 
-\Phi_w^\prime {\bf P}^\prime \Lambda_X
,
\ee
with matrix of derivatives
\bea
\Phi^\prime_{w} (k,l,i; x,y) &=&
\frac{\partial \phi (x,y,w)}{\partial w_{kl,i}}
\label{backprop}
\\
&=&
\sigma^\prime (v_n) \sum_{l_{n-1}} w_{l_{n-1},n}
\sigma^\prime (v_{l_{n-1},n-1}) \sum_{l_{n-2}} w_{l_{n-1}l_{n-2},n-1}
\nonumber\\
&&\cdots\;
\sum_{l_{i+1}} w_{l_{i+2}l_{i+1},i+2}
\sigma^\prime (v_{l_{i+1},i+1})
w_{l_{i+1}k,i+1}
\sigma^\prime (v_{l_{i},i}) z_{l,i-1} 
\nonumber
,
\eea
and $\sigma^\prime (v)$ = $d\sigma(v)/dv$.
While $\phi (x,y,w)$ is calculated by forward propagating
$z$ = $(x,y)$ through the net defined by
weight vector $w$
according to 
Eq.\ (\ref{multi-layer-neural-net}) the derivatives
$\Phi^\prime$ can efficiently be calculated by back--propagation
according to Eq.\ (\ref{backprop}).
Notice that even for diagonal ${\bf P}^\prime$
the derivatives are not needed only at data points
but the prior and normalization term require derivatives 
at all $x$, $y$.
Thus, in practice terms like $\Phi^\prime {\bf K} \phi$
have to be calculated in a relatively poor discretization.
Notice, however, that regularization is here
not only due to the prior term but 
follows also from the restrictions implicit 
in a chosen neural network architecture.
In many practical cases 
a relatively poor discretization of the prior term may thus be sufficient.

Table \ref{summary-param} summarizes the discussed approaches.

\begin{table}
\begin{center}
\begin{tabular}{ | l | l | l|}
\hline
Ansatz  & Functional form & to be optimized\\
\hline
\hline\rule[-3mm]{0mm}{9mm}
linear ansatz&$\phi(z) = \sum_l \xi_l B_l(z)$    & $\xi_l$\\
\hline\rule[0mm]{0mm}{6mm}
linear model&$\phi(z) = \xi_0 + \sum_l \xi_l z_l$    & $\xi_0$, $\xi_l$\\
\rule[-3mm]{0mm}{0mm}
$\;$with interaction
   & $\qquad\quad + \sum_{mn} \xi_{mn} z_m z_n +\cdots$ 
   & $\xi_{mn},\cdots$\\
\hline\rule[-3mm]{0mm}{9mm}
mixture model & $\phi(z) = \sum \xi_{0,l} B_l(\xi_l,z)$ 
    & $\xi_{0,l}$, $\xi_l$\\
\hline\rule[0mm]{0mm}{6mm}
additive model &$\phi(z) = \sum_l\phi_l(z_l)$ & $\phi_l(z_l)$\\
\rule[-3mm]{0mm}{0mm}
$\;$with interaction
   &$ \qquad\quad+ \sum_{mn} \phi_{mn}(z_m z_n) +\cdots$ 
   & $\phi_{mn}(z_mz_n),\cdots$\\
\hline\rule[-3mm]{0mm}{9mm}
product ansatz &$\phi(z) = \prod_l\phi_l(z_l)$ & $\phi_l(z_l)$\\
\hline\rule[-3mm]{0mm}{9mm}
decision trees 
   & $\phi(z) = \sum_l \xi_{l} \prod_k \Theta(z_{\xi_{lk}}-\xi_{0,lk})$ 
   & $\xi_l$, $\xi_{0,lk}$, $\xi_{lk}$\\
\hline\rule[-3mm]{0mm}{9mm}
projection pursuit 
   & $\phi(z)=\xi_0+\sum_l \phi_l(\xi_{0,l}+\sum_l\xi_l z_l)$
   & $\phi_l$, $\xi_0$, $\xi_{0,l}$, $\xi_l$\\
\hline\rule[-3mm]{0mm}{9mm}
neural net (2 lay.)
   & $\phi(z)=\sigma\!\left(\sum_{l} \xi_{l}\,
              \sigma\!\left(\sum_k  \xi_{lk} z_k\right)\right)$ 
   & $\xi_l$, $\xi_{lk}$\\
\hline
\end{tabular}
\end{center}
\caption{Some possible parameterizations.}
\label{summary-param}
\end{table}

\section{Parameterizing priors: Hyperparameters}
\label{hyperparameters}

\subsection{Prior normalization}
\label{prior-normalization}

In Chapter \ref{variational}.
parameterization of $\phi$ have been studied.
This section now discusses parameterizations
of the prior density $p(\phi|D_0)$.
For Gaussian prior densities
that means parameterization of mean and/or covariance.
The parameters of the prior functional, 
which we will denote by $\theta$,
are in a Bayesian context also known as {\it hyperparameters}.
Hyperparameters $\theta$ can be considered
as part of the hidden variables. 
 
In a full Bayesian approach the ${h}$--integral 
therefore has to be completed
by an integral over the additional hidden variables $\theta$.
Analogously, the prior densities
can be supplemented by priors for $\theta$,
also be called {\it hyperpriors},
with corresponding energies $E_\theta$.

In saddle point approximation
thus an additional stationarity equation 
will appear,
resulting from the derivative
with respect to $\theta$.
The saddle point approximation of the $\theta$--integration
(in the case of uniform hyperprior $p(\theta)$
and with the ${h}$--integral being calculated exactly or by approximation)
is also known as ML--II prior \cite{Berger-1980}
or evidence framework
\cite{Gull-1988,Gull-1989,Skilling-1991,MacKay-1992c,MacKay-1992d,MacKay-1994d,Bishop-1995b}.

There are some cases where it is convenient 
to let the likelihood $p(y|x,h)$ 
depend, besides on a function $\phi$,
on a few additional parameters.
In regression such a parameter
can be the variance of the likelihood.
Another example is the inverse temperature $\beta$ introduced in 
Section \ref{mixtures-for-regression},
which, like $\phi$ also appears in the prior.
Such parameters may formally be added to 
the ``direct'' hidden variables 
$\phi$ yielding an enlarged $\tilde \phi$.
As those ``additional likelihood parameters''
are like other hyperparameters typically
just real numbers, and not functions like $\phi$,
they can often be treated analogously to hyperparameters.
For example, they may also be determined by cross--validation (see below)
or by a low dimensional integration.
In contrast to pure prior parameters, however,
the functional derivatives with respect to such
``additional likelihood parameters''
contain terms arising from the derivative of the likelihood.

Within the Frequentist interpretation of error minimization
as empirical risk minimization
hyperparameters $\theta$ can be determined by 
minimizing the {\it empirical generalization error} 
on a new set of test or validation data $D_T$ 
being independent from the training data $D$. 
Here the empirical generalization error is meant to be
the pure data term 
$E_D(\theta)$ = $E_D(\phi^*(\theta))$ 
of the error functional 
for $\phi^*$ 
being the optimal $\phi$ for the full regularized $E_\phi(\theta)$
at $\theta$ and for given training data $D$.
Elaborated techniques include
cross--validation and bootstrap methods
which have been mentioned in Sections \ref{empirical-risk}
and  \ref{neural}.

Within the Bayesian interpretation
of error minimization as posterior maximization
the introduction of hyperparameters
leads to a new difficulty.
The problem arises from the fact
that it is usually desirable
to interpret the error term $E_\theta$
as prior energy for $\theta$, meaning that
\be
p(\theta) = \frac{e^{-E_\theta}}{Z_\theta}
\label{prior-factorization}
,
\ee
with normalization
\be
Z_\theta
=
{\int\!d\theta\, e^{-E_\theta}}
,
\ee
represents the prior density for $\theta$.
Because the joint prior factor for $\phi$ and $\theta$ is given by  
the product
\be
p(\phi,\theta) = p(\phi|\theta) p(\theta)
,
\ee
one finds
\be
p(\phi|\theta) = \frac{e^{-E (\phi|\theta)}}{Z_\phi(\theta)}
.
\ee
Hence, the $\phi$--dependent part of the energy represents
a {\it conditional prior energy} denoted here $E(\phi|\theta)$. 
As this conditional normalization
\be
Z_\phi(\theta)
=
\int\!d\phi\, e^{-E(\phi|\theta)}
,
\ee
is in general $\theta$--dependent a normalization term
\be
E_N (\theta) = \ln Z_\phi(\theta)
\ee
must therefore be included
in the error functional when minimizing with respect to $\theta$.

It is interesting to look what happens
if $p(\phi,\theta)$ 
of Eq.\ (\ref{prior-factorization}) 
is expressed
in terms of {\it joint energy} $E(\phi,\theta)$ as follows
\be
p(\phi,\theta) = \frac{e^{-E (\phi,\theta)}}{Z_{\phi,\theta}}
.
\ee
Then the joint normalization 
\be
Z_{\phi,\theta}
=
\int\!d\phi\, d\theta\, e^{-E(\phi,\theta)}
,
\ee
is independent of $\phi$ and $\theta$ and could be skipped from the functional.
However, in that case the term $E_\theta$
cannot easily be related to the prior $p(\theta)$.

Notice especially, 
that this discussion also applies to the case 
where $E_\theta$ is assumed to be uniform 
so it does not have to appear explicitly in the error functional.
The two ways of expressing $p(\phi,\theta)$ 
by a joint or conditional energy, respectively, 
are equivalent
if the joint density  factorizes. 
In that case, however,
$\theta$ and $\phi$ are independent,
so $\theta$ 
cannot be used to parameterize the density of $\phi$.

Numerically the need to calculate $Z_\phi(\theta)$
can be disastrous
because normalization factors $Z_\phi (\theta)$
represent often an extremely high dimensional (functional) integral
and are, in contrast to the normalization of $P$ over $y$,
very difficult to calculate.

There are, however, 
situations for which $Z_\phi (\theta)$ remains $\theta$--independent.
Let $p(\phi,\theta)$ stand for example
for a Gaussian specific prior $p(\phi,\theta|\tilde D_0)$
(with the normalization condition factored out 
as in Eq.\ (\ref{specific-prior})). 
Then, because the normalization of a Gaussian is independent
of its mean, parameterizing the mean $t$ = $t(\theta)$ results in
a $\theta$--independent $Z_\phi (\theta)$.

Besides their mean, Gaussian processes are  
characterized by their covariance operators ${{\bf K}}^{-1}$.
Because the normalization only depends on $\det {{\bf K}}$
a second possibility 
yielding $\theta$--dependent $Z_\phi (\theta)$ 
are parameterized transformations of the form 
${{\bf K}} \rightarrow {\bf O}{{\bf K}}{\bf O}^{-1}$
with orthogonal ${\bf O}$ = ${\bf O}(\theta)$.
Indeed, such transformations do not change the determinant $\det {{\bf K}}$.
They are only non--trivial for multi--dimensional Gaussians.

For general parameterizations of density estimation problems, however,
the normalization term 
$\ln Z_\phi(\theta)$ must be included. 
The only way to get rid of that normalization term
would be to assume a {\it compensating hyperprior} 
\be
p(\theta)\propto Z_\phi (\theta)
, 
\ee
resulting in an error term 
$E(\theta)$ = $-\ln Z_\phi (\theta)$ 
compensating $E_N (\theta)$.

Thus, in the general case we have to consider the functional
\be
E_{\theta,\phi} =
-(\ln P(\phi ),\,N)
+ (P(\phi),\, \Lambda_X )
+ E_\phi (\theta)
+ E_\theta
+\ln Z_\phi (\theta) 
.
\label{adapEgeneral}
\ee
writing $E(\phi|\theta)$ = $E_\phi$ and $E(\theta)$ = $E_\theta$.
The stationarity conditions have the form
\bea
\frac{\delta E_\phi}{\delta \phi} &=&
{\bf P}^\prime(\phi) {\bf P}^{-1}(\phi) N  
- {\bf P}^\prime (\phi) \Lambda_X
,
\label{adam1g}
\\
\frac{\partial E_\phi}{\partial \theta} 
&=& 
-{\bf Z}^\prime Z_\phi^{-1}(\theta)
-E_{\theta}^\prime 
,
\label{adam2g}
\eea
with
\be
{\bf Z}^\prime (l,k)
= \delta (l-k) \frac{\partial Z_\phi(\theta)}{d\theta_l}
,\quad
E_{\theta}^\prime (l)
=
\frac{\partial E_\theta}{\partial \theta_l}
.
\ee
For compensating hyperprior 
$ E_\theta = -\ln Z_\phi (\theta)$
the right hand side of Eq.\ (\ref{adam2g})
vanishes.

Finally, we want to remark that
in case function evaluation of $p(\phi,\theta)$
is much cheaper than calculating the gradient (\ref{adam2g}),
minimization methods not using the gradient
should be considered, like for example the
downhill simplex method
\cite{Press-Teukolsky-Vetterling-Flannery-1992}.

\subsection{Adapting prior means}
\label{adaptingpm}

\subsubsection{General considerations}

A prior mean or template function $t$
represents a prototype, reference function or base line
for $\phi$.
It may be a typical expected pattern in time series prediction
or a reference image in image reconstruction.
Consider, for example, the task of completing an image $\phi$
given some pixel values (training data)
\cite{Lemm-1999a}.
Expecting the image to be that of a face the 
template function $t$ may be chosen 
to be some prototypical image of a face.
We have seen in Section \ref{Non--zero-means}
that a single template $t$ could be eliminated
for Gaussian (specific) priors
by solving for $\phi-t$ instead for $\phi$.
Restricting, however, to only a single template may be a very bad choice.
Indeed, faces for example appear on images in many variations,
like
in different scales, translated, rotated, various illuminations, 
and other kinds of deformations.
We may now describe such variations
by a family of templates $t(\theta)$,
the parameter $\theta$ describing scaling, translations, 
rotations, and more general deformations.
Thus, 
we expect a function to be similar to
only one of the templates $t(\theta)$
and want to implement a (soft, probabilistic) OR,
approximating
$t(\theta_1)$ OR $t(\theta_2)$ OR $\cdots$
(See also \cite{Lemm-1998,Lemm-1998a,Lemm-1998b,Lemm-1998c}).

A (soft, probabilistic)  AND of approximation conditions,
on the other hand, 
is implemented by adding error terms.
For example, 
classical error functionals
where data and prior terms are added
correspond to an approximation of training data AND {\it a priori} data.

Similar considerations apply for {\it model selection}.
We could for example expect $\phi$ to be well approximated by a 
neural network or a decision tree.
In that case $t(\theta)$ spans, for example, a space of neural networks
or decision trees.
Finally, let us emphasize again that
the great advantage and practical feasibility of adaptive templates
for regression problems
comes from the fact that no additional normalization terms have to be
added to the error functional.

\subsubsection{Density estimation}

The general case with adaptive means for Gaussian prior factors
and hyperparameter energy $E_\theta$ yields an error functional
\be
E_{\theta,\phi} =
-(\ln P(\phi),\,N)
+\frac{1}{2} \Big(\phi-t(\theta) ,\,{{\bf K}}\,(\phi-t(\theta))\Big)
+ (P(\phi),\, \Lambda_X )+ E_\theta
.
\label{adapE}
\ee
Defining
\be
{\bf t}^\prime (l;x,y)
= \frac{\partial t(x,y;\theta)}{\partial \theta_l}
,
\ee
the stationarity equations of (\ref{adapE})
obtained from the functional derivatives with
respect to $\phi$ and hyperparameters $\theta$
become
\bea
{{\bf K}}(\phi-t) &=&
{\bf P}^\prime (\phi) {\bf P}^{-1}(\phi) N  
- {\bf P}^\prime(\phi) \Lambda_X
,
\label{adam1}
\\
{\bf t}^\prime {{\bf K}} (\phi-t) 
&=& 
-E_{\theta}^\prime 
.
\label{adam2}
\eea
Inserting
Eq.\ (\ref{adam1}) in Eq.\ (\ref{adam2}) gives
\be
{\bf t}^\prime{\bf P}^\prime (\phi) {\bf P}^{-1}(\phi) N  
= {\bf t}^\prime{\bf P}^\prime(\phi) \Lambda_X
-E_{\theta}^\prime 
.
\label{adam3}
\ee
Eq.(\ref{adam3}) becomes equivalent to the parametric
stationarity equation
(\ref{parametricSE}) with vanishing prior term
in the deterministic limit of vanishing prior covariances ${\bf K}^{-1}$,
i.e., under the assumption $\phi=t(\theta)$,
and for vanishing $E_\theta^\prime$.
Furthermore, a non--vanishing prior term in (\ref{parametricSE}) can be 
identified with the term $E_\theta$.
This shows, that parametric methods can be considered
as deterministic limits of (prior mean) hyperparameter approaches.
{\it In particular, a parametric solution can thus 
serve as reference template $t$,
to be used within a specific prior factor.}
Similarly,
such a parametric solution 
is a natural initial guess for a nonparametric $\phi$
when solving a stationarity equation by iteration.

If working with parameterized $\phi(\xi)$
extra prior terms Gaussian in some function $\psi(\xi)$ can be included
as discussed in Section \ref{gauss-param}.
Then, analogously to templates $t$ for $\phi$, also
parameter templates $t_\psi$ can be made adaptive 
with hyperparameters $\theta_\psi$.
Furthermore, prior terms 
$E_\theta$ and $E_{\theta_\psi}$ 
for the hyperparameters $\theta$, $\theta_\psi$
can be added.
Including such additional error terms yields
\bea
E_{\theta,\theta_\psi,\phi(\xi),\psi(\xi)} 
&=&
-(\ln P(\,\phi(\xi)\,),\,N) + (P(\,\phi(\xi)\,),\, \Lambda_X )
\nonumber\\&&
+\frac{1}{2} \Big(\phi(\xi)-t(\theta) ,\,{{\bf K}}\,(\phi(\xi)-t(\theta))\Big)
\nonumber\\&&
+\frac{1}{2}
\Big(\psi(\xi)-t_\psi(\theta_\psi),
     \,{{\bf K}}_\psi\,(\psi(\xi)-t_\psi(\theta_\psi))\Big)
\nonumber\\&&
+E_{\theta} + E_{\theta_\psi} 
,
\label{generalE}
\eea
and Eqs.(\ref{adam1}) and (\ref{adam1}) change to
\bea
\Phi^\prime {{\bf K}}(\phi-t) 
+\Psi^\prime {{\bf K}}_\psi (\psi-t_\psi)
 &=&
{\bf P}_\xi^\prime  {\bf P}^{-1}N  
- {\bf P}_\xi^\prime  \Lambda_X
,
\label{adam1xi}
\\
{\bf t}^\prime {{\bf K}} (\phi-t) 
&=& 
-E_{\theta}^\prime 
,
\\
\label{adam2xi}
{\bf t}_\psi^\prime {{\bf K}}_\psi (\psi-t_\psi)
&=&- E_{\theta_\psi}^\prime 
\label{adam3xi}
,
\eea
where ${\bf t}_\psi^\prime $,
$E_{\theta_\psi}^\prime$,
$E_{\theta}^\prime$ ,  
denote derivatives with respect
to the parameters $\theta_\psi$ or $\theta$, respectively.
Parameterizing $E_{\theta}$ and $E_{\theta_\psi}$ the process 
of introducing hyperparameters can be iterated.

\subsubsection{Unrestricted variation}
\label{unrestricted-variation}

To get a first understanding of the approach (\ref{adapE})
let us consider the extreme example of
completely unrestricted $t$--variations.
In that case the template function $t(x,y)$ itself
represents the hyperparameter.
(Such function hyperparameters or hyperfields
are also discussed in Sect.\  \ref{hyperfields}.)
Then, ${\bf t}^\prime = {\bf I}$ and
Eq.\ (\ref{adam2}) gives
${{\bf K}}(\phi -t)$ = $0$
(which for invertible ${{\bf K}}$
is solved uniquely by $t=\phi$),
resulting according to Eq.\ (\ref{lambdatemp}) in
\be
\Lambda_X = N_X
.
\ee
The case of a completely free prior mean $t$
is therefore equivalent to a situation without prior.
Indeed, for invertible ${\bf P}^\prime$, 
projection of Eq.\ (\ref{adam3}) into the $x$--data space
by ${\bf I}_D$ of Eq.\ (\ref{dataprojector})
yields
\be
P_D =  {\bf \Lambda}_{X,D}^{-1} N
,
\ee
where 
${\bf \Lambda}_{X,D}$ 
= ${\bf I}_D {\bf \Lambda}_{X}{\bf I}_D $ 
is invertible 
and $P_D = {\bf I}_D P$.
Thus for $x_i$ for which $y_i$ are available
\be
P(x_i,y_i) 
= \frac{N(x_i,y_i)}{N_X (x_i)}
\ee
is concentrated on the data points.
Comparing this with solutions of Eq.\ (\ref{lambdain})
for fixed $t$
we see that adaptive means tend to lower the influence
of prior terms.

\subsubsection{Regression}

Consider now the case of regression
according to functional (\ref{regression-functional})
with an adaptive template $t_0(\theta)$.
The system of stationarity equations for 
the regression function $h(x)$ (corresponding to $\phi (x,y)$)
and $\theta$ becomes
\bea
{{\bf K}}_0({h}-t_0) &=& {{\bf K}_D}(t_D-{h})
,
\label{adamreg1}
\\
{\bf t}_0^\prime {{\bf K}}_0 ({h} -t_0) &=& 0
.
\label{adamreg2}
\eea
It will also be useful
to insert Eq.\ (\ref{adamreg1}) in Eq.\ (\ref{adamreg2}), 
yielding
\be
0 = {\bf t}_0^\prime {{\bf K}_D}({h}-t_D)
.
\label{adamreg3}
\ee
For fixed $t$ Eq.\ (\ref{adamreg1}) is solved by 
the template average $t$
\be
{h} = t = \left({{\bf K}}_0 + {{\bf K}}_D\right)^{-1}
       \left({{\bf K}}_0 t_0 + {{\bf K}}_D t_D\right)
,
\label{total-template}
\ee
so that
Eq.\ (\ref{adamreg2}) or Eq.\ (\ref{adamreg3}), respectively, become
\be
0={\bf t}_0^\prime {{\bf K}}_0 
(t-t_0)
,
\ee
\be
0 = {\bf t}_0^\prime {{\bf K}}_D (t-t_D)
.
\label{adamregres}
\ee
It is now interesting to note
that if we replace in Eq.\ (\ref{adamregres}) the full template average $t$
by $t_0$ we get 
\be
0 = {\bf t}_0^\prime {{\bf K}}_D (t_0-t_D)
,
\label{comp1}
\ee
which is equivalent to the stationarity equation
\be
0 = {{\bf H}}^\prime {{\bf K}}_D ({h}-t_D),
\label{comp2}
\ee
(the derivative matrix ${{\bf H}}^\prime$ being the analogue to $\Phi^\prime$ 
for $h$)
of an error functional
\be
E_{D,{h}(\xi)} 
= \frac {1}{2} (\,{h}(\xi) - t_D,\, {{\bf K}}_D ({h}(\xi)-t_D)\,)
\label{D-phi-xi-functional}
\ee
without prior terms
but with parameterized ${h}(\xi)$, e.g., a neural network.
The approximation
${h}$ = $t$ = $t_0$ can, for example,
be interpreted as limit $\lambda\rightarrow \infty$,
\be
\lim_{\lambda\rightarrow \infty} {h} =
\lim_{\lambda\rightarrow \infty} t = t_0
,
\ee
after replacing ${{\bf K}}_0$ by $\lambda {{\bf K}}_0$
in Eq.\ (\ref{total-template}).
The setting ${h}$ = $t_0$ can then be used
as initial guess ${h}^0$ for an iterative solution for ${h}$.
For existing ${{\bf K}}_0^{-1}$ 
${h}$ = $t_0$ is also obtained after one iteration step
of the iteration scheme
${h}^{i} = t_0 + {{\bf K}}_0^{-1} {{\bf K}}_D (t_D-{h}^{i-1})$
starting with initial guess ${h}^0=t_D$.

For comparison with Eqs.(\ref{adamregres},\ref{comp1},\ref{comp2})
we give the stationarity equations 
for parameters $\xi$ 
for a parameterized regression functional 
including an additional prior term
with hyperparameters
\be
E_{\theta,{h}(\xi)} = 
\frac {1}{2} (\,{h}(\xi) - t_D,\, {{\bf K}}_D ({h}(\xi)-t_D)\,)
+\frac {1}{2} (\,{h}(\xi) - t_0(\theta),\, {{\bf K}}_0(\theta)
 ({h}(\xi)-t_0(\theta))\,)
,
\label{theta-phi-xi-functional}
\ee
which are
\be
0 = {{\bf H}}^\prime {{\bf K}}_D ({h}-t_D) + {h}^\prime {{\bf K}}_0 ({h}-t_0)
.
\ee

Let us now compare 
the various regression functionals 
we have met up to now.
The non--parameterized and regularized regression 
functional $E_{{h}}$ (\ref{regression-functional})
implements prior information
explicitly by a regularization term.

A parameterized and regularized functional
$E_{{h}(\xi)}$ of the form (\ref{error-xi-h})
corresponds to a functional of the form (\ref{theta-phi-xi-functional})
for $\theta$ fixed. 
It imposes restrictions on the regression function $h$
in two ways, by choosing a specific parameterization 
and by including an explicit prior term.
If the number of data is large enough,
compared to the flexibility of the parameterization,
the data term of $E_{{h}(\xi)}$ alone
can have a unique minimum. 
Then, at least technically, no
additional prior term would be required.
This corresponds to the classical 
error minimization methods 
used typically for parametric approaches.
Nevertheless, also in such situations
the explicit prior term can be useful
if it implements useful prior knowledge over $h$.

The regularized functional with prior-- or hyperparameters
$E_{\theta,{h}}$ (\ref{adapE})
implements, compared to $E_{{h}}$,
effectively weaker prior restrictions.
The prior term corresponds to  a {\it soft restriction}
of ${h}$ to the space spanned by the parameterized $t(\theta )$.
In the limit where the parameterization of 
$t(\theta)$ is rich enough to
allow $t(\theta^*)$ = ${h}^*$ at the stationary point
the prior term vanishes completely.

The parameterized and regularized functional 
$E_{\theta,{h}(\xi)}$ (\ref{theta-phi-xi-functional}),
including prior parameters $\theta$,
implements prior information explicitly by a regularization term
and implicitly by the parameterization of ${h}(\xi)$.
The explicit prior term vanishes
if $t(\theta^*)$ = ${h} (\xi^*)$ at the stationary point.
The functional combines a {\it hard restriction}
of ${h}$ with respect to the space spanned by the parameterization ${h}(\xi)$
and a {\it soft restriction} of ${h}$ with respect to 
the space spanned by the parameterized $t(\theta)$.
Finally, the parameterized and non--regularized functional
$E_{D,{h}(\xi)} $ (\ref{D-phi-xi-functional})
implements prior information only implicitly by parameterizing ${h}(\xi)$.
In contrast to the functionals 
$E_{\theta,{h}}$ and $E_{\theta,{h}(\xi)}$ 
it implements only a {\it hard restriction} for ${h}$.
The following table summarizes the discussion:

\begin{center}
\begin{tabular}{ | r | c | l|}
\hline
Functional & Eq. & prior implemented \\
\hline
$E_{{h}}$             & (\ref{regression-functional})
                       &explicitly\\
$E_{{h}(\xi)}$        &(\ref{error-xi-h})           &explicitly and implicitly\\
$E_{\theta,{h}}$      &(\ref{adapE})              &explicitly\\
                       &&no prior for $t(\theta^*) = {h}^*$ \\
$E_{\theta,{h}(\xi)}$ &(\ref{theta-phi-xi-functional})
                       & explicitly and implicitly\\
                       &&no expl.\ prior for $t(\theta^*) = {h} (\xi^*)$ \\
$E_{D,{h}(\xi)} $     &(\ref{D-phi-xi-functional})&implicitly\\
\hline
\end{tabular}
\end{center}

\subsection{Adapting prior covariances}
\subsubsection{General case}

Parameterizing covariances ${{\bf K}}^{-1}$
is often desirable in practice.
It includes for example adapting
the trade--off between data and prior terms
(i.e., the determination of the regularization factor),
the selection between different symmetries, smoothness measures,
or in the multidimensional situation the determination
of directions with low variance.
As far as the normalization depends on ${{\bf K}}(\theta)$
one has to consider the error functional
\be
E_{\theta,\phi} =
-(\ln P(\phi),\,N)
+\frac{1}{2} \Big(\phi-t ,\,{{\bf K}} (\theta)\,(\phi-t)\Big)
+ (P(\phi),\, \Lambda_X )
+\ln Z_\phi (\theta)
+E_\theta
,
\label{adapO}
\ee
with 
\be
Z_\phi (\theta) = 
(2\pi)^\frac{d}{2}(\det {{\bf
K}}(\theta))^{-\frac{1}{2}}
,
\ee
for a $d$--dimensional Gaussian specific prior,
and
stationarity equations
\bea
{{\bf K}}(\phi-t) &=& 
{\bf P}^\prime(\phi) {\bf P}^{-1}(\phi) N  
- {\bf P}^\prime(\phi) \Lambda_X
,
\label{adaO1}
\\
\frac{1}{2} 
\Big(\phi-t ,\,\frac{\partial {{\bf K}}(\theta)}{\partial \theta}\,(\phi-t)\Big)
&=& 
{\rm Tr} \left({\bf K}^{-1}(\theta)
 \frac{\partial {{\bf K}}(\theta)}{\partial \theta}\right)
-E_\theta^\prime
.
\label{adaO2}
\eea
Here we used
\be
\frac{\partial}{\partial \theta} \ln \det {{\bf K}}
=\frac{\partial}{\partial \theta} {\rm Tr}\, \ln {{\bf K}}
={\rm Tr} 
\left( {{\bf K}}^{-1} \frac{\partial {{\bf K}}}{\partial \theta} \right)
.
\ee
In case of an unrestricted variation of the matrix elements
of ${{\bf K}}$
the hyperparameters become 
$\theta_l$ = $\theta (x,y;x^\prime,y^\prime)$
= ${{\bf K}} (x,y;x^\prime,y^\prime)$.
Then, using
\be
\frac{\partial {{\bf K}}(x,y;x^\prime,y^\prime)}
     {\partial \theta (x^{\prime\prime},y^{\prime\prime};
                    x^{\prime\prime\prime},y^{\prime\prime\prime})}
=
\delta (x-x^{\prime\prime}) \delta (y-y^{\prime\prime}) 
\delta (x^{\prime}-x^{\prime\prime\prime}) 
\delta (y^{\prime}-y^{\prime\prime\prime})
,
\ee
Eqs.(\ref{adaO2}) becomes the inhomogeneous equation
\be
\frac{1}{2} (\phi-t) \, (\phi-t)^T
= 
{\rm Tr} \left({\bf K}^{-1}(\theta)
 \frac{\partial {{\bf K}}(\theta)}{\partial \theta}\right)
-E_\theta^\prime
.
\ee

We will in the sequel consider the two special cases where 
the determinant of the covariance is $\theta$--independent
so that the trace term vanishes,
and where $\theta$ is just a multiplicative factor 
for the specific prior energy, i.e., a so called regularization parameter.

\subsubsection{Automatic relevance detection}

A useful application of hyperparameters 
is the identification of sensible directions
within the space of $x$ and $y$ variables.
Consider the general case of a covariance, decomposed into components
${\bf K}_0$ = $\sum_i \theta_i {\bf K}_i$.
Treating the coefficient vector $\theta$ (with components $\theta_i$)
as hyperparameter with hyperprior $p(\theta)$ 
results in a prior energy (error) functional
\be
\frac{1}{2} 
 \big(\phi-t,\; (-\sum_i \theta_i {\bf K}_i)(\phi-t)\,\big)
-\ln p(\theta) +\ln Z_\phi (\theta)
.
\ee
The $\theta$--dependent normalization $\ln Z_\phi(\theta)$ 
has to be included to obtain the correct
stationarity condition for $\theta$.
The components ${\bf K}_i$
can be the components of a negative Laplacian,
for example, 
${\bf K}_i$ = $-\partial_{x_i}^2$
or
${\bf K}_i$ = $-\partial_{y_i}^2$.
In that case adapting the hyperparameters
means searching for sensible directions in the
space of $x$ or $y$ variables.
This technique has been called
{\it Automatic Relevance Determination}
by MacKay and Neal \cite{Neal-1996}.
The positivity constraint for $a$
can be implemented explicitly,
for example by using
${\bf K}_0$ = $\sum_i \theta_i^2 {\bf K}_i$
or
${\bf K}_0$ = $\sum_i \exp(\theta_i) {\bf K}_i$.

\subsubsection{Local smoothness adaption}
\label{local-smoothness}

Similarly, the regularization factor of a
smoothness related covariance operator may be adapted locally.
Consider, for example, a prior energy for $\phi(x,y)$
\be
E(\phi|\theta)=\frac{1}{2} 
 \big(\phi-t,\; {\bf K}(a,b)(\phi-t)\,\big)
,
\ee
with a Laplacian prior
\be
{\bf K}(x,x^\prime,y,y^\prime;\theta) = 
-\sum_i^{m_x} e^{\theta_{x,i}(x)}\,\delta(x_i-x_i^\prime)\,\partial_{x_i}^2
-\sum_i^{m_y} e^{\theta_{y,i}(y)}\,\delta(y-y_i^\prime) \,\partial_{y_i}^2
,
\ee
for
$m_x$--dimensional vector $x$ 
and $m_y$--dimensional vector $y$
depending on functions $\theta_{x,i}(x)$ and $\theta_{y,i}(y)$
(or more general $\theta_{x,i}(x,y)$ and $\theta_{y,i}(x,y)$)
collectively denoted by $\theta$.
Expressing the coefficient functions as exponentials 
$\exp(\theta_{x,i})$, $\exp(\theta_{y,i})$ 
is one possibility to enforce their positivity.
Typically, one might impose a smoothness hyperprior
on the functions $\theta_{x,i}(x)$ and $\theta_{y,i}(y)$, 
for example by using an energy functional
\be
E(\phi,\theta)
+\frac{1}{2}\sum_i^{m_x} (\theta_{x,i},\; {\bf K}_{\theta,x} \theta_{x,i})
+\frac{1}{2}\sum_i^{m_y} (\theta_{y,i},\; {\bf K}_{\theta,y} \theta_{y,i})
+\ln Z_\phi (\theta)
,
\ee
with smoothness related ${\bf K}_{\theta,x}$, ${\bf K}_{\theta,y}$.
The stationarity equation for a functions $\theta_{x,i}(x)$ reads
\bea
0 &=& ({\bf K}_{\theta,x}\theta_{x,i})(x) 
- \left(\phi(x,y)-t(x,y)\right) 
  \left( \partial_{x_i}^2 
  \left(\phi(x,y)-t(x,y)\right)\right) e^{\theta_{x,i}(x)}
\nonumber
\\&&
+\partial_{\theta_{x,i}(x)} \ln Z_\phi(\theta) 
.
\eea
The functions $\theta_{x,i}(x)$ and $\theta_{y,i}(y)$
are examples of function hyperparameters
(see Sect.\ \ref{hyperfields}).

\subsubsection{Local masses and gauge theories}
\label{local-masses}

The Bayesian analog of a mass term in quantum field theory
is a term 
proportional to the identity matrix ${\bf I}$
in the inverse prior covariance ${\bf K}_0$. 
Consider, for example,
\be
{\bf K}_0 = \theta^2 \, {\bf I} - \Delta 
,
\ee
with $\theta$ real (so that $\theta^2\ge 0$) 
representing a mass parameter.
For large masses $\phi$ 
tends to copy the template $t$ locally,
and longer range effects of data points following from
smoothness requirements become less important.
Similarly to Sect.\ \ref{local-smoothness}
a constant mass can be replaced by a mass function $\theta(x)$.
This allows to adapt locally that interplay between ``template copying'' 
and smoothness related influence of training data.
As hyperprior, one may
use a smoothness constraint on the mass function $\theta(x)$,
e.g.,
\be
\frac{1}{2}
(\phi-t,\; {\bf M}^2 (\phi-t) )
-\frac{1}{2}
(\phi-t,\; \Delta (\phi-t) )
+\lambda\, (\theta,\; {\bf K}_\theta \theta)
+\ln Z_\phi(\theta)
,
\ee
where ${\bf M}$ denotes the diagonal mass operator 
with diagonal elements $\theta(x)$.

Functional hyperparameters 
like $\theta(x)$ represent, 
in the language of physicists,
additional fields entering the problem
(see also Sect.\ \ref{hyperfields}).
There are similarities  for example to gauge fields in physics.
In particular, a gauge theory--like formalism can be constructed
by decomposing $\theta(x)$ = $\sum_i \theta_i(x)$, 
so that the inverse covariance
\be
{\bf K}_0 
= \sum_i \left( {\bf M}^2_i - \partial_i^2\right) 
= \sum_i \left( {\bf M}_i + \partial_i\right) 
         \left( {\bf M}_i - \partial_i\right) 
= \sum_i D^\dagger_i D_i
,
\ee
can be expressed in terms of a ``covariant derivative''
$D_i$ = $\partial_i + \theta_i$.
Next, one may choose as hyperprior for $\theta_i(x)$
\be
\frac{1}{2}\left(
\sum_i^{m_x} \big(\theta_i,\; -\Delta \,\theta_i\big)
-\big(\sum_i^{m_x} \partial_{x_i} \theta_i,\; 
      \sum_j^{m_x} \partial_{x_j} \theta_j \big)
\right)
=
\frac{1}{4}\sum_{ij}^{m_x} F_{ij}^2
\ee
which can be expressed in terms of 
a ``field strength tensor'' (for Abelian fields),
\be
F_{ij} = \partial_i \theta_j - \partial_j \theta_i
,
\ee
like, for example,
the Maxwell tensor in quantum electrodynamics.
(To relate this, as in electrodynamics, 
to a local $U(1)$ gauge symmetry $\phi\rightarrow e^{i\alpha}\phi$
one can consider complex functions $\phi$, 
with the restriction that their phase cannot be measured.)
Notice, that, due to the interpretation of the prior as
product $p(\phi|\theta) p(\theta)$, an additional $\theta$--dependent
normalization term
$\ln Z_\phi(\theta)$ enters the energy functional.
Such a term is not present in quantum field theory,
where one relates the prior functional directly to
$p(\phi,\theta)$, so the norm is independent of $\phi$ and $\theta$.

\subsubsection{Invariant determinants}

In this section we discuss parameterizations of the covariance
of a Gaussian specific prior which leave the determinant invariant.
In that case no $\theta$--dependent normalization factors
have to be included which are usually very difficult to calculate.
We have to keep in mind, however, 
that in general a large freedom for ${{\bf K}}(\theta)$ 
effectively diminishes the influence of the parameterized prior term.

A determinant is, for example, 
invariant under general similarity transformations,
i.e., $\det \tilde {{\bf K}}$ = $\det {{\bf K}}$
for
${{\bf K}} \rightarrow \tilde {{\bf K}}$ 
= ${\bf O}{{\bf K}}{\bf O}^{-1}$
where ${\bf O}$ could be any element of the general linear group.
Similarity transformations do not change the eigenvalues,
because from
${{\bf K}}\psi$ = $\lambda \psi$ 
follows
${\bf O}{{\bf K}}{\bf O}^{-1}{\bf O}\psi$
= $\lambda {\bf O}\psi$.
Thus, if ${{\bf K}}$ is positive definite
also $\tilde {{\bf K}}$ is.
The additional constraint that $\tilde {{\bf K}}$ 
has to be real symmetric,
\be
\tilde {{\bf K}} = \tilde {{\bf K}}^T = \tilde {{\bf K}}^\dagger 
,
\ee
requires ${\bf O}$ to be real and orthogonal
\be
{\bf O}^{-1} = {\bf O}^T = {\bf O}^\dagger 
.
\ee
Furthermore, as an overall factor of ${\bf O}$ does not change 
$\tilde {{\bf K}}$ one can restrict ${\bf O}$ 
to a special orthogonal group $SO(N)$ with $\det {\bf O}=1$.
If ${{\bf K}}$ has degenerate eigenvalues 
there exist orthogonal transformations with ${{\bf K}}$ = $\tilde {{\bf K}}$.

While in one dimension
only the identity remains as transformation,
the condition of an invariant determinant
becomes less restrictive in higher dimensions.
Thus, especially for large dimension $d$ 
of ${{\bf K}}$ (infinite for continuous $x$)
there is a great freedom to adapt covariances
without the need to calculate normalization factors,
for example to adapt the sensible directions
of a multivariate Gaussian.

A positive definite ${{\bf K}}$
can be diagonalized by an orthogonal matrix ${\bf O}$ 
with $\det {\bf O}$ = $1$, i.e.,
${{\bf K}}  = {\bf O}{\bf D}{\bf O}^{T}$.
Parameterizing ${\bf O}$ the specific prior term becomes
\be
\frac{1}{2} \Big(\phi-t ,\,{{\bf K}} (\theta)\,(\phi-t)\Big)
=
\frac{1}{2} 
\Big(\phi-t ,\,{\bf O} (\theta){\bf D}{\bf O}^{T}(\theta)\,(\phi-t)\Big)
,
\ee
so the stationarity Eq.\ (\ref{adaO2}) reads
\be
\Big(\phi-t ,\,
   \frac{\partial {\bf O}}{\partial \theta}{\bf D}{\bf O}^{T}
\,(\phi-t)\Big)
=
-E_\theta^\prime
.
\label{rotation}
\ee 
Matrices ${\bf O}$ from $SO(N)$
include rotations and inversion.
For a Gaussian specific prior
with nondegenerate eigenvalues Eq.\ (\ref{rotation})
allows therefore to adapt the `sensible' directions 
of the Gaussian.

There are also transformations which can change eigenvalues,
but leave eigenvectors invariant.
As example, consider a diagonal matrix ${\bf D}$
with diagonal elements (and eigenvalues) $\lambda_i\ne 0$, i.e.,
$\det {\bf D}$ = $\prod_i \lambda_i$.
Clearly, any permutation of the eigenvalues $\lambda_i$
leaves the determinant invariant and transforms a positive definite matrix
into a positive definite matrix.
Furthermore, one may introduce 
continuous parameters $\theta_{ij}>0$ with $i<j$
and transform ${\bf D} \rightarrow \tilde{\bf D}$ 
according to
\be
\lambda_i \rightarrow 
\tilde \lambda_i = \lambda_i \theta_{ij}
,\quad
\lambda_j \rightarrow 
\tilde \lambda_j = \frac{\lambda_j}{\theta_{ij}}
,
\ee
which leaves the product
$\lambda_i\lambda_j$ 
= $\tilde\lambda_i\tilde\lambda_j$
and therefore also the determinant invariant
and transforms a positive definite matrix into a positive definite matrix.
This can be done with every pair of eigenvalues
defining a set of continuous parameters 
$\theta_{ij}$ with $i<j$
($\theta_{ij}$ can be completed to a symmetric matrix)
leading to 
\be
\lambda_i \rightarrow 
\tilde \lambda_i
=
\lambda_i \frac{\prod_{j>i} \theta_{ij}}{\prod_{j<i} \theta_{ji}}
,
\label{prodinvariants}
\ee
which also leaves the determinant invariant
\be
\det \tilde {\bf D} = \prod_i \tilde \lambda_i
=
\prod_i
\left(
\lambda_i \frac{\prod_{j>i} \theta_{ij}}{\prod_{j<i} \theta_{ji}}
\right)
=
\left( \prod_i \lambda_i \right)
\frac{\prod_i\prod_{j>i} \theta_{ij}}{\prod_i\prod_{j<i} \theta_{ji}}
= \prod_i \lambda_i 
= \det {\bf D}
.
\ee
A more general transformation with unique parameterization
by $\theta_i>0$, $i\ne i^*$, 
still leaving the eigenvectors unchanged, would be 
\be
\tilde \lambda_i = \lambda_i \theta_i,\; i\ne i^*
;\quad
\tilde \lambda_{i^*} = \lambda_{i^*} \prod_{i\ne i^*} \theta_i^{-1}
\label{prodinvariants2}
.
\ee 
This techniques can be applied to a general positive definite ${{\bf K}}$
after diagonalizing 
\be
{{\bf K}}  = {\bf O}{\bf D}{\bf O}^{T}
\rightarrow
\tilde {{\bf K}} = {\bf O}\tilde {\bf D}{\bf O}^{T}
\Rightarrow
 \det {{\bf K}} = \det \tilde{{\bf K}}
.
\ee
As example consider the transformations 
(\ref{prodinvariants}, \ref{prodinvariants2})
for which the specific prior term becomes
\be
\frac{1}{2} \Big(\phi-t ,\,{{\bf K}} (\theta)\,(\phi-t)\Big)
=
\frac{1}{2} \Big(\phi-t ,\,{\bf O}{\bf D}(\theta){\bf O}^{T}\,(\phi-t)\Big)
,
\ee
and stationarity Eq.\ (\ref{adaO2}) 
\be
\frac{1}{2} 
\Big(\phi-t ,\,
   {\bf O}\frac{\partial {\bf D}}{\partial \theta}{\bf O}^{T}
\,(\phi-t)\Big)
=
-E_\theta^\prime
,
\ee
and for (\ref{prodinvariants}), 
with $k<l$,
\be
\frac{\partial {\bf D}(i,j)}
     {\partial \theta_{kl}}
=
\delta (i-j) \, 
\Bigg(
\delta (k-i) \, \lambda_k 
   \frac{\prod_{l\ne n>k} \theta_{kn}}{\prod_{n<k} \theta_{nk}}
+\;
\delta (l-i) \, \lambda_l 
   \frac{\prod_{n>l} \theta_{ln}}{\prod_{k\ne n<l} \theta_{nl}}
\Bigg)
,
\ee
or, for (\ref{prodinvariants2}), 
with $k\ne i^*$,
\be
\frac{\partial {\bf D}(i,j)}
     {\partial \theta_{k}}
=
\delta (i-j) \, 
\Bigg(
\delta (k-i) \, \lambda_k 
+\;
\delta (i-i^*) \, \lambda_{i^*} 
   \frac{1}{\theta_k \prod_{l\ne i^*} \theta_{l}}
\Bigg)
.
\ee

If, for example, ${{\bf K}}$ is a translationally invariant operator
it is diagonalized in a basis of plane waves.
Then also $\tilde {{\bf K}}$ is translationally invariant,
but its sensitivity to certain frequencies has changed.
The optimal sensitivity pattern is determined
by the given stationarity equations.

\subsubsection{Regularization parameters}

Next we consider the example
${{\bf K}}(\gamma)$ = $\gamma {{\bf K}}_0$ 
where $\theta \ge 0$ has been denoted $\gamma$,
representing a regularization parameter or
an inverse temperature variable for the specific prior.
For a $d$--dimensional Gaussian integral
the normalization factor becomes
$Z_\phi (\gamma)$ 
= $(\frac{2\pi}{\gamma})^\frac{d}{2}(\det {{\bf K}}_0)^{-1/2}$.
For positive (semi)definite ${{\bf K}}$ the dimension $d$ is
given by the rank of ${{\bf K}}$ under a chosen discretization.
Skipping constants results in a normalization energy
$E_N(\gamma)$ = $-\frac{d}{2}\ln \gamma$.
With
\be
\frac{\partial {{\bf K}}}
     {\partial \gamma}
=
{{\bf K}}_0
\ee
we obtain the stationarity equations
\bea
\gamma {{\bf K}}_0(\phi-t) &=& 
{\bf P}^\prime (\phi) {\bf P}^{-1}(\phi) N  
- {\bf P}^\prime (\phi) \Lambda_X
,
\label{adaT1}
\\
\frac{1}{2} 
(\phi-t ,\,{{\bf K}}_0\,(\phi-t))
&=& 
\frac{d}{2\, \gamma}
-E_\gamma^\prime
.
\label{adaT2}
\eea
For compensating hyperprior
the right hand side of Eq.\ (\ref{adaT2}) vanishes, 
giving thus no stationary point for $\gamma$. 
Using however the condition $\gamma\ge 0$
one sees that for positive definite ${{\bf K}}_0$ Eq.\ (\ref{adaT1})
is minimized for $\gamma$ = $0$
corresponding to the `prior--free' case.
For example, in the case of Gaussian regression the solution 
would be the data template $\phi$ = $h$ = $t_D$. 
This is also known as ``$\delta$--catastrophe''.
To get a nontrivial solution for $\gamma$
a noncompensating hyperparameter energy $E_\gamma$ = $E_\theta$ 
must be used
so that $\ln Z_\phi + E_N$ is nonuniform
\cite{Berger-1980,Bishop-1995b}.

The other limiting case is 
a vanishing $E_\gamma^\prime$ for which
Eq.\ (\ref{adaT2}) becomes
\be
\gamma = \frac{d}{ (\phi-t ,\,{{\bf K}}_0\,(\phi-t)) }.
\ee
For $\phi\rightarrow t$
one sees that $\gamma\rightarrow \infty$.
Moreover, in case $P[t]$ represents a normalized probability,
$\phi=t$ is also a solution of the first stationarity equation (\ref{adaT1})
in the limit $\gamma\rightarrow \infty$.
Thus, for vanishing $E_\gamma^\prime$ 
the `data--free' solution $\phi=t$
is a selfconsistent solution
of the stationarity equations (\ref{adaT1},\ref{adaT2}).

\begin{figure}[tb]
\begin{center}
\epsfig{file=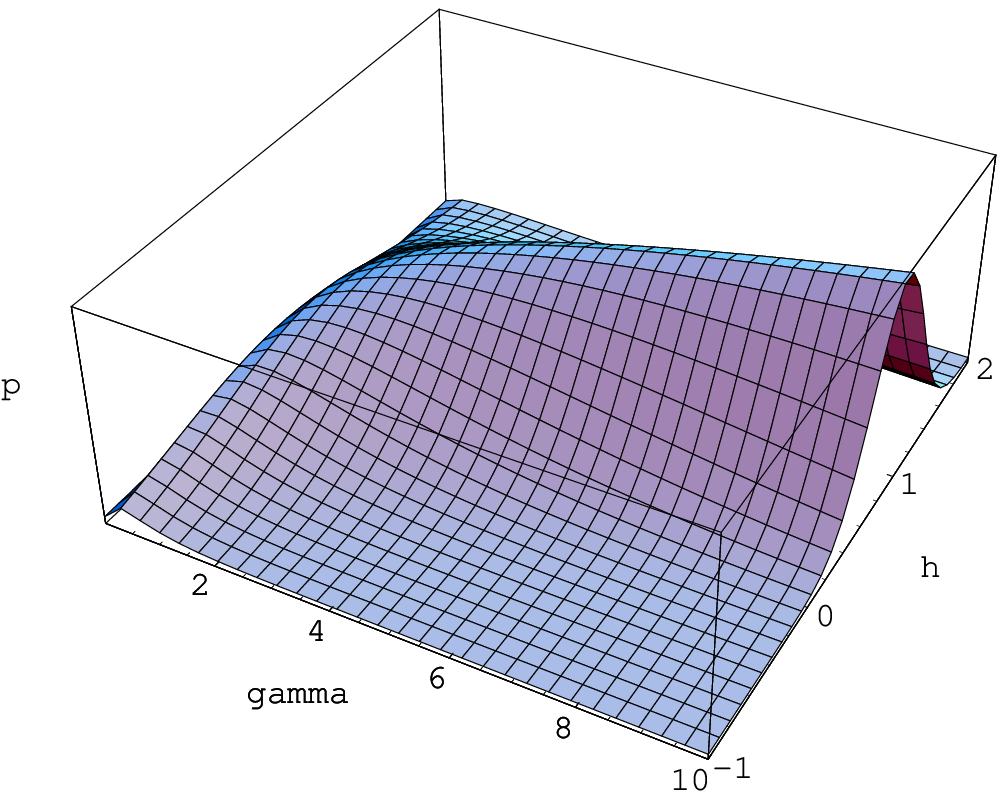, width= 65mm}    
\epsfig{file=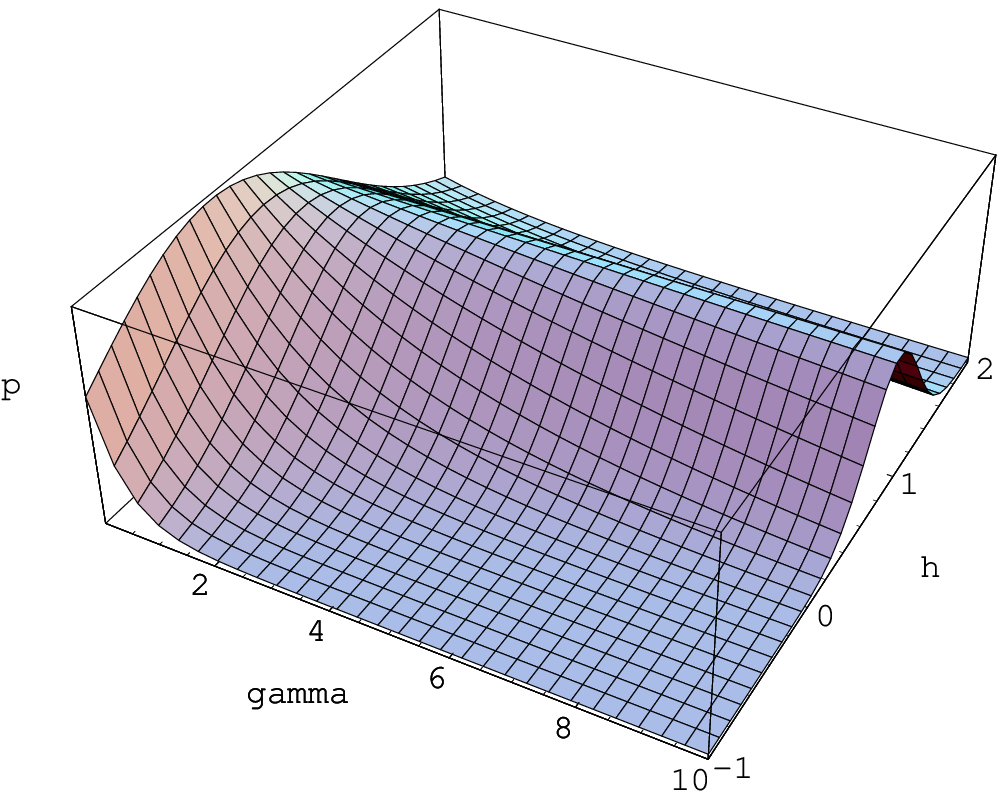, width= 65mm}    
\end{center}
\caption{Shown is the joint posterior density of $h$ and $\gamma$, i.e.,
$p({h},\gamma|D,D_0)$
$\propto p(y_D|{h})p({h}|\gamma,D_0)p(\gamma)$
for a zero--dimensional example of Gaussian regression
with training data $y_D=0$ and prior data $y_{D_0}=1$.
L.h.s: 
For uniform prior $p(\gamma) \propto 1$ 
so that the joint posterior becomes
$p \propto e^{-\frac{1}{2} {h}^2 -\frac{\gamma}{2} ({h}-1)^2 
+\frac{1}{2}\ln \gamma}$, 
having its maximum is at $\gamma$ = $\infty$, ${h}=1$.
R.h.s.:
For compensating hyperprior 
$p(\gamma) \propto 1/\protect\sqrt{\gamma}$ so that
$p \propto e^{-\frac{1}{2} {h}^2 -\frac{\gamma}{2} ({h}-1)^2}$
having its maximum is at $\gamma$ = $0$, ${h}=0$.
}
\label{priorT}
\end{figure}

Fig.\ref{priorT} shows a posterior surface 
for uniform and for compensating hyperprior
for a one--dimensional regression example.
The Maximum A Posteriori Approximation
corresponds to the highest point of the joint posterior
over $\gamma$, $h$ in that figures.
Alternatively one can treat
the $\gamma$--integral 
by Monte--Carlo--methods \cite{Williams-Barber-1998}.

Finally we remark that
in the setting of empirical risk minimization,
due to the different interpretation of the error functional,
regularization parameters are usually determined by
cross--validation or similar techniques
\cite{Mosteller-Wallace-1963,Allen-1974,Wahba-Wold-1975,Stone-1974,Stone-1977,Golup-Heath-Wahba-1979,Craven-Wahba-1979,Silverman-1986,Wahba-1990,Efron-Tibshirani-1993,Green-Silverman-1994}.

\subsection{Exact posterior for hyperparameters}

In the previous sections we have studied saddle point approximations
which lead us to maximize  the joint posterior $p(h,\theta|D,D_0)$
simultaneously with respect to the hidden variables $h$ and $\theta$
\be
p(y|x,D,D_0) =
p(y_D|x_D,D_0)^{-1} 
\!\!\int \!\!dh\,  
\int \!\!d\theta \,
p(y|x,h) 
\underbrace{
  p(y_D|x_D,h) p(h|D_0,\theta) p(\theta)
}_{\propto p(h,\theta|D,D_0),\; \rm max\, w.r.t.\, \theta\; and\; h}
,
\label{spa1}
\ee
assuming for the maximization with respect to $h$
a slowly varying $p(y|x,h)$ at the stationary point.

This simultaneous maximization with respect to both variables
is consistent with the usual asymptotic
justification of a saddle point approximation.
For example, for a function $f(h,\theta)$ 
of two 
(for example, one--dimensional) 
variables $h$, $\theta$
\be
\int \!dh\, d\theta\, e^{-\beta f(h,\theta)}
\approx
e^{-\beta f(h^*,\theta^*)
-\frac{1}{2}\ln \det (\beta {\bf H}/2\pi)
}
\label{spaex1}
\ee
for large enough $\beta$ (and a unique maximum).
Here $f(h^*,\theta^*)$ denotes the joint minimum
and ${\bf H}$ the Hessian of $f$ with respect to $h$ and $\theta$.
For $\theta$--dependent determinant of the covariance 
and the usual definition of $\beta$,
results in a function $f$ of the form
$f(h,\theta)$ 
= $E(h,\theta) + (1/2\beta)\ln \det (\beta {\bf K}(\theta)/2 \pi)$,
where both terms are  relevant for
the minimization of $f$ with respect to $\theta$.
For large $\beta$, however, the second term
becomes small compared to the first one.
(Of course, there is the possibility that a saddle point approximation 
is not adequate for the $\theta$ integration.
Also, we have seen 
that the condition of a positive definite covariance 
may lead to a solution for $\theta$ on the boundary 
where the (unrestricted) stationarity equation is not fulfilled.)

Alternatively,
one might think of performing the two integrals stepwise.
This seems especially useful if one integral
can be calculated analytically.
Consider, for example
\be
\int \!dh\, d\theta\, e^{-\beta f(h,\theta)}
\approx
\int \!d\theta\, e^{
-\beta f(\theta,h^*(\theta))
-\frac{1}{2}\ln \det (\frac{\beta}{2\pi}
\frac{\partial^2 f(h^*(\theta))}{\partial h^2})
}
\label{spaex2}
\ee
which would be exact for a Gaussian $h$--integral.
One sees now that 
minimizing the complete negative exponent
$\beta f(\theta,h^*)$
+ $\frac{1}{2}\ln \det (\beta (\partial^2 f/\partial h^2)/2\pi)$
with respect to $\theta$
is different from minimizing only $f$ in (\ref{spaex1}),
if the second derivative of $f$ with respect to $h$ depends on $\theta$
(which is not the case for a Gaussian $\theta$ integral).
Again this additional term becomes negligible for large enough $\beta$. 
Thus, at least asymptotically, 
this term may be altered or even be skipped,
and differences in the results of 
the variants of saddle point approximation
will be expected to be small.

Stepwise approaches like (\ref{spaex2}) can be used, for example
to perform Gaussian integrations analytically,
and lead to somewhat simpler
stationarity equations for $\theta$--dependent covariances 
\cite{Williams-Barber-1998}.

In particular, 
let us look at 
the case of Gaussian regression in a bit more detail.
The following discussion, however, also applies 
to density estimation if, as in (\ref{spaex2}), 
the Gaussian first step integration is replaced
by a saddle point approximation including the normalization factor. 
(This requires the calculation of the determinant of the Hessian.)
Consider the two step procedure for Gaussian regression
\bea
p(y|x,D,D_0)\! &=& \!
p(y_D|x_D,D_0)^{-1} 
\!\!\int \!\!d\theta 
\underbrace{
 p(\theta)
 \underbrace{
  \!\!\int \!\!dh\,  p(y|x,h) 
     p(y_D|x_D,h) p(h|D_0,\theta) 
}_{\rm exact}
}_{p(\theta)p(y,y_D|x,x_D,D_0,\theta) 
\propto p(y,\theta|x,D,D_0)
{\rm\, max\, w.r.t.\,} \theta}
,
\nonumber\\&=&\!
\int \! d\theta\,   
\underbrace{
\underbrace{
 p(\theta|D,D_0)
}_{\propto\rm exact} 
\underbrace{
 p(y|x,D,D_0,\theta)
}_{\rm exact}
}_{p(y,\theta|x,D,D_0),\; \rm max\, w.r.t.\, \theta}
\label{spa3}
\eea
where in a first step $p(y,y_D|x,x_D,D_0,\theta)$ can be calculated 
analytically and in a second step the $\theta$ integral is performed
by Gaussian approximation around a stationary point.
Instead of maximizing
the joint posterior $p(h,\theta|D,D_0)$ with respect to $h$ and $\theta$
this approach performs the $h$--integration analytically and maximizes
$p(y,\theta|x,D,D_0)$ with respect to $\theta$.
The disadvantage of this approach 
is the $y$--, and $x$--dependency of the resulting solution.

Thus, 
assuming a slowly varying
$p(y|x,D,D_0,\theta)$ at the stationary point
it appears simpler to
maximize the $h$--marginalized
posterior, $p(\theta|D,D_0)$ = $\int dh\, p(h,\theta|D,D_0)$,
if the $h$--integration can be performed exactly,
\be
p(y|x,D,D_0) =
\int \! d\theta\,   
\underbrace{
\underbrace{
 p(\theta|D,D_0) 
}_{\rm exact}
}_{\rm max\, w.r.t.\, \theta }
\underbrace{
p(y|x,D,D_0,\theta)
}_{\rm exact}
.
\label{spa2}
\ee
Having found a maximum posterior solution $\theta^*$ 
the corresponding analytical solution
for $p(y|x,D,D_0,\theta^*)$
is then given by Eq.\ (\ref{predictive-dens}).
The posterior density $p(\theta|D,D_0)$
can be obtained from the likelihood of $\theta$
and a specified prior $p(\theta)$
\be
p(\theta|D,D_0) =
\frac{p(y_D|x_D,D_0,\theta)
p(\theta)}{p(y_D|x_D,D_0)} 
.
\ee
Thus, in case the $\theta$--likelihood can be calculated 
analytically,
the $\theta$--integral is calculated in saddle point approximation
by maximizing the posterior for $\theta$
with respect to $\theta$.
In the case of a uniform $p(\theta)$
the optimal $\theta^*$ is obtained
by maximizing the $\theta$--likelihood.
This is also known as {\it empirical Bayes approach}
\cite{Carlin-Louis-1996}.
As $h$ is integrated out in $p(y_D|x_D,D_0,\theta)$
the $\theta$--likelihood is also called
{\it marginalized likelihood}.

Indeed, for Gaussian regression,
the $\theta$-likelihood 
can be integrated analytically,
analogously to Section \ref{exact},
yielding \cite{Wahba-1990,Williams-Rasmussen-1996,Williams-Barber-1998},
\bea
p(y_D|x_D,D_0,\theta) 
&=&
\int\!dh\, p(y_D|x_D,h) \, p(h|D_0,\theta)
\nonumber\\&=&
\int\!dh\, 
e^{-\frac{1}{2} \sum_{i=0}^{n}
           \big( {h}-t_i,\, {\bf K}_i ({h}-t_i) \big) 
    +\frac{1}{2} \sum_{i=0}^{n} \ln \det_i ({\bf K}_i/2\pi)
}
\nonumber\\
&=&
e^{
 -\frac{1}{2}\sum_{i=0}^n \big( t_i,\, {\bf K}_i t_i \big) 
 +\frac{1}{2}\big( t,\, {\bf K} t \big) 
 +\frac{1}{2}\ln \det_D(\widetilde {\bf K} /2\pi)
}
\nonumber\\
&=&
e^{-\frac{1}{2}
  \Big( t_D - t_0,\, \widetilde {\bf K}
      ( t_D - t_0) \Big) 
 +\frac{1}{2}\ln\det_D \widetilde {\bf K} 
 -\frac{\tilde n}{2}\ln(2\pi) 
}
\nonumber\\
&=&
e^{-\widetilde E
 +\frac{1}{2}\ln\det_D \widetilde {\bf K} 
 -\frac{\tilde n}{2}\ln(2\pi) 
}
,
\label{anal-likel}
\eea
where
$\widetilde E$ = $\frac{1}{2}
\Big( t_D - t_0,\, \widetilde {\bf K}
      ( t_D - t_0) \Big)$,
$\widetilde {\bf K}$ 
= $({\bf K}_D^{-1}+{\bf K}_{0,DD}^{-1}(\theta))^{-1}$
= ${\bf K}_D + {\bf K}_D{\bf K}^{-1}{\bf K}_D$, 
$\det_D$ the determinant in data space,
and we used that
from
${\bf K}_i^{-1}{\bf K}_j$ = $\delta_{ij}$ for  $i,j>0$
follows
$\sum_{i=0}^n \big( t_i,\, {\bf K}_i t_i \big)$
= $\big( t_D,\, {\bf K}_D t_D \big)$
+ $\big( t_0,\, {\bf K}_0 t_0 \big)$
= $\big( t_D,\, {\bf K} t\big)$,
with
${\bf K}$  = $\sum_{i=0}^{n} {\bf K}_i$.
In cases where the marginalization over $h$, 
necessary to obtain the evidence,
cannot be performed analytically
and all $h$--integrals are calculated in saddle point approximation,
we get the same result as 
for a direct simultaneous MAP for $h$ and $\theta$
for the predictive density
as indicated in (\ref{spa1}).

Now we are able to compare the three
resulting stationary equations for
$\theta$--dependent mean $t_0(\theta)$,
covariance ${\bf K}_0(\theta)$ and prior $p(\theta)$.
Setting the derivative of the joint posterior $p(h,\theta|D,D_0)$
with respect to $\theta$ 
to zero yields 
\bea
0&=&
\left( \frac{\partial t_0}{\partial \theta},\; {\bf K}_0(t_0-{h})\right)\
+\frac{1}{2} 
\Big({h}-t_0 ,
   \,\frac{\partial {{\bf K}_0}(\theta)}{\partial \theta}\,({h}-t_0)\Big)
\nonumber\\&&
-{\rm Tr} \left({\bf K}_0^{-1}
 \frac{\partial {{\bf K}_0}}{\partial \theta}\right)
-\frac{1}{p(\theta)} \frac{\partial p(\theta)}{\partial \theta}
.
\label{firstStEq}
\eea
This equation which we have already discussed
has to be solved simultaneously with the stationarity equation
for $h$.
While this approach is easily adapted
to general density estimation problems,
its difficulty for $\theta$--dependent covariance determinants
lies in calculation of the derivative of the determinant of ${\bf K}_0$.
Maximizing the $h$--marginalized posterior $p(\theta|D,D_0)$, 
on the other hand,
only requires 
the calculation of the derivative of the determinant of the 
$\tilde n \times \tilde n$ matrix 
$\widetilde {\bf K}$
\bea
0 &=& 
\left( \frac{\partial t_0}{\partial \theta},\; 
\widetilde {\bf K}(t_0-t_D)\right)\
+\frac{1}{2} \left((t_D-t_0),\, 
\frac{\partial \widetilde {\bf K}}{\partial \theta}(t_D-t_0)\right)
\nonumber\\&&
-{\rm Tr}\left(\widetilde {\bf K}^{-1}
  \frac{\partial \widetilde {\bf K}}{\partial \theta}\right)
-\frac{1}{p(\theta)} \frac{\partial p(\theta)}{\partial \theta}
.
\label{ex-theta1}
\eea
Evaluated
at the stationary 
$h^*$ = $t_0+{\bf K}_0^{-1} \widetilde {\bf K} (t_D-t_0)$,
the first term of Eq.\ (\ref{firstStEq}),
which does not contain derivatives of the covariances, 
becomes equal to the first term of Eq.\ (\ref{ex-theta1}).
The last terms of 
Eqs.\ (\ref{firstStEq}) and (\ref{ex-theta1})
are always identical.
Typically, the data--independent ${\bf K}_0$ 
has a more regular structure
than the data--dependent $\widetilde {\bf K}$.
Thus,
at least for one or two dimensional $x$,
a straightforward numerical solution of Eq.\ (\ref{firstStEq})
by discretizing $x$
can also be a good choice for Gaussian regression problems.

Analogously, from Eq.\ (\ref{predictive-dens}) follows
for maximizing $p(y,\theta|x,D,D_0)$ with respect to $\theta$
\bea
0 &=& 
\left( \frac{\partial t}{\partial\theta},\;{\bf K}_{y}(t - y)\right)\
+\frac{1}{2} \left((y-t),\, 
\frac{\partial {\bf K}_{y}}{\partial \theta}(y-t)\right)
\nonumber\\&&
-{\rm Tr}\left({\bf K}_{y}^{-1}
  \frac{\partial {\bf K}_{y}}{\partial \theta}\right)
-\frac{1}{p(\theta|D,D_0)} \frac{\partial p(\theta|D,D_0)}{\partial \theta}
,
\eea
which is $y$--, and $x$--dependent.
Such an approach may be considered
if interested only in specific test data $x$, $y$.

We may remark that also in Gaussian regression the $\theta$--integral
may be quite different from a Gaussian integral,
so a saddle point approximation 
does not necessarily have to give satisfactory results.
In cases one encounters problems one can, for example, try 
variable transformations 
$\int f(\theta)d\theta$ = 
$\int \det (\partial \theta/\partial \theta^\prime) 
f(\theta(\theta^\prime))d\theta^\prime$ 
to obtain a more Gaussian shape of the integrand.
Due to the presence of the Jacobian determinant, however,
the asymptotic interpretation of the corresponding saddle point approximation
is different for the two integrals.
The variability of saddle point approximations
results from the freedom to add terms which vanish 
asymptotically but remains finite in the nonasymptotic region.
Similar effects are known in quantum many body theory
(see for example \cite{Negele-Orland-1988}, chapter 7.)
Alternatively, the $\theta$--integral can be solved numerically
by Monte Carlo methods\cite{Williams-Rasmussen-1996,Williams-Barber-1998}.

\subsection{Integer hyperparameters}

The hyperparameters $\theta$ considered up to now
have been real numbers, or vector of real numbers.
Such hyperparameters can describe continuous transformations,
like the translation, rotation or scaling of template functions
and the scaling of covariance operators.
For real $\theta$ and differentiable posterior,
stationarity conditions can be found by differentiating
the posterior with respect to $\theta$.

Instead of a class of continuous transformations
a finite number of alternative template functions or covariances
may be given.
For example, an image to be reconstructed 
might be expected to show a digit between zero and nine, 
a letter from some alphabet,
or the face of someone
who is a member of known group of people.
Similarly,
a particular times series may 
be expected to be either in a high or in a low variance regime.
In all these cases, 
there exist a finite number of classes $i$
which could be represented by specific templates $t_i$
or covariances ${\bf K}_i$.
Such ``class'' variables $i$ are 
nothing else than hyperparameters 
$\theta$ with integer values.

Binary parameters, for example,
allow to select from two reference functions or two covariances
that one which fits the data best.
E.g., for $i$ = $\theta\in \{0,1\}$ one can write
\bea
t(\theta) &=& (1-\theta) t_1 + \theta t_2,\label{integer-hyper-t}\\
{\bf K}(\theta) &=& (1-\theta) {\bf K}_1 + \theta {\bf K}_2
\label{integer-hyper-K}
.
\eea

For integer $\theta$ the integral
$\int\! d\theta$ 
becomes a sum $\sum_\theta$ 
(we will also use the letter $i$ and write $\sum_i$ 
for integer hyperparameters),
so that prior, posterior, and predictive density
have the form of a {\it finite mixture} 
with components $\theta$.

For a moderate number of components
one may be able to include all of the mixture components.
Such prior mixture models will be studied in Section \ref{non-Gaussian}.

If the number of mixture components is too large to 
include them all explicitly,
one again must restrict to some of them.
One possibility is to select a random sample
using Monte--Carlo methods. 
Alternatively, one may search for the $\theta^*$
with maximal posterior.
In contrast to typical optimization problems for real variables,
the corresponding integer optimization problems
are usually not very smooth with respect to $\theta$
(with smoothness defined in terms of differences instead of derivatives),
and are therefore often much harder to solve.

There exists, however,
a variety of deterministic and stochastic integer optimization algorithms,
which may be combined with ensemble methods like genetic algorithms
\cite{Holland-1975,Goldberg-1989,Davis-1991,
Michalewicz-1992,Koza-1992,Schwefel-1995,Mitchell-1996},
and with homotopy methods, like simulated annealing
\cite{Kirkpatrick-Gelatt-Vecchi-1983,Mezard-Parisi-Virasoro-1987,Ripley-1987,Davis-1987,Aarts-Korts-1989,Rose-Gurewitz-Fox-1990,Yuille-1990,Gelfand-Mitter-1993,Yuille-Kosowski-1994,Yuille-Stolorz-Utans-1994}.
Annealing methods are similar to (Markov chain) Monte--Carlo methods,
which aim in sampling many points 
from a specific distribution
(i.e., for example at fixed temperature).
For them it is important to have (nearly) independent samples
and the correct limiting distribution of the Markov chain.
For annealing methods
the aim is to find the correct minimum 
(i.e., the ground state having zero temperature)
by smoothly changing the temperature from a finite value to zero.
For them it is less important to model the distribution
for nonzero temperatures exactly, but
it is important to use an adequate
cooling scheme for lowering the temperature.

Instead of an integer optimization problem 
one may also try to solve a similar problem
for real $\theta$.
For example, 
the binary $\theta\in \{0,1\}$ 
in Eqs.\ (\ref{integer-hyper-t}) and (\ref{integer-hyper-K})
may be extended to real $\theta\in [0,1]$.
By smoothly increasing an appropriate additional hyperprior $p(\theta)$
one can finally enforce again binary hyperparameters $\theta\in\{0,1\}$.

\subsection{Local hyperfields}
\label{hyperfields}

Most, but not all hyperparameters $\theta$ 
considered so far have been real or integer {\it numbers}, 
or {\it vectors} with real or integer components $\theta_i$.
With the unrestricted template functions of Sect.\ \ref{unrestricted-variation}
or the functions parameterizing the covariance
in Sections \ref{local-smoothness} and \ref{local-masses},
we have, however, also already encountered 
{\it function hyperparameters} or {\it hyperfields}.
In this section we will now discuss 
function hyperparameters in more detail.

Functions can be seen as continuous vectors,
the function values $\theta(u)$
being the (continuous) analogue of vector components $\theta_i$.
In numerical calculations, in particular, 
functions usually have to be discretized,
so, numerically, functions stand for high dimensional vectors.

Typical arguments of function hyperparameters
are the independent variables $x$ 
and, for general density estimation, also the dependent variables $y$. 
Such functions $\theta(x)$ or $\theta(x,y)$ will be called
{\it local hyperparameters} or {\it local hyperfields}.
Local hyperfields $\theta(x)$ 
can be used, for example,
to adapt templates or covariances locally.
(For general density estimation problems
replace here and in the following $x$ by $(x,y)$.)

The price to be paid for the additional flexibility
of function hyperparameters
is a large number of additional degrees of freedom. 
This can considerably complicate calculations and,
requires a sufficient number of training data
and/or  a sufficiently restrictive hyperprior
to be able to determine the hyperfield
and not to make the prior useless.

To introduce local hyperparameters $\theta (x)$
we express real symmetric, positive (semi--) definite inverse covariances
by square roots or ``filter operators'' ${\bf W}$,
${\bf K}$ = ${\bf W}^T{\bf W}$
=
$\int \!dx\; {W}_x{W}^T_x$
where
${W}_x$ represents the vector ${\bf W}(x,\cdot)$.
Thus, in components
\be
{\bf K}(x,x^\prime) 
= \int \!dx^{\prime\prime}\; 
   {\bf W}^T(x,x^{\prime\prime}){\bf W}(x^{\prime\prime},x^{\prime})
,
\ee
and therefore
\bea
\Big( \phi-t\, , \,{\bf K} (\phi-t) \Big)
&=&
\int\! dx\,dx^\prime\, 
[\phi(x)-t(x)]
\,  {\bf K}^T(x,x^{\prime}) \,
[\phi(x^{\prime})-t(x^{\prime})]
\nonumber\\
&=&
\int\! dx\,dx^\prime\, dx^{\prime\prime}\,
[\phi(x)-t(x)]
 {\bf W}^T(x,x^{\prime})
\nonumber\\
&&\qquad\times\;
{\bf W}(x^{\prime},x^{\prime\prime})
[\phi(x^{\prime\prime})-t(x^{\prime\prime})]
\nonumber\\
&=&
\int \! dx\, |\omega (x)|^2
,
\eea
where we defined the  ``filtered differences''
\be
\omega (x)
=
\big( \, W_x\, ,\, \phi-t \, \big)
=
\int \!dx^\prime \, {\bf W}(x,x^\prime) 
[\phi(x^\prime)-t(x^\prime )]
.
\label{filtered-diff}
\ee
Thus, for a Gaussian prior for $\phi$ we have
\be
p(\phi) \propto
e^{- \frac{1}{2} \big( \phi-t\, , \,{\bf K} (\phi-t) \big)}
=
e^{- \frac{1}{2}\int \!dx \, |\omega(x)|^2}
.
\ee
A real local hyperfield
$\theta (x)$
mixing, for instance,  
locally two alternative filtered differences
may now be introduced as follows
\be
p(\phi|\theta) =
e^{-\frac{1}{2}
\int \!dx |\omega (x;\theta)|^2
-\ln Z_\phi(\theta)}
=
e^{-\frac{1}{2} \int \!dx \, 
 \left| [1-\theta(x) ] \omega_1(x)
 +\theta(x) \omega_2(x)
 \right|^2
-\ln Z_\phi(\theta)}
\label{local-hyper-p-r}
,
\ee
where
\be
\omega (x;\theta) 
= [1-\theta(x)] \, \omega_1(x) + \theta(x) \,\omega_2(x)
\label{hyper-function-omega}
,
\ee
and, say, $\theta(x)\in [0,1]$.
For unrestricted real $\theta(x)$
an arbitrary real $\omega (x;\theta)$ can be obtained.
For a binary local hyperfield
with $\theta(x)\in \{0,1\}$
we have
$\theta^2$ = $\theta$,
$(1-\theta)^2$ = $(1-\theta)$,
and 
$\theta(1-\theta)$ = $0$,
so
Eq.~(\ref{local-hyper-p-r})  
becomes
\be
p(\phi|\theta) =
e^{-\frac{1}{2}
\int \!dx |\omega (x;\theta)|^2
-\ln Z_\phi(\theta)}
=
e^{-\frac{1}{2} \int \!dx \, 
 \left( [1-\theta(x)]|\omega_1(x)|^2
 +\theta(x) |\omega_2(x)|^2
 \right)
-\ln Z_\phi(\theta)}
\label{local-hyper-p}
.
\ee
For real $\theta(x)$
in Eq.\ (\ref{hyper-function-omega}) 
terms with $\theta^2(x)$, 
$[1-\theta(x)]^2$, and $[1-\theta(x)]\theta(x)$
would appear in Eq.\ (\ref{local-hyper-p}).
A binary $\theta$ variable can
be obtained from a real
$\theta$ 
by replacing
\be
B_\theta(x) = \Theta(\theta(x)-\vartheta)
\rightarrow  \theta(x)
.
\label{def-B}
\ee
Clearly, if both prior and hyperprior are formulated 
in terms of such $B_\theta(x)$
this is equivalent to using directly a binary hyperfield.
 
For a local hyperfield $\theta(x)$
a local adaption of the functions $\omega (x;\theta)$ 
as in Eq.~(\ref{hyper-function-omega})
can be achieved by switching
locally between alternative templates 
or alternative filter operators ${\bf W}$ 
\bea
t_x(x^\prime;\theta) 
&=& 
[1-\theta(x)] \, t_{1,x}(x^\prime)
 + \theta(x)\, t_{2,x}(x^\prime),
\label{hyper-function-t}
\\
{\bf W}(x,x^\prime; \theta) &=& 
[1-\theta(x)] \, {\bf W}_{1}(x,x^\prime)
+ \theta(x) \,{\bf W}_{2}(x,x^\prime)
\label{hyper-function-W}
.
\eea
In Eq.~(\ref{hyper-function-t})
it is important to notice that ``local'' templates 
$t_x(x^\prime; \theta)$ 
for fixed $x$ 
are still functions 
of an $x^\prime$ variable.
Indeed, to obtain $\omega(x;\theta)$,
the function $t_x$ is needed for all $x^\prime$ for which ${\bf W}$
has nonzero entries. 
\be
\omega(x;\theta )
=
\int\!dx^\prime\, 
{\bf W}(x,x^\prime) 
[\phi(x^\prime)-t_x(x^\prime;\theta)]
.
\ee
That means that the template
is adapted individually for every local filtered difference.
Thus, Eq.~(\ref{hyper-function-t})
has to be distinguished from the choice
\be
t(x^\prime;\theta) 
=
[1-\theta(x^\prime )] \, t_{1}(x^\prime)
 + \theta(x^\prime )\, t_{2}(x^\prime)
\label{mixing-t}
.
\ee
The unrestricted adaption of templates
discussed in Sect.\ \ref{unrestricted-variation},
for instance,
can be seen as an approach of the form of
Eq.~(\ref{mixing-t}) 
with an unbounded real hyperfield $\theta(x)$.

Eq.~(\ref{hyper-function-W}) 
corresponds
for binary $\theta$ to an inverse covariance
\be
{\bf K}(\theta) 
= \int \!dx\; {\bf K}_x(\theta) 
= \int \!dx\; 
\left(
[1-\theta(x)] {W}_{1,x}{W}^T_{1,x}
+  \theta(x)  {W}_{2,x}{W}^T_{2,x}
\right)
,
\ee
where
${\bf K}_x(\theta)$ = ${W}_{x}(\theta){W}^T_{x}(\theta)$
and 
$W_{i,x}$ = ${\bf W}_i(x,\cdot)$,
$W_{x}(\theta )$ = ${\bf W}(x,\cdot\,;\theta)$.
We remark that $\theta$--dependent covariances
require to include the normalization factors
when integrating over $\theta$ or 
solving for the optimal $\theta$ in MAP.
If we consider
two binary hyperfields $\theta$, $\theta^\prime$,
one for $t$ and one for ${\bf W}$,
we find a prior
\be
p(\phi|\theta,\theta^\prime)
\propto
e^{-E(\phi|\theta,\theta^\prime)}
=
e^{-\frac{1}{2}
\int\!dx\, 
\big( \phi - t_x(\theta)\, , \,
   {\bf K}_x(\theta^\prime)\,[\phi-t_x(\theta)]\big)
}
.
\ee
Up to a $\phi$--independent constant 
(which still depends on $\theta$, $\theta^\prime$) 
the corresponding prior energy
 can again be written in the form
\be
E(\phi|\theta,\theta^\prime)
=
\frac{1}{2}
\Big( \phi-t(\theta,\theta^\prime)\, ,\, 
    {\bf K}(\theta^\prime ) [\phi-t(\theta,\theta^\prime)] \Big)
.
\ee
Indeed, 
the corresponding effective template $t(\theta,\theta^\prime)$
and effective covariance ${\bf K}(\theta^\prime)$
are according to Eqs.\ (\ref{regression-functional},\ref{Ereg2})
given by
\bea
t (\theta ,\theta^\prime)
&=&
{\bf K}(\theta^\prime)^{-1}
\int\!dx\, {\bf K}_x(\theta^\prime ) \,  t_x(\theta)
,
\\
{\bf K}(\theta^\prime)
&=&
\int \! dx\, {\bf K}_x(\theta^\prime) 
.
\eea
Hence, one may rewrite
\bea
\int\! dx \, |\omega(x;\theta,\theta^\prime)|^2 
&=&
\Big( \phi-t(\theta,\theta^\prime),\; 
{\bf K}(\theta^\prime) \, [\phi-t(\theta,\theta^\prime)] \Big)
\label{eff-E}
\\
&&+
\sum_x \Big( t_x(\theta),\; 
  {\bf K}_x(\theta^\prime) \, t_x(\theta) \Big)
-
\Big( t(\theta,\theta^\prime),\; 
  {\bf K} \,t(\theta,\theta^\prime) \Big)
.
\nonumber
\eea
The MAP solution of Gaussian regression
for a prior corresponding to (\ref{eff-E})
at optimal $\theta^*$, ${\theta^\prime}^*$
is according to Section \ref{regression}
therefore 
given by 
\be
\phi^* (\theta^*,{\theta^\prime}^*)
= 
[{\bf K}_D+{\bf K}({\theta^\prime}^*)]^{-1} 
\left({\bf K}_D t_D + {\bf K}({\theta^\prime}^*)\, 
t(\theta^*,{\theta^\prime}^*)\right)
.
\ee

One may avoid dealing with
``local'' templates $t_x(\theta)$
by adapting templates
in prior terms where ${\bf K}$ is equal to the identity ${\bf I}$.
In that case $(t_{0})_{x}(x^\prime;\theta)$ 
is only needed for $x$ = $x^\prime$
and we may thus directly write
$(t_{0})_{x}(x^\prime;\theta)$ 
=
$t_{0}(x^\prime;\theta)$.
As example, consider the following  prior energy,
where the $\theta$--dependent template
is located in a term with ${\bf K}$ = ${\bf I}$
and another, say smoothness, prior is added
with zero template
\be
E(\phi|\theta) =
\frac{1}{2}
\Big( \phi-t_0(\theta),\; (\phi-t_0(\theta)) \Big)
+\frac{1}{2}
\Big( \phi,\; {\bf K}_0 \,\phi \Big)
.
\ee
Combining both terms yields
\be
E(\phi|\theta) 
=
\frac{1}{2} 
\bigg( 
\Big(\phi-t(\theta),\; {\bf K}[\phi-t(\theta)] \Big)
+
\Big( t_0(\theta),\;       
  \left({\bf I}-{\bf K}^{-1}\right) t_0(\theta) \Big)
\bigg)
,
\ee
with effective template and effective inverse covariance
\be
t(\theta) = {\bf K}^{-1} t_0(\theta)
,\quad 
{\bf K} = {\bf I}+ {\bf K}_0
.
\ee
For differential operators ${\bf W}$
the effective $t(\theta)$ 
is thus a smoothed version of $t_0(\theta)$.

The extreme case would be to treat
$t$ and ${\bf W}$ itself as unrestricted hyperparameters.
Notice, however, that increasing flexibility tends to lower
the influence of the corresponding prior term.
That means,
using completely free templates and covariances
without introducing additional restricting hyperpriors,
just eliminates the corresponding prior term
(see Section \ref{unrestricted-variation}).

Hence, to restrict the flexibility,
typically a smoothness hyperprior may be imposed
to prevent highly oscillating functions $\theta (x)$.
For real $\theta(x)$, for example, a smoothness prior
like $(\theta, -\Delta \theta)$ can be used 
in regions where it is defined. 
(The space of $\phi$--functions
for which a smoothness prior $(\phi-t,\,{\bf K}(\phi-t))$ 
with discontinuous $t(\theta)$ is defined
depends on the locations of the discontinuities.)
An example of a non--Gaussian hyperprior is,
written here for a one--dimensional $x$,
\be
p(\theta) \propto 
e^{-\frac{\kappa}{2} \int\!dx \, C_\theta(x)}
,
\label{hyperprior-C}
\ee
where $\kappa$ is some constant 
and 
\be
C_\theta(x) = 
\Theta \left( \left(\frac{\partial \theta}{\partial x}\right)^2 
  - \vartheta_\theta\right)
.
\label{jumps}
\ee
is zero at locations where the square of the first derivative
is smaller than a certain 
threshold $0\le \vartheta_\theta < \infty$,
and one otherwise.
(The step function $\Theta$ is defined as $\Theta(x)$ = 0 for $x\le 0$
and  $\Theta(x)$ = 1 for $1<x\le\infty$.)
To enable differentiation
the step function $\Theta$ could be replaced by a sigmoidal function.
For discrete $x$ one can count analogously the number of jumps 
larger than a given threshold.
Similarly, one may penalize the number $N_d(\theta)$ 
of discontinuities
where $\left(\frac{\partial \theta}{\partial x}\right)^2$ = $\infty$
and use 
\be
p(\theta) \propto e^{-\frac{\kappa}{2} N_d(\theta)}
.
\label{hyperprior-Nd}
\ee
In the case of a binary field
this corresponds 
to counting
the number of times the field changes its value.

The expression $C_\theta$
of Eq. (\ref{jumps})
can be generalized to
\be
C_\theta(x)
= \Theta\left( |\omega_\theta(x)|^2-\vartheta_\theta\right)
,
\label{Cdef}
\ee
where,
analogously to Eq.~(\ref{filtered-diff}), 
\be
\omega_\theta(x)
=
\int \!dx^\prime \, 
{\bf W}_\theta(x,x^\prime) 
[\theta(x^\prime)-t_\theta(x^\prime)]
,
\ee
and
${\bf W}_\theta$
is some filter operator acting on the hyperfield
and
$t_\theta(x^\prime)$
is a template for the hyperfield.

Discontinuous functions $\phi$ can either be approximated
by using discontinuous templates $t(x;\theta)$
or by eliminating matrix elements of the inverse covariance
which connect the two sides of the discontinuity.
For example, consider the discrete version 
of a negative Laplacian
with periodic boundary conditions,
\be
{\bf K} = {\bf W}^T {\bf W} =
\left(
\begin{tabular}{    c     c     c     c     c    c }
                    2  & $-1$&  0  &  0  &  0  &$-1$\\
                   $-1$&  2  & $-1$&  0  &  0  & 0  \\
                    0  & $-1$&  2  & $-1$&  0  & 0  \\
                    0  &  0  & $-1$&  2  & $-1$&  0  \\
                    0  &  0  &  0  & $-1$&  2  & $-1$ \\
                   $-1$&  0  &  0  &  0  & $-1$&  2  \\
\end{tabular}
\right),
\ee
and square root,
\be
{\bf W} =
\left(
\begin{tabular}{    c     c     c     c     c    c }
                    1  & $-1$&  0  &  0  &  0  & 0  \\
                    0  &  1  & $-1$&  0  &  0  & 0  \\
                    0  &  0  &  1  & $-1$&  0  &  0 \\
                    0  &  0  &  0  &  1  & $-1$&  0  \\
                    0  &  0  &  0  &  0  &  1  & $-1$ \\
                   $-1$&  0  &  0  &  0  &  0  &  1  \\
\end{tabular}
\right)
.
\ee
The first three points
can be disconnected from the last three points
by setting 
${\bf W}(3)$ and ${\bf W}(6)$ to zero, namely,
\be
{\bf W} = 
\left(
\begin{tabular}{    c     c     c  |  c     c    c }
                    1  & $-1$&  0  &  0  &  0  & 0  \\
                    0  &  1  & $-1$&  0  &  0  & 0  \\
                    0  &  0  &  0  &  0  &  0  & 0  \\
\hline
                    0  &  0  &  0  &  1  & $-1$&  0  \\
                    0  &  0  &  0  &  0  &  1  & $-1$ \\
                    0  &  0  &  0  &  0  &  0  &  0  \\
\end{tabular}
\right)
\ee
so that
the smoothness prior is ineffective
between points from different regions,
\be
{\bf K} = {\bf W}^T {\bf W} =
\left(
\begin{tabular}{    c     c     c  |  c     c    c }
                    1  & $-1$&  0  &  0  &  0  & 0  \\
                  $-1$ &  2  & $-1$&  0  &  0  & 0  \\
                    0  & $-1$&  1  &  0  &  0  & 0  \\
\hline
                    0  &  0  &  0  &  1  & $-1$&  0  \\
                    0  &  0  &  0  & $-1$&  2  & $-1$ \\
                    0  &  0  &  0  &  0  & $-1$&  1  \\
\end{tabular}
\right)
.
\ee
In contrast to using discontinuous templates,
the height of the jump at the discontinuity 
has not to be given in advance
when
using such disconnected Laplacians (or other inverse covariances).
On the other hand
training data are then required for all separated regions
to determine the free constants 
which correspond to the zero modes of the Laplacian.

Non--Gaussian priors, 
which will be discussed in more detail in the next Section, 
often provide an alternative 
to the use of function hyperparameters.
Similarly to Eq.\ (\ref{jumps}) 
one may for example 
define a binary function $B(x)$ in terms of $\phi$, 
\be
B(x) = 
\Theta \left(|\omega_1(x)|^2-|\omega_2(x)|^2 - \vartheta \right)
,
\label{jumps2}
\ee
like, for a negative Laplacian prior,
\be
B(x) = 
\Theta \left( 
\left|\frac{\partial (\phi-t_1)}{\partial x}\right|^2
-\left|\frac{\partial (\phi-t_2)}{\partial x}\right|^2
            - \vartheta
\right)
.
\label{jumps2b}
\ee
Here $B(x)$ is directly determined by $\phi$
and is not considered as an independent hyperfield.
Notice also that the functions $\omega_i(x)$
and $B(x)$ may be nonlocal with respect to $\phi(x)$,
meaning they may depend on more than one $\phi(x)$ value.
The threshold $\vartheta$ has to be related to
the prior expectations on $\omega_i$.
A possible non--Gaussian prior for $\phi$ formulated in terms of $B$
can be,
\be
p(\phi) \propto
e^{-\frac{1}{2} \int\!dx\, 
\left(
|\omega_1(x)|^2 (1-B(x))
+
|\omega_2(x)|^2 B(x)
-\frac{\kappa}{2} N_d(B)
\right)
}
,
\label{omega-B-energy}
\ee
with
$N_d(B)$ counting the number of discontinuities of $B(x)$.
Alternatively to $N_d$ one may for a real $B$ define, 
similarly to (\ref{Cdef}),
\be
C(x)
= \Theta\left( |\omega_B(x)|^2-\vartheta_B\right)
,
\ee
with 
\be
\omega_B(x)
=
\int \!dx^\prime \, 
{\bf W}_B(x,x^\prime) 
[B(x^\prime)-t_B(x^\prime)]
,
\ee
and some filter operator
${\bf W}_B$
and template
$t_B$.
Similarly to the introduction of hyperparameters,
one can treat
$B(x)$ formally as an independent function
by including a term 
$
\lambda
\left(
B(x)-\Theta \left(|\omega_1(x)|^2-|\omega_2(x)|^2 - \vartheta 
\right)
\right)$
in the prior energy
and taking the limit $\lambda\rightarrow\infty$.

Eq.\ (\ref{omega-B-energy}) looks similar to
the combination of the prior (\ref{local-hyper-p})
with the hyperprior (\ref{hyperprior-Nd}),
\be
p(\phi,\theta) \propto
e^{-\frac{1}{2} \int\!dx\, 
\left(
|\omega_1(x)|^2 (1-B_\theta(x))
+
|\omega_2(x)|^2 B_\theta(x)
-\frac{\kappa}{2} N_d(B_\theta)
-\ln Z_\phi(\theta)
\right)
}
.
\label{omega-theta-energy}
\ee
Notice, however, that the definition (\ref{def-B}) of 
the hyperfield $B_\theta$ 
(and $N_d(B_\theta)$ or $C_\theta$, respectively),
is different from that of $B$ (and $N_d(B)$ or $C$),
which are direct functions of $\phi$.
If the $\omega_i$ differ only in their templates,
the normalization term can be skipped.
Then, identifying $B_\theta$ in (\ref{omega-theta-energy})
with a binary $\theta$ and assuming 
$\vartheta$ = $0$,
$\vartheta_\theta$ = $\vartheta_B$, 
${\bf W}_\theta$  = ${\bf W}_B$,
the two equations are equivalent
for 
$\theta$ = $\Theta\left(|\omega_1(x)|^2-|\omega_2(x)|^2\right)$.
In the absence of hyperpriors,
it is indeed easily seen 
that this is a selfconsistent solution for $\theta$,
given $\phi$.
In general, however, when 
hyperpriors are included,
another solution for $\theta$ 
may have a larger posterior.
Non--Gaussian priors will be discussed
in Section \ref{other-non-gaussian}.

Hyperpriors or non--Gaussian prior terms
are useful to enforce specific
global constraints for $\theta(x)$ or $B(x)$.
In images, for example, discontinuities 
are expected to form closed curves.
Hyperpriors, organizing discontinuities along lines or closed curves, 
are thus important for image segmentation
\cite{Geman-Geman-1984,Marroquin-Mitter-Poggio-1987,Geiger-Girosi-1991,Geiger-Yuille-1991,Winkler-1995,Zhu-Yuille-1996}.

\section{Non--Gaussian prior factors}
\label{non-Gaussian}

\subsection{Mixtures of Gaussian prior factors}
\label{mix-of-G}

Complex, non--Gaussian prior factors,
for example being multimodal,
may be constructed or approximated 
by using mixtures of simpler prior components.
In particular, it is convenient
to use as components or ``building blocks''
Gaussian densities,
as then many useful results obtained for Gaussian processes
survive the generalization to mixture models
\cite{Lemm-1996,Lemm-1998,Lemm-1998a,Lemm-1998b,Lemm-1998c}.
We will therefore in the following discuss
applications of mixtures of Gaussian priors.
Other implementations of non--Gaussian priors
will be discussed in Section \ref{other-non-gaussian}.

In Section \ref{prior-normalization} we have seen that 
hyperparameters label components of mixture densities.
Thus, if $j$ labels the components of a mixture model,
then $j$ can be seen as hyperparameter.
In Section \ref{hyperparameters} 
we have treated the corresponding hyperparameter integration
completely in saddle point approximation.
In this section we will assume the hyperparameters $j$ 
to be discrete and try to calculate 
the corresponding summation exactly.

Hence, consider
a discrete hyperparameter $j$,
possibly in addition to continuous hyperparameters $\theta$.
In contrast to the $\theta$--integral
we aim now in treating the analogous sum over $j$ exactly,
i.e., we want to study {\it mixture models}
\be
p(\phi,\theta |\tilde D_0)
= \sum_j^m p(\phi,\theta,j |\tilde D_0)
= \sum_j^m p(\phi|\tilde D_0,\theta,j) p(\theta,j)
\label{hiddenj}
.
\ee
In the following we concentrate on
{\it mixtures of Gaussian specific priors}.
Notice that such models do {\it not} correspond to
Gaussian mixture models for $\phi$ 
as they are often used in density estimation.
Indeed, the form of $\phi$ may be completely unrestricted,
it is only its prior or posterior density 
which is modeled by a mixture. 
We also remark that a strict asymptotical justification
of a saddle point approximation would require the introduction of a parameter
$\tilde \beta$ so that
$p(\phi,\theta |\tilde D_0)\propto e^{\tilde\beta \ln \sum_j p_j}$.
If the sum is reduced to a single term,
then $\tilde\beta$ corresponds to $\beta$.

We already discussed shortly in Section \ref{adaptingpm} that,
in contrast to a product of probabilities or a sum of error terms
implementing a probabilistic AND of approximation conditions,
a sum over $j$ implements a probabilistic OR.
Those alternative approximation conditions will in the sequel
be represented by alternative templates $t_j$ and covariances ${{\bf K}}_j$.
A prior (or posterior) density in form of a probabilistic OR means that 
the optimal solution does not necessarily have to approximate
all but possibly only one of the $t_j$
(in a metric defined by ${{\bf K}}_j$). 
For example, we may expect in an image reconstruction task
blue or brown eyes whereas a mixture
between blue and brown might not be as likely.
Prior mixture models are potentially useful for

\begin{itemize}
\item[1.] Ambiguous (prior) data.
Alternative templates can for example
represent different expected trends for a time series.
\item[2.] Model selection.
Here templates represent alternative reference models 
(e.g., different neural network architectures, decision trees)
and determining the optimal $\theta$
corresponds to training of such models.
\item[3.] Expert knowledge. 
Assume {\it a priori} knowledge to be formulated 
in terms of conjunctions and disjunctions 
of simple components or building blocks
(for example verbally).
E.g., an image of a face is expected to contain 
certain constituents (eyes, mouth, nose; AND)
appearing in various possible variants (OR).
Representing the simple components/building blocks 
by Gaussian priors centered around 
a typical example (e.g.,of an eye)
results in Gaussian mixture models.
This constitutes a possible interface between 
symbolic and statistical methods.
Such an application of prior mixture models 
has some similarities with the
quantification of ``linguistic variables'' by fuzzy methods
\cite{Klir-Yuan-1995,Klir-Yuan-1996}.
\end{itemize}

For a discussion of possible applications of prior mixture models
see also
\cite{Lemm-1996,Lemm-1998,Lemm-1998a,Lemm-1998b,Lemm-1998c}.
An application of prior mixture models to image completion
can be found in \cite{Lemm-1999a}.

\subsection{Prior mixtures for density estimation}
\label{mix-de}

The mixture approach (\ref{hiddenj}) 
leads in general to non--convex  error functionals.
For Gaussian components
Eq.\ (\ref{hiddenj}) results in an error functional
\bea
E_{\theta,\phi} 
&=&
-(\ln P(\phi),\,N)
+ (P(\phi),\, \Lambda_X )
\nonumber\\
&&-
\ln \sum_j e^{-\left(
\frac{1}{2} 
  \left(\phi-t_j(\theta),\,{{\bf K}}_j (\theta)\,(\phi-t_j(\theta))\right)
+\ln Z_\phi (\theta,j)+E_{\theta,j}
\right)}
,
\label{adapmix}
\\&=&
-\ln \sum_j e^{-E_{\phi,j}-E_{\theta,j}+c_j},
\eea
where
\be
E_{\phi,j} =
-(\ln P(\phi),\,N)
+ (P(\phi),\, \Lambda_X )
+\frac{1}{2} 
  \Big(\phi-t_j(\theta),\,{{\bf K}}_j (\theta)\,(\phi-t_j(\theta))\Big)
,
\ee
and 
\be
c_j
=
-\ln Z_\phi (\theta,j)
.
\ee
The stationarity equations for $\phi$ and $\theta$
\bea
0 &=& \sum_j^m \frac{\delta E_{\phi,j}}{\delta \phi}\, 
e^{-E_{\phi,j}-E_{\theta,j}+c_j},
\\
0 &=& \sum_j^m \left( \frac{\partial E_{\phi,j}}{\partial \theta} 
                    +\frac{\partial E_{\theta,j}}{\partial \theta} 
                    +{\bf Z}_j^\prime Z_\phi^{-1}(\theta,j) \right)
e^{-E_{\phi,j}-E_{\theta,j}+c_j}
,
\eea
can also be written 
\bea
0 &=& \sum_j^m \frac{\delta E_{\phi,j}}{\delta \phi}\,
p(\phi,\theta,j|\tilde D_0),
\\
0 &=& \sum_j^m \left( \frac{\partial E_{\phi,j}}{\partial \theta} 
                    +\frac{\partial E_{\theta,j}}{\partial \theta} 
                    +{\bf Z}_j^\prime Z_\phi^{-1}(\theta,j) \right)
p(\phi,\theta,j|\tilde D_0)
.
\eea
Analogous equations are obtained
for parameterized $\phi(\xi)$.

\subsection{Prior mixtures for regression}
\label{mixtures-for-regression}

For regression it is especially useful to introduce
an inverse temperature multiplying the terms
depending on $\phi$, i.e., likelihood and prior.\footnote{As also 
the likelihood term depends on $\beta$
it may be considered part of a $\tilde \phi$
together regression function $h(x)$. 
Due to its similarity to a regularization factor
we have included 
$\beta$ in this chapter about hyperparameters.}
As in regression $\phi$ is represented by the regression function $h(x)$
the temperature--dependent error functional becomes
\be
E_{\theta,{h}} 
= -\ln \sum_j^m e^{ - \beta E_{{h},j} - E_{\theta,\beta,j} + c_j}
= -\ln \sum_j^m e^{ -E_j + c_j}
,
\label{adapmixregr}
\ee
with
\be
E_j = E_D + E_{0,j}+E_{\theta,\beta,j}
,
\label{adapmixE}
\ee
\be
E_D =
\frac{1}{2} \left({h}-t_D ,\,{{\bf K}}_D\,({h}-t_D)\right)
,\quad
E_{0,j} =
\frac{1}{2} 
  \left({h}-t_j(\theta),\,{{\bf K}}_j (\theta)\,({h}-t_j(\theta))\right)
,
\ee
some hyperprior energy $E_{\theta,\beta,j}$,
and
\bea
c_j (\theta,\beta)
&=& -\ln Z_{h} (\theta,j,\beta)
+\frac{n}{2}\ln \beta -\frac{\beta}{2} V_D -c
\nonumber\\
&=& \frac{1}{2}\ln 
\det \big({{\bf K}}_j(\theta )\big)
+\frac{d+n}{2}\ln \beta
-\frac{\beta}{2} V_D
\eea
with some constant $c$.
If we also maximize with respect to $\beta$
we have to include the (${h}$--independent)
training data variance $V_D=\sum_i^n V_i$
where
$V_i$ = $\sum_k^{n_i} y(x_k)^2/n_{i} - t_D^2(x_i)$
is the variance of the $n_i$ training data at $x_i$.
In case every $x_i$ appears only once $V_D$ vanishes.
Notice that $c_j$ includes a contribution from the $n$ data points
arising from the $\beta$--dependent normalization
of the likelihood term. 
Writing the stationarity equation
for the hyperparameter $\beta$ separately,
the corresponding three stationarity conditions
are found as
\bea
0
&=& \sum_j^m 
\Big({{\bf K}}_D \,({h}-t_D)+{{\bf K}}_j \,({h}-t_j)\Big)
\, e^{-\beta E_{{h},j}-E_{\theta,\beta,j}+c_j},
\label{regr1}
\\
0 &=& \sum_j^m 
\left( 
      E_{{h},j}^\prime + E_{\theta,\beta,j}^\prime
      +{\rm Tr}\, \left(
                      {{\bf K}}_j^{-1}\frac{\partial {{\bf K}}_j}{\partial \theta} 
                  \right)
\right)
e^{-\beta E_{{h},j}-E_{\theta,\beta,j}+c_j}
,
\label{regr2}
\\
0 &=& \sum_j^m 
\left( 
       E_{0,j} + \frac{\partial E_{\theta,\beta,j}}{\partial \beta}
       + \frac{d+n}{2\beta}
\right)
e^{-\beta E_{{h},j}-E_{\theta,\beta,j}+c_j}
.
\label{regr3}
\eea
As $\beta$ is only a one--dimensional parameter
and its density can be quite non--Gaussian
it is probably most times more informative
to solve for varying values of $\beta$
instead to restrict to a single 
`optimal' $\beta^*$.
Eq.\ (\ref{regr1}) 
can also be written
\be
{h}
=
\left( {{\bf K}}_D + \sum_j^m a_j {{\bf K}}_j \right)^{-1}
\left( {{\bf K}}_D t_D + \sum_l^m a_j {{\bf K}}_j t_j \right)
\label{regr1a}
,
\label{regr-h}
\ee
with
\bea
a_j 
&=& p(j|h,\theta,\beta,D_0)
= \frac{e^{-E_j+c_j}}{\sum_k^m e^{-E_k+c_k}}
= \frac{e^{-\beta E_{0,j}-E_{\theta,\beta,j}+\frac{1}{2}\ln\det{{\bf K}}_j}}
{\sum_k^me^{-\beta E_{0,k}
  -E_{\theta,\beta,k}+\frac{1}{2}\ln\det{{\bf K}}_k}}
\nonumber\\
&=&
\frac{p(h|j,\theta,\beta,D_0)p(j|\theta,\beta,D_0)}
{p(h|\theta,\beta,D_0)}
=
\frac{p(h|j,\theta,\beta,D_0)p(j,\theta|\beta,D_0)}
{p(h,\theta|\beta,D_0)}
,
\label{full-aj}
\eea
being thus still a nonlinear equation for ${h}$.

\subsubsection{High and low temperature limits}

It are the limits of large and small $\beta$
which make the introduction of this additional parameter useful.
The reason being that the high temperature limit 
$\beta\rightarrow 0$ gives the convex case,
and statistical mechanics provides us 
with high and low temperature expansions.
Hence, we study the high temperature and low temperature limits
of Eq.\ (\ref{regr1a}).

In the {\it high temperature limit}
$\beta \rightarrow 0$
the exponential factors $a_j$ become ${h}$--independent
\be
a_j 
\stackrel{\beta\rightarrow0}{\longrightarrow}
a^0_j =
\frac{e^{-E_{\theta,\beta,j}+\frac{1}{2}\ln\det{{\bf K}}_j}}
{\sum_k^me^{-E_{\theta,\beta,k}+\frac{1}{2}\ln\det{{\bf K}}_k}}
.
\label{highTa}
\ee
In case one chooses 
$E_{\theta,\beta,j}$ = $E_{\beta,j} + \beta E_{\theta}$
one has to replace
$E_{\theta,\beta,j}$ by $E_{\beta,j}$.
The high temperature solution becomes
\be
{h} = \bar t
\ee
with (generalized) `complete template average'
\be
\bar t =
\left( {{\bf K}}_D + \sum_j^m a^0_j {{\bf K}}_j \right)^{-1}
\left( {{\bf K}}_D t_D + \sum_l^m a^0_j {{\bf K}}_j t_j \right)
.
\label{completeT}
\ee
Notice that $\bar t$ corresponds to the minimum of the quadratic
functional
\be
E_{(\beta = \infty)} =
 \Big( {h} - t_D,\,{{\bf K}}_D ({h}-t_D)\Big)
+\sum_j^m a^0_j 
 \Big( {h} - t_j,\,{{\bf K}}_j ({h}-t_j)\Big)
.
\ee
Thus, in the infinite temperature limit
a combination of quadratic priors by OR is 
effectively replaced by a combination by AND.

In the {\it low temperature limit} $\beta\rightarrow \infty$
we have,
assuming 
$E_{\theta,\beta,j}$ = $E_{\beta}+E_{j} + \beta E_{\theta}$,
\be
\sum_j e^{-\beta (E_{0,j}+E_\theta) - E_{\beta}-E_{j}}
=
e^{-\beta (E_{0,j^*}+E_\theta)-E_{\beta}} 
\sum_j e^{-\beta (E_{0,j}-E_{0,{j^*}})-E_{j}}
\ee
\be
\stackrel{\beta\rightarrow\infty}{\longrightarrow}
e^{-\beta (E_{0,j^*}+E_\theta) -E_{\beta}-E_j}
\qquad 
{\rm for}
\quad 
E_{0,{j^*}}<E_{0,j}, \; \forall j\ne j^*,
\; p(j^*) \ne 0,
,
\ee
meaning that
\be
a_j \stackrel{\beta\rightarrow\infty}{\longrightarrow}
a^\infty_j =
\left\{ \begin{array}{r@{\quad : \quad}l}
1 & j = {\rm argmin}_j E_{0,j} = {\rm argmin}_j E_{{h},j}
\\
0 & j \ne  {\rm argmin}_j E_{0,j}= {\rm argmin}_j E_{{h},j}
\end{array} \right.
.
\label{lowTa}
\ee
Henceforth, 
all (generalized) `component averages'  $\bar t_j$ become solutions
\be
{h} = 
\bar t_{j}
,
\ee
with 
\be
\bar t_{j}
=
\left( {{\bf K}}_D + {{\bf K}}_{j} \right)^{-1}
\left( {{\bf K}}_D t_D + {{\bf K}}_j t_{j} \right)
,
\label{componentT}
\ee
provided the
$\bar t_{j}$ fulfill the stability condition 
\be
 E_{{h},j} ({h}=\bar t_j)
<E_{{h},j^\prime} ({h}=\bar t_j)
, \quad \forall j^\prime \ne j
,
\ee
i.e.,
\be
 V_{j}
<
\frac{1}{2} \Big( \bar t_{j} - \bar t_{j^\prime},\, 
            \left( {{\bf K}}_D + {{\bf K}_{j^\prime}} \right)
            ( \bar t_{j} - \bar t_{j^\prime})\Big)
 + V_{j^\prime}, 
\quad \forall j^\prime \ne j
,
\ee
where
\be
V_j = 
\frac{1}{2} \Bigg(
\Big(t_D,\,{{\bf K}}_D\,t_D\Big)
+\Big(t_j,\,{{\bf K}}_j\,t_j\Big)
-\Big(\bar t_j,\,({{\bf K}}_D+{{\bf K}}_j) \,\bar t_j\Big)
\Bigg)
.
\label{Vterm}
\ee
That means 
single components become solutions at zero temperature $1/\beta$
in case their (generalized) `template variance' $V_j$, 
measuring the discrepancy
between data and prior term, is not too large.
Eq.\ (\ref{regr-h})
for $h$ can also be expressed by the 
(potential) low temperature solutions $\bar t_j$
\be
h = \left( \sum_j^m a_j ({\bf K}_D +{\bf K}_j)\right)^{-1} 
\sum_j^m a_j \,({\bf K}_D +{\bf K}_j)\,\bar t_j
.
\label{regr-h2}
\ee

Summarizing, in the high temperature limit
the stationarity equation (\ref{regr1})
becomes linear with a single solution being essentially
a (generalized) average of all template functions.  
In the low temperature limit the single component solutions become stable 
provided their (generalized) variance corresponding to their minimal error 
is small enough.

\subsubsection{Equal covariances}

Especially interesting is the case
of $j$--independent 
${{\bf K}}_j(\theta)$ = ${{\bf K}}_0 (\theta )$
and $\theta$--independent
$\det {{\bf K}}_0 (\theta)$.
In that case the often difficult to obtain
determinants of ${{\bf K}}_j$ do not have to be calculated.

For $j$--independent covariances 
the high temperature solution 
is
according to
Eqs.(\ref{completeT},\ref{componentT})
a linear combination
of the (potential) low temperature solutions
\be
\bar t = \sum_j^m a^0_j \bar t_j
.
\ee
It is worth to emphasize that, as the solution
$\bar t$ is {\it not} a mixture of the component templates $t_j$
but of component solutions $\bar t_j$,
even poor choices for the template functions $t_j$
can lead to good solutions, if enough data are available.
That is indeed the reason why the most common choice $t_0\equiv 0$
for a Gaussian prior can be successful.

Eqs.(\ref{regr-h2}) simplifies to
\be
{h} 
= \frac{\sum_j^m \bar t_j e^{-\beta E_{{h},j} ({h})-E_{\theta,\beta,j}+c_j}}
       {\sum_j^m e^{-\beta E_{{h},j} ({h})-E_{\theta,\beta,j}+c_j}}
=\sum_j^m a_j \bar t_j
=\bar t + \sum_j^m (a_j-a_j^0) \, \bar t_j
,
\label{equalcov}
\ee
where
\be
\bar t_j 
= \left( {{\bf K}}_D + {{\bf K}}_0 \right)^{-1}
\left(
  {{\bf K}}_D t_D + {{\bf K}}_0 t_j 
\right)
,
\ee
and (for $j$--independent $d$)
\be
a_j 
= \frac{e^{-E_j}}
           {\sum_k e^{-E_k}}
= \frac{e^{-\beta E_{{h},j} -E_{\theta,\beta,j}}}
           {\sum_k e^{-\beta E_{{h},k} -E_{\theta,\beta,k}}}
= \frac{e^{-\frac{\beta}{2}a {B}_j a+d_j}}
       {\sum_k e^{-\frac{\beta}{2} a {B}_k a+d_k}}
,
\label{stata}
\ee
introducing
vector $a$ with components $a_j$,
$m\times m$ matrices
\be
{B}_j (k,l) = 
\Big(\bar t_k-\bar t_j,\,\left( {{\bf K}}_D + {{\bf K}}_0\right)
\,(\bar t_l-\bar t_j)\Big)
\ee
and constants
\be
d_j= 
-\beta V_j-E_{\theta,\beta,j}
,
\ee
with $V_j$ given in (\ref{Vterm}).
Eq.\ (\ref{equalcov}) is still a nonlinear equation
for ${h}$, it shows however that the solutions
must be convex combinations of the ${h}$--independent $\bar t_j$.
Thus, it is sufficient to solve
Eq.\ (\ref{stata}) for $m$ mixture coefficients $a_j$
instead of Eq.\ (\ref{regr1}) for the function ${h}$. 

The high temperature relation
Eq.\ (\ref{highTa}) becomes
\be
a_j 
\stackrel{\beta\rightarrow 0}{\longrightarrow}
a^0_j =
\frac{e^{-E_{\theta,\beta,j}}}
     {\sum_k^me^{-E_{\theta,\beta,k}}}
,
\ee
or $a^0_j = 1/m$ for a hyperprior $p(\theta,\beta,j)$
uniform with respect to $j$.
The low temperature relation Eq.\ (\ref{lowTa})
remains unchanged.

\begin{figure}[tb]
\vspace{-5cm}
\begin{center}
\epsfig{file=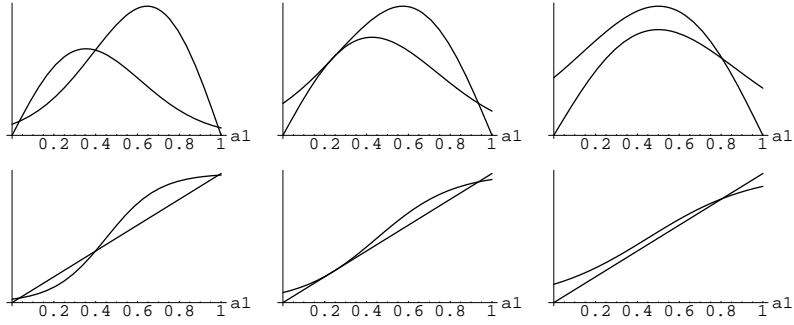, width=110mm}
\end{center}
\vspace{-6.0cm}
\caption{The solution of stationary equation
Eq.\ (\ref{stata}) is given by the point where 
$a_1 e^{-\frac{\beta}{2} b a_1^2 + d_2}$ = 
$(1-a_1) e^{-\frac{\beta}{2} b (1-a_1)^2+d_1}$ 
(upper row) or, equivalently,
$a_1$ = $\frac{1}{2} \left(\tanh \Delta + 1\right)$ (lower row).
Shown are, from left to right, a situation
at high temperature and one stable solution ($\beta$ = $2$), 
at a temperature ($\beta$ = $2.75$) near the bifurcation, 
and at low temperature with
two stable and one unstable solutions $\beta$ = $4$. 
The values of $b$ = $2$, $d_1 = -0.2025 \beta$ and
$d_2 = -0.3025 \beta$  used for the plots
correspond for example to the one--dimensional 
model of Fig.\ref{priorM} with $t_1=1$, $t_2=-1$, $t_D = 0.1$.
Notice, however, that the shown relation is valid
for $m=2$ at arbitrary dimension.}
\label{b2pic}
\end{figure}

\begin{figure}[tb]
\vspace{-5cm}
\begin{center}
$\!\!\!\!\!\!\!\!\!\!$
\epsfig{file=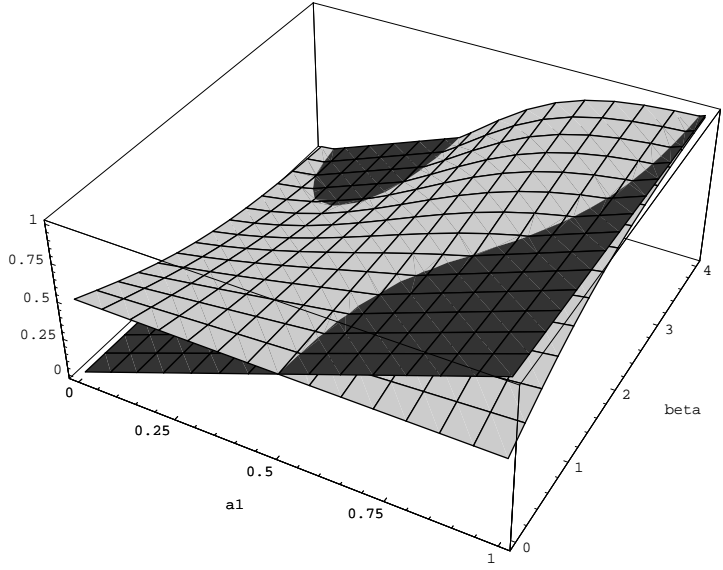, width=95mm}
\end{center}
\vspace{-3.5cm}
\caption{
As in Fig.\ref{b2pic}
the plots of $f_1(a_1)=a_1$ 
and $f_2(a_1)=\frac{1}{2} \left(\tanh \Delta + 1\right)$
are shown 
within the inverse temperature range $0\le\beta\le 4$.
}
\label{tpic}
\end{figure}

For $m=2$ Eq.\ (\ref{equalcov}) becomes
\be
{h} 
=\sum_j^2 a_j \bar t_j
=\frac{\bar t_1 + \bar t_2}{2} 
+ (a_1-a_2) \frac{\bar t_1 - \bar t_2}{2} 
=\frac{\bar t_1 + \bar t_2}{2} 
+ \left(\tanh \Delta\right)  \frac{\bar t_1 - \bar t_2}{2} 
,
\ee
with
$(\bar t_1+\bar t_2)/2$ = $\bar t$ 
in case $E_{\theta,\beta,j}$ is uniform in $j$
so that $a_j^0$ = $0.5$, and 
\bea
\Delta &=&
\frac{E_2-E_1}{2} 
\; = \;  \beta \frac{E_{{h},2}-E_{{h},1}}{2}
  +\frac{E_{\theta,\beta,2}-E_{\theta,\beta,1}}{2}
\nonumber\\ 
&=&
-\frac{\beta}{4} a(B_1-B_2)a +\frac{d_1-d_2}{2}
\; =\;  \frac{\beta}{4} \,b(2 a_1-1) + \frac{d_1-d_2}{2}
,
\eea
because
the matrices $B_j$ are in this case zero except
$B_1(2,2) = B_2(1,1) = b$.
The stationarity Eq.\ (\ref{stata}) can be solved
graphically (see Figs.\ref{b2pic}, \ref{tpic}),
the solution being given by the point
where 
$
a_1 e^{-\frac{\beta}{2} b a_1^2 + d_2}
= 
(1-a_1) e^{-\frac{\beta}{2} b (1-a_1)^2+d_1}
$,
or, alternatively,
\be
a_1 = \frac{1}{2} \left(\tanh \Delta + 1\right)
.
\ee
That equation is analogous to
the celebrated mean field equation of the
ferromagnet.

We conclude that in the case of equal component covariances,
in addition to the linear low--temperature equations,
only a $m-1$--dimensional nonlinear equation has to be solved
to determine the
`mixing coefficients' $a_1,\cdots , a_{m-1}$.

\begin{figure}[tb]
\begin{center}
\epsfig{file=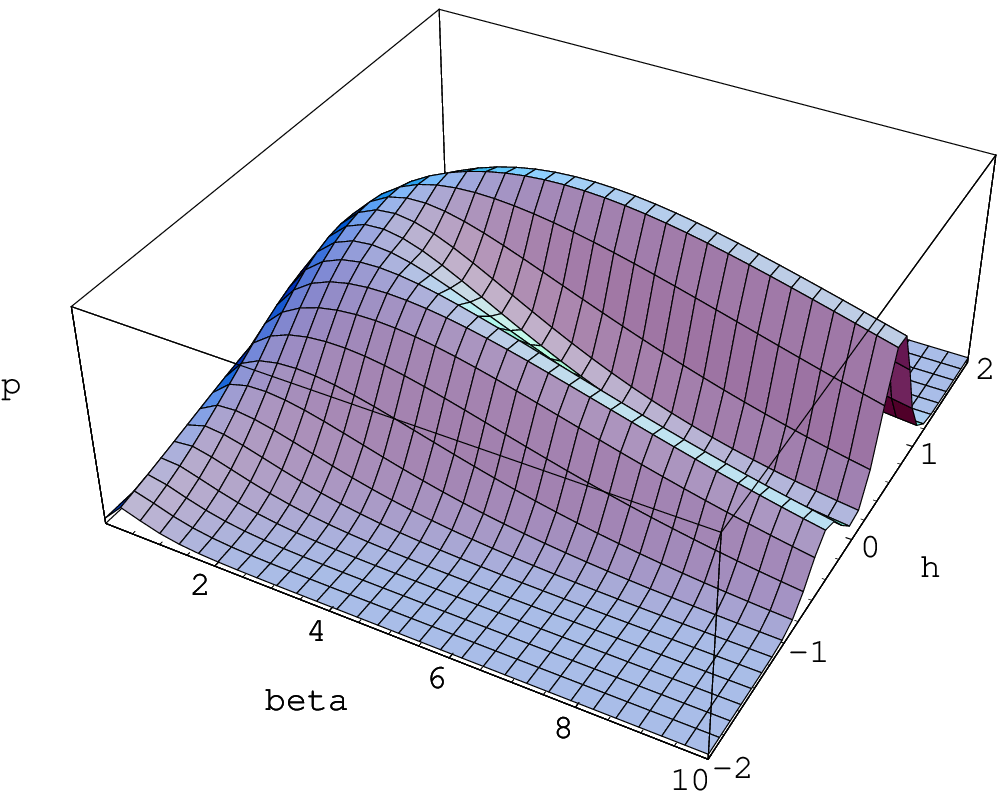, width= 65mm}    
\epsfig{file=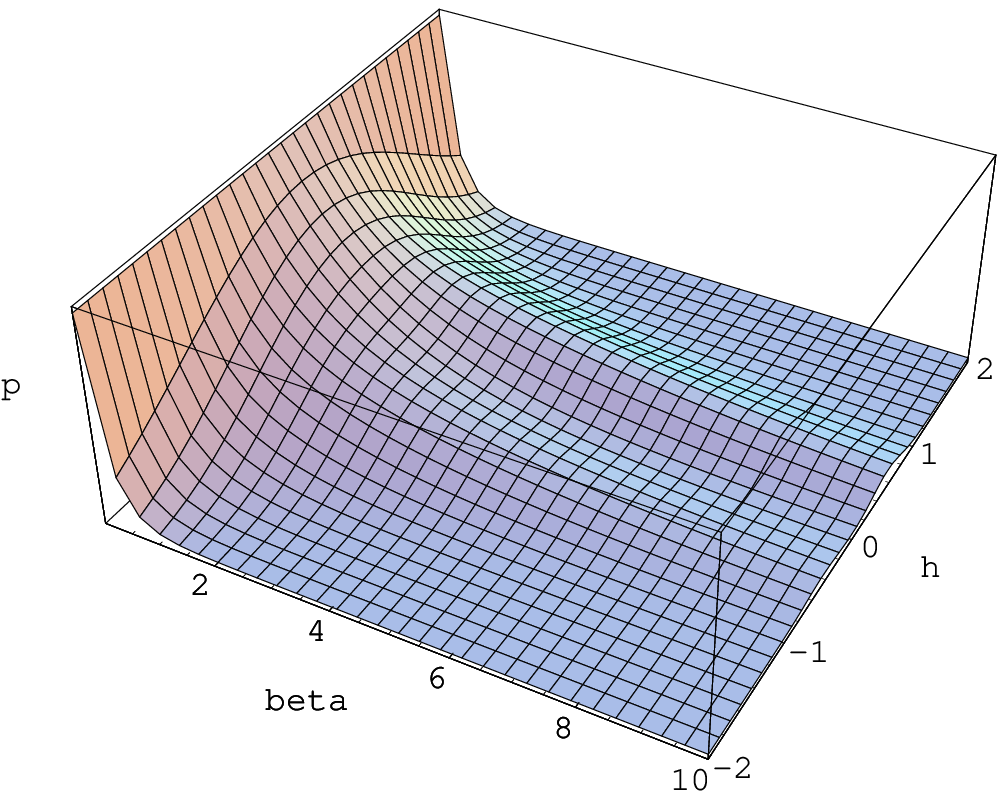, width= 65mm}    
\end{center}
\caption{Shown is the joint posterior density of $h$ and $\beta$, i.e.,
$p({h},\beta|D,D_0)$
$\propto p(y_D|{h},\beta )p({h}|\beta,D_0)p(\beta)$
for a zero--dimensional example 
of a Gaussian prior mixture model
with training data $y_D=0.1$ and prior data $y_{D_0}=\pm 1$
and inverse temperature $\beta$.
L.h.s.:
For uniform prior (middle) $p(\beta) \propto 1$ with
joint posterior $p \propto$ 
$e^{-\frac{\beta}{2} {h}^2 + \ln \beta}$ 
$\left(e^{-\frac{\beta}{2} ({h}-1)^2} 
+ e^{-\frac{\beta}{2} ({h}+1)^2}\right)$ 
the maximum appears at finite $\beta$.
(Here no factor $1/2$ appears in front of $\ln\beta$ 
because normalization constants for prior and likelihood term have 
to be included.)
R.h.s.:
For compensating hyperprior $p(\beta) \propto 1/\protect\sqrt{\beta}$ with
$p \propto$ 
$e^{-\frac{\beta}{2} {h}^2 -\frac{\beta}{2} ({h}-1)^2}$
$+$ $e^{-\frac{\beta}{2} {h}^2 -\frac{\beta}{2} ({h}+1)^2}$
the maximum is at $\beta$ = $0$.  
}
\label{priorM}
\end{figure}

\begin{figure}[tb]
\begin{center}
\epsfig{file=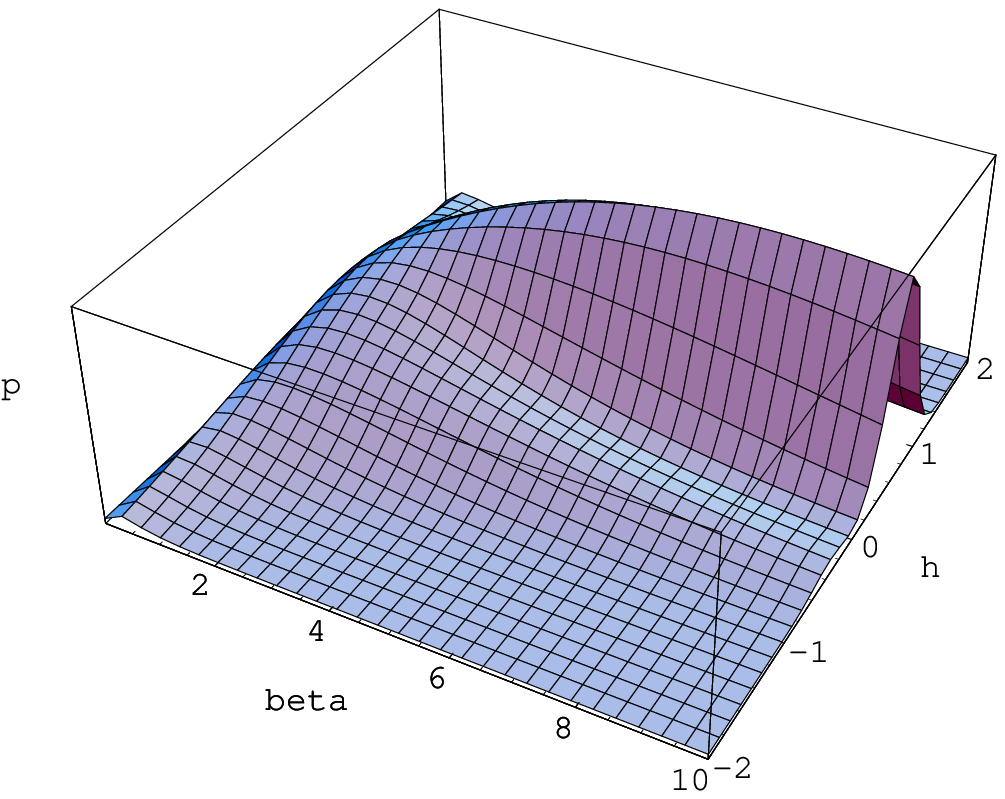, width= 65mm}    
\epsfig{file=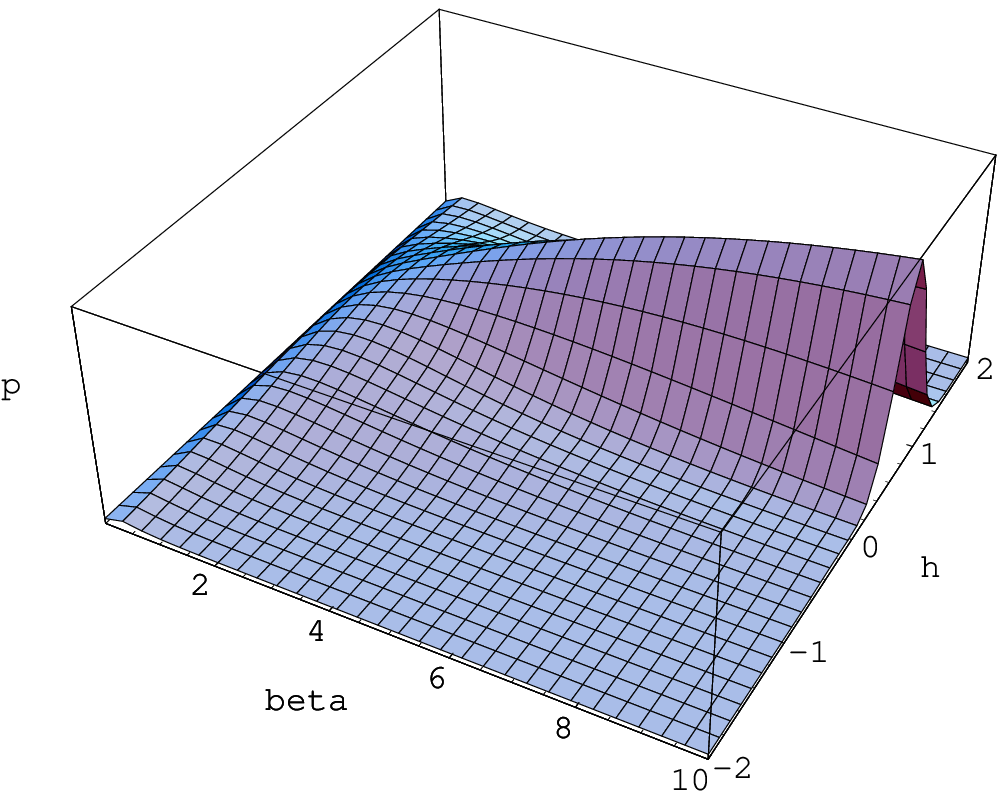, width= 65mm}    
\end{center}
\caption{
Same zero--dimensional prior mixture model 
for uniform hyperprior on $\beta$
as in Fig.\ref{priorM},
but for varying data
$x_d=0.3$ (left),
$x_d=0.5$ (right).
}
\label{priorMU}
\end{figure}

\subsubsection{Analytical solution of mixture models}

For regression under a Gaussian mixture model
the predictive density can be calculated analytically
for fixed $\theta$.
This is done by expressing the predictive density
in terms of the likelihood of 
$\theta$ and $j$,
marginalized over $h$
\be
p(y|x,D,D_0) 
=
\sum_j \int \!dh\,d\theta\,
\frac{p(\theta,j)\,p(y_D|x_D,D_0,\theta,j)}
     {\sum_j \int \!d\theta\, p(\theta,j) p(y_D|x_D,D_0,\theta,j)}
p(y|x,D,D_0,\theta,j)
.
\ee
(Here we concentrate on $\theta$.
The parameter $\beta$ can be treated analogously.)
According to Eq.\ (\ref{anal-likel}) the likelihood can be written
\be
p(y_D|x_D,D_0,\theta,j)
=
e^{-\beta \widetilde E_{0,j}(\theta)
+\frac{1}{2}\ln \det (\frac{\beta}{2\pi} \widetilde {\bf K}_{j} (\theta))
}
,
\ee
with 
\be
\widetilde E_{0,j}(\theta)
= 
\frac{1}{2}  
\big( t_D - t_j(\theta),\, 
\widetilde {\bf K}_j (\theta) (t_D - t_j(\theta ) )\big)
=V_j
,
\ee
and $\widetilde {\bf K}_{j}(\theta)$ 
= $({\bf K}_D^{-1}+{\bf K}_{j,DD}^{-1}(\theta))^{-1}$
being a $\tilde n \times \tilde n$--matrix in data space.
The equality of $V_j$ and $\widetilde E_{0,j}$
can be seen using
${\bf K}_j-{\bf K}_j ({\bf K}_D + {\bf K}_j)^{-1}{\bf K}_j$
=
${\bf K}_D-{\bf K}_D ({\bf K}_D + {\bf K}_{j,DD})^{-1}{\bf K}_D$
=
${\bf K}_{j,DD}-{\bf K}_{j,DD }({\bf K}_D 
+ {\bf K}_{j,DD)}^{-1}{\bf K}_{j,DD}$
=
$\widetilde {\bf K}$.
For the predictive mean,
being the optimal solution under squared--error loss
and log--loss (restricted to Gaussian densities with fixed variance)
we find therefore
\be
\bar y (x)
= \int\!dy\, y\, p(y|x,D,D_0)
= \sum_j  
\int d\theta \; b_j(\theta )\, \bar t_j(\theta)
,
\ee
with, according to Eq.\ (\ref{pmean2}),
\be
\bar t_j(\theta)
=
t_j + {\bf K}_j^{-1} \widetilde {\bf K}_j(t_D-t_j)
,
\ee 
and mixture coefficients
\bea
b_j(\theta)
&=&
p(\theta,j|D)
=
\frac{p(\theta,j) p(y_D|x_D,D_0,\theta,j)} 
                    {\sum_j\int d\theta p(\theta,j) p(y_D|x_D,D_0,\theta,j)} 
\nonumber\\
&\propto& e^{-\beta \widetilde E_j (\theta) -E_{\theta,j}
+\frac{1}{2} \ln \det (\widetilde K_j (\theta))}
,
\label{exact-mix-mean}
\eea
which defines $\widetilde E_j$ = $ \beta \widetilde E_{0,j}$ + $E_{\theta,j}$.
For solvable $\theta$--integral
the coefficients can therefore be obtained exactly.

If $b_j$ is calculated in saddle point
approximation at $\theta$ = $\theta^*$
it has the structure of $a_j$
in (\ref{full-aj})
with $E_{0,j}$ replaced by 
$\widetilde E_j$ and 
${\bf K}_j$ by $\widetilde {\bf K}_j$.
(The inverse temperature $\beta$ could be treated analogously to $\theta$.
In that case $E_{\theta,j}$ would have to be replaced by $E_{\theta,\beta,j}$.)

Calculating also the likelihood for $j$, $\theta$
in Eq.\ (\ref{exact-mix-mean}) 
in saddle point approximation, i.e.,
$p(y_D|x_D,D_0,\theta^*,j)
\approx 
p(y_D|x_D,h^*) p(h^*|D_0,\theta^*,j)$,
the terms $p(y_D|x_D,h^*)$ in numerator and denominator cancel,
so that, skipping $D_0$ and $\beta$,
\be
b_j(\theta^*)
= \frac{p(h^*|j,\theta^*)p(j,\theta^*)}{p(h^*,\theta^*)}
= a_j(h^*,\theta^*)
,
\ee
becomes equal to
the $a_j(\theta^*)$ 
in Eq.\ (\ref{full-aj}) at 
$h$ = $h^*$.

Eq.\ (\ref{exact-mix-mean}) yields
as stationarity equation for $\theta$,
similarly to Eq.\ (\ref{ex-theta1})
\bea
0 &=& 
\sum_j b_j
\left( \frac{\partial \widetilde E_j}{\partial \theta}
-{\rm Tr}\left(\widetilde {\bf K}_{j}^{-1}
  \frac{\partial \widetilde {\bf K}_{j}}{\partial \theta}\right)
\right) 
\\&=&
\sum_j b_j
\Bigg(
\left( \frac{\partial t_j(\theta)}{\partial \theta},\; 
     \widetilde {\bf K}_{j}(\theta)(t_j(\theta)-t_D)\right)\
\nonumber\\&&
+\frac{1}{2} \left((t_D-t_j(\theta)),\, 
\frac{\partial \widetilde {\bf K}_{j}(\theta)}
     {\partial \theta}(t_D-t_j(\theta))\right)
\nonumber\\&&
-{\rm Tr}\left(\widetilde {\bf K}_{j}^{-1}(\theta)
  \frac{\partial \widetilde {\bf K}_{j}(\theta)}{\partial \theta}\right)
-\frac{1}{p(\theta,j)} \frac{\partial p(\theta,j)}{\partial \theta}
\Bigg)
.
\eea

\begin{figure}
\begin{center}
\epsfig{file=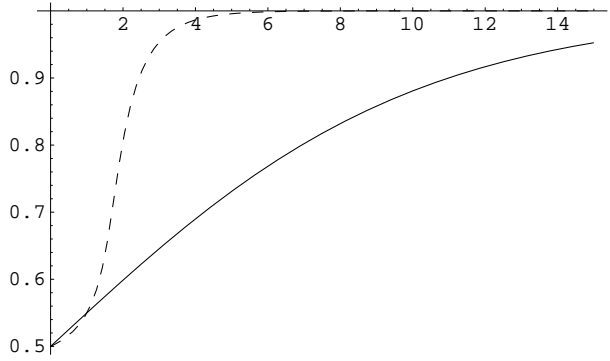, width=80mm}
\end{center}
\vspace{-0.7cm}
\caption{
Exact $b_1$ and $a_1$ (dashed) vs. $\beta$ 
for two mixture components with equal covariances 
and $B_1(2,2)$ = $b$ = 2,
$\widetilde E_1$ = 0.405,
$\widetilde E_2$ = 0.605.}
\label{cmp-pic}
\end{figure}

For fixed $\theta$ and $j$--independent covariances
the high temperature solution
is a mixture of component solutions
weighted by their prior probability 
\be
\bar y \stackrel{\beta\rightarrow 0}{\longrightarrow} 
\sum_j p(j)\;\bar t_{j}
= \sum_j a_j^0 \;\bar t_{j}
= \bar t
.
\ee
The low temperature solution becomes the 
component solution $\bar t_j$ with minimal
distance between data and prior template
\be
\bar y \stackrel{\beta\rightarrow \infty}{\longrightarrow} \bar t_{j^*}
\quad {\rm with} \;
j^* = \;{\rm argmin}_j \big( t_D - t_j,\, 
 \widetilde {\bf K}_{j} (t_D - t_j )\big)
.
\ee
Fig.\ref{cmp-pic} compares
the exact mixture coefficient $b_1$
with the dominant solution of the maximum 
posterior coefficient $a_1$ (see also \cite{Lemm-1996})
which are related according to (\ref{stata})
\be
a_j= \frac{e^{-\frac{\beta}{2}a {B}_j a-\widetilde E_j}}
       {\sum_k e^{-\frac{\beta}{2} a {B}_k a-\widetilde E_k}}
= \frac{b_j \,e^{-\frac{\beta}{2}a {B}_j a}}
       {\sum_k b_k\, e^{-\frac{\beta}{2} a {B}_k a}}
.
\ee

\subsection{Local mixtures}
\label{local-mix}

Global mixture components can be obtained
by combining local mixture components.
Predicting a time series, for example,
one may allow to switch locally (in time)
between two or more possible regimes,
each corresponding to a different local covariance or template.

The problem which arises when combining local alternatives
is the fact that 
the total number of mixture components
grows exponentially in the number local components 
which have to be combined 
for a global mixture component.

Consider a local prior mixture model,
similar to Eq.\ (\ref{omega-B-energy}),
\be
p(\phi|\theta) 
= 
e^{-\int \!dx; |\omega(x;\theta(x))|^2-\ln Z_\phi(\theta)}
\ee
where $\theta(x)$ may be a binary or an integer variable.
The local mixture variable $\theta(x)$
labels local alternatives for  filtered differences
$\omega(x;\theta(x))$ which may differ in
their templates $t(x;\theta(x))$ and/or 
their local filters ${\bf W}(x;\theta(x))$.
To avoid infinite products,
we choose a discretized $x$ variable (which may include
the $y$ variable for general density estimation problems),
so that
\be
p(\phi)
= 
\sum_\theta
p(\theta)e^{-\sum_x |\omega(x;\theta(x))|^2-\ln Z_\phi(\theta)}
,
\ee
where the sum $\sum_\theta$ 
is over all local integer variables $\theta(x)$, i.e.,
\be
\sum_\theta
= \sum_{\theta(x_1)} \cdots \sum_{\theta(x_l)}
= \left( \prod_x \sum_{\theta(x_1)}\right)
.
\ee

Only for factorizing hyperprior
$p(\theta)$ = $\prod_x p(\theta(x))$
the complete posterior factorizes
\bea
p(\phi)
&=& 
\left(\prod_{x^\prime} \sum_{\theta(x^\prime)}\right)
\prod_x 
\left(
p(\theta(x))e^{- |\omega(x;\theta(x))|^2-\ln Z_\phi(x,\theta(x))}
\right)
\nonumber\\
&=& 
\prod_{x} \sum_{\theta(x)}
\left(
p(\theta(x))e^{- |\omega(x;\theta(x))|^2-\ln Z_\phi(x,\theta(x))}
\right)
,
\eea
because
\be
Z_\phi
=
\prod_x \sum_{\theta(x)}
\left(e^{-|\omega(x;\theta(x))|^2}
\right)
=
\prod_x Z_\phi(x,\theta(x))
.
\ee

Under that condition the mixture coefficients $a_{\theta}$
of Eq.\ (\ref{full-aj})
can be obtained from the equations,
local in $\theta(x)$,
\be
a_{\theta} = a_{\theta(x_1)\cdots \theta(x_l)} 
= p(\theta|\phi)
=\prod_x a_{\theta(x)}
\ee
with
\be
a_{\theta(x)} 
= 
\frac{p(\theta(x))e^{-|\omega(x;\theta(x))|^2-\ln Z_\phi(x;\theta(x))}}
     {\sum_{\theta^\prime(x)} 
      p(\theta^\prime(x))
      e^{-|\omega(x;\theta^\prime(x))|^2-\ln Z_\phi(x;\theta^\prime(x))}}
.
\ee
For equal covariances this is a nonlinear equation
within a space of dimension equal to the number of local components.
For non--factorizing hyperprior
the equations for different $\theta(x)$
cannot be decoupled.

\subsection{Non--quadratic potentials}
\label{other-non-gaussian}

Solving learning problems numerically
by discretizing the $x$ and $y$ variables
allows in principle to deal with 
arbitrary non--Gaussian priors.
Compared to Gaussian priors, however,
the resulting stationarity equations are intrinsically nonlinear.

As a typical example let us formulate a prior
in terms of nonlinear and non--quadratic
``potential'' functions $\psi$
acting on ``filtered differences'' 
$\omega$ = ${\bf W}(\phi -t)$,
defined with respect to some 
positive (semi--)definite inverse covariance
${\bf K}$ = ${\bf W}^T{\bf W}$.
In particular, consider a prior factor 
of the following form
\be
p(\phi) 
=
e^{-\int \! dx\, \psi(\omega(x))-\ln Z_\phi}
=
\frac{e^{-E(\phi)}}{Z_\phi}
,
\label{nonlinear-filter-prior}
\ee
where
$E(\phi)$ = $\int \! dx \,\psi(\omega(x))$.
For general density estimation problems
we understand $x$ to stand for a pair $(x,y)$.
Such priors are for example used for image restoration
\cite{Geman-Geman-1984,Blake-Zisserman-1987,Mumford-Shah-1989,Geman-Reynoids-1992,Zhu-Wu-Mumford-1997,Zhu-Mumford-1997}.

For differentiable $\psi$ function
the functional derivative with respect to $\phi(x)$
becomes
\be
\delta_{\phi(x)} p(\phi) 
=
-e^{-\int \! dx^\prime\, \psi(\omega(x^\prime))-\ln Z_\phi}
\int \! dx^{\prime\prime}\, \psi^\prime(\omega(x^{\prime\prime}))
{\bf W}(x^{\prime\prime},x)
,
\ee
with $\psi^\prime(s)$ = $d\psi(z)/dz$,
from which follows 
\be
\delta_{\phi} E(\phi)  
=
-\delta_{\phi} \ln p(\phi)  = {\bf W}^T \psi^\prime
.
\ee
For nonlinear filters acting on $\phi-t$,
${\bf W}$ in Eq.\ (\ref{nonlinear-filter-prior})
must be replaced by 
$\omega^\prime(x)$
=
$\delta_\phi(x)\omega(x)$.
Instead of one ${\bf W}$
a ``filter bank'' ${\bf W}_\alpha$
with corresponding
${\bf K}_\alpha$,
$\omega_\alpha$,
and $\psi_\alpha$ may be used, so that
\be
e^{-\sum_\alpha 
   \int \! dx\, \psi_\alpha(\omega_\alpha(x))-\ln Z_\phi}
\label{nonlinear-filter-bank}
,
\ee
and
\be
\delta_{\phi} E(\phi)  
= \sum_\alpha {\bf W}_\alpha^T \psi_\alpha ^\prime
\label{nonlinear-filter-bank-stat-eq}
.
\ee

\begin{figure}
\vspace{-0.5cm}
\begin{center}
\raisebox{2.2cm}{(a)$\;$}
\epsfig{file=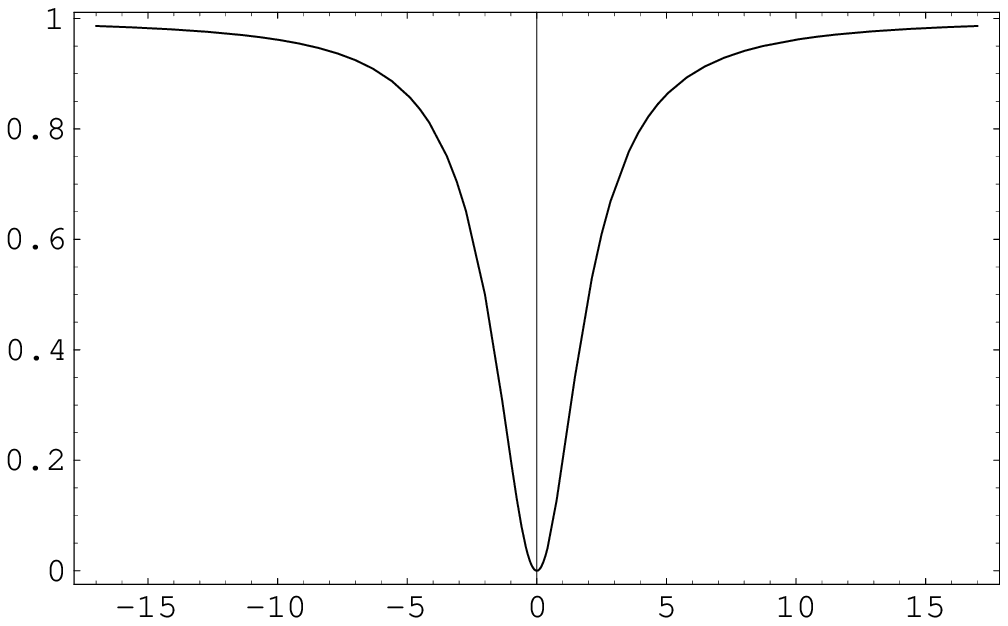, width=71mm}\\
\raisebox{2.2cm}{(b)$\;$}
\epsfig{file=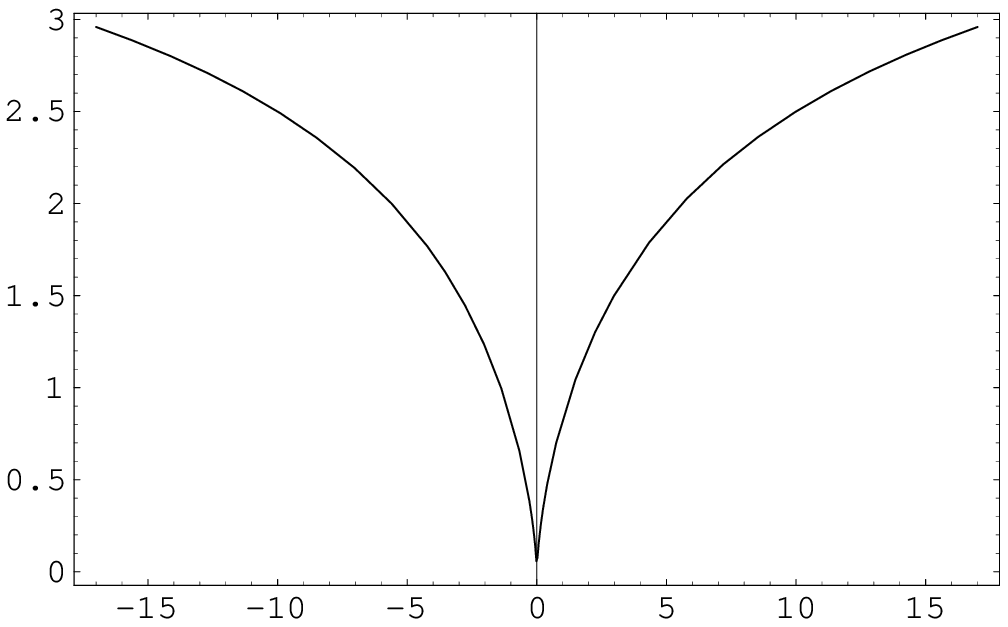, width=71mm}\\
\raisebox{2.2cm}{(c)$\,$}
\epsfig{file=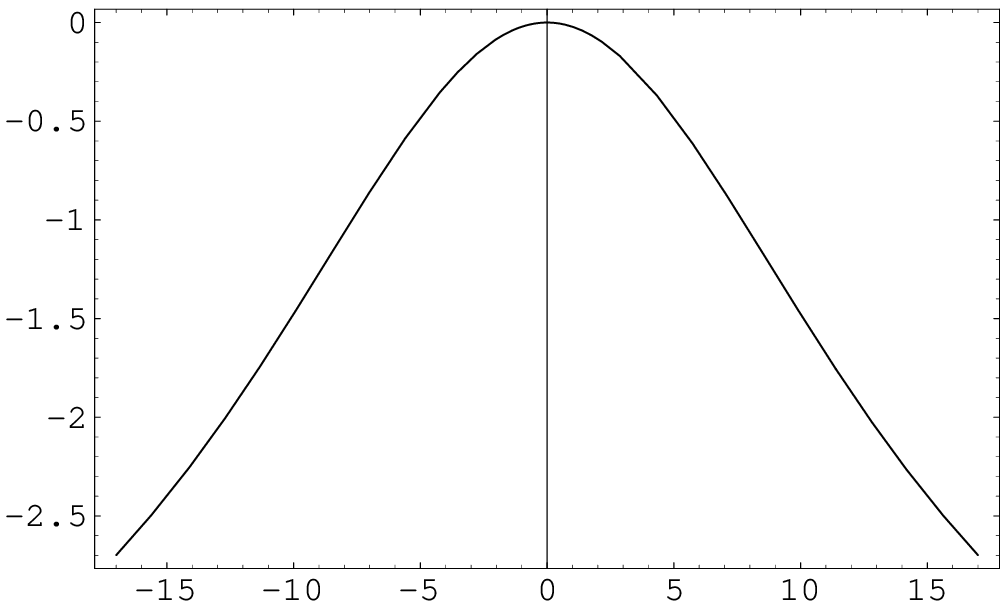, width=71mm}
\end{center}
\vspace{-0.5cm}
\caption{Non--quadratic
potentials of the form
$\psi(x)$ = $a( 1.0 - 1/(1+(|x-x0|/b)^\gamma))$, 
\cite{Zhu-Mumford-1997}:
``Diffusion terms'':
(a)
Winkler's cup function \cite{Winkler-1995} 
($a$= $5$,
$b$ = $10$,
$\gamma$ = $0.7$,
$x_0$ = $0$),
(b)
with cusp
($a$= $1$,
$b$ = $3$,
$\gamma$ = $2$,
$x_0$ = $0$),
(c)
``Reaction term''
($a$ = $-4.8$,
$b$ = $15$,
$\gamma$ = $2.0$
$x_0$ = $0$).
}
\label{Zhu-Mum-pic}
\end{figure}

The potential functions $\psi$ 
may be fixed in advance for a given problem.
Typical choices to allow discontinuities
are symmetric ``cup'' functions 
with minimum at zero and flat tails 
for which one large step is cheaper than many small ones
\cite{Winkler-1995}).
Examples are shown in Fig.\ \ref{Zhu-Mum-pic} (a,b).
The cusp in (b), where the derivative does not exist,
requires special treatment \cite{Zhu-Mumford-1997}. 
Such functions can also be interpreted in the sense of robust statistics
as flat tails reduce
the sensitivity with respect to outliers
\cite{Huber-1979,Huber-1981,Geiger-Yuille-1991,Black-Rangarajan-1996}. 

Inverted ``cup'' functions, like those
shown in Fig.\ \ref{Zhu-Mum-pic} (c),
have been obtained by optimizing a set of $\psi_\alpha$
with respect to a sample of natural images
\cite{Zhu-Mumford-1997}.
(For statistics of natural images 
their relation to wavelet--like filters and sparse coding 
see also
\cite{Olshausen-Field-1995,Olshausen-Field-1996}.)

While, for ${\bf W}$ which are differential operators, 
cup functions promote smoothness,
inverse cup functions can be used to implement
structure.
For such ${\bf W}$ 
the gradient algorithm for minimizing $E(\phi)$, 
\be
\phi_{\rm new} 
= \phi_{\rm old} - \eta \delta_\phi E(\phi_{\rm old})
,
\ee
becomes in the continuum limit
a nonlinear parabolic partial differential equation,
\be
\phi_\tau 
= -\sum_\alpha {\bf W}_\alpha^T 
\psi_\alpha^\prime ({\bf W}_\alpha(\phi-t))
.
\ee
Here a formal time variable $\tau$ have been introduced
so that
$(\phi_{\rm new}-\phi_{\rm old})/\eta\rightarrow \phi_\tau = d\phi/d\tau$.
For cup functions this equation is of diffusion type 
\cite{Nitzberg-Shiota-1992,Perona-Malik-1990},
if also inverted cup functions are included
the equation is of reaction--diffusion type
\cite{Zhu-Mumford-1997}.
Such equations are known to generate
a great variety of patterns.

Alternatively to fixing $\psi$ in advance
or, which is sometimes possible for low--dimensional 
discrete function spaces like images,
to approximate $\psi$ by sampling from the prior distribution,
one may also introduce hyperparameters
and adapt potentials $\psi(\theta)$ to the data.

For example, attempting to adapt a unrestricted function
$\psi(x)$ with hyperprior $p(\psi)$ by 
Maximum A Posteriori Approximation
one has to solve the stationarity condition
\be
0 =
\delta_{\psi(s)} \ln p(\phi,\psi)
= 
\delta_{\psi(s)} \ln p(\phi|\psi)
+\delta_{\psi(s)} \ln p(\psi)
.
\ee
From
\be
\delta_{\psi(s)} p(\phi|\psi)
=
-p(\phi|\psi) \int \!dx\, \delta \left(s-\omega(x) \right)
- \frac{1}{Z_\phi^2} \delta_{\psi(s)} Z_\phi
,
\ee
it follows
\be
-\delta_{\psi(s)} \ln p(\phi|\psi)
=
n(s) - <n(s)>
,
\label{psi-hyper-stat-eq}
\ee
with integer
\be
n(s) = \int \!dx\, \delta \left(s-\omega(x) \right)
,
\ee
being the histogram of the filtered differences,
and average histogram
\be
<n(s)> \; = \int\! d\phi \, p(\phi|\psi) \, n(s)
.
\ee
The right hand side of Eq.\ (\ref{psi-hyper-stat-eq})
is zero at $\phi^*$
if, e.g., $p(\phi|\psi)$ = $\delta(\phi-\phi^*)$,
which is the case for
$\psi(\omega(x;\phi))$ 
= 
$\beta \left(\omega(x;\phi)-\omega(x;\phi^*)\right)^2$ 
in the 
${\beta\rightarrow\infty}$
limit. 

Introducing hyperparameters
one has to keep in mind that
the resulting additional flexibility 
must be balanced 
by the number of training data 
and the hyperprior
to be useful in practice.

\section{Iteration procedures: Learning}
\label{learning}
\subsection{Numerical solution of stationarity equations}

Due to the presence of the logarithmic
data term $-(\ln P, N)$
and the normalization constraint in density estimation problems
the stationary equations are in general nonlinear,
even for Gaussian specific priors.
An exception are Gaussian regression problems 
discussed in Section \ref{regression}
for which $-(\ln P, N)$ becomes quadratic 
and the normalization constraint can be skipped.
However, the nonlinearities 
arising from the data term $-(\ln P, N)$
are restricted to a finite number of 
training data points
and for Gaussian specific priors one may expect them, 
like those arising from the normalization constraint,
to be numerically not very harmful.
Clearly, severe nonlinearities can appear
for general non--Gaussian specific priors
or general nonlinear parameterizations $P(\xi)$.

As nonlinear equations the stationarity conditions
have in general to be solved by iteration.
In the context of empirical learning iteration procedures
to minimize an error functional represent possible {\it learning algorithms}.

In the previous sections we have encountered 
stationarity equations 
\be
0 = \frac{\delta (-E_\phi)}{\delta \phi} = G(\phi)
,
\ee
for error functionals $E_\phi$, e.g., $\phi$ = $L$ or $\phi$ = $P$,
written in a form
\begin{equation}
{{\bf K}} \phi  = T.
\label{OV=T}
\end{equation}
with $\phi$--dependent $T$ (and possibly ${\bf K}$).
For the stationarity Eqs.\ (\ref{OL=T}), (\ref{OP=T}),
and (\ref{Ophi=Tphi})
the operator ${{\bf K}}$ is a 
$\phi$--independent inverse covariance
of a Gaussian specific prior.
It has already been mentioned that for existing 
(and not too ill--conditioned)
${{\bf K}}^{-1}$
(representing the covariance of the prior process)
Eq.\ (\ref{OV=T})
suggests an iteration scheme
\be
\phi^{(i+1)} = {{\bf K}}^{-1} T(\phi^{(i)})
,
\label{phiKTphi}
\ee
for discretized $\phi$ starting from some initial guess $\phi^{(0)}$.
In general, like for the non--Gaussian specific priors discussed in 
Section \ref{non-Gaussian}, 
${{\bf K}}$ can be $\phi$--dependent.
Eq.\ (\ref{Oxi=Txi}) shows that general nonlinear parameterizations $P(\xi)$ 
lead to nonlinear operators ${{\bf K}}$.

Clearly, 
if allowing $\phi$--dependent $T$,
the form (\ref{OV=T}) is no restriction of generality.
One always can choose an arbitrary invertible (and not too ill--conditioned) 
${\bf A}$,
define 
\be
T_{\bf A}= G(\phi) +  {\bf A} \phi,
\ee
write a stationarity equation as
\begin{equation}
{\bf A} \phi\ = T_{\bf A},
\end{equation}
discretize and iterate with ${\bf A}^{-1}$.
To obtain a numerical iteration scheme 
we will choose  
a linear, positive definite {\it learning matrix} ${\bf A}$.
The learning matrix may depend on $\phi$ 
and may also change during iteration.

To connect a stationarity equation given in form (\ref{OV=T})
to an arbitrary iteration scheme with 
a learning matrix ${\bf A}$  
we define
\be
{\bf B} = {{\bf K}} - {\bf A},
\quad
{\bf B}_\eta ={{\bf K}} - \frac{1}{\eta}{\bf A},
\ee
i.e., we split ${{\bf K}}$ into two parts
\be
{{\bf K}}={\bf A} + {\bf B} = \frac{1}{\eta}{\bf A} + {\bf B}_\eta,
\ee
where we introduced $\eta$ for later convenience.
Then we obtain from the stationarity equation (\ref{OV=T})
\be
\phi = \eta {\bf A}^{-1} (T - {\bf B}_\eta \phi ) .
\ee
To iterate we start by inserting an approximate solution 
$\phi^{(i)}$ to the right--hand side
and obtain a new  $\phi^{(i+1)}$ by calculating the left hand side.
This can be written in one of the following equivalent forms
\begin{eqnarray}
\phi^{(i+1)} &=& 
\eta {\bf A}^{-1} \left( T^{(i)} - {\bf B}_\eta \phi^{(i)} \right)
\label{iter-1}\\
&=& 
(1-\eta) \phi^{(i)} + \eta {\bf A}^{-1} \left( T^{(i)} - {\bf B}  \phi^{(i)} \right) 
\label{iter-2}\\
&=& 
\phi^{(i)} + \eta {\bf A}^{-1} \left( T^{(i)} -  {{\bf K}} \phi^{(i)} \right),
\label{iter-3}
\end{eqnarray}
where $\eta$ plays the role of a learning rate
or step width,
and ${\bf A}^{-1}$ = $\left({\bf A}^{(i)}\right)^{-1}$ 
may be iteration dependent.
The update equations (\ref{iter-1}--\ref{iter-3}) can be written
\begin{equation}
\Delta \phi^{(i)} = \eta {\bf A}^{-1} G (\phi^{(i)}),
\end{equation}
with $\Delta \phi^{(i)}$ = $\phi^{(i+1)} - \phi^{(i)}$.
Eq.\ (\ref{iter-3}) does not require
the calculation of ${\bf B}$ or ${\bf B}_\eta$
so that instead of ${\bf A}$
directly ${\bf A}^{-1}$ can be given
without the need to calculate its inverse.
For example operators approximating ${{\bf K}}^{-1}$ and being easy to
calculate may be good choices for ${\bf A}^{-1}$.

For positive definite ${\bf A}$
(and thus also positive definite inverse)
convergence can be guaranteed, at least theoretically.
Indeed,
multiplying with $(1/\eta ) {\bf A}$ and projecting onto
an infinitesimal $d\phi$        
gives
\begin{equation}
\frac{1}{\eta} (\, d\phi,\, {\bf A}\,\Delta \phi\,) = 
 \Big(\,d\phi,\,\frac{\delta (-E_\phi)}{\delta \phi}\Bigg|_{\phi=\phi^{(i)}}\,\Big) 
= d(-E_\phi).
\end{equation}
In an infinitesimal neighborhood of $\phi^{(i)}$
where ${\Delta \phi^{(i)}}$ becomes equal to $d\phi$ in first order 
the left--hand side is for positive (semi) definite ${\bf A}$
larger (or equal) to zero.
This shows that at least for $\eta$ small  enough
the posterior log--probability $-E_\phi$ increases
i.e., the differential $dE_\phi$ is smaller or equal to zero
and the value of the error functional $E_\phi$ decreases.

Stationarity equation
(\ref{stat-eq-bra-ket}) for minimizing $E_L$ 
yields for (\ref{iter-1},\ref{iter-2},\ref{iter-3}),
\begin{eqnarray}
L^{(i+1)} &=& 
\eta {\bf A}^{-1} \left( N - {\bf \Lambda}_X^{(i)} e^{L^{(i)}} 
 - {{\bf K}} L^{(i)} + \frac{1}{\eta} {\bf A} L^{(i)}\right)\\
&=& 
(1-\eta) L^{(i)} + \eta {\bf A}^{-1} \left( N - {\bf \Lambda}_X^{(i)} e^{L^{(i)}} 
- {{\bf K}} L^{(i)} +{\bf A} L^{(i)} \right)\\
&=& 
L^{(i)} + \eta {\bf A}^{-1} \left( N - {\bf \Lambda}_X^{(i)} e^{L^{(i)}} 
  - {{\bf K}} L^{(i)} \right).
\label{L3}
\end{eqnarray}
The function $\Lambda_X^{(i)}$ is also unknown
and is part of the variables
we want to solve for.
The normalization conditions 
provide the necessary additional equations,
and the matrix ${\bf A}^{-1}$ can be extended
to include the iteration procedure for $\Lambda_X$.
For example, we can insert the stationarity equation for $\Lambda_X$
in (\ref{L3}) to get
\begin{equation}
L^{(i+1)} =
L^{(i)} + \eta {\bf A}^{-1} 
\left[ N - e^{{\bf L}^{(i)}} (N_X-{\bf I}_X {{\bf K}} L) 
- {{\bf K}} L^{(i)} \right] .
\end{equation}
If normalizing $L^{(i)}$ at each iteration this
corresponds to an iteration procedure for $g = L + \ln Z_X$.

Similarly, for the functional $E_P$
we have to solve (\ref{stat-eq-P})
and obtain for (\ref{iter-3}),
\begin{eqnarray}
P^{(i+1)}
&=& P^{(i)} + \eta {\bf A}^{-1} \left( T^{(i)}_P-{{\bf K}} P^{(i)} \right)\\
&=& P^{(i)} + \eta {\bf A}^{-1} \left( {\bf P^{(i)}}^{-1}  N   
   -  \Lambda_X^{(i)}  -{{\bf K}} P^{(i)} \right)\\
&=& P^{(i)} + \eta  {\bf A}^{-1} \left(
{\bf P^{(i)}}^{-1} N  -  N_X - {\bf I}_X {\bf P}^{(i)} {{\bf K}} P^{(i)}
          -{{\bf K}} P^{(i)} \right).
\end{eqnarray}
Again, normalizing $P$ at each iteration this is equivalent to
solving for $z = Z_X P$,
and the update procedure for $\Lambda_X$ can be varied.

\subsection{Learning matrices}
\subsubsection{Learning algorithms for density estimation}

There exists a variety of well developed
numerical methods for unconstraint as well as for constraint optimization
\cite{Pierre-1986,Fletcher-1987,Hackbusch-1989,Press-Teukolsky-Vetterling-Flannery-1992,Hackbusch-1993,Bazaraa-Sherali-Shetty-1993,Bertsekas-1995,Golden-1996,Polak-1997}.
Popular examples include 
conjugate gradient,
Newton, and quasi--Newton methods,
like the variable metric methods 
DFP (Davidon--Fletcher--Powell) or
BFGS (Broyden--Fletcher--Goldfarb--Shanno).

All of them correspond to the choice of specific,
often iteration dependent, learning matrices ${\bf A}$
defining the learning algorithm.
Possible simple choices are:
\begin{eqnarray}
{\bf A} = {\bf I} &:& {\rm Gradient}\label{a-1}\\
{\bf A} = {\bf D} &:& {\rm Jacobi}\label{a-1a}\\
{\bf A} = {\bf L} + {\bf D} 
&:& \mbox{\rm Gauss--Seidel}\label{a-1b}\\
{\bf A} = {{\bf K}} &:& \mbox{prior relaxation}\label{a-2}
\end{eqnarray}
where ${\bf I}$ stands for the identity operator,
${\bf D}$ for a diagonal matrix,
e.g.\ the diagonal part of ${{\bf K}}$,
and ${\bf L}$ for a lower triangular matrix,
e.g.\ the lower triangular part of ${{\bf K}}$.
In case ${{\bf K}}$ represents the 
operator of the prior term in an error functional
we will call iteration with 
${{\bf K}}^{-1}$ (corresponding to the covariance of the prior process)
{\it prior relaxation}.
For $\phi$--independent ${{\bf K}}$ and $T$,
$\eta=1$ with invertible ${{\bf K}}$
the corresponding linear equation is solved by 
prior relaxation in one step.
However, also linear equations are solved
by iteration if the size of ${{\bf K}}$ is too large to be inverted.
Because of ${\bf I}^{-1}$ =  ${\bf I}$
the gradient algorithm does not require inversion.

On one hand, density estimation is a rather general problem
requiring the solution of constraint, 
inhomogeneous, nonlinear (integro--)differential equations.
On the other hand, density estimation problems are,
at least for Gaussian specific priors and non restricting parameterization,
typically ``nearly'' linear 
and have only a relatively simple non--negativity and normalization constraint.
Furthermore, the inhomogeneities are commonly restricted
to a finite number of discrete training data points.
Thus, we expect the inversion of ${{\bf K}}$ to be the essential
part of the solution for density estimation problems.
However, ${{\bf K}}$ is not necessarily invertible
or may be difficult to calculate.
Also, inversion of ${{\bf K}}$ is not exactly what
is optimal and there are improved methods.
Thus, we will discuss in the following basic optimization
methods adapted especially to the situation of density estimation.

\subsubsection{Linearization and Newton algorithm}

For linear equations ${{\bf K}} \phi$ = $T$
where $T$ and ${{\bf K}}$
are no functions of $\phi$
a spectral radius $\rho ({\bf M})<1$
(the largest modulus of the eigenvalues)
of the {\it iteration matrix} 
\begin{equation}
{\bf M}
= - \eta {\bf A}^{-1} {\bf B}_\eta
=(1-\eta) {\bf I} - \eta {\bf A}^{-1} {\bf B}
= {\bf I} - \eta {\bf A}^{-1} {{\bf K}}
\end{equation}
would guarantee convergence of the iteration scheme.
This is easily seen by solving the linear equation 
by iteration according to (\ref{iter-1})
\bea
\phi^{(i+1)} &=& \eta {\bf A}^{-1}T + {\bf M}\phi^{(i)}
\\ &=&
\eta {\bf A}^{-1}T + \eta {\bf M} {\bf A}^{-1}T + {\bf M}^2 \phi^{(i-1)}
\\ &=&
\eta \sum_{n=0}^\infty {\bf M}^n {\bf A}^{-1} T
.
\eea
A zero mode of ${{\bf K}}$,
for example a constant function for 
differential operators without boundary conditions,
corresponds to an eigenvalue $1$ of ${\bf M}$
and would lead to divergence of the sequence $\phi^{(i)}$.
However, a nonlinear $T(\phi)$ or ${{\bf K}}(\phi)$,
like the nonlinear normalization constraint
contained in $T(\phi)$,
can then still lead to a unique solution.

A convergence analysis for nonlinear equations can be done 
in a linear approximation around a fixed point.
Expanding the gradient at $\phi^*$ 
\begin{equation}
G(\phi) 
=
\frac{\delta (-E_\phi) }{\delta \phi}\Bigg|_{\phi^*}
+ (\phi-\phi^*) \,\, {\bf H} (\phi^*)
+ \cdots
\end{equation}
shows that the factor of the linear term
is the Hessian.
Thus in the vicinity of $\phi^*$
the spectral radius of the iteration matrix 
\begin{equation}
{\bf M} = {\bf I} + \eta {\bf A}^{-1} {\bf H},
\end{equation}
determines the rate of convergence.
The Newton algorithm uses the negative
Hessian  $-{\bf H}$ as learning matrix provided it exists
and is positive definite. Otherwise it must resort to other methods.
In the linear approximation (i.e., for quadratic energy)
the Newton algorithm 
\begin{eqnarray}
{\bf A} = -{\bf H} &:& {\rm Newton}\label{a-N}
\end{eqnarray}
is optimal.
We have already seen in Sections \ref{Hessians-L} and \ref{Hessians-P}
that the inhomogeneities
generate in the Hessian in addition to ${{\bf K}}$ a diagonal part 
which can remove zero modes of ${{\bf K}}$.

\subsubsection{Massive relaxation}

We now consider methods to construct a 
positive definite or at least invertible learning matrix.
For example, far from a minimum the Hessian ${\bf H}$ 
may not be positive definite
and like a differential operator ${{\bf K}}$
with zero modes, not even invertible.
Massive relaxation 
can transform a non--invertible or not positive definite operator ${\bf A}_0$,
e.g.\ ${\bf A}_0 = {{\bf K}}$ or ${\bf A}_0 = - {\bf H}$,
into an invertible or positive definite operators:
\begin{eqnarray}
{\bf A} = {\bf A}_0 + m^2 {\bf I} &:& {\rm Massive \,\, relaxation}
\label{a-3}
\end{eqnarray}
A generalization would be to allow $m = m(x,y)$.
This is, for example, used
in some realizations of Newton`s method for minimization
in regions where ${\bf H}$ is not positive definite
and a diagonal operator is added to $-{\bf H}$, 
using for example a modified Cholesky factorization
\cite{Bertsekas-1995}. 
The mass term removes the zero modes of ${{\bf K}}$
if $-m^2$ is not in the spectrum of ${\bf A}_0$
and makes it positive definite if $m^2$ 
is larger than the smallest eigenvalue of ${\bf A}_0$.
Matrix elements 
$(\, \phi,\,  ({\bf A}_0 - z {\bf I})^{-1} \,\phi\,)$ 
of the resolvent ${\bf A}^{-1} (z)$,
$z=-m^2$ representing in this case a complex variable,
have poles at discrete eigenvalues of ${\bf A}_0$
and a cut at the continuous spectrum
as long as $\phi$ has a non--zero  overlap with
the corresponding eigenfunctions.
Instead of multiples of the identity,
also other operators may be added to remove zero modes.
The Hessian ${\bf H}_L$ in (\ref{H_L}), for example,
adds a $x$--dependent mass--like, but not necessarily positive definite
term to ${{\bf K}}$.
Similarly, for example ${\bf H}_P$ in (\ref{H_P}) has
$(x,y)$--dependent mass ${\bf P}^{-2} {\bf N}$ 
restricted to data points.

While full relaxation is the massless limit $m^2\rightarrow 0$ of massive
relaxation, a gradient algorithm with $\eta^\prime $ can be obtained
as infinite mass limit $m^2\rightarrow \infty$
with $\eta \rightarrow \infty$
and $m^2/\eta = 1/\eta^\prime$.

Constant functions are typical zero modes, i.e.,
eigenfunctions with zero eigenvalue,
for differential operators
with periodic boundary conditions.
For instance for a common smoothness term 
$ -\Delta $ (kinetic energy operator) 
as regularizing operator ${{\bf K}}$
the inverse of ${\bf A} = {{\bf K}} +m^2 {\bf I}$ has the form
\begin{equation}
{\bf A}^{-1} (x^\prime , y^\prime ;  x,y ) =
\, \, \frac{1}{-\Delta + m^2}.
\end{equation}
\begin{equation}
=\int_{-\infty}^\infty \! \frac{d^{d_X}\!k_x \, d^{d_Y}\!k_y}{(2 \pi)^d} \, 
\frac{e^{i k_x (x-x^\prime) + ik_y (y-y^\prime )}
}{k_x^2 + k_y^2+m^2},
\end{equation}
with $d=d_X+d_Y$, $d_X$ = dim($X$), $d_Y$ = dim($Y$).
This Green`s function or matrix element of the resolvent kernel 
for ${\bf A}_0$ is analogous to the (Euclidean) propagator
of a free scalar field with mass $m$,
which is its two--point correlation function 
or  matrix element of the covariance operator.
According to $1/x = \int_0^\infty \! dt\, e^{-xt}$
the denominator can be brought into the exponent
by introducing an additional integral.
Performing the resulting Gaussian integration over $k = (k_x,k_y)$
the inverse can be expressed as
\[
{\bf A}^{-1} (x^\prime , y^\prime ;  x,y ; m)
=
m^{d-2} {\bf A}^{-1} (m (x-x^\prime) , m (y-y^\prime) ; 1)
\]
\begin{equation}
=(2 \pi)^{-d/2} \left( \frac{m}{|x-x^\prime|+|y-y^\prime|}\right)^{(d-2)/2}
\!\!K_{(d-2)/2}( m |x-x^\prime|+ m |y-y^\prime|  ),
\end{equation}
in terms of the 
modified Bessel functions $K_\nu (x)$ 
which 
have the following integral representation
\begin{equation}
K_\nu (2 \sqrt{\beta \gamma}) =
\frac{1}{2} \left(\frac{\gamma}{\beta}\right)^{\frac{\nu}{2}}
\int_0^\infty \! dt\, t^{\nu-1} e^{\frac{\beta}{t}-\gamma t}.
\label{modBessK}
\end{equation}
Alternatively, the same result can be obtained
by switching to $d$--dimensional spherical coordinates,
expanding the exponential in ultra-spheric harmonic functions
and performing the integration over the angle-variables
\cite{Kleinert-1993}.
For the example $d=1$ this corresponds to Parzen´s kernel
used in density estimation
or for $d=3$ 
\begin{equation}
{\bf A}^{-1} (x^\prime , y^\prime ;  x,y ) 
= \frac{1}{4 \pi |x-x^\prime|+ 4 \pi |y-y^\prime|}
e^{- m |x-x^\prime| - m |y-y^\prime| }.
\end{equation}

The Green's function for periodic, Neumann, or Dirichlet boundary conditions
can be expressed  by sums over ${\bf A}^{-1} (x^\prime , y^\prime ;  x,y )$
\cite{Glimm-Jaffe-1987}.

The lattice version 
of the Laplacian with lattice spacing $a$ reads
\begin{equation}
\hat \Delta f(n) =
\frac{1}{a^2} \sum_j^d 
[  f(n-a_j)-2 f(n) + f(n+a_j) ],
\end{equation}
writing $a_j$ for a vector in direction $j$ and length $a$.
Including a mass term 
we get as lattice approximation for ${\bf A}$
\[
\hat{\bf A} (n_x,n_y ;  m_x,m_y ) =
-\frac{1}{a^2}\sum_{i=1}^{d_X} 
\delta_{n_y,m_y}
(\delta_{n_x+a_i^x,m_x} 
-2 \delta_{n_x,m_x}
+ \delta_{n_x-a^x_i,m_x})
\]
\begin{equation}
-\frac{1}{a^2}\sum_{j=1}^{d_Y} 
\delta_{n_x,m_x}
(\delta_{n_y+a^y_j,m_y}
-2 \delta_{n_y,m_y}
+\delta_{n_y-a^y_j,m_y})
+m^2 \delta_{n_x,m_x} \delta_{n_y,m_y}
\end{equation}
Inserting the Fourier representation (\ref{fourierdelta})
of $\delta (x)$ gives 
\[
\hat{\bf A} (n_x,n_y ;  m_x,m_y ) 
= \frac{2d}{a^2}
\int_{-\pi}^\pi \!\! 
\frac{d^{d_X}\!k_x \, d^{d_Y}\!k_y}{(2\pi)^d}
e^{ik_x (n_x-m_x) + ik_y (n_y-m_y)}
\]
\begin{equation}
\times \left(
1 + \frac{m^2 a^2}{2d}
-\frac{1}{d} \sum_{i=1}^{d_X} \cos k_{x,i} 
-\frac{1}{d} \sum_{j=1}^{d_Y} \cos k_{y,j}
\right),
\end{equation}
with $k_{x,i}$ = $k_x a^x_i $, $\cos k_{y,j}$ = $\cos k_y a^y_j$
and inverse
\[
\hat {\bf A}^{-1} (n_x , n_y ; m_x,m_y) =
\int_{-\pi}^\pi \!\! 
\frac{d^{d_X}\!k_x \, d^{d_Y}\!k_y}{(2\pi)^d}
\hat {\bf A}^{-1} (k_x , k_y)
e^{ik_x (n_x-m_x) + ik_y (n_y-m_y)}
\]
\begin{equation}
=\frac{a^2}{2d}
\int_{-\pi}^\pi \!\! 
\frac{d^{d_X}\!k_x \, d^{d_Y}\!k_y}{(2\pi)^d}
\frac{e^{ik_x (n_x-m_x) + ik_y (n_y-m_y)}}
{   
1 \!+\! \frac{m^2a^2}{2d}  \!-\! 
\frac{1}{d} \sum_{i=1}^{d_X} \cos k_{x,i} 
\!-\!
\frac{1}{d} \sum_{j=1}^{d_Y} \cos k_{y,j}
}.
\end{equation}
(For $m=0$ and $d\le 2$ the integrand diverges
for $k\rightarrow 0$ (infrared divergence).
Subtracting formally the also infinite
$\hat {\bf A}^{-1} (0,0 ; 0,0)$
results in finite difference.
For example in $d=1$ one finds
$\hat {\bf A}^{-1} (n_y ; m_y) - 
\hat {\bf A}^{-1} (0;0)$ = $-(1/2) |n_y-m_y |$
\cite{Itzykson-Drouffe-1989}.
Using $1/x = \int_0^\infty \! dt\, e^{-xt}$ 
one obtains \cite{Pordt-1998}
\begin{equation}
\hat {\bf A}^{-1} (k_x , k_y) =
\frac{1}{2} \int_0^\infty \! dt\,
e^{ -\mu t + a^{-2} t 
\left( \sum_i^{d_X} \cos k_{x,i} + \sum_j^{d_Y} \cos k_{y,j} \right) },
\end{equation}
with $\mu = d/a^2 + m^2/2$.
This allows to express the inverse $\hat {\bf A}^{-1}$
in terms of the modified Bessel functions $I_\nu (n)$
which have for integer argument $n$ the integral representation 
\begin{equation}
I_\nu (n) =
\frac{1}{\pi} \int_0^\pi \!d\Theta\, e^{n \cos \Theta}\cos (\nu \Theta).
\label{modBessI}
\end{equation}
One finds
\begin{equation}
\hat {\bf A}^{-1} (n_x , n_y ; m_x,m_y) =
\frac{1}{2} \int_0^\infty e^{-\mu t} 
\prod_{i =1}^{d_X} K_{|n_{x,i} - n^\prime_{x,i}|} (t/a^2)
\prod_{j =1}^{d_Y} K_{|m_{y,j} - m^\prime_{y,j}|} (t/a^2).
\end{equation}

It might be interesting to remark that the matrix elements of the 
inverse learning matrix or free massive propagator  
on the lattice $\hat {\bf A}^{-1}(x^\prime,y^\prime;x,y)$
can be given an interpretation
in terms of (random) walks connecting the two points
$(x^\prime, y^\prime)$ and $(x,y)$
\cite{Fernandez-Froehlich-Sokal-1992,Pordt-1998}.
For that purpose
the lattice Laplacian is splitted into a diagonal and a nearest neighbor part 
\begin{equation}
- \hat \Delta  = \frac{1}{a^2} \left( 2d {\bf I} - {\bf W} \right),
\end{equation}
where the nearest neighbor matrix
${\bf W}$ has matrix elements equal one for nearest neighbors
and equal to zero otherwise.
Thus, 
\begin{equation}
\left( - \hat \Delta + m^2\right)^{-1}
=
\frac{1}{2 \mu} \left( {\bf I} - \frac{1}{2 \mu a^2} {\bf W}\right)^{-1}
=\frac{1}{2 \mu} \sum_{n=0}^\infty
\left(\frac{1}{2 \mu a^2}\right)^n {\bf W}^n ,
\end{equation}
can be written as geometric series.
The matrix elements ${\bf W}^n (x^\prime,y^\prime;x,y)$
give the number of walks $w[(x^\prime,y^\prime)\rightarrow (x,y)]$ 
of length  $|w|$  = $n$ connecting
the two points $(x^\prime,y^\prime)$ and $(x,y)$.
Thus, one can write
\begin{equation}
\left( - \hat \Delta + m^2\right)^{-1}(x^\prime,y^\prime;x,y)
= \frac{1}{2 \mu} \sum_{w[(x^\prime,y^\prime)\rightarrow (x,y)]}
\left(\frac{1}{2 \mu a^2}\right)^{|w|} .
\end{equation}

\subsubsection{Gaussian relaxation}

As Gaussian kernels are often used in density estimation and
also in function approximation 
(e.g.\ for radial basis functions \cite{Poggio-Girosi-1990})
we consider the example
\begin{eqnarray}
{\bf A} = 
\sum_{k=0}^\infty \frac{1}{k!}
 \left( \frac{{\bf M}^2 }{2\tilde\sigma^2}\right)^k 
=e^{\frac{{\bf M}^2}{2\tilde\sigma^2}}
&:& {\rm Gaussian}
\label{a-3a}
\end{eqnarray} 
with positive semi--definite ${\bf M}^2$.
The contribution for $k=0$ corresponds to a mass term
so for positive semi--definite  ${\bf M}$ this ${\bf A}$ 
is positive definite and therefore invertible
with inverse 
\begin{equation}
{\bf A}^{-1} 
= e^{-\frac{{\bf M}^2}{2\tilde\sigma^2}},
\label{inva-gauss}
\end{equation} 
which is diagonal and Gaussian in ${\bf M}$--representation.
In the limit $\tilde \sigma\rightarrow \infty$
or for zero modes of ${\bf M}$ the Gaussian 
${\bf A}^{-1}$ becomes the identity ${\bf I}$,
corresponding to the gradient algorithm.
Consider
\begin{equation}
{\bf M}^2 (x^\prime , y^\prime ; x, y)  
= - \delta (x-x^\prime )\delta (y-y^\prime ) \Delta 
\end{equation}
where the $\delta$--functions 
are usually skipped from the notation,
and
\[
\Delta =
\frac{\partial^2}{\partial x^2} + \frac{\partial^2}{\partial y^2},
\]
denotes the Laplacian.
The kernel of the inverse is diagonal in Fourier representation
\begin{equation}
A(k_x^\prime,k_y^\prime;,k_x,k_y)
= \delta (k_x-k_x^\prime )\delta (k_y-k_y^\prime ) 
e^{-\frac{k_x^2+k_y^2}{2\tilde\sigma^2}}
\end{equation}
and non--diagonal, but also  Gaussian in $(x,y)$--representation
\begin{equation}
{\bf A}^{-1} (x^\prime , y^\prime ;  x,y )
= e^{-\frac{\Delta}{2 \tilde\sigma^2}}
= \int\! \frac{dk_x dk_y}{(2 \pi)^d} \, 
e^{-\frac{k_x^2 + k_y^2}{2 \tilde\sigma^2}
+i k_x (x-x^\prime ) +i k_y (y-y^\prime )}
\end{equation} 
\begin{equation}
=\left(\frac{\tilde\sigma}{ \sqrt{2 \pi}}\right)^d
e^{-\tilde\sigma^2((x-x^\prime )^2 + (y-y^\prime )^2)}
= \frac{1}{ \left(\sigma \sqrt{2 \pi} \right)^d}
\, \, e^{-\frac{(x-x^\prime )^2 + (y-y^\prime )^2}{2\sigma^2}},
\end{equation}
with $\sigma = 1/{\tilde\sigma}$
and  $d=d_X+d_Y$, $d_X$ = dim($X$), $d_Y$ = dim($Y$).

\subsubsection{Inverting in subspaces}

Matrices considered as learning matrix
have to be invertible.
Non-invertible matrices can only be inverted in
the subspace which is the complement of its zero space.
With respect to a symmetric 
${\bf A}$ we define 
the projector 
${\bf Q}_0 = {\bf I} - \sum_i \psi_i^T\,\psi_i$ into its zero space
(for the more general case of a normal ${\bf A}$
replace $\psi_i^T$ by the hermitian conjugate $\psi_i^ \dagger$)
and its complement ${\bf Q}_1 = {\bf I} - {\bf Q}_0 = \sum_i \psi_i^T\,\psi_i$
with $\psi_i$ denoting orthogonal eigenvectors with eigenvalues $a_i\ne 0$
of ${\bf A}$, i.e.,
${\bf A} \psi_i = a_i \psi_i \ne 0$.
Then, denoting projected sub-matrices by 
${\bf Q}_i {\bf A} {\bf Q}_j$ = ${\bf A}_{ij}$
we have 
${\bf A}_{00}$ = ${\bf A}_{10}$ = ${\bf A}_{01}$ = $0$, 
i.e.,
\begin{equation}
{\bf A}  = 
{\bf Q}_1 {\bf A} {\bf Q}_1
= 
{\bf A}_{11}.
\end{equation}
and in the update equation
\begin{equation}
{\bf A} \Delta \phi^{(i)} = \eta  \, G
\end{equation}
only ${\bf A}_{11}$ can be inverted.
Writing
${\bf Q}_j \phi$ = $\phi_j$ for a projected vector, 
the iteration scheme acquires the form
\begin{eqnarray}
\Delta \phi^{(i)}_1 &=& \eta {\bf A}^{-1}_{11} G_1,\\
0 &=& \eta \, G_0.
\end{eqnarray}
For positive semi--definite
${\bf A}$ the sub-matrix ${\bf A}_{11}$ is positive definite.
If the second equation is already fulfilled
or its solution is postponed to a later iteration step we have
\begin{eqnarray}
\phi^{(i+1)}_1 &=& \phi^{(i)}_1 + \eta {\bf A}_{11}^{-1}
\left( T_1^{(i)} - {{\bf K}}_{11}^{(i)} \phi_1^{(i)} -{{\bf K}}_{10}^{(i)} \phi_0^{(i)}\right),\\
\phi^{(i+1)}_0 &=& \phi^{(i)}_0.
\end{eqnarray}
In case the projector
${\bf Q}_0 = {\bf I}_0$ is diagonal in the chosen representation
the projected equation 
can directly be solved by skipping the corresponding components.
Otherwise one can use the Moore--Penrose inverse ${\bf A}^\#$ of ${\bf A}$ to 
solve the projected equation
\begin{equation}
\Delta \phi^{(i)} = \eta {\bf A}^\# G.
\end{equation}
Alternatively, an invertible operator $\tilde {\bf A}_{00}$  
can be added to ${\bf A}_{11}$ to obtain a complete iteration scheme
with ${\bf A}^{-1}$ = ${\bf A}_{11}^{-1}$ + $\tilde {\bf A}_{00}^{-1}$ 
\begin{eqnarray}
\phi^{(i+1)} &=& \phi^{(i)} 
+\eta {\bf A}_{11}^{-1}
\left( T_1^{(i)} - {{\bf K}}_{11}^{(i)} \phi_1^{(i)} -{{\bf K}}_{10}^{(i)} \phi_0^{(i)}\right)
\nonumber\\
&&+\eta 
\tilde {\bf A}_{00}^{-1}\left( T_0^{(i)} - {{\bf K}}_{01}^{(i)} \phi_1^{(i)} 
     -{{\bf K}}_{00}^{(i)} \phi_0^{(i)}\right).
\end{eqnarray}

The choice
${\bf A}^{-1}$ = 
$\left( {\bf A}_{11} + {\bf I}_{00} \right)^{-1}$ 
= ${\bf A}_{11}^{-1} + {\bf I}_{00}$, 
= ${\bf A}_{11}^{-1} + {\bf Q}_{0}$, 
for instance, 
results in a gradient algorithm on the zero space
with additional coupling between the two subspaces.

\subsubsection{Boundary conditions}

For a differential operator 
invertability can be achieved
by adding an operator 
restricted to a subset $B \subset X\times Y$ (boundary).
More general, we consider an projector 
${\bf Q}_B$ on a space which we will call boundary
and the projector on the interior ${\bf Q}_I$ = ${\bf I} - {\bf Q}_B$.
We write ${\bf Q}_k {{\bf K}} {\bf Q}_l$ = ${{\bf K}}_{kl}$
for $k,l\in\{I,B\}$,
and require ${{\bf K}}_{BI} = 0$.
That means ${{\bf K}}$ is not symmetric,
but ${{\bf K}}_{II}$ can be,
and we have
\begin{equation}
{{\bf K}} 
= ({\bf I} - {\bf Q}_B) {{\bf K}} + {\bf Q}_{B} {{\bf K}} {\bf Q}_{B} 
= 
{{\bf K}}_{II} + {{\bf K}}_{IB} + {{\bf K}}_{BB}.
\label{boundaryEQ}
\end{equation}
For such an ${{\bf K}}$
an equation of the form
${{\bf K}} \phi = T$ can be decomposed into
\begin{eqnarray}
{{\bf K}}_{BB} \phi_B    &=& T_B, 
\label{boundary1}\\
{{\bf K}}_{IB} \phi_B + {{\bf K}}_{II} \phi_I &=& T_I, 
\label{boundary2}
\end{eqnarray}
with projected 
$\phi_k = {\bf Q}_k \phi$,
$T_k = {\bf Q}_k T$,
so that
\begin{eqnarray}
\phi_B  &=& {{\bf K}}_{BB}^{-1} T_B, \\
\phi_I  &=& {{\bf K}}_{II} ^{-1} 
\left( T_I - {{\bf K}}_{IB} {{\bf K}}_{BB}^{-1}T_B \right). 
\end{eqnarray}
The boundary part is independent of the interior,
however, the interior can depend on the boundary.
A basis can be chosen so that the projector onto the boundary is diagonal,
i.e.,
\[
{\bf Q}_B =
{\bf I}_B 
= \sum_{j:(x_j,y_j)\in B} 
 (\delta(x_j) \otimes \delta(y_j)) \otimes (\delta(x_j) \otimes \delta(y_j))^T
.
\]
Eliminating the boundary
results in an  equation for the interior
with adapted inhomogeneity.
The special case ${{\bf K}}_{BB}$ =  ${\bf I}_{BB}$,
i.e.,  $\phi_B = T_B$ on the boundary,
is known as Dirichlet boundary conditions.

As trivial example of an equation ${\bf K}\phi$ = $T$
with boundary conditions,
consider a one--dimensional finite difference approximation
for a negative Laplacian ${\bf K}$, 
adapted to include boundary conditions
as in Eq.\ (\ref{boundaryEQ}),
\be
\left(
\begin{tabular}{    c     c     c     c     c    c }
                    1  &  0 &  0  &  0  &  0   &  0   \\
                   $-1$&  2  & $-1$&  0  &  0  &  0   \\
                    0  & $-1$&  2  & $-1$&  0  &  0   \\
                    0  &  0  & $-1$&  2  & $-1$&  0   \\
                    0  &  0  &  0  & $-1$&  2  & $-1$ \\
                    0  &  0  &  0  &  0  &  0  &  1   \\
\end{tabular}
\right)
\left(
\begin{tabular}{    c   }
                    $\phi_1$ \\
                    $\phi_2$ \\
                    $\phi_3$ \\
                    $\phi_4$ \\
                    $\phi_5$ \\
                    $\phi_6$ 
\end{tabular}
\right)
=
\left(
\begin{tabular}{    c   }
                    $b$ \\
                    $T_2$ \\
                    $T_3$ \\
                    $T_4$ \\
                    $T_5$ \\
                    $b$  
\end{tabular}
\right)
.
\ee
Then Eq.\ (\ref{boundary1})
is equivalent to the boundary conditions,
$\phi_1$ = $b$, $\phi_6$ = $b$,
and the interior equation Eq.\ (\ref{boundary2}) reads
\be
\left(
\begin{tabular}{    c     c     c    c }
                    2  & $-1$&  0  &  0    \\
                   $-1$&  2  & $-1$&  0    \\
                    0  & $-1$&  2  & $-1$  \\
                    0  &  0  & $-1$&  2  
\end{tabular}
\right)
\left(
\begin{tabular}{    c   }
                    $\phi_2$ \\
                    $\phi_3$ \\
                    $\phi_4$ \\
                    $\phi_5$ 
\end{tabular}
\right)
=
\left(
\begin{tabular}{    c   }
                    $T_2$ \\
                    $T_3$ \\
                    $T_4$ \\
                    $T_5$ \\
\end{tabular}
\right)
+
\left(
\begin{tabular}{    c   }
                    $b$ \\
                    $0$ \\
                    $0$ \\
                    $b$ \\
\end{tabular}
\right)
.
\ee
(Useful references dealing with the
numerical solution of partial differential equations are, for example,
\cite{Ames-1977,Mitchell-Griffiths-1980,Hackbusch-1985,Press-Teukolsky-Vetterling-Flannery-1992,Grossmann-Roos-1994}.)

Similarly to boundary conditions for ${\bf K}$,
we may use a learning matrix ${\bf A}$ 
with boundary conditions
(corresponding for example to those used for ${\bf K}$):
\begin{eqnarray}
{\bf A} = 
{\bf A}_{II} + {\bf A}_{IB} +{\bf A}_{BB} 
&:& \mbox{Boundary}
\label{a-4a}\\
{\bf A} = {\bf A}_{II} + {\bf A}_{IB}  + {\bf I}_{BB} 
&:& {\rm Dirichlet \,\,boundary}
\label{a-5}
\end{eqnarray}
For linear ${\bf A}_{BB}$ the form (\ref{a-4a})
corresponds to general linear boundary conditions.
(It is also possible to require nonlinear boundary conditions.)
${\bf A}_{II}$ can be chosen symmetric,
and therefore positive definite,
and the boundary of ${\bf A}$ can 
be changed during iteration.
Solving ${\bf A} (\phi^{(i+1)} - \phi^{(i)})$ = $\eta (T^{(i)} - {{\bf K}}^{(i)} \phi^{(i)})$
gives on the boundary and for the interior
\begin{equation}
\phi_B^{(i+1)} = \phi_B^{i} +
\eta {\bf A}_{BB}^{-1}
\left(
T_B^{(i)} - {{\bf K}}_{BB}^{(i)} \phi_B^{(i)} - {{\bf K}}_{BI}^{(i)} \phi_I^{(i)}
\right),  
\end{equation}
\begin{equation}
\phi_I^{(i+1)} = \phi_I^{i} +
\eta {\bf A}_{II}^{-1}
\left(
T_I^{(i)} - {{\bf K}}_{II}^{(i)} \phi_I^{(i)} - {{\bf K}}_{IB}^{(i)} \phi_B^{(i)}
\right)
- {\bf A}_{II}^{-1}
{\bf A}_{IB}
\left( \phi_B^{(i+1)} - \phi_B^{(i)}\right),
\end{equation}
For fulfilled boundary conditions with
$\phi_B^{(i)} = \left({{\bf K}}_{BB}^{(i)}\right)^{-1} T_B^{(i)}$
and ${{\bf K}}_{BI}^{(i)} = 0$,
or for $\eta {\bf A}_{BB}^{-1} \rightarrow 0$
so the boundary is not updated,
the term $\phi_B^{(i+1)} - \phi_B^{(i)}$
vanishes.  
Otherwise, inserting the first in the second equation gives
\bea
\phi_I^{(i+1)} &=& \phi_I^{i} +
\eta {\bf A}_{II}^{-1}
\left(
T_I^{(i)} - {{\bf K}}_{II}^{(i)} \phi_I^{(i)} - {{\bf K}}_{IB}^{(i)} \phi_B^{(i)}
\right)
\\
&&-\eta {\bf A}_{II}^{-1}
{\bf A}_{IB}
{\bf A}_{BB}^{-1}
\left(
T_B^{(i)} - {{\bf K}}_{BB}^{(i)} \phi_B^{(i)} - {{\bf K}}_{BI}^{(i)} \phi_I^{(i)}
\right)
.
\nonumber\eea
Even if ${{\bf K}}$ is not defined with boundary conditions,
an invertible ${\bf A}$ can be obtained from
${{\bf K}}$ by introducing a boundary for ${\bf A}$.
The updating process is then
restricted to the interior.
In such cases the boundary should be systematically changed
during iteration.
Block--wise updating of $\phi$ represent a special case
of such learning matrices with variable boundary.

The following table summarizes the learning matrices
we have discussed in some detail for the setting of density estimation
(for conjugate gradient and quasi--Newton methods
see, for example, \cite{Press-Teukolsky-Vetterling-Flannery-1992}):

\begin{center}
\begin{tabular}{ | r | l |}
\hline
Learning algorithm & Learning matrix\\
\hline
Gradient            & ${\bf A} = {\bf I}$\\
Jacobi              & ${\bf A} = {\bf D}$\\
Gauss--Seidel       & ${\bf A} = {\bf L} + {\bf D}$\\ 
Newton              & ${\bf A} = -{\bf H} $\\
prior relaxation& ${\bf A} = {{\bf K}} $\\
massive relaxation  & ${\bf A} = {\bf A}_0 + m^2 {\bf I}$\\
linear boundary& 
${\bf A} = {\bf A}_{II}+ {\bf A}_{IB}  + {\bf A}_{BB} $\\
Dirichlet boundary  & 
${\bf A} = {\bf A}_{II}+ {\bf A}_{IB}  + {\bf I}_{BB} $\\
Gaussian            & ${\bf A} = \sum_{k=0}^\infty \frac{1}{k!}
                  \left( \frac{{\bf M}^2 }{2\tilde\sigma^2}\right)^k 
                  =e^{\frac{{\bf M}^2}{2\tilde\sigma^2}}$\\
\hline
\end{tabular}
\end{center}

\subsection{Initial configurations and kernel methods}
\subsubsection{Truncated equations}

To solve the nonlinear Eq.\ (\ref{phiKTphi})
by iteration
one has to begin with an initial configuration  $\phi^{(0)}$.
In principle any easy to use technique for density estimation 
could be chosen to construct starting guesses $\phi^{(0)}$.

One possibility to obtain initial guesses 
is to neglect some terms of the full stationarity equation
and solve the resulting simpler (ideally linear)
equation first.
The corresponding solution may be taken as initial guess
$\phi^{(0)}$ for solving the full equation.

Typical error functionals
for statistical learning problems 
include a term $( L,\, N)$ consisting
of a discrete sum over a finite number $n$ of training data.
For diagonal ${\bf P}^\prime$ those contributions
result (\ref{varlag}) in 
$n$  $\delta$--peak contributions to the inhomogeneities $T$
of the stationarity equations,
like $\sum_i \delta (x-x_i)\delta (y-y_i)$
in Eq.\ (\ref{OL=T})
or 
$\sum_i \delta (x-x_i)\delta (y-y_i)/P(x,y)$
in Eq.\ (\ref{OP=T}).
To find an initial guess, one can 
now keep only that $\delta$--peak contributions $T_\delta$ 
arising from the training data
and ignore the other, typically continuous parts of $T$.
For (\ref{OL=T}) and (\ref{OP=T})
this means setting $\Lambda_X = 0$
and yields a truncated equation 
\be
{{\bf K}} \phi
= {\bf P}^\prime {\bf P}^{-1} N
= T_{\delta}
.
\ee
Hence, $\phi$ can for diagonal ${\bf P}^\prime$
be written as a sum of $n$ terms
\begin{equation}
 \phi(x,y) =  \sum_{i=1}^n {{\bf C}} (x,y;x_i ,y_i) 
\frac{P^\prime (x_i,y_i)}{P(x_i,y_i)},
\label{vtrunc}
\end{equation}
with ${{\bf C}} = {{\bf K}}^{-1}$,
provided the inverse ${{\bf K}}^{-1}$ exists.
For $E_L$ the resulting truncated equation is linear in $L$.
For $E_P$, however, the truncated equations remains nonlinear.
Having solved the truncated equation
we restore the necessary constraints for $\phi$,
like normalization and non--negativity for $P$
or normalization of the exponential for $L$.

In general, a ${{\bf C}}\ne {{\bf K}}^{-1}$ can be chosen.
This is necessary if ${{\bf K}}$ is not invertible
and can also be useful if its inverse is difficult to calculate.
One possible choice for the kernel is the inverse negative Hessian
${{\bf C}} = - {\bf H}^{-1}$ evaluated 
at some initial configuration $\phi^{(0)}$
or an approximation of it.
A simple possibility to construct an invertible operator from a
noninvertible ${{\bf K}}$ would be
to add a mass term 
\begin{equation}
{{\bf C}}
= 
\left( {{\bf K}} + m_C^2 {\bf I} \right)^{-1}
,
\end{equation}
or to impose additional boundary conditions.

Solving a truncated equation of the form (\ref{vtrunc})
with ${{\bf C}}$ 
means skipping the term 
$-{{\bf C}}({\bf P}^\prime \Lambda_X+({{\bf K}}-{{\bf C}}^{-1})\phi)$
from the exact relation
\be
\phi = {{\bf C}} {\bf P}^\prime {\bf P}^{-1} N
     -{{\bf C}}({\bf P}^\prime \Lambda_X+({{\bf K}}-{{\bf C}}^{-1})\phi)
.
\label{nontrunc}
\ee
A kernel used to create an initial guess $\phi^{(0)}$
will be called an {\it initializing kernel}.

A similar possibility is to start with
an ``empirical solution'' 
\be
\phi^{(0)} = \phi_{\rm emp},
\ee
where $\phi_{\rm emp}$ is defined as
a $\phi$ which reproduces the conditional empirical density $P_{\rm emp}$ of 
Eq.\ (\ref{cond-P-emp1}) obtained from the training data,
i.e.,
\be
P_{\rm emp} = P (\phi_{\rm emp}).
\ee
In case, there are not data points for every $x$--value,
a correctly normalized initial solution 
would for example be given by
$\tilde P_{\rm emp}$ defined in Eq.\ (\ref{tilde-P-emp}).
If zero values of the empirical density
correspond to infinite values for $\phi$, like in the case $\phi$ = $L$, 
one can use $P^\epsilon_{\rm emp}$ as defined in Eq.\ (\ref{Pepsilon}),
with small $\epsilon$, 
to obtain an initial guess.

Similarly to Eq.\ (\ref{vtrunc}),
it is often also useful to choose a (for example smoothing) kernel ${\bf C}$
and use as initial guess
\be
\phi^{(0)} = {\bf C} \phi_{\rm emp}
,
\label{smoothed-empirical-initialization}
\ee
or a 
properly normalized version thereof.
Alternatively, 
one may also let the (smoothing) operator ${\bf C}$
directly act on $P_{\rm emp}$
and use a corresponding $\phi$ as initial guess, 
\be
\phi^{(0)} = (\phi)^{(-1)}{\bf C} P_{\rm emp})
,
\label{smoothed-empirical-initialization2}
\ee
assuming an inverse $(\phi)^{(-1)}$ 
of the mapping $P(\phi)$ exists.

We will now discuss the cases $\phi=L$ and $\phi=P$ in some more detail.

\subsubsection{Kernels for $L$}

For $E_L$ we have the truncated equation
\begin{equation}
L 
= {{\bf C}} N.
\label{Lkernel}
\end{equation}
Normalizing  the exponential of the solution gives
\begin{equation}
L(x,y) = \sum_i^n {{\bf C}} (x,y;x_i,y_i) 
- \ln \int \! dy^\prime \,  e^{\sum_i^n {{\bf C}} (x,y^\prime ;x_i,y_i)} , 
\label{L-kernel}
\end{equation}
or 
\be
L = {\bf C} N -\ln {\bf I}_X {\bf e}^{{\bf C}N}
\label{L-kernel2}
.
\ee
Notice that normalizing $L$ 
according to Eq.\ (\ref{L-kernel})
after each iteration
the truncated equation (\ref{Lkernel}) 
is equivalent to a one--step iteration with uniform $P^{(0)}$ = $e^{L^{(0)}}$
according to
\begin{equation}
L^1 = {{\bf C}} N + {\bf C}{\bf P}^{(0)} \Lambda_X,
\end{equation}
where only $({\bf I}-{\bf C}{\bf K})L$ is missing from the 
nontruncated equation (\ref{nontrunc}),
because the additional $y$--independent term ${\bf C}{\bf P}^{(0)} \Lambda_X$
becomes inessential if $L$ is normalized afterwards.

Lets us consider as example the choice 
${{\bf C}}$ = $-{\bf H}^{-1}(\phi^{(0)})$
for uniform initial $L^{(0)}=c$
corresponding to a normalized $P$
and ${{\bf K}}L^{(0)}$ = $0$
(e.g., a differential operator).
Uniform $L^{(0)}$ means uniform $P^{(0)}=1/v_y$,
assuming that $v_y = \int \!dy$ exists
and, according to Eq.\ (\ref{lambda}),
$\Lambda_X$ = $N_X$ for ${{\bf K}}L^{(0)}$ = $0$.
Thus, the Hessian (\ref{L-Hessian}) at $L^{(0)})$
is found as
\begin{equation}
{\bf H}(L^{(0)})
= 
- \left( {\bf I} - \frac{{\bf I}_X}{v_y} \right)
{{\bf K}} \left( {\bf I} - \frac{{\bf I}_X}{v_y} \right)
- \left( {\bf I} - \frac{{\bf I}_X}{v_y} \right) 
\frac{{\bf N}_X}{v_y}
= -{{\bf C}}^{-1}
, 
\end{equation}
which can be invertible due to the presence of the second term.

Another possibility is to start with an
approximate empirical log--density, defined as
\be
L^\epsilon_{\rm emp} = \ln P^\epsilon_{\rm emp}
,
\ee
with $P^\epsilon_{\rm emp}$ given in  Eq.\ (\ref{Pepsilon}).
Analogously to Eq.\ (\ref{smoothed-empirical-initialization}), 
the empirical log--density may for example also be smoothed 
and correctly normalized again,
resulting in an initial guess,
\be
L^{(0)} =
{\bf C} L^\epsilon_{\rm emp}
- \ln {\bf I}_X {\bf e}^{{\bf C} L^\epsilon_{\rm emp}}
.
\label{condLemp-initialization}
\ee
Similarly, one may let a kernel ${\bf C}$, 
or its normalized version $\tilde {\bf C}$ 
defined below in Eq.\ (\ref{P-kernel}),
act on $P_{\rm emp}$ first and then take the logarithm
\be
L^{(0)} =
\ln (\tilde {\bf C} P^\epsilon_{\rm emp})
.
\label{condLemp-initialization2}
\ee
Because already $\tilde {\bf C} P_{\rm emp}$ is typically
nonzero it is most times not necessary to work here
with $P^\epsilon_{\rm emp}$.
Like in the next section
$P_{\rm emp}$ may be also be replaced by 
$\tilde P_{\rm emp}$ as defined in Eq.\ (\ref{tilde-P-emp}).

\subsubsection{Kernels for $P$}

For $E_P$ the truncated equation
\be
P = {{\bf C}}{\bf P}^{-1}N  
,
\label{pkernel}
\ee
is still nonlinear in $P$.
If we solve this equation approximately by a one--step iteration
$P^1$ = ${{\bf C}}({\bf P}^{(0)})^{-1}N$  
starting from a uniform initial $P^{(0)}$ and normalizing afterwards 
this corresponds for a single $x$--value 
to the   
classical kernel methods commonly used in density estimation.
As normalized density results
\begin{equation}
P(x,y) 
= \frac{\sum_i {{\bf C}} (x,y ;x_i,y_i)}
{\int \! dy^\prime \, \sum_i {{\bf C}} (x,y^\prime ;x_i,y_i)}
=  \sum_i \bar {{\bf C}} (x,y ;x_i,y_i),
\label{smoothedNinitialization}
\end{equation}
i.e.,
\begin{equation}
P 
= {\bf N}_{K,X}^{-1} {{\bf C}} N
= \bar {{\bf C}} N,
\label{P-kernel}
\end{equation}
with (data dependent) normalized kernel 
$\bar {{\bf C}}$ = ${\bf N}_{{C},X}^{-1} {{\bf C}}$
and ${\bf N}_{{C},X}$ the diagonal matrix with diagonal elements
${\bf I}_X {{\bf C}} N$.
Again ${{\bf C}} = {{\bf K}}^{-1}$ 
or similar invertible choices can be used
to obtain a starting guess for $P$.
The form of the Hessian (\ref{H_P}) suggests in particular to include
a mass term on the data.

It would be interesting to interpret Eq.\ (\ref{P-kernel}) 
as stationarity equation of a functional $\hat E_P$
containing the usual data term $\sum_i \ln P(x_i,y_i)$.
Therefore, to obtain the derivative ${\bf P}^{-1}N$
of this data term we multiply for existing $\bar {\bf C}^{-1}$ 
Eq.\ (\ref{P-kernel}) by 
${\bf P}^{-1} \bar {{\bf C}}^{-1}$, 
where $P\ne 0$ at data points,  
to obtain
\begin{equation} 
{\widetilde {{\bf C}}^{-1} P} 
= {\bf P}^{-1}N
,
\end{equation}
with data dependent
\begin{equation}
{\widetilde {{\bf C}}^{-1}} (x,y;x^\prime ,y^\prime)
=\frac{ {\bar {{\bf C}}}^{-1} (x,y;x^\prime ,y^\prime) }
{\sum_i {\bar {{\bf C}}} (x,y;x_i ,y_i) }.
\end{equation}
Thus, Eq.\ (\ref{P-kernel})
is the stationarity equation of the functional
\begin{equation}
\hat E_P =
-(\,N,\,\ln P\,)
+\frac{1}{2}\,
(\,P,\, {\widetilde {{\bf C}}^{-1}} \, P \,)
.
\label{kernelfunc}
\end{equation}

To study the dependence on the number $n$ of training data 
for a given ${{\bf C}}$ consider a normalized kernel with 
$\int\! dy \, {{\bf C}} (x,y;x^\prime,y^\prime ) = \lambda$,
$\forall x,x^\prime ,y^\prime$.
For such a kernel the denominator of $\bar {{\bf C}}$
is equal to $n\lambda$ so we have
\begin{equation}
\bar {{\bf C}} = \frac{{{\bf C}}}{n\lambda} 
,\quad
P= \frac{{{\bf C}} N}{n \lambda }
\end{equation}
Assuming that for large $n$ the 
empirical average $(1/n)\sum_i {{\bf C}}(x,y;x_i ,y_i)$
in the denominator of $\widetilde {{\bf C}}^{-1}$
becomes $n$ independent,
e.g., converging to the true average 
$n \!\int \!\!dx^\prime dy^\prime \,  p(x^\prime,y^\prime) 
{{\bf C}}(x,y;x^\prime ,y^\prime)$,
the regularizing term in functional (\ref{kernelfunc})
becomes proportional to $n$
\be
{\widetilde {{\bf C}}^{-1}} \propto n \lambda^2
,
\ee
According to Eq.\ (\ref{n-asym})
this would allow to relate a
saddle point approximation to a large $n$--limit.

Again, a similar possibility
is to start with the
empirical density
$\tilde P_{\rm emp}$ defined in  Eq.\ (\ref{tilde-P-emp}).
Analogously to Eq.\ (\ref{smoothed-empirical-initialization}), 
the empirical density can for example also be smoothed 
and correctly normalized again, so that
\be
P^{(0)} 
=  \tilde {\bf C} \tilde P_{\rm emp}
.
\label{condPinitialization}
\ee
with $\tilde {\bf C}$
defined in Eq.\ (\ref{P-kernel}).

\begin{figure}[ht]
\begin{center}
\epsfig{file=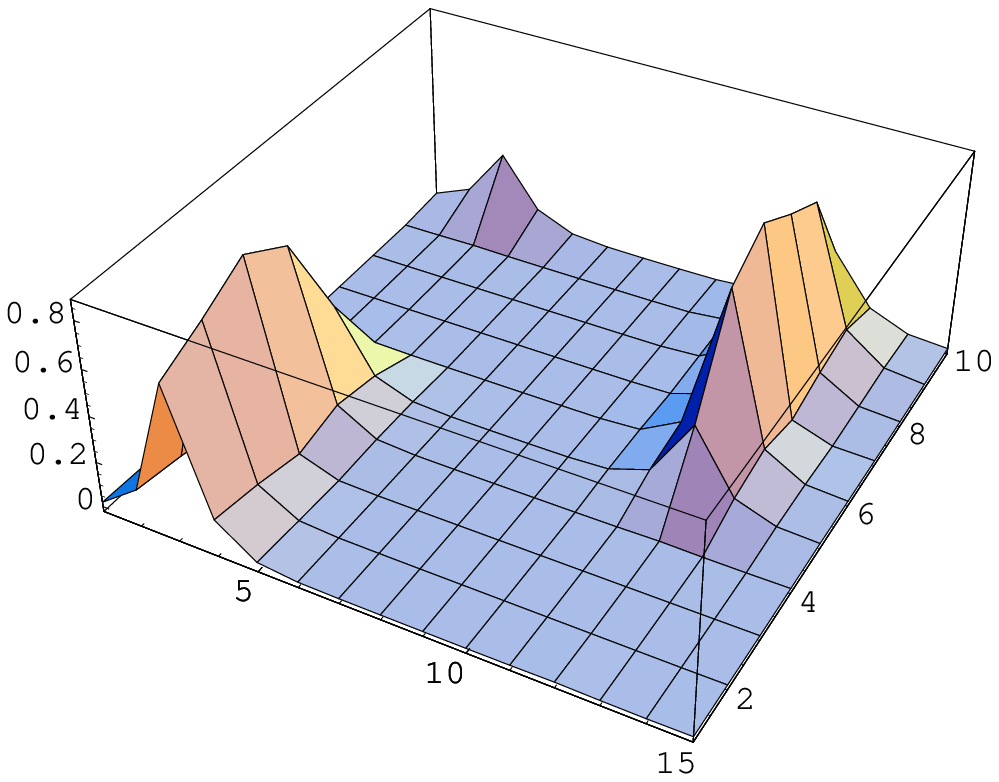, width= 65mm}    
\epsfig{file=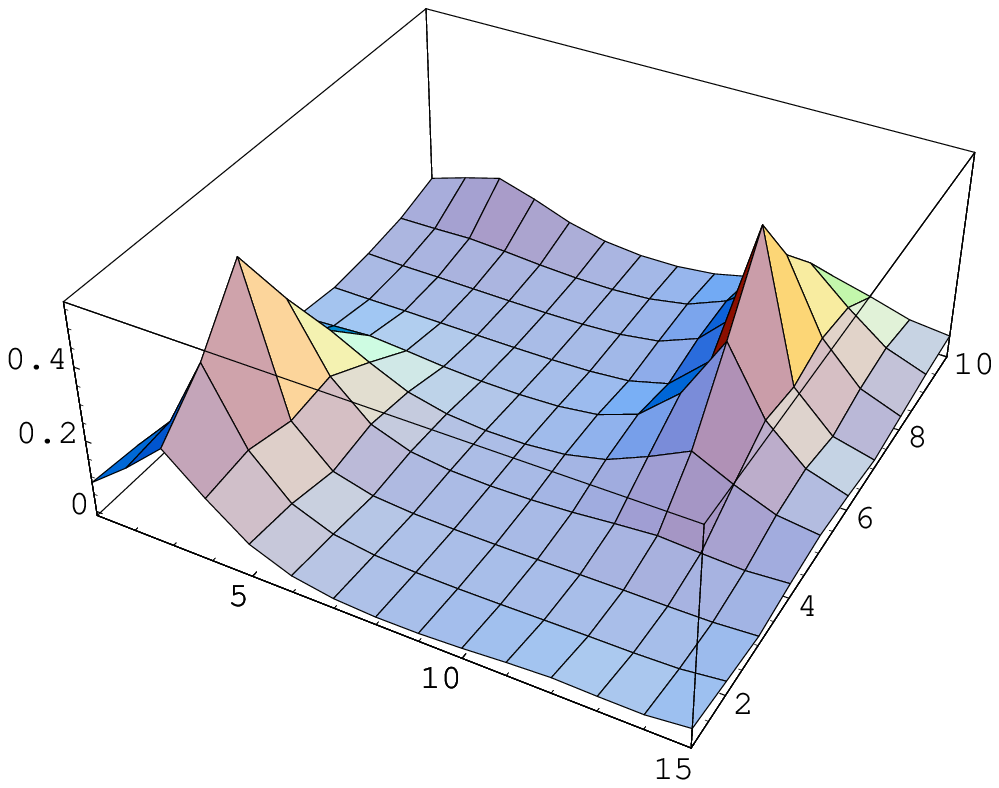, width= 65mm}\\    
\epsfig{file=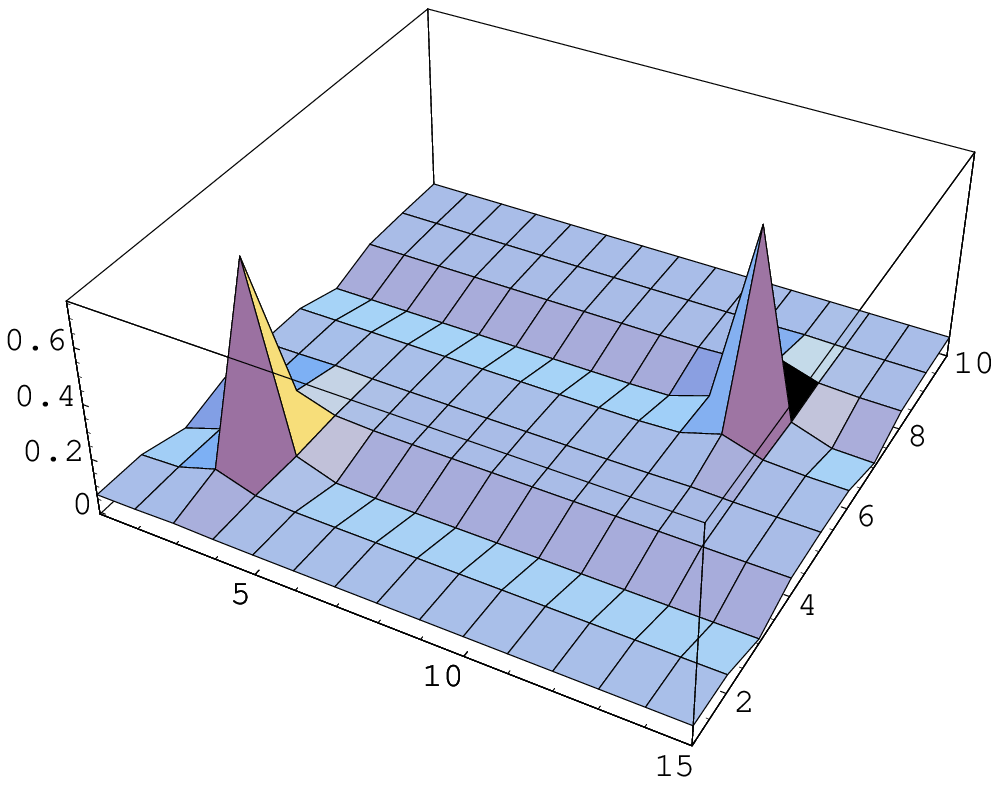, width= 65mm}  
\epsfig{file=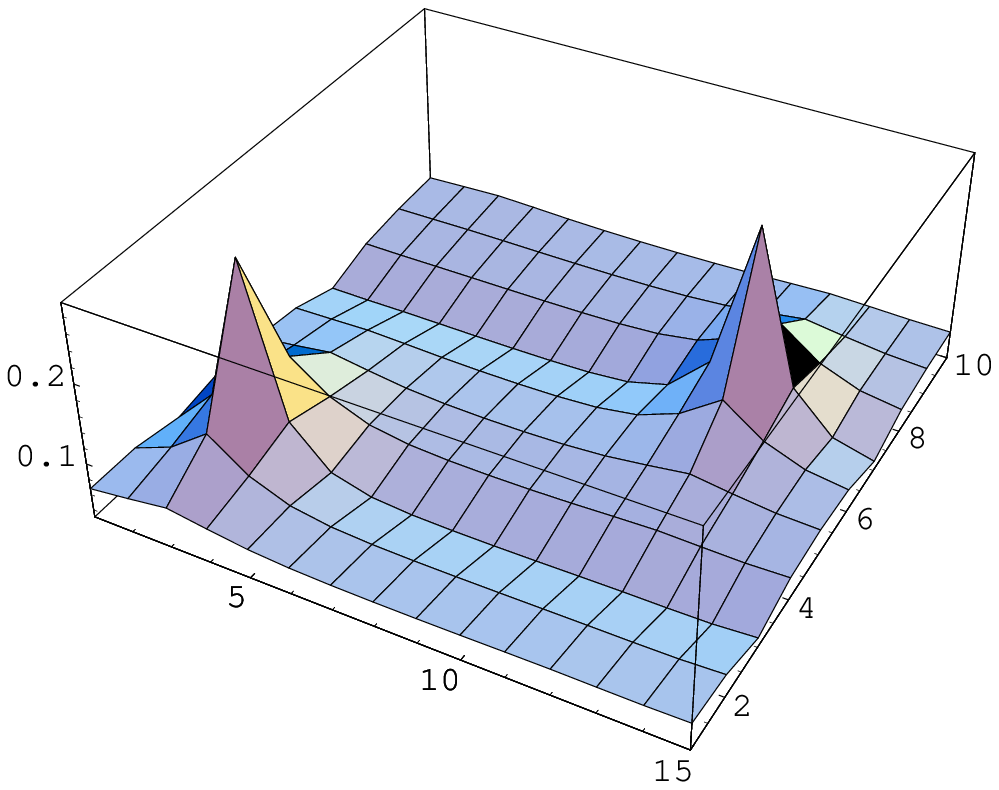, width= 65mm}    
\end{center}
\caption{Comparison of initial guesses $P^{(0)}(x,y)$
for a case with two data points located
at $(3,3)$ and $(7,12)$
within the intervals $y\in[1,15]$
and $x\in[1,10]$ 
with periodic boundary conditions.
First row: $P^{(0)}$ = $\tilde {\bf C} N$.
(The smoothing operator
acts on the unnormalized $N$.
The following conditional normalization changes the
shape more drastically than in the example shown in the second row.)
Second row: $P^{(0)}$ = $\tilde {\bf C}\tilde P_{\rm emp}$.
(The smoothing operator
acts on the already conditionally normalized $\tilde P_{\rm emp}$.)
The kernel $\tilde {\bf C}$ is given by Eq.\ (\ref{P-kernel})
with  ${\bf C}$ = $({\bf K}+m_C^2{\bf I})$,
$m_C^2$ = $1.0$,
and a ${\bf K}$ of the form of Eq.\ (\ref{smoothness-prior-for-examples})
with
$\lambda_0$ = $\lambda_4$ = $\lambda_6$ = 0,
and
$\lambda_2$ = $0.1$ (figures on the l.h.s.)
or
$\lambda_2$ = $1.0$ (figures on the r.h.s.),
respectively.
}
\label{densI}
\end{figure}

Fig.\ \ref{densI}
compares the initialization according to 
Eq. (\ref{smoothedNinitialization}),
where the smoothing operator $\tilde C$ acts on $N$,
with an initialization according to
Eq. (\ref{condPinitialization}),
where the smoothing operator $\tilde C$
acts on the correctly normalized $\tilde P_{\rm emp}$.

\subsection{Numerical examples}

\subsubsection{Density estimation with Gaussian specific prior}

In this section we look at some numerical examples
and discuss implementations of the nonparametric learning algorithms 
for density estimation
we have discussed in this paper.

As example,
consider a problem 
with a one--dimensional $X$--space
and  a one--dimensional $Y$--space,
and a smoothness prior 
with inverse covariance
\begin{equation}
{{\bf K}} = \lambda_x \left( {{\bf K}}_X\otimes 1_Y \right)
         + \lambda_y \left( 1_X\otimes {{\bf K}}_Y \right) , 
\label{smoothness-prior-for-examples}
\end{equation}
where
\begin{eqnarray}
{{\bf K}}_X &=& 
\lambda_0 {\bf I}_X- \lambda_2 {\Delta}_x 
+ \lambda_4 {\Delta}_x^2 - \lambda_6 {\Delta}_x^3\\ 
{{\bf K}}_Y &=& 
\lambda_0 {\bf I}_Y- \lambda_2 {\Delta}_y 
+ \lambda_4 {\Delta}_y^2 - \lambda_6 {\Delta}_y^3,
\end{eqnarray}
and Laplacian
\begin{equation}
{\Delta}_x (x,x^\prime ) 
= \delta^{\prime\prime} (x-x^\prime)
= \delta (x-x^\prime)\frac{d^2}{dx^2},
\end{equation}
and analogously for ${\Delta}_y$.
For
$\lambda_2\ne0$ = $\lambda_0$ = $\lambda_4$ = $\lambda_6$
this corresponds to the two Laplacian prior factors
$\Delta_x$ for $x$ and $\Delta_y$ for $y$.
(Notice that also for $\lambda_x$ = $\lambda_y$
the  $\lambda_4$-- and $\lambda_6$--terms
do not include all terms of 
an iterated 2--dimensional Laplacian, 
like $\Delta^2$ = $(\Delta_x+\Delta_y)^2$ or $\Delta^4$, 
as the mixed derivatives $\Delta_x\Delta_y$ are missing.)

We will now study nonparametric density estimation
with prior factors being Gaussian with respect to $L$ 
as well as being Gaussian with respect to $P$.

The error or energy functional for a Gaussian prior factor in $L$
is given by Eq.\ (\ref{L-functional}).
The corresponding iteration procedure is 
\begin{equation}
L^{(i+1)} =
L^{(i)} + \eta {\bf A}^{-1} 
\left(N - {{\bf K}} L^{(i)} 
-{\bf e}^{{\bf L}^{(i)}} \left[ 
N_X-{\bf I}_X {{\bf K}} L^{(i)} 
\right]\right) 
.
\label{L-iter-eq}
\end{equation}
Written explicitly
for $\lambda_2=1$, $\lambda_0$ = $\lambda_4$ = $\lambda_6$ = $0$
Eq.\ (\ref{L-iter-eq}) reads,
\begin{equation}
L^{(i+1)}(x,y) = L^{(i)}(x,y) + 
\eta \sum_j {\bf A}^{-1} (x,y;x_j ,y_j )
\end{equation}
\[
+\eta \int \! dx^\prime dy^\prime {\bf A}^{-1} (x,y;x^\prime ,y^\prime )
\left[
 \frac{d^2}{d(x^\prime )^2} L^{(i)} (x^\prime ,y^\prime )
+\frac{d^2}{d(y^\prime )^2} L^{(i)} (x^\prime ,y^\prime )
\right.
\]
\[
\left.
\!-\!
\left(
\sum_j \! \delta(x^\prime\! -\!x_j) 
\!+\!\int \!dy^{\prime\prime}\! 
\frac{d^2}{d(x^\prime )^2} L^{(i)} (x^\prime ,y^{\prime\prime} )
\!+\!\int \!\!dy^{\prime\prime}
\frac{d^2}{d(y^{\prime\prime} )^2} L^{(i)} (x^\prime ,y^{\prime\prime} )
\right)
\!e^{L^{(i)}(x^\prime ,y^\prime )}
\right].
\]
Here 
\[
\int_{y_A}^{y_B} \!dy^{\prime\prime}\,
\frac{d^2}{d(y^{\prime\prime} )^2} L^{(i)} (x^\prime ,y^{\prime\prime} )
=
\frac{d}{d(y^{\prime\prime} )} 
L^{(i)} (x^\prime ,y^{\prime\prime} )\Bigg|_{y_A}^{y_B}
\]
vanishes if the first derivative $\frac{d}{dy} L^{(i)} (x,y)$ 
vanishes at the boundary or if periodic.

Analogously, 
for error functional $E_P$ (\ref{P-functional}) the iteration procedure
\begin{equation}
P^{(i+1)} =
P^{(i)} + \eta {\bf A}^{-1} 
\left[
({\bf P}^{(i)})^{-1} N -  
N_X - {\bf I}_X {\bf P}^{(i)} {{\bf K}} P^{(i)}
- {{\bf K}}  P^{(i)}   
\right].
\end{equation}
becomes
for 
$\lambda_2=1$, $\lambda_0$ = $\lambda_4$ = $\lambda_6$ = $0$
\begin{equation}
P^{(i+1)}(x,y) = P^{(i)}(x,y) + 
\eta \sum_j \frac{{\bf A}^{-1} (x,y;x_j ,y_j )}
{P^{(i)} (x_j,y_j)}
\end{equation}
\[
+\eta \int \! dx^\prime dy^\prime {\bf A}^{-1} (x,y;x^\prime ,y^\prime )
\left[
 \frac{d^2}{d(x^\prime )^2} P^{(i)} (x^\prime ,y^\prime )
+\frac{d^2}{d(y^\prime )^2} P^{(i)} (x^\prime ,y^\prime )
\right.
\]
\[
\left.
-\left(
\sum_j \delta(x^\prime\!\! -\!x_j) 
+\!\!\int \!\!dy^{\prime\prime} P^{(i)} (x^\prime ,y^{\prime\prime} )
\frac{d^2 P^{(i)} (x^\prime ,y^{\prime\prime} )}{d(x^\prime )^2} 
\right.\right.
\]\[
\left.\left.
+\int \!dy^{\prime\prime} P^{(i)} (x^\prime ,y^{\prime\prime} )
\frac{d^2 P^{(i)} (x^\prime ,y^{\prime\prime} )}{d(y^{\prime\prime} )^2} 
\right)
\right].
\]
Here
\[
\int_{y_A}^{y_B} \!\!\!dy^{\prime\prime} P^{(i)} (x^\prime ,y^{\prime\prime} )
\frac{d^2 P^{(i)} (x^\prime ,y^{\prime\prime} )}{d(y^{\prime\prime} )^2} 
=
\]
\be
P^{(i)} (x^\prime ,y^{\prime\prime} )
\frac{d P^{(i)} (x^\prime ,y^{\prime\prime} )}{d(y^{\prime\prime} )} 
\Bigg|_{y_A}^{y_B}
-
\int_{y_A}^{y_B} \!\!\! dy^{\prime\prime} 
\left(\!
\frac{d P^{(i)} (x^\prime ,y^{\prime\prime} )}{dy^{\prime\prime}}
\! \right)^2\!, 
\ee
where the first term vanishes for $P^{(i)}$
periodic or vanishing at the boundaries.
(This has to be the case for $i\, d/dy$ to be hermitian.)

We now study density estimation problems
numerically.
In particular, we want 
to check the influence of the nonlinear normalization constraint.
Furthermore, we want to compare models
with Gaussian prior factors for $L$ 
with models with Gaussian prior factors for $P$.

The following numerical calculations
have been performed on a mesh
of dimension 10$\times$ 15, i.e., $x\in [1,10]$ and  $y\in [1,15]$,
with periodic boundary conditions on $y$
and sometimes also in $x$.
A variety of different iteration and initialization methods 
have been used.

\begin{figure}[ht]
\vspace{-4cm}
\begin{center}
\epsfig{file=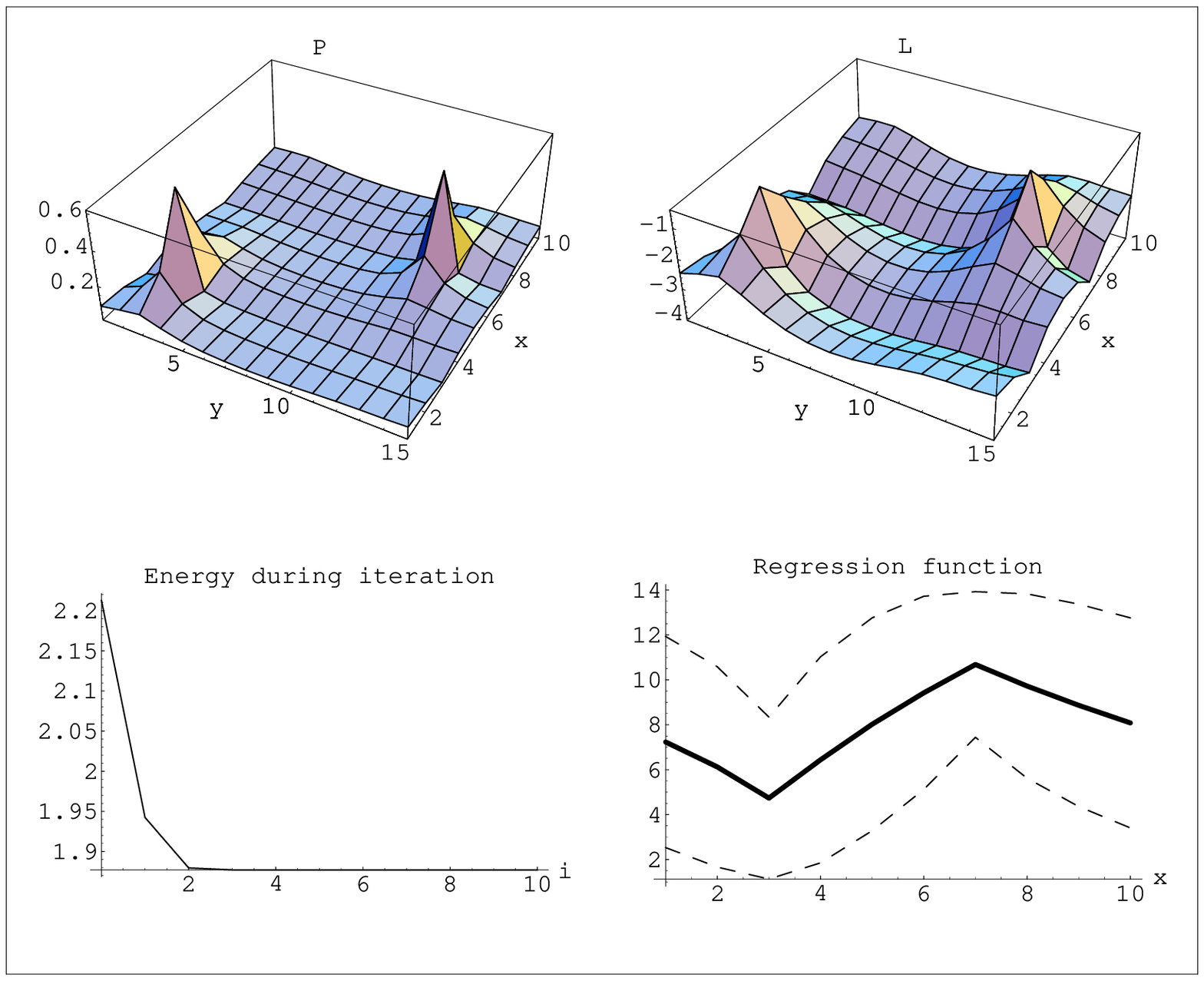, width=132mm}
\end{center}
\vspace{-4.1cm}
\caption{
Density estimation with 2 data points
and a Gaussian prior factor
for the log--probability $L$.
First row: Final $P$ and $L$.
Second row: The l.h.s.\ shows the energy $E_L$ (\ref{L-functional})
during iteration,
the r.h.s.\ the 
regression function 
$h(x)$ 
= $\int \!dy \, y p(y|x,h_{\rm true})$
= $\int \!dy \, y P_{\rm true}(x,y)$.
The dotted lines indicate the range of 
one standard deviation above and below the regression function
(ignoring periodicity in $x$).
The fast convergence shows that the problem is nearly linear.
The asymmetry of the solution
between the $x$-- and $y$--direction 
is due to the normalization constraint, only required for $y$.
(Laplacian smoothness prior ${\bf K}$ 
as given in Eq.\ (\ref{smoothness-prior-for-examples})
with
$\lambda_x$ = 
$\lambda_y$ = 1,
$\lambda_0$ = 0,
$\lambda_2$ = 0.025,
$\lambda_4$ = 
$\lambda_6$ = 0.
Iteration with
negative Hessian ${\bf A}$ = $-{\bf H}$ 
if positive definite, 
otherwise with the gradient algorithm, i.e., ${\bf A}$ = ${\bf I}$.
Initialization with $L^{(0)}$ = $\ln (\tilde {\bf C} \tilde P_{\rm emp})$,
i.e., $L^{(0)}$ normalized to $\int \!dy\,e^L$ = $1$,
with $\tilde {\bf C}$ of Eq.\ (\ref{P-kernel}) and
${\bf C}$ = $({\bf K}+m_C^2{\bf I})$, $m_C^2$ = $0.1$.
Within each iteration step the optimal step width $\eta$ has been found
by a line search.
Mesh with $10$ points in $x$-direction and 
$15$ points in $y$--direction,
periodic boundary conditions in $x$ and $y$. 
The $2$ data points are $(3,3)$ and $(7,12)$.)
}
\label{dens2Laa}
\end{figure}

\begin{figure}[ht]
\vspace{-2cm}
\begin{center}
\epsfig{file=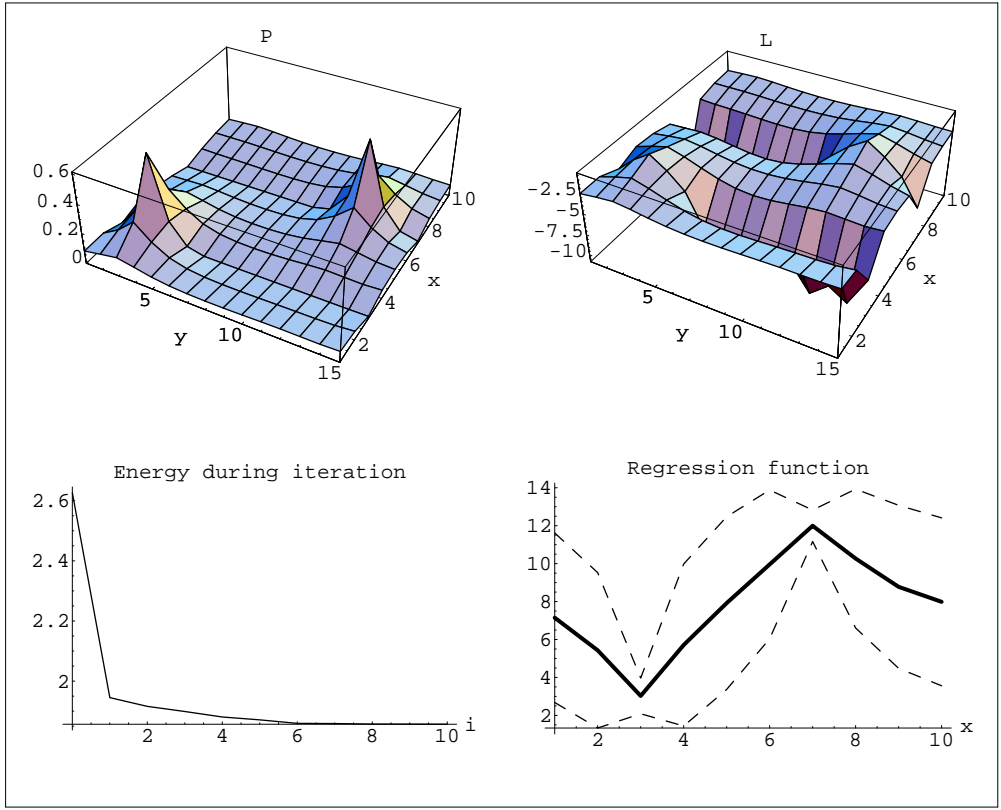, width=132mm}
\end{center}
\vspace{-4.1cm}
\caption{
Density estimation with 2 data points, 
this time with a Gaussian prior factor
for the probability $P$,
minimizing the energy functional $E_P$ (\ref{P-functional}).
To make the figure comparable with Fig.\ \ref{dens2Laa}
the parameters have been chosen so that 
the maximum of the solution $P$ 
is the same in both figures ($\max P$ = 0.6).
Notice, that compared to Fig.\ \ref{dens2Laa}
the smoothness prior is less effective
for small probabilities.
(Same data, mesh
and periodic boundary conditions 
as for Fig.\ \ref{dens2Laa}.
Laplacian smoothness prior ${\bf K}$ 
as in Eq.\ (\ref{smoothness-prior-for-examples})
with
$\lambda_x$ = 
$\lambda_y$ = 1,
$\lambda_0$ = 0,
$\lambda_2$ = 1,
$\lambda_4$ = 
$\lambda_6$ = 0.
Iterated using massive prior relaxation, i.e.,
${\bf A}$ = ${\bf K}+m^2{\bf I}$ with  $m^2$ = 1.0.
Initialization with  $P^{(0)}$ = $\tilde {\bf C} \tilde P_{\rm emp}$,
with 
$\tilde {\bf C}$ of Eq.\ (\ref{P-kernel})
so $P^{(0)}$ is correctly normalized,
and ${\bf C}$ = $({\bf K}+m_C^2{\bf I})$,
$m_C^2$ = $1.0$.
Within each iteration step the optimal factor $\eta$ has been found
by a line search algorithm.)
}
\label{dens2Paa}
\end{figure}

\begin{figure}[ht]
\vspace{-2cm}
\begin{center}
\epsfig{file=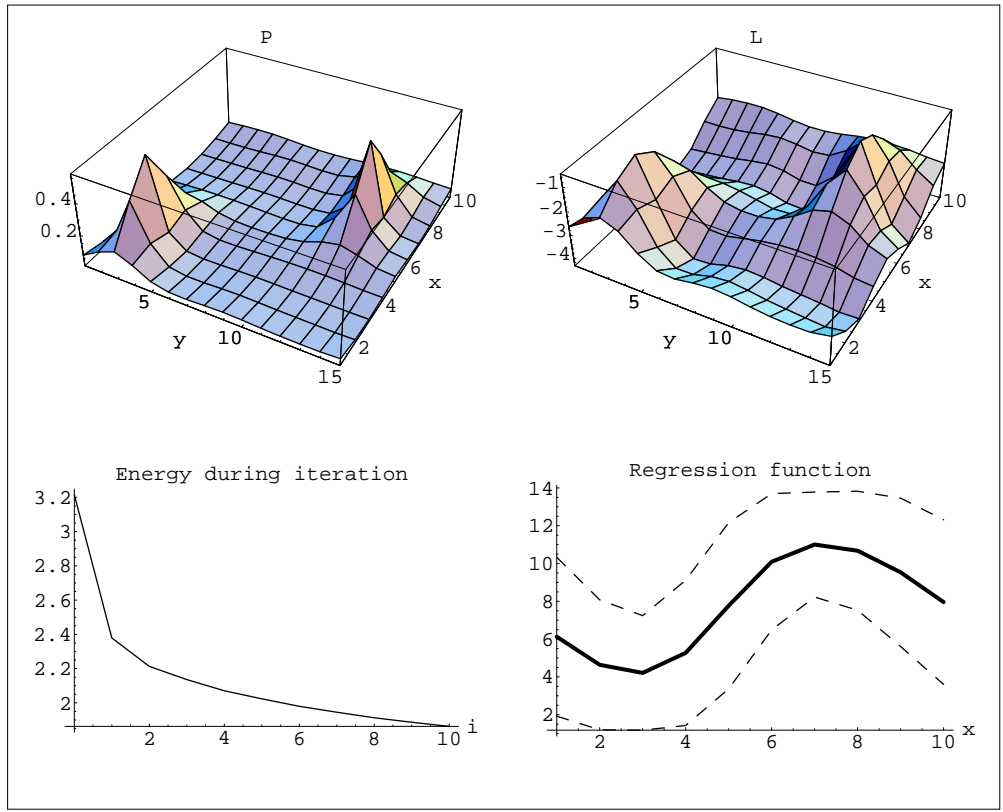, width=132mm}
\end{center}
\vspace{-4cm}
\caption{
Density estimation with a $(-\Delta^3)$ Gaussian prior factor
for the log--probability $L$. 
Such a prior favors probabilities of Gaussian shape.
(Smoothness prior ${\bf K}$ 
of the form of Eq.\ (\ref{smoothness-prior-for-examples})
with
$\lambda_x$ = 
$\lambda_y$ = 1,
$\lambda_0$ = 0,
$\lambda_2$ = 0,
$\lambda_4$ = 0,
$\lambda_6$ = 0.01.
Same iteration procedure, initialization, data, mesh
and periodic boundary conditions 
as for Fig.\ \ref{dens2Laa}.)
}
\label{dens2Lgauss}
\end{figure}

\begin{figure}[ht]
\vspace{-2cm}
\begin{center}
\epsfig{file=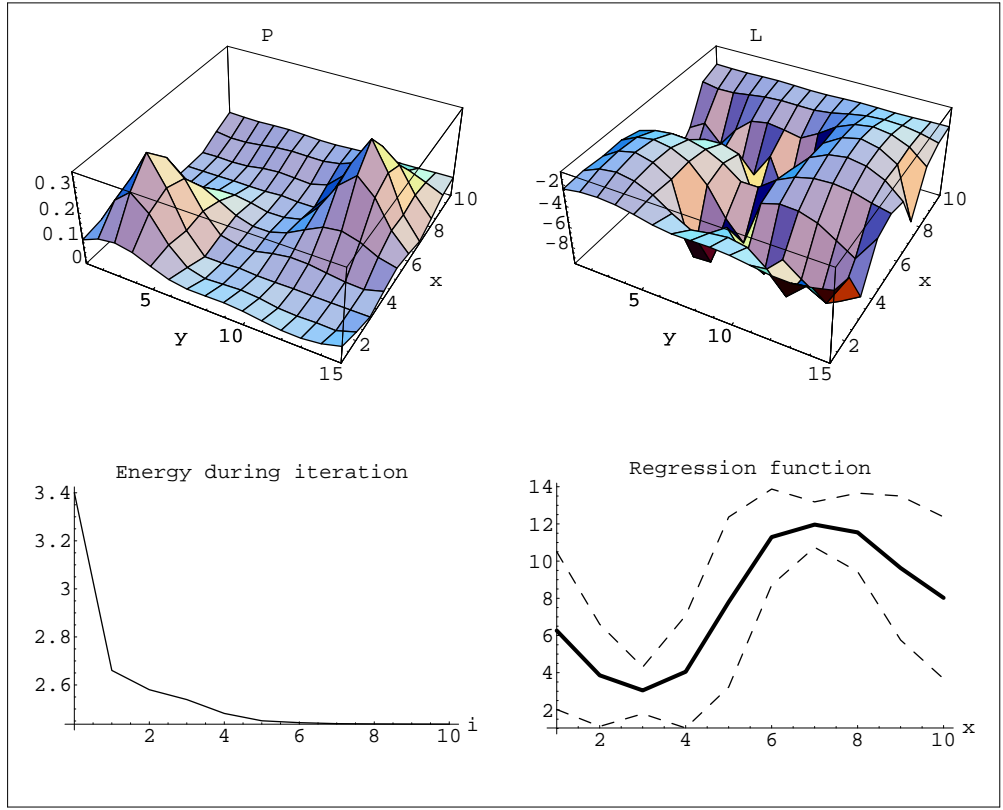, width=132mm}
\end{center}
\vspace{-4cm}
\caption{
Density estimation with a $(-\Delta^3)$ Gaussian prior factor
for the probability $P$. 
As the variation of $P$ is smaller than that of $L$,
a smaller $\lambda_6$ has been chosen than in  Fig.\ \ref{dens2Pgauss}.
The Gaussian prior in $P$ is also relatively less effective 
for small probabilities than a comparable Gaussian prior in $L$. 
(Smoothness prior ${\bf K}$ 
of the form of Eq.\ (\ref{smoothness-prior-for-examples})
with
$\lambda_x$ = 
$\lambda_y$ = 1,
$\lambda_0$ = 0,
$\lambda_2$ = 0,
$\lambda_4$ = 0,
$\lambda_6$ = 0.1.
Same iteration procedure, initialization, data, mesh
and periodic boundary conditions 
as for Fig.\ \ref{dens2Paa}.)
}
\label{dens2Pgauss}
\end{figure}

Figs.\
\ref{dens2Laa} --
\ref{dens2Pgauss}
summarize results for density estimation problems
with only two data points,
where differences in the effects of varying smoothness priors 
are particularly easy to see.
A density estimation with
more data points can be found in
Fig.\ \ref{mix50g}.

For Fig.\ \ref{dens2Laa}
a  Laplacian smoothness prior on $L$ has been implemented.
The solution has been obtained
by iterating with the negative Hessian,
as long as positive definite.
Otherwise the gradient algorithm has been used.
One iteration step means one iteration according to Eq.\ (\ref{iter-3})
with the optimal $\eta$. 
Thus, each iteration step includes the optimization of $\eta$
by a line search algorithm. 
(For the figures the Mathematica function FindMinimum
has been used to optimize $\eta$.)

As initial guess in Fig.\ \ref{dens2Laa}
the kernel estimate
$L^{(0)}$ = 
$\ln (\tilde {\bf C} \tilde P_{\rm emp})$
has been employed, 
with $\tilde {\bf C}$ defined in Eq.\ (\ref{P-kernel}) and
${\bf C}$ = $({\bf K}+m_C^2{\bf I})$
with squared mass $m_C^2$ = $0.1$.
The fast drop--off
of the energy $E_L$ within the first two iterations
shows the quality of this initial guess.
Indeed, this fast convergence seems to indicate that the problem
is nearly linear,
meaning that the influence of the 
only nonlinear term in the stationarity equation,
the normalization constraint,
is not too strong.
Notice also,
that the reconstructed regression shows the
typical piecewise linear approximations well known
from  one--dimensional (normalization constraint free) regression problems
with Laplacian prior.

Fig.\ \ref{dens2Paa} shows a density estimation
similar to Fig.\ \ref{dens2Laa},
but for a Gaussian prior factor in $P$
and thus also with  different $\lambda_2$,
different initialization,
and slightly different iteration procedure.
For Fig.\ \ref{dens2Paa}
also a  kernel estimate
$P^{(0)}$ = 
$(\tilde {\bf C} \tilde P_{\rm emp})$
has been used as initial guess, again  
with $\tilde {\bf C}$ as defined in Eq.\ (\ref{P-kernel}) and
${\bf C}$ = $({\bf K}+m_C^2{\bf I})$
but with squared mass $m_C^2$ = $1.0$.
The solution has been obtained by
prior relaxation 
${\bf A}$ = ${\bf K}+m^2{\bf I}$ 
including a mass term with $m^2$ = 1.0
to get for a Laplacian ${\bf K}$ = $-\Delta$
and periodic boundary conditions
an invertible  ${\bf A}$.
This iteration scheme
does not require to calculate the Hessian ${\bf H}_P$
at each iteration step.
Again the quality of the initial guess (and the iteration scheme) 
is indicated by the fast drop--off
of the energy $E_P$ during the first iteration.

Because the range of $P$--values, being between zero and one,
is smaller than that of $L$--values, being between minus infinity and zero,
a larger Laplacian smoothness factor $\lambda_2$
is needed for Fig.\ \ref{dens2Paa}
to get similar results than for Fig.\ \ref{dens2Laa}.
In particular, such $\lambda_2$ values have been chosen 
for the two figures
that the maximal values
of the the two reconstructed probability densities $P$
turns out to be nearly equal.

Because the logarithm particularly expands the distances
between small probabilities
one would expect
a Gaussian prior for $L$ to be 
especially effective for small probabilities.
Comparing Fig.\ \ref{dens2Laa} and Fig.\ \ref{dens2Paa}
this effect can indeed be seen.
The deep valleys appearing in the $L$--landscape 
of Fig.\ \ref{dens2Paa}
show that small values of $L$ are not smoothed out
as effectively as in Fig.\ \ref{dens2Laa}.
Notice, that therefore
also the variance of the solution $p(y|x,h)$ 
is much smaller for a Gaussian prior in $P$
at those $x$ which are in the training set.

Fig.\ \ref{dens2Lgauss} resumes results 
for a model similar to that presented in 
Fig.\ \ref{dens2Laa}, but with a
$(-\Delta^3)$--prior replacing the 
Laplacian $(-\Delta)$--prior.
As all quadratic functions have zero third derivative
such a prior favors, applied to $L$, 
quadratic log--likelihoods,
corresponding to Gaussian probabilities $P$.
Indeed,
this is indicated by the striking difference 
between the regression functions
in Fig.\ \ref{dens2Lgauss} and
in Fig.\ \ref{dens2Laa}:
The $(-\Delta^3)$--prior produces a much rounder
regression function,
especially at the $x$ values which appear in the data.
Note however,
that in contrast to a pure Gaussian regression problem,
in density estimation an additional
non--quadratic normalization constraint is present.

In Fig.\ \ref{dens2Pgauss}
a similar prior has been applied, 
but this time being Gaussian in $P$
instead of $L$.
In contrast to a $(-\Delta^3)$--prior for $L$,
a $(-\Delta^3)$--prior for $P$ implements a tendency to quadratic $P$.
Similarly to the difference between
Fig.\ \ref{dens2Laa} and Fig.\ \ref{dens2Lgauss},  
the regression function in Fig.\ \ref{dens2Pgauss}
is also rounder than that in Fig.\ \ref{dens2Paa}.
Furthermore,
smoothing in Fig.\ \ref{dens2Pgauss} 
is also less effective for smaller probabilities
than it is in Fig.\ \ref{dens2Lgauss}. 
That is the same result we have found
comparing the two priors for $L$
shown in Fig.\ \ref{dens2Paa} and Fig.\ \ref{dens2Laa}. 
This leads to deeper valleys in the $L$--landscape
and to a smaller variance 
especially at $x$ which appear in the training data.

Fig.\ \ref{mix50g} depicts the results of a density estimation 
based on more than two data points.
In particular, fifty training data have been obtained
by sampling with uniform $p(x)$
from the ``true'' density 
\be
P_{\rm true}(x,y) 
= p(y|x,h_{\rm true}) 
= \frac{1}{2\sqrt{2\pi}\sigma_0}
 \left(
                 e^{-\frac{(y-h_a(x))^2}{2\sigma_0^2}}
 +
                 e^{-\frac{(y-h_b(x))^2}{2\sigma_0^2}}
 \right)
,
\label{trueP}
\ee
with $\sigma_0$ = $1.5$,
$h_a(x)$ = $125/18+(5/9)x$,
$h_b(x)$ = $145/18-(5/9)x$,
shown in the top row of Fig.\ \ref{mixTrue}.
The sampling process has been implemented 
using the transformation method
(see for example \cite{Press-Teukolsky-Vetterling-Flannery-1992}).
The corresponding empirical density $N/n$
(\ref{jointPemp})
and conditional empirical density
$P_{\rm emp}$ of Eq.\ (\ref{cond-P-emp1}),
in this case equal to the extended $\tilde P_{\rm emp}$
defined in Eq.\ (\ref{tilde-P-emp}),
can be found in Fig.\ \ref{mixData}.

Fig.\ \ref{mix50g} shows
the maximum posterior solution $p(y|x,h^*)$
and its logarithm, 
the energy $E_L$ during iteration,
the {\it regression function}
\be
h(x) 
= \int \!dy \, y \, p(y|x,h_{\rm true})
= \int \!dy \, y \, P_{\rm true}(x,y)
,
\ee
(as reference, the regression function for the 
true likelihood $p(y|x,h_{\rm true})$ 
is given in Fig.\ \ref{mixTrueR}),
the {\it average training error} (or {\it empirical (conditional) log--loss})
\be
<-\ln p(y|x,h) >_D  
= -\frac{1}{n}\sum_{i=1}^n\log p(y_i|x_i,h)
,
\ee
and 
the {\it average test error} 
(or {\it true expectation of (conditional) log--loss})
for uniform $p(x)$
\be
<-\ln p(y|x,h) >_{P_{\rm true}}
=
-\int \!dy\,dx\,  p(x)p(y|x,h_{\rm true}) \ln p(y|x,h)
,
\ee
which is, up to a constant, equal to the
expected Kullback--Leibler distance
between the actual solution and the true likelihood,
\be
{\rm KL}\Big(p(x,y|h_{\rm true}),p(y|x,h)\Big)
=
-\int \!dy\,dx\, p(x,y|h_{\rm true}) 
\ln \frac{p(y|x,h)}{p(y|x,h_{\rm true}) }
.
\ee
The test error
measures the quality of the achieved solution.
It has, in contrast to the energy and training error, 
of course not been available to the learning algorithm.

The maximum posterior solution 
of Fig.\ \ref{mix50g} 
has been calculated
by minimizing $E_L$ using massive prior iteration with
${\bf A}$ = ${\bf K}+m^2 {\bf I}$, a squared mass $m^2$ = $0.01$,
and a (conditionally) normalized, constant $L^{(0)}$ as initial guess.
Convergence has been fast,
the regression function is similar to the true one
(see Fig.\ \ref{mixTrueR}).

Fig.\ \ref{mix50comp}
compares some iteration procedures and initialization methods
Clearly, all methods do what they should do,
they decrease the energy functional.
Iterating with the negative Hessian 
yields the fastest convergence.
Massive prior iteration
is nearly as fast, even for uniform initialization,
and does not require calculation of the Hessian
at each iteration. 
Finally, the slowest iteration method, but the easiest to implement,
is the gradient algorithm.

Looking at Fig.\ \ref{mix50comp}
one can distinguish data--oriented from prior--oriented initializations.
We understand 
data--oriented initial guesses to be those for which the
training error is smaller at the beginning of the iteration
than for the final solution and 
prior--oriented initial guesses to be those for which the opposite is true.
For good initial guesses the difference is small.
Clearly, the uniform initializations is prior--oriented,
while an empirical log--density $\ln(N/n+\epsilon)$
and the shown kernel initializations are data--oriented.

The case where the test error grows
while the energy is decreasing
indicates a misspecified prior
and is typical for overfitting.
For example, in the fifth row of Fig.\ \ref{mix50comp}
the test error (and in this case also the average training error)
grows again after having reached a minimum
while the energy is steadily decreasing.

\subsubsection{Density estimation with Gaussian mixture prior}

Having seen Bayesian field theoretical models working
for Gaussian prior factors
we will study in this section the 
slightly more complex prior mixture models.
Prior mixture models
are an especially useful tool
for implementing complex and
unsharp prior knowledge.
They may be used, for example, 
to translate verbal statements of experts
into quantitative prior densities
\cite{Lemm-1996,Lemm-1998,Lemm-1998a,Lemm-1998b,Lemm-1998c},
similar to the
quantification of ``linguistic variables'' by  fuzzy methods
\cite{Klir-Yuan-1995,Klir-Yuan-1996}.

We will now study a prior mixture with
Gaussian prior components in $L$.
Hence, consider the following energy functional with mixture prior
\be
E_L 
= -\ln \sum_j p_j e^{-E_j}
= - (L,N) +(e^L,\Lambda_X)
-\ln \sum_j p_j e^{-\lambda E_{0,j}}
\label{L-mixture-functional1}
\ee
with mixture components 
\be
E_j = -(L,N) + \lambda E_{0,j}
+(e^L,\Lambda_X)
.
\ee
We choose Gaussian component prior factors
with equal covariances but differing means
\be
E_{0,j} = \frac{1}{2} \Big(L-t_j,\,{{\bf K}} (L-t_j)\Big)
.
\ee
Hence, the stationarity equation for Functional (\ref{L-mixture-functional1})
becomes
\be
0 = N - \lambda {\bf K} \left(L - \sum_j a_j t_j \right) 
      - {\bf e}^{\bf L} \Lambda_X
,
\ee
with Lagrange multiplier function
\be
\Lambda_X 
= N_X - \lambda{\bf I}_X{\bf K} \left(L - \sum_j a_j t_j \right) 
,
\ee
and mixture coefficients
\be
a_j = \frac{p_j e^{-\lambda E_{0,j}}}{\sum_k p_k e^{-\lambda E_{0,k}}}
.
\ee
The parameter $\lambda$ 
plays here a similar role as the inverse temperature $\beta$ 
for prior mixtures in regression
(see Sect.\ \ref{mixtures-for-regression}).
In contrast to the $\beta$--parameter in regression, however,
the ``low temperature'' solutions for
$\lambda\rightarrow\infty$
are the pure prior templates $t_j$,
and for $\lambda\rightarrow 0$
the prior factor is switched off.

\begin{figure}[ht]
\vspace{-2cm}
\begin{center}
\epsfig{file=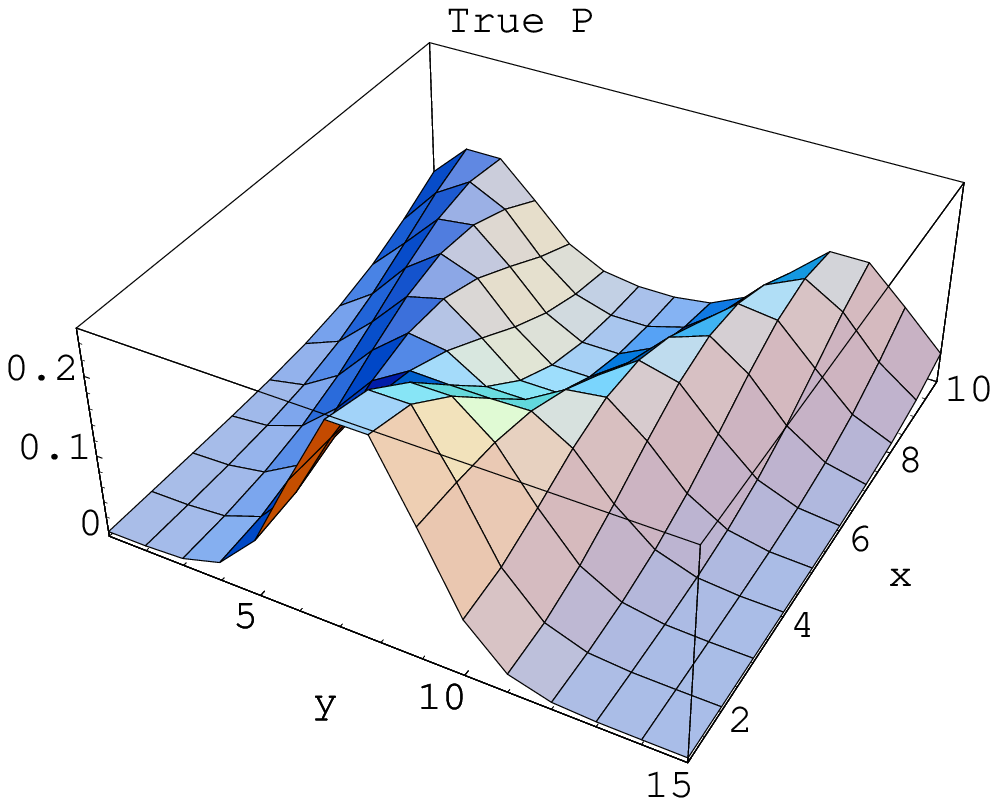, width= 65mm}
\epsfig{file=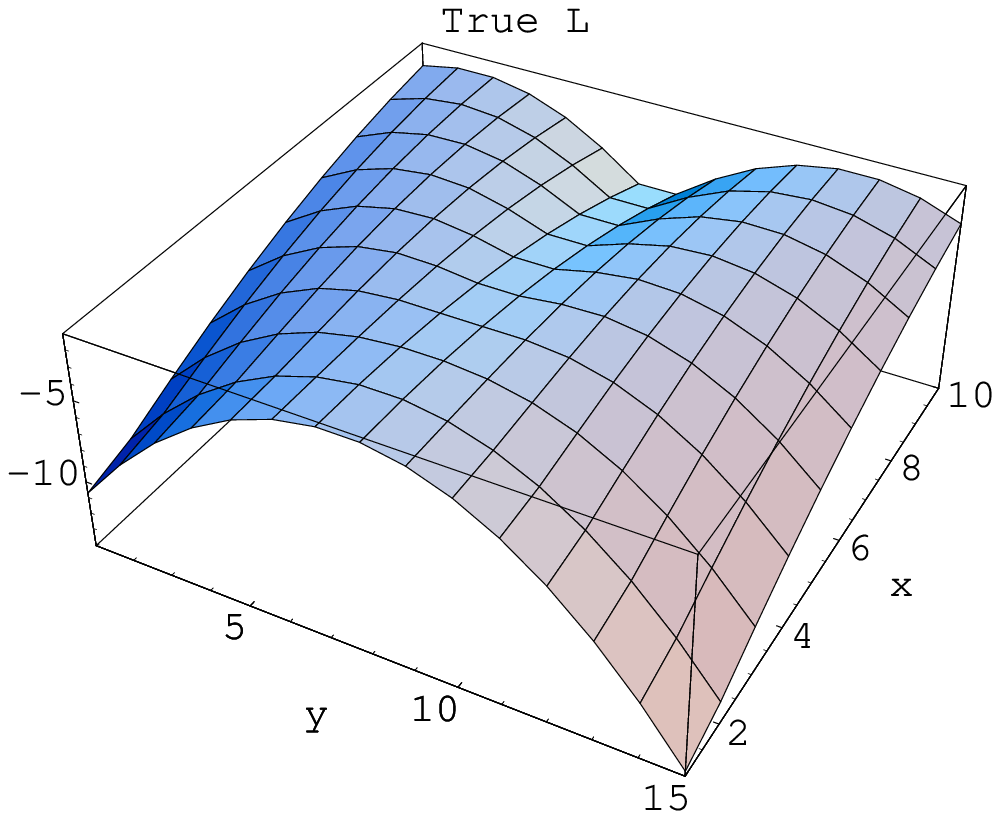, width= 65mm}\\
\epsfig{file=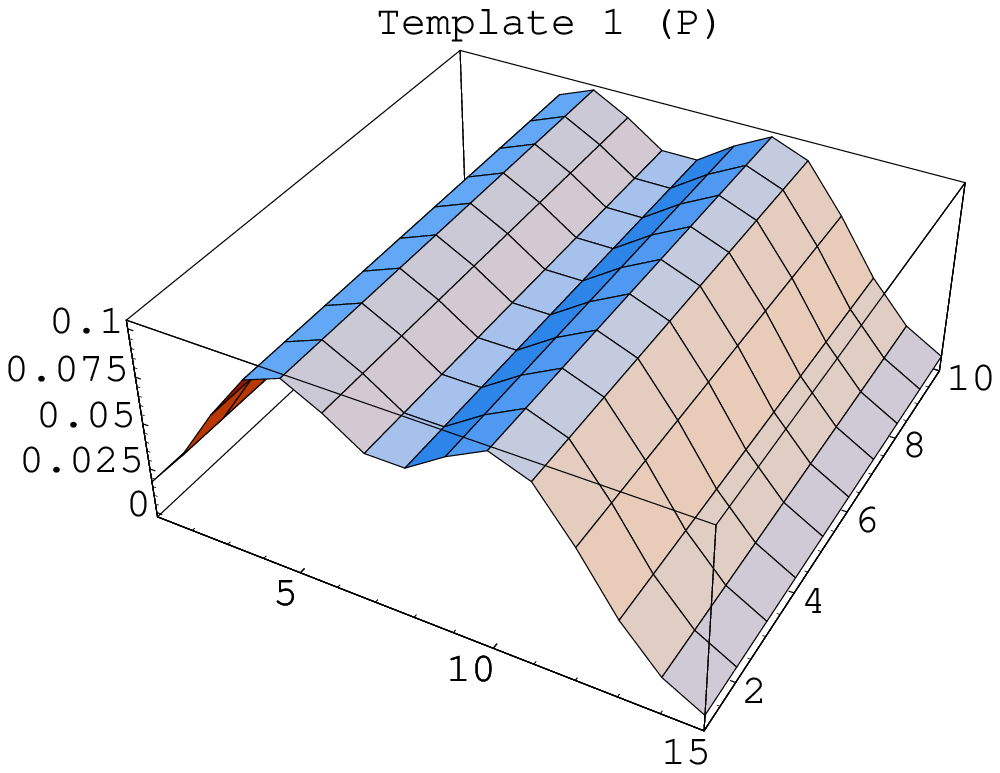, width= 65mm}
\epsfig{file=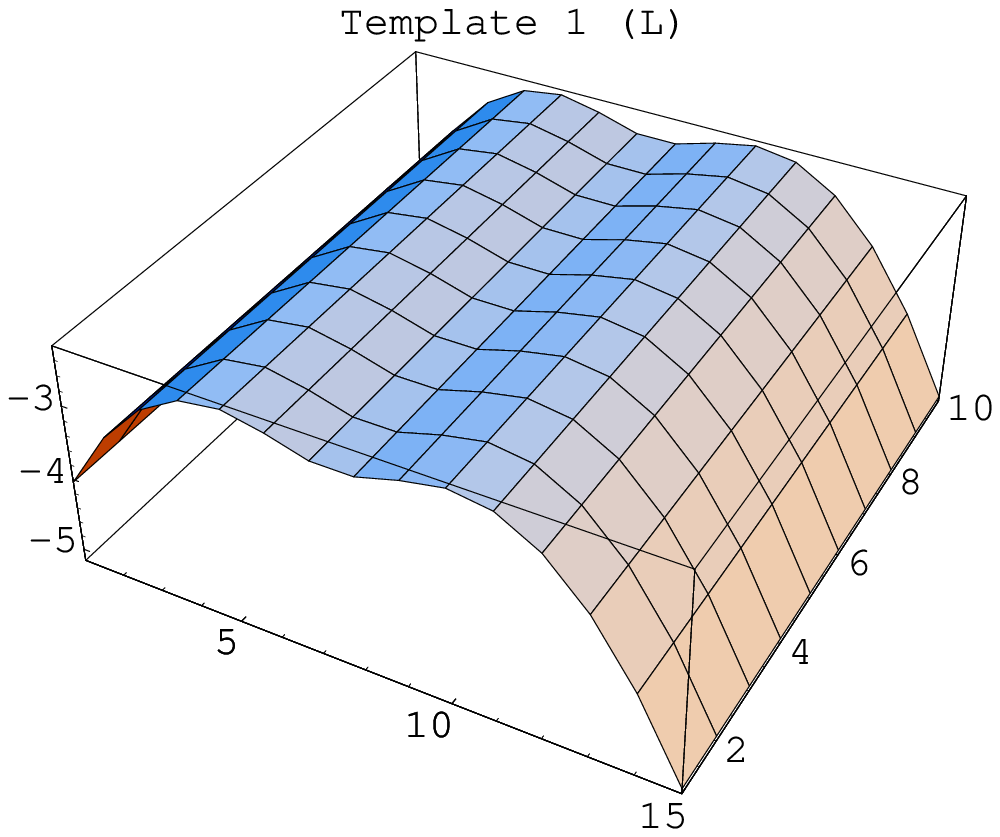, width= 65mm}\\
\epsfig{file=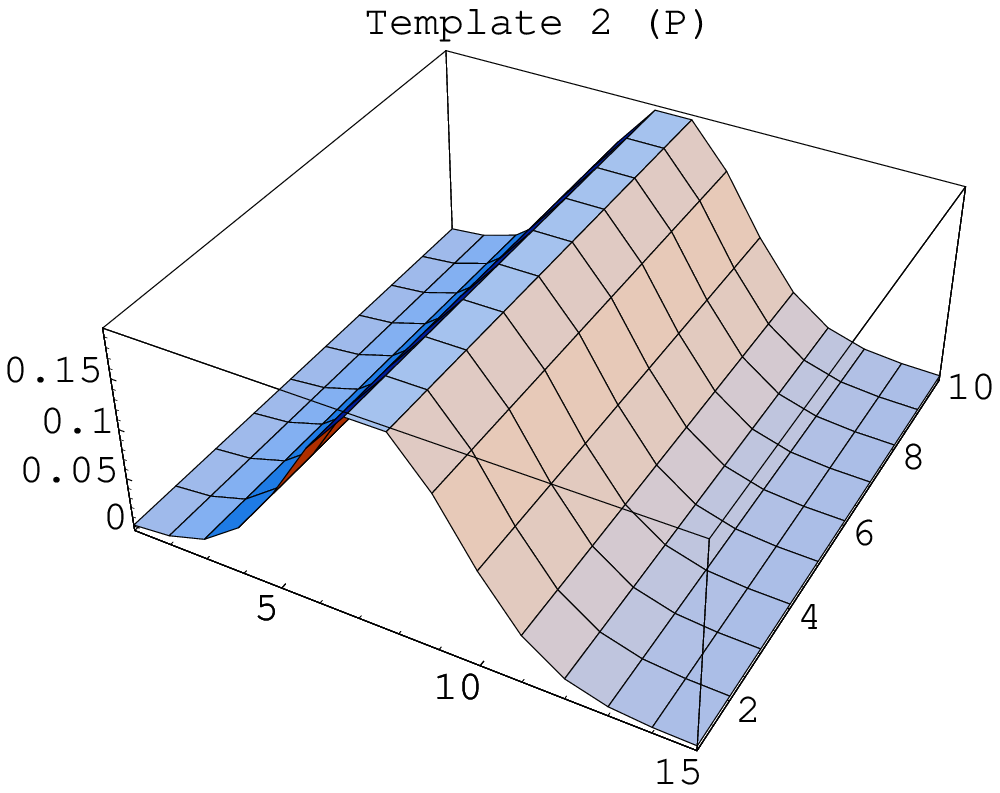, width= 65mm}
\epsfig{file=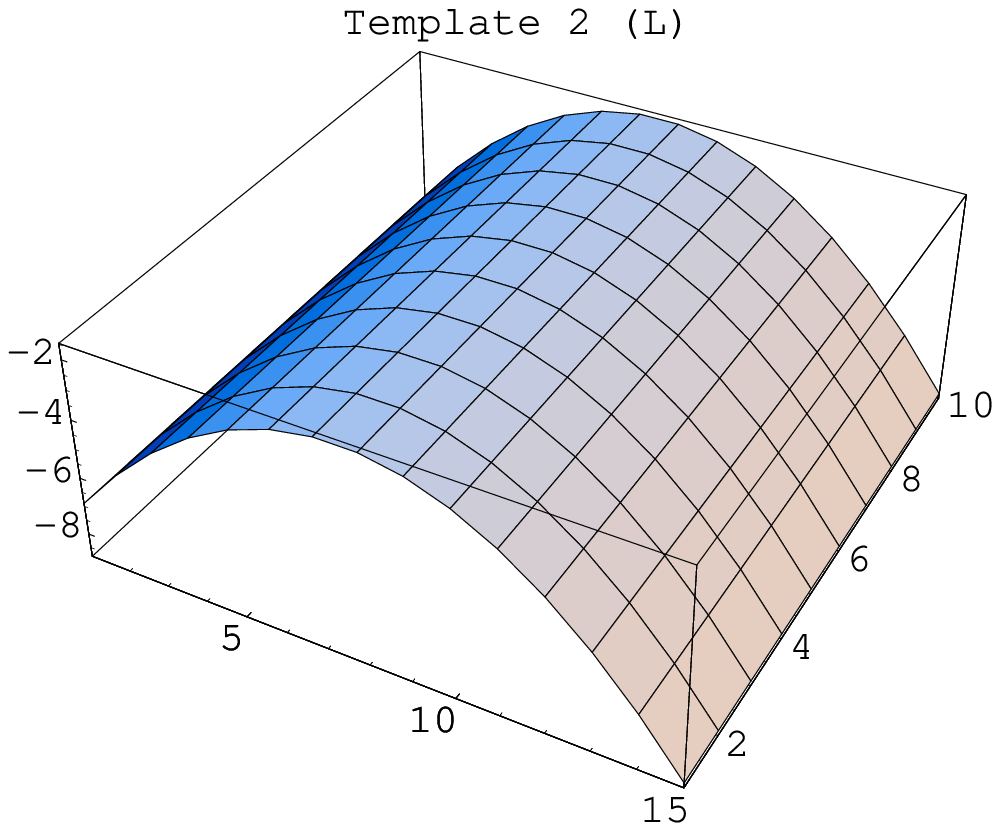, width= 65mm}
\end{center}
\caption{First row:
True density $P_{\rm true}$ (l.h.s.)
true log--density $L_{\rm true}$  = $\log P_{\rm true}$ (r.h.s.)
used for Figs.\ \ref{mix50g}--\ref{mix50f}.
Second and third row: 
The two templates $t_1$ and $t_2$ 
of Figs.\ \ref{mix50a}--\ref{mix50f}
for $P$ ($t_i^P$, l.h.s.)
or for $L$ ($t_i^L$, r.h.s.), respectively,
with $t_i^L$ = $\log t_i^P$.
As reference for the following figures
we give the expected test error
$\int \!dy\,dx\,  p(x)p(y|x,h_{\rm true}) \ln p(y|x,h)$
under the true $p(y|x,h_{\rm true})$ for uniform $p(x)$. 
It is for $h_{\rm true}$ equal to  2.23 
for template $t_1$ equal to 2.56,
for template $t_2$ equal  2.90 
and for a uniform $P$ equal to 2.68.
}
\label{mixTrue}
\end{figure}

\begin{figure}[ht]
\begin{center}
\epsfig{file=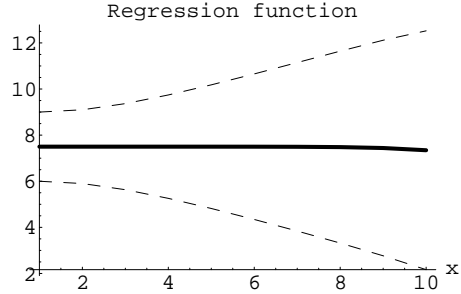, width= 60mm}
\end{center}
\caption{
Regression function $h_{\rm true}(x)$ for the true density $P_{\rm true}$
of Fig.\ \ref{mixTrue},
defined as
$h(x)$ 
= $\int \!dy \, y p(y|x,h_{\rm true})$
= $\int \!dy \, y P_{\rm true}(x,y)$.
The dashed lines indicate the range of 
one standard deviation above and below the regression function.
}
\label{mixTrueR}
\end{figure}

\begin{figure}[ht]
\begin{center}
\epsfig{file=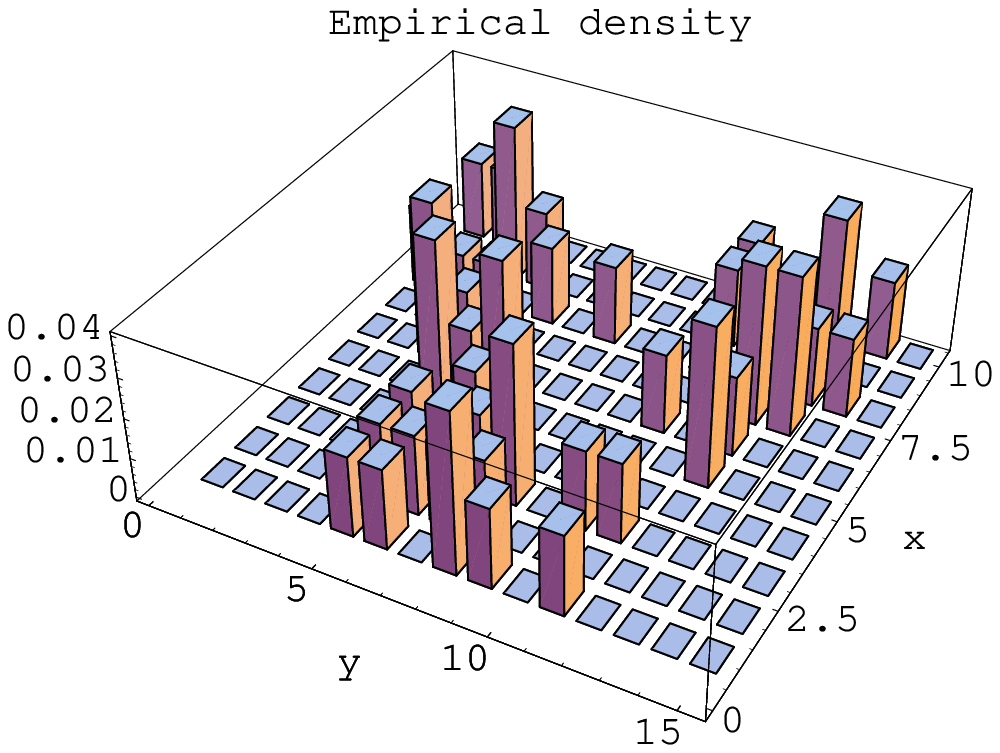, width= 65mm}
\epsfig{file=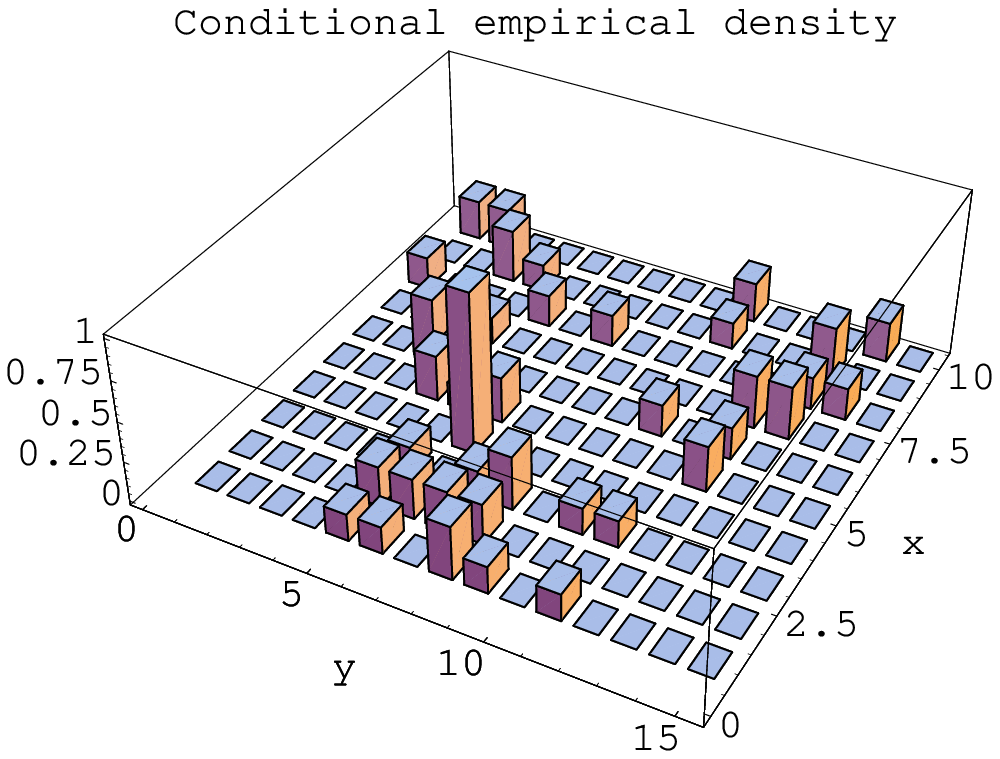, width= 65mm}
\end{center}
\caption{L.h.s.:
Empirical density 
$N(x,y)/n$
=  
\mbox{$\sum_i\delta(x-x_i)\delta(y-y_i)/\sum_i 1$}.
sampled from 
$p(x,y|h_{\rm true})$ 
=
$p(y|x,h_{\rm true}) p(x)$ 
with uniform $p(x)$.
R.h.s.:
Corresponding conditional empirical density 
$P_{\rm emp}(x,y)$
= $({\bf N}_X^{-1} N)(x,y)$
= \mbox{$\sum_i\delta(x-x_i)\sum_i\delta(y-y_i)\sum_i/\sum_i \delta(x-x_i)$}.
Both densities are obtained from the 50 data points 
used for Figs.\ \ref{mix50g}--\ref{mix50f}.
}
\label{mixData}
\end{figure}

\begin{figure}[ht]
\begin{center}
\epsfig{file=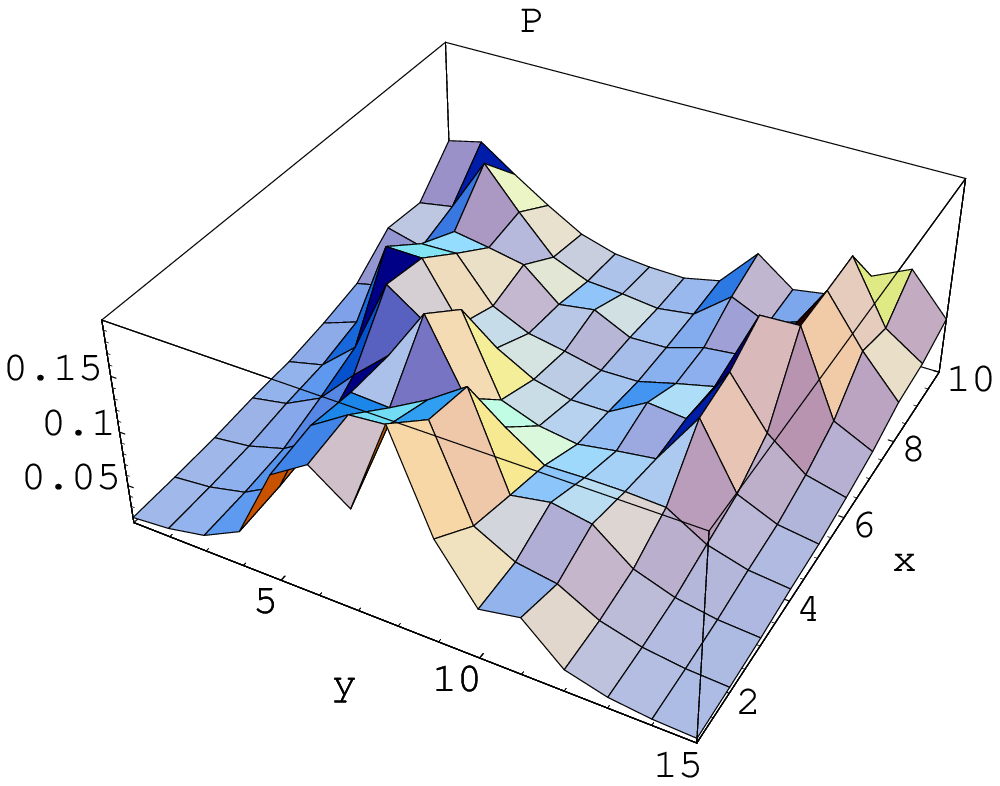, width= 65mm}
\epsfig{file=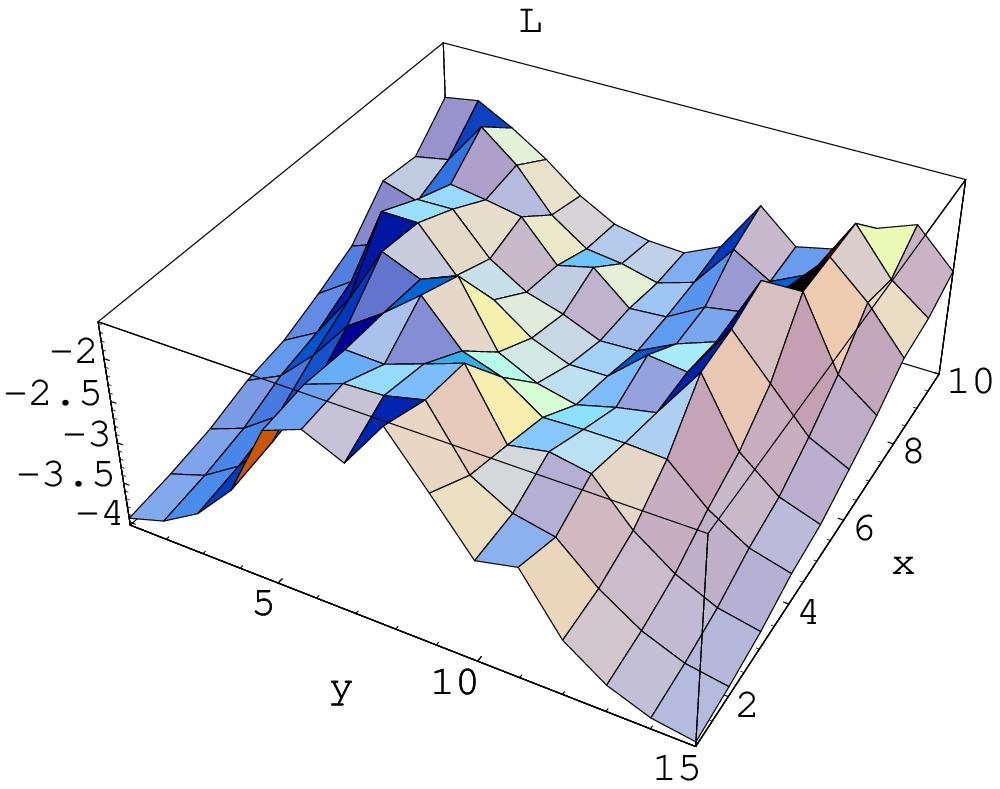, width= 65mm}\\
\vspace{0.5cm}
\epsfig{file=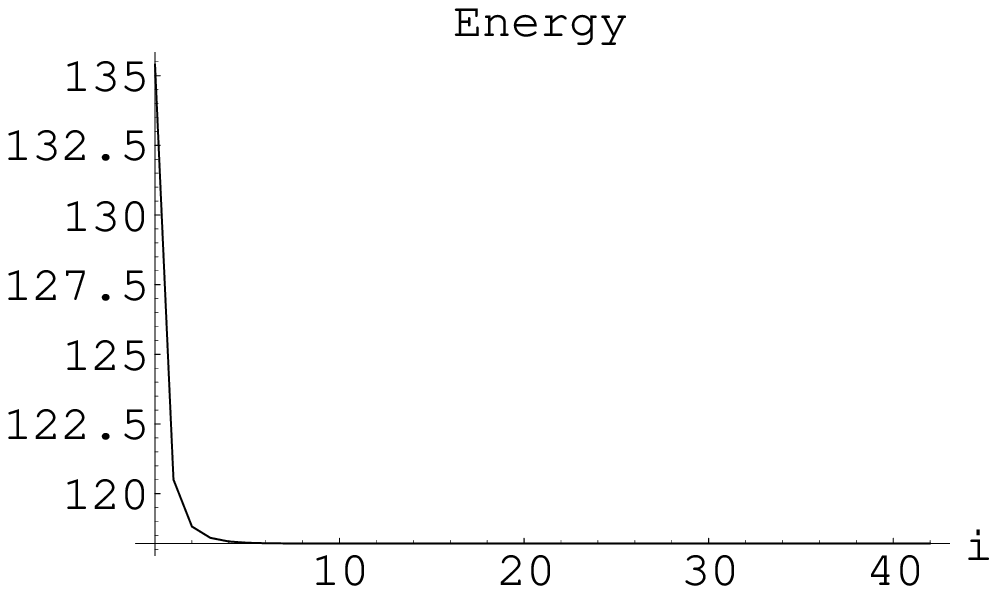, width= 50mm}
\epsfig{file=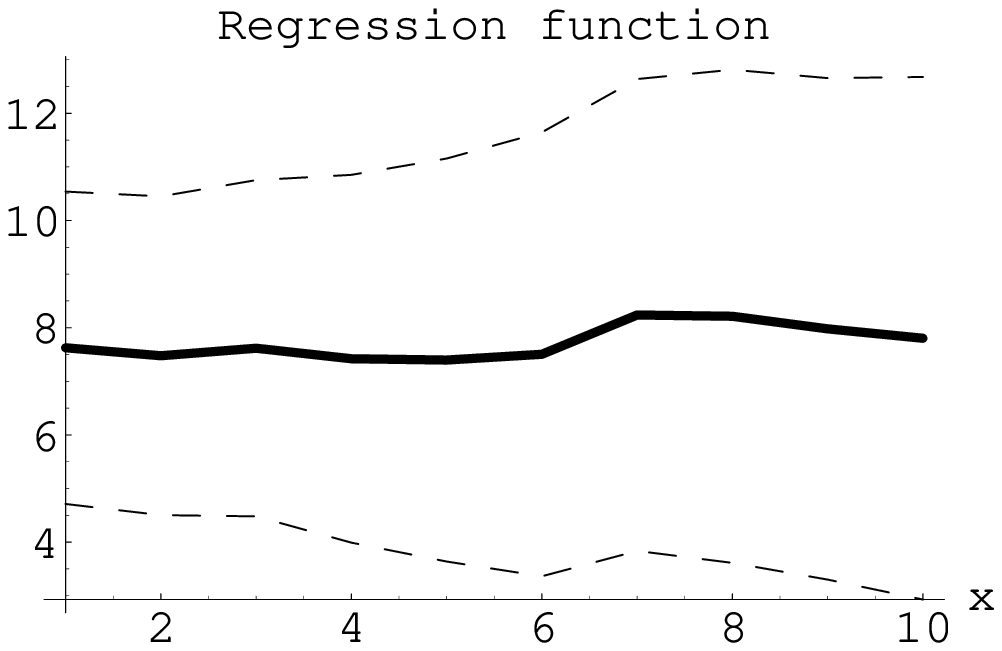, width= 50mm}\\
\epsfig{file=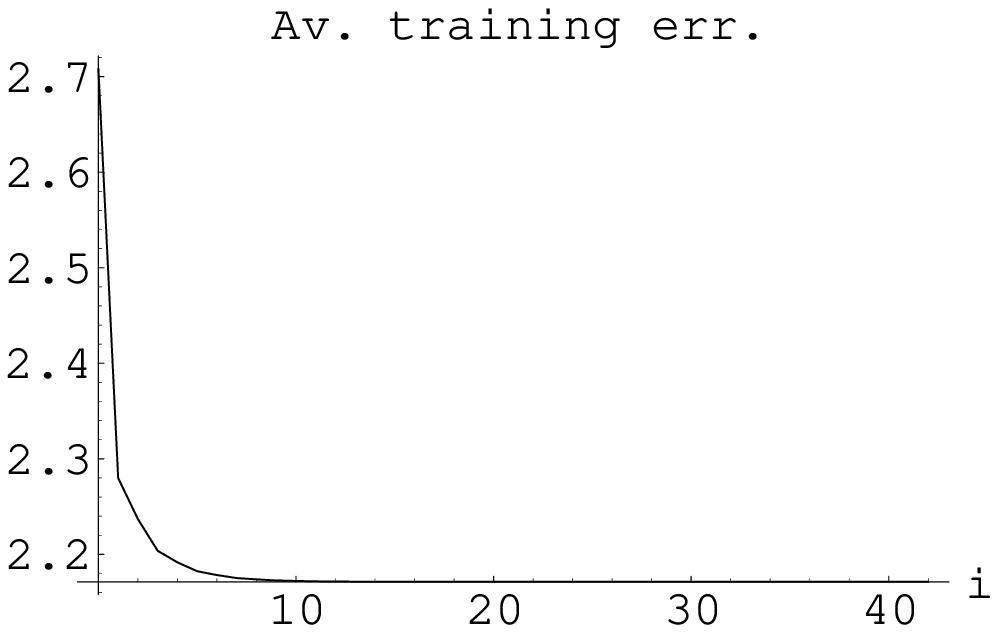, width= 50mm}
\epsfig{file=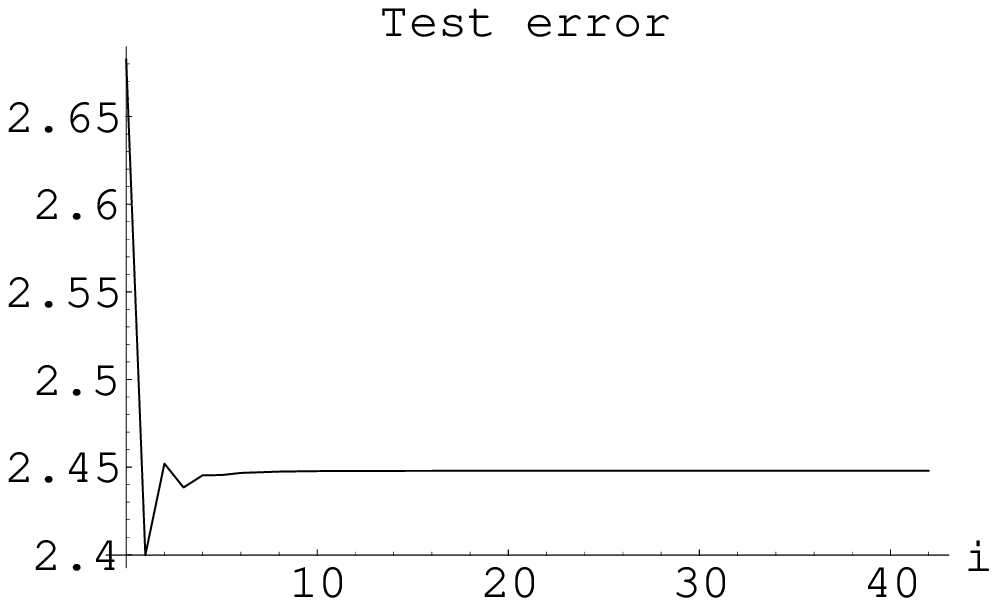, width= 50mm}
\end{center}
\vspace{-0.2cm}
\caption{
Density estimation with Gaussian prior factor
for log--probability $L$  with 50 data points
shown in Fig.\  \ref{mixData}.
Top row: Final solution $P(x,y)$ = $p(y|x,h)$ and $L=\log P$.
Second row: Energy $E_L$ (\ref{L-functional})
during iteration and final regression function.
Bottom row: 
Average training error
$-(1/n)\sum_{i=1}^n\log p(y_i|x_i,h)$ during iteration
and
average test error 
$-\int \!dy\,dx\,  p(x)p(y|x,h_{\rm true}) \ln p(y|x,h)$
for uniform $p(x)$. 
(Parameters: Zero mean Gaussian 
smoothness prior with 
inverse covariance $\lambda {\bf K}$,
$\lambda$ = 0.5  
and ${\bf K}$  
of the form (\ref{smoothness-prior-for-examples})
with
$\lambda_x$ = 2,
$\lambda_y$ = 1,
$\lambda_0$ = 0,
$\lambda_2$ = 1,
$\lambda_4$ = 
$\lambda_6$ = 0,
massive prior iteration with
${\bf A}$ = ${\bf K}+m^2 {\bf I}$ and squared mass $m^2$ = $0.01$.
Initialized with normalized constant $L$.
At each iteration step the factor $\eta$ 
has been adapted by a line search algorithm.
Mesh with $10$ points in $x$-direction and 
$15$ points in $y$--direction, 
periodic boundary conditions in $y$.)
}
\label{mix50g}
\end{figure}

\begin{figure}[ht]
\vspace{-1.4cm}
\begin{center}
\epsfig{file=ps/mix50jEne.eps, width= 39mm}
\epsfig{file=ps/mix50jED.eps, width= 39mm}
\epsfig{file=ps/mix50jErr.eps, width= 39mm}
\\
\epsfig{file=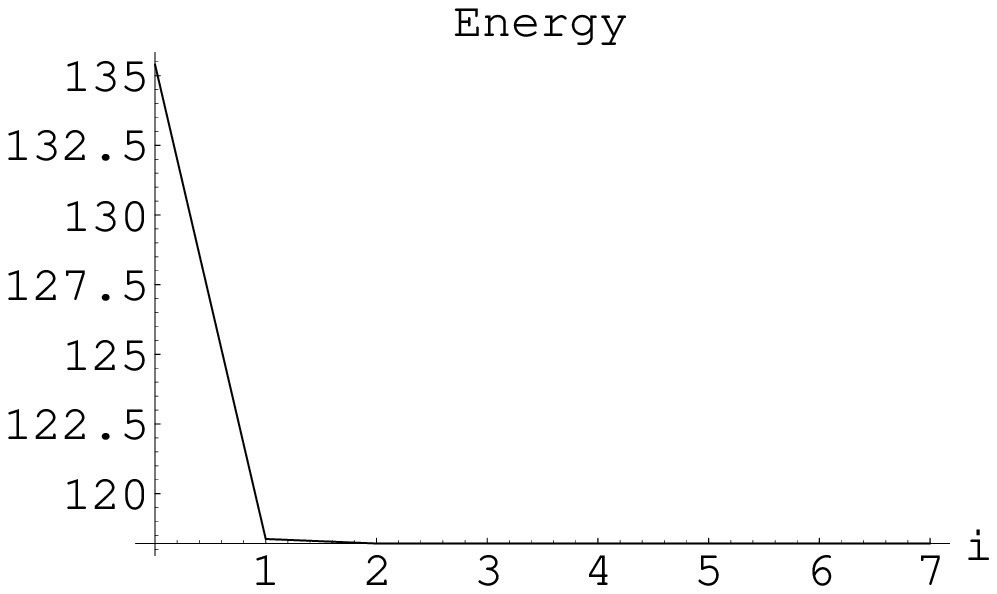, width= 39mm}
\epsfig{file=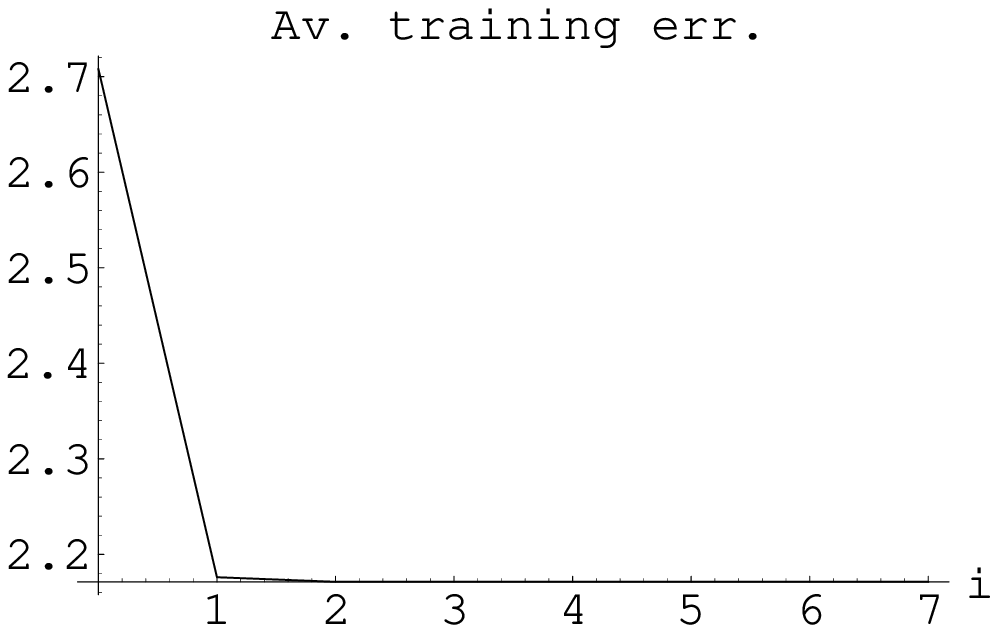, width= 39mm}
\epsfig{file=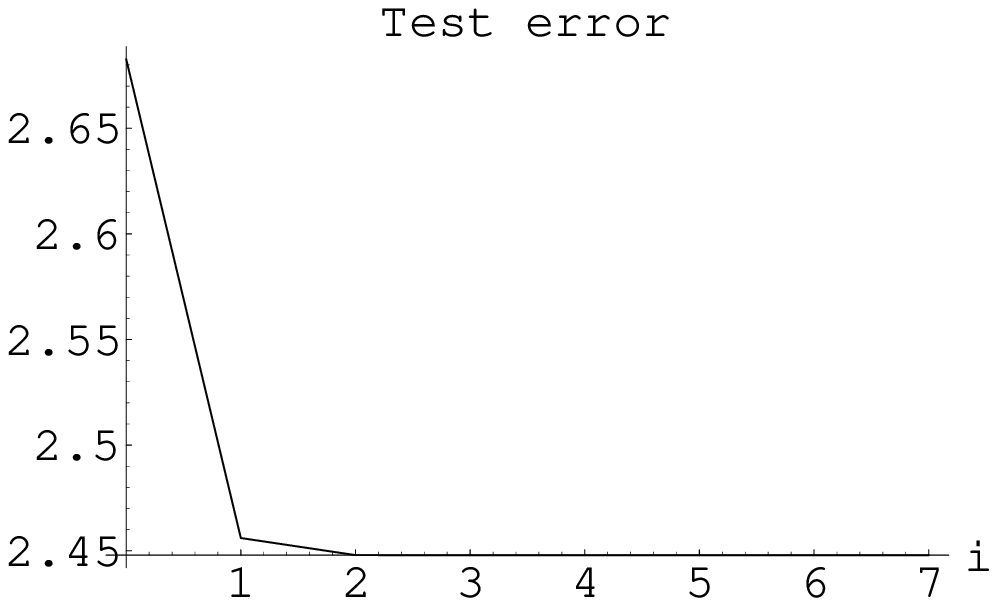, width= 39mm}
\\
\epsfig{file=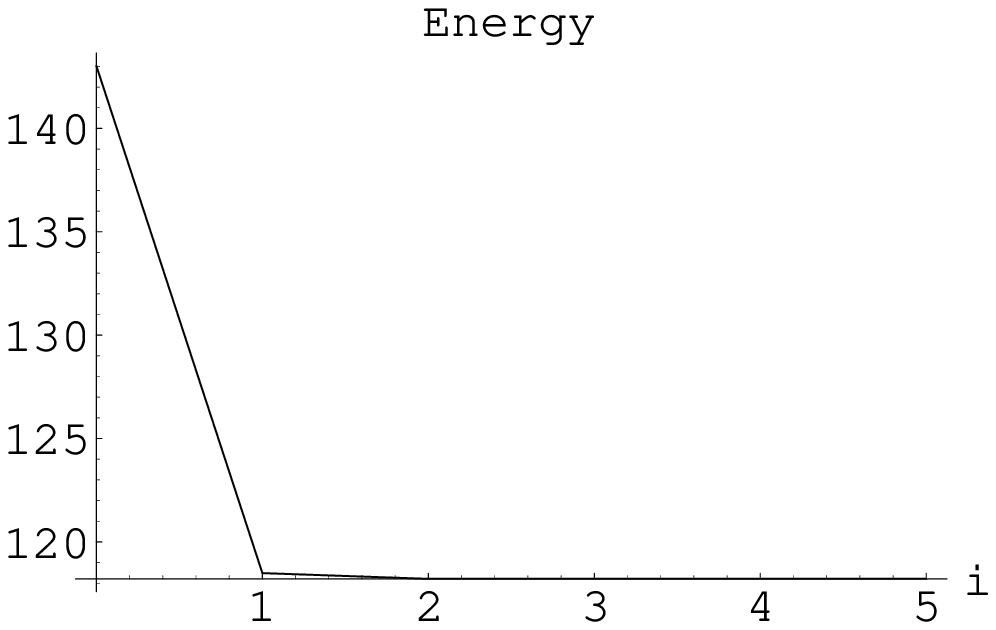, width= 39mm}
\epsfig{file=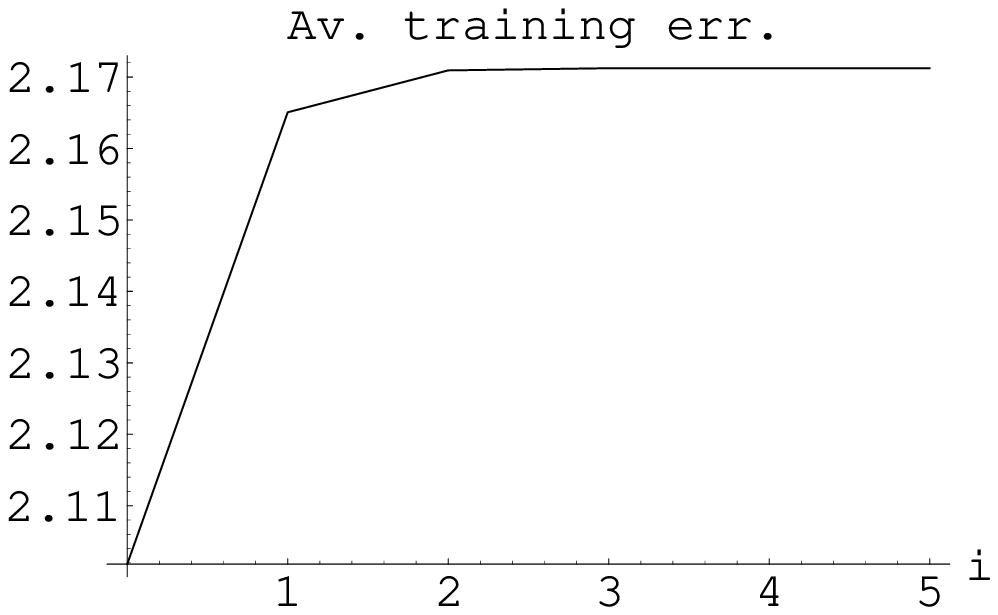, width= 39mm}
\epsfig{file=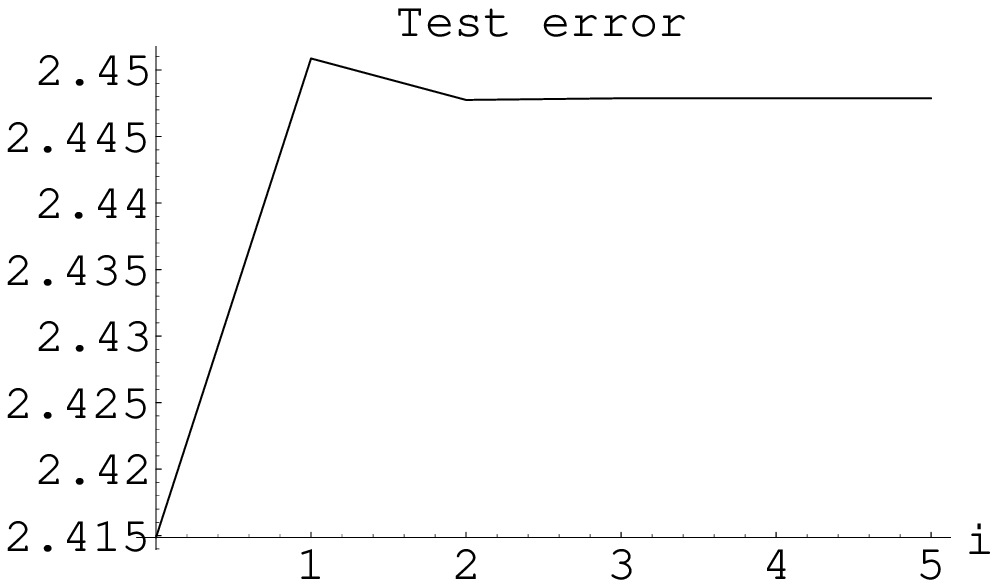, width= 39mm}
\\
\epsfig{file=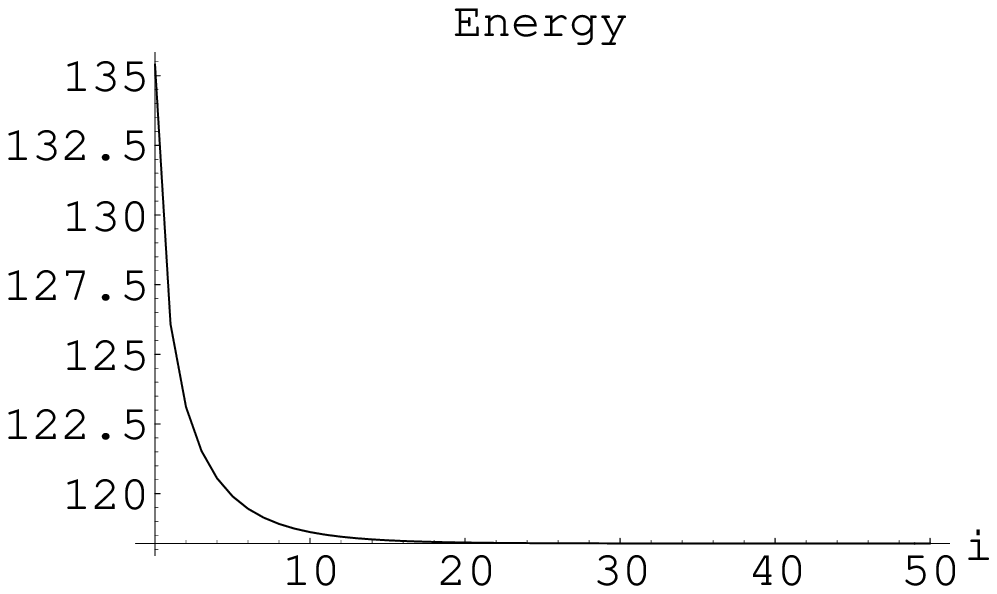, width= 39mm}
\epsfig{file=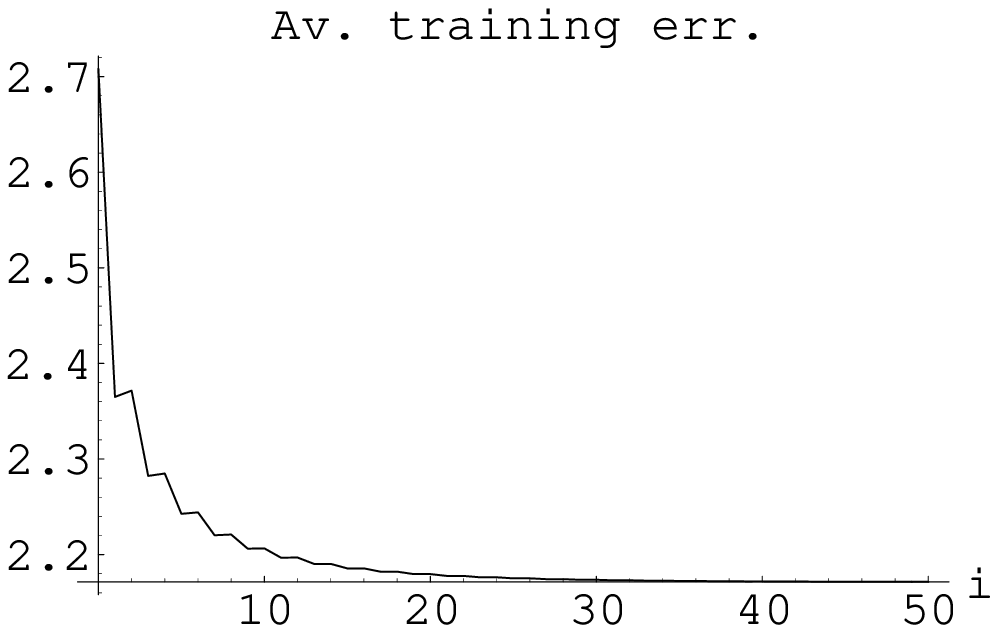, width= 39mm}
\epsfig{file=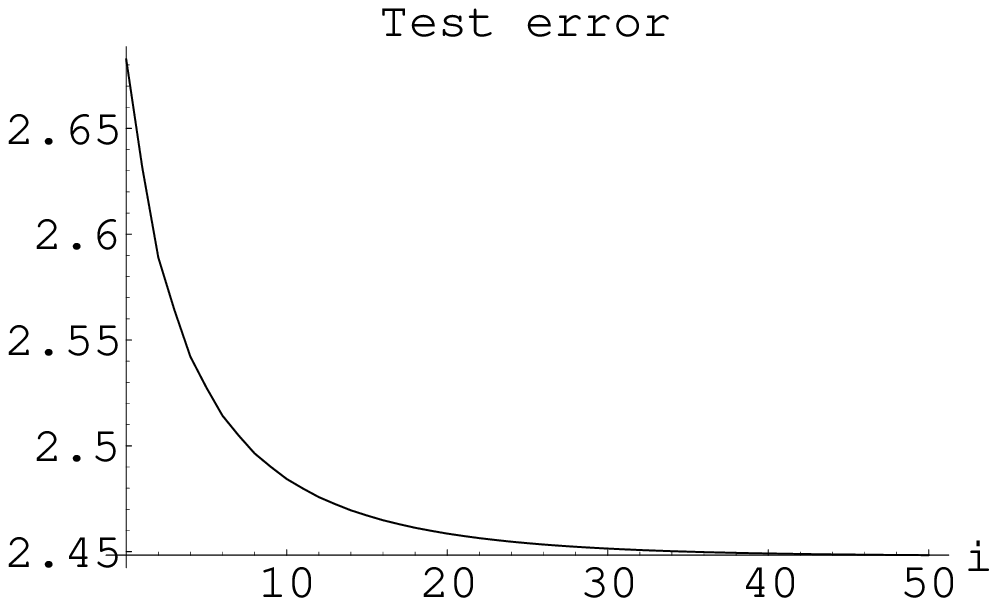, width= 39mm}
\\
\epsfig{file=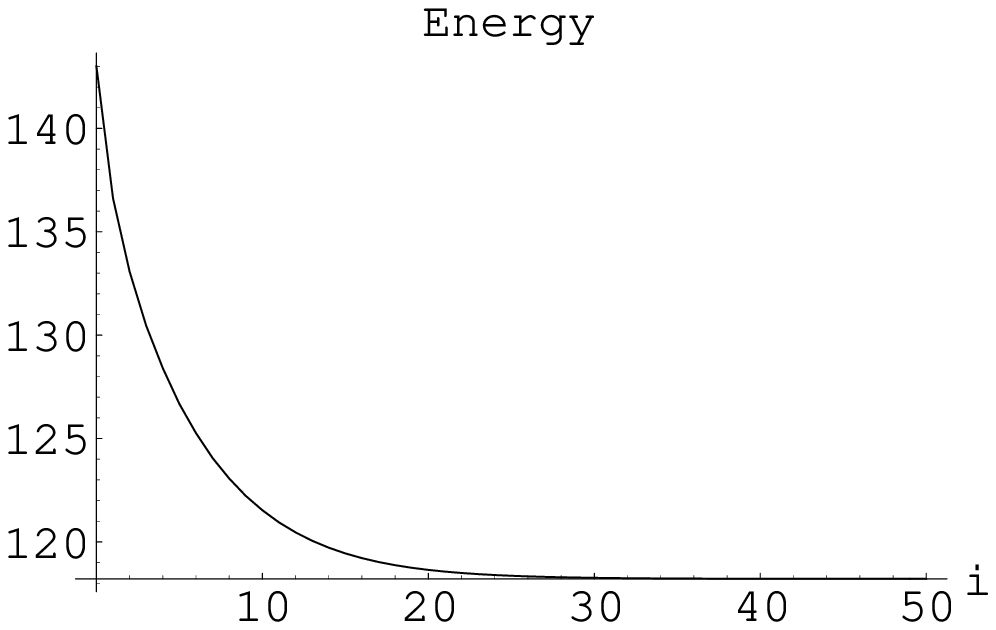, width= 39mm}
\epsfig{file=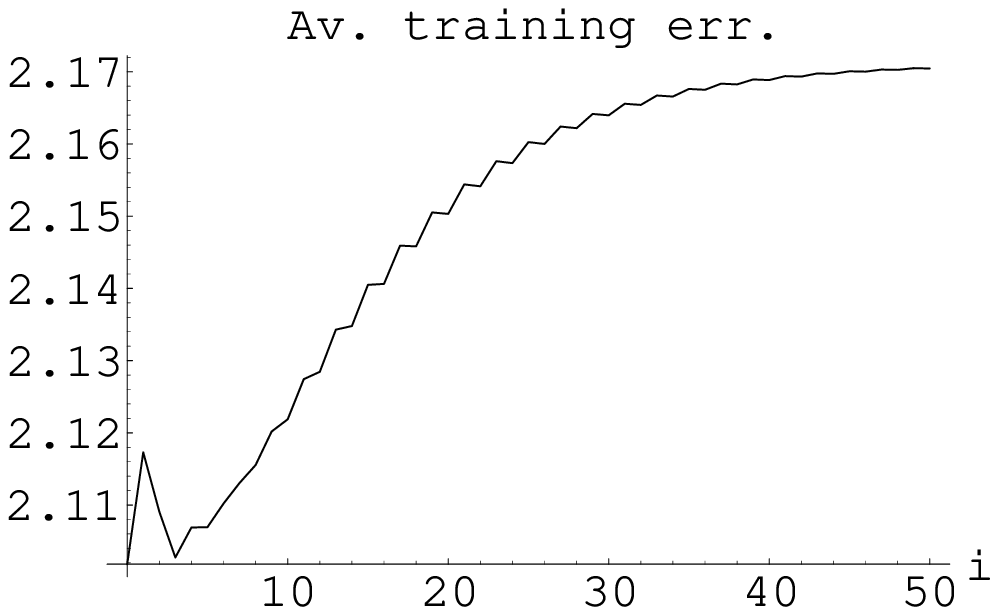, width= 39mm}
\epsfig{file=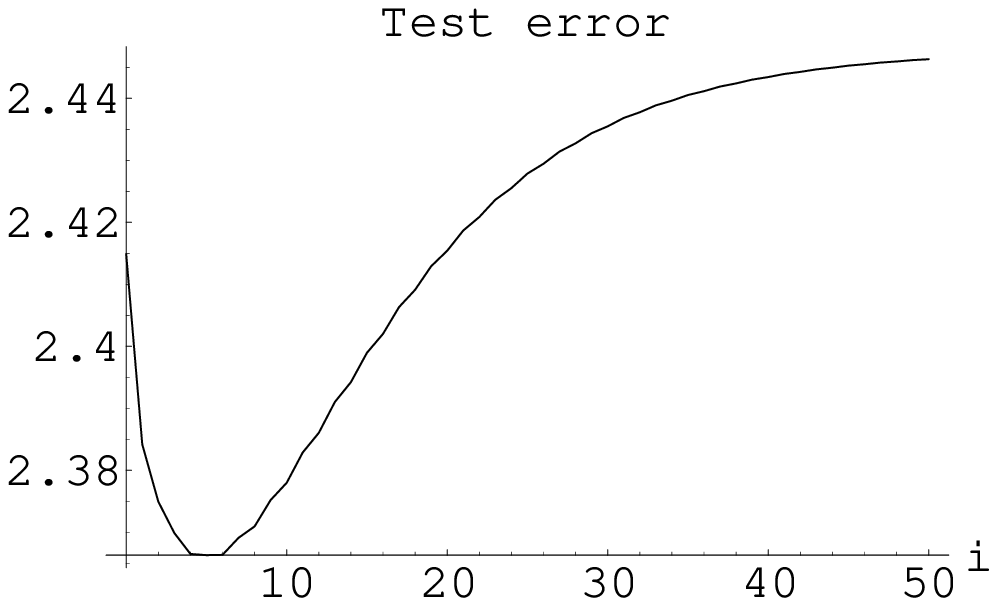, width= 39mm}
\\
\epsfig{file=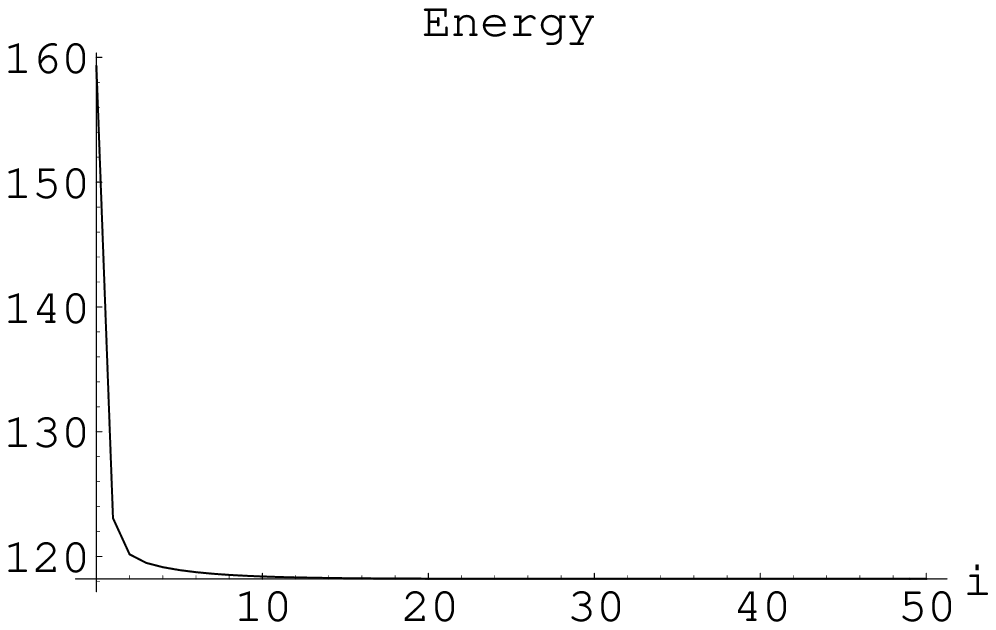, width= 39mm}
\epsfig{file=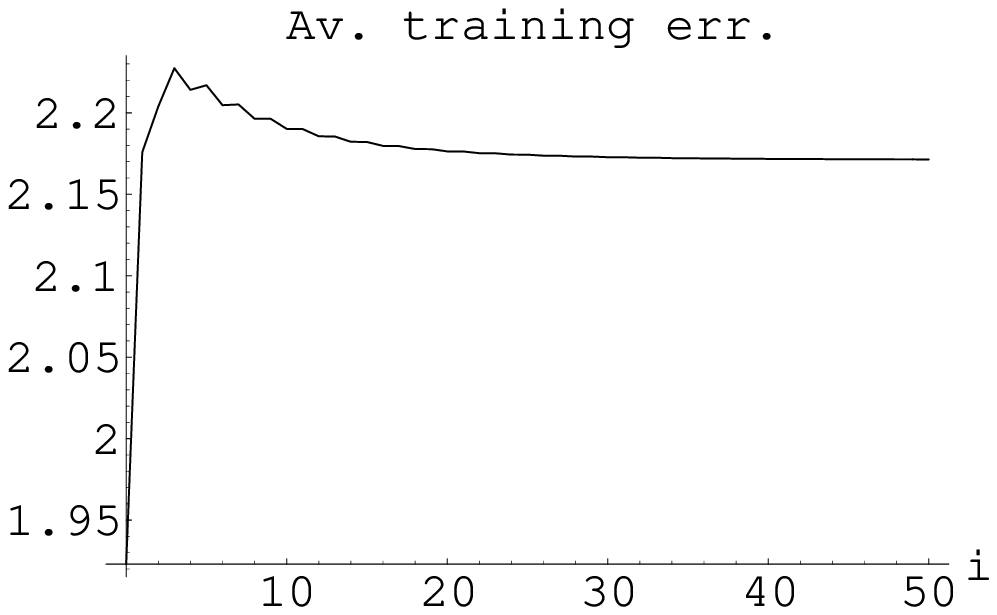, width= 39mm}
\epsfig{file=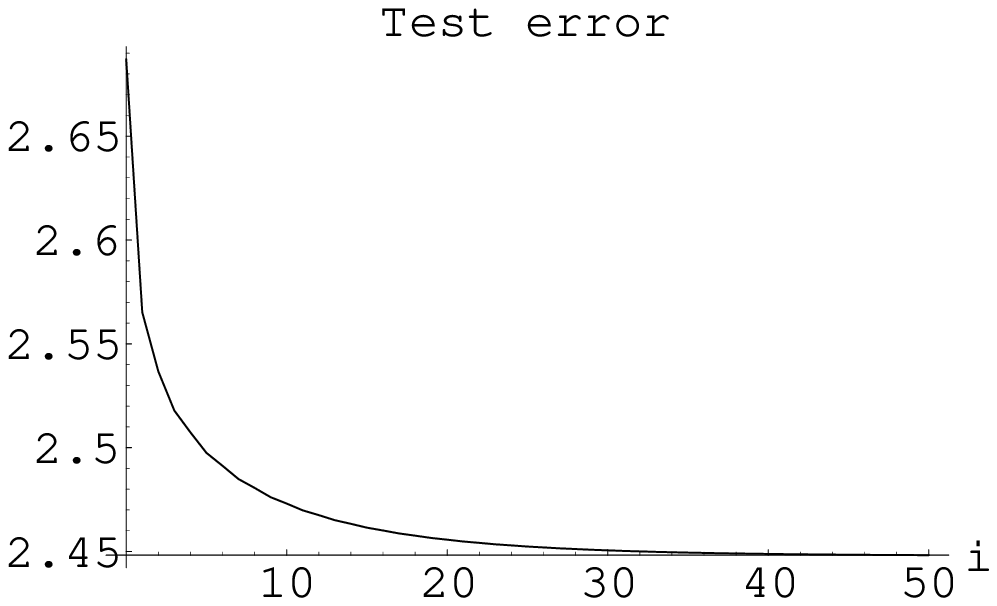, width= 39mm}
\end{center}
\vspace{-0.5cm}
\caption{
Comparison of iteration schemes and initialization.
First row: Massive prior iteration 
(with ${\bf A}$ = ${\bf K}+m^2 {\bf I}$, $m^2$ = $0.01$)
and uniform initialization.
Second row: Hessian iteration (${\bf A}$ = $-{\bf H}$) 
and uniform initialization.
Third row: Hessian iteration and kernel initialization 
(with ${\bf C}$ = ${\bf K}+m_C^2 {\bf I}$, $m_C^2$ = $0.01$ 
and normalized afterwards).
Forth row: Gradient (${\bf A}$ = ${\bf I}$) with uniform initialization.
Fifth row: Gradient with kernel initialization.
Sixth row: Gradient 
with delta--peak initialization.
(Initial $L$ equal to $\ln (N/n+\epsilon)$, $\epsilon$ = $10^{-10}$,
conditionally normalized.
For $N/n$ see Fig.\ \ref{mixData}).
Minimal number of iterations 4,
maximal number of iterations 50,
iteration stopped if $|L^{(i)}-L^{(i-1)}|<10^{-8}$.
Energy functional and parameters as for Fig.\ \ref{mix50g}.
}
\label{mix50comp}
\end{figure}

\begin{figure}[ht]
\vspace{-1.cm}
\begin{center}
\epsfig{file=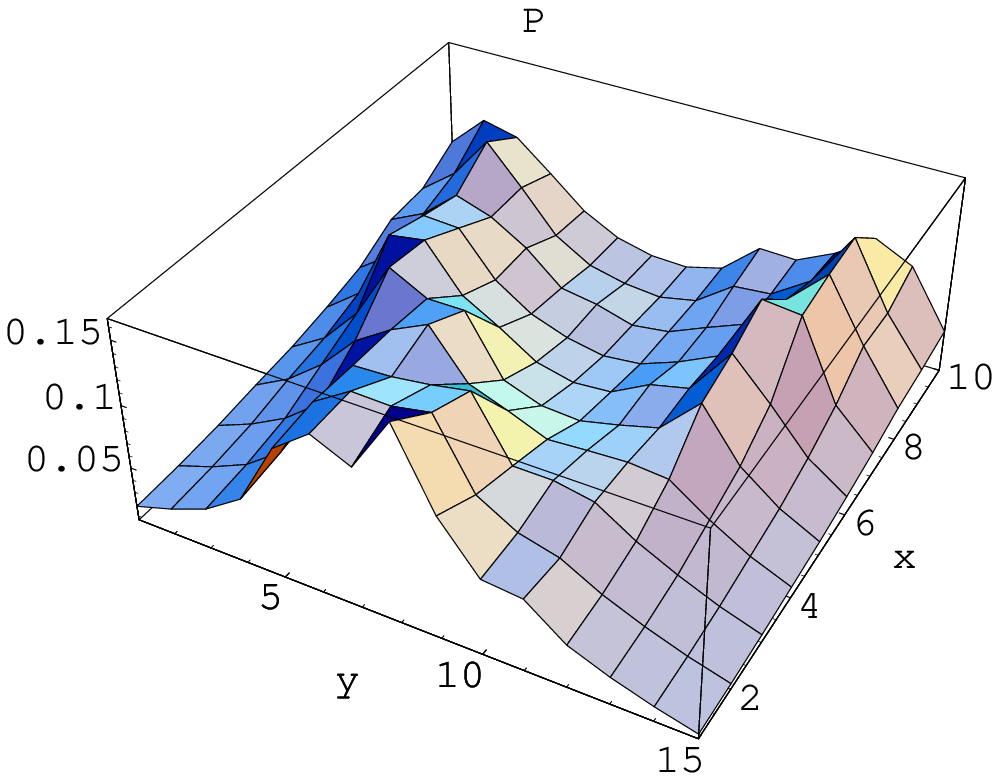, width= 65mm}
\epsfig{file=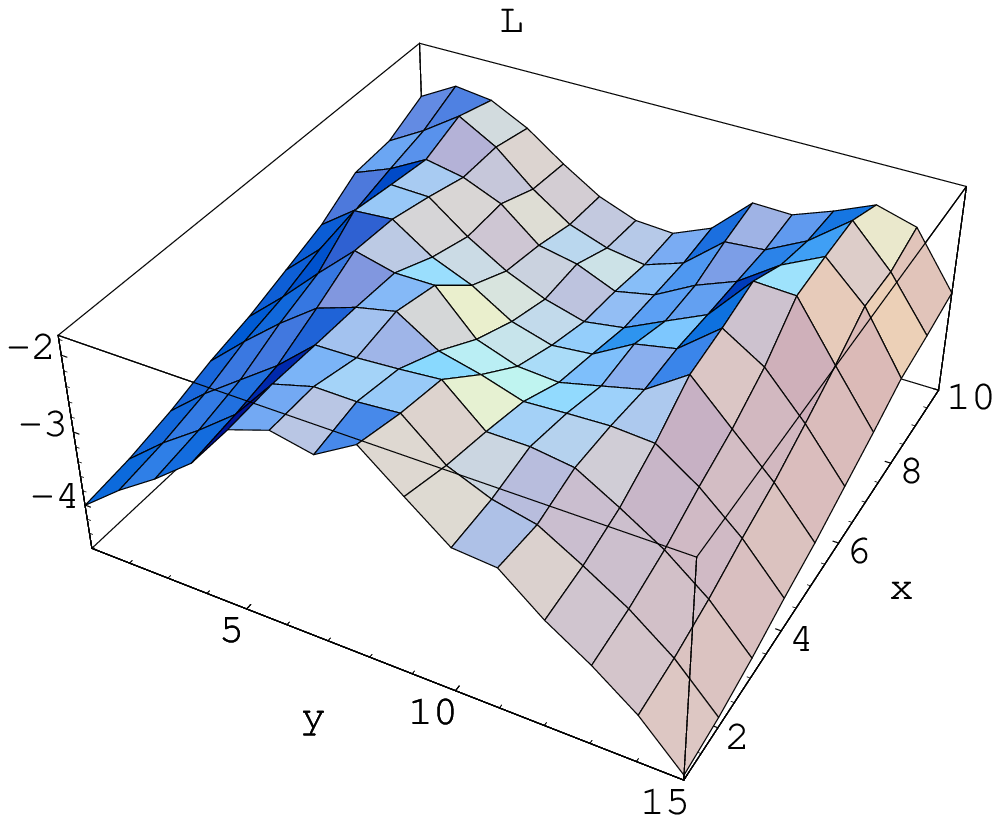, width= 65mm}\\
\vspace{0.5cm}
\epsfig{file=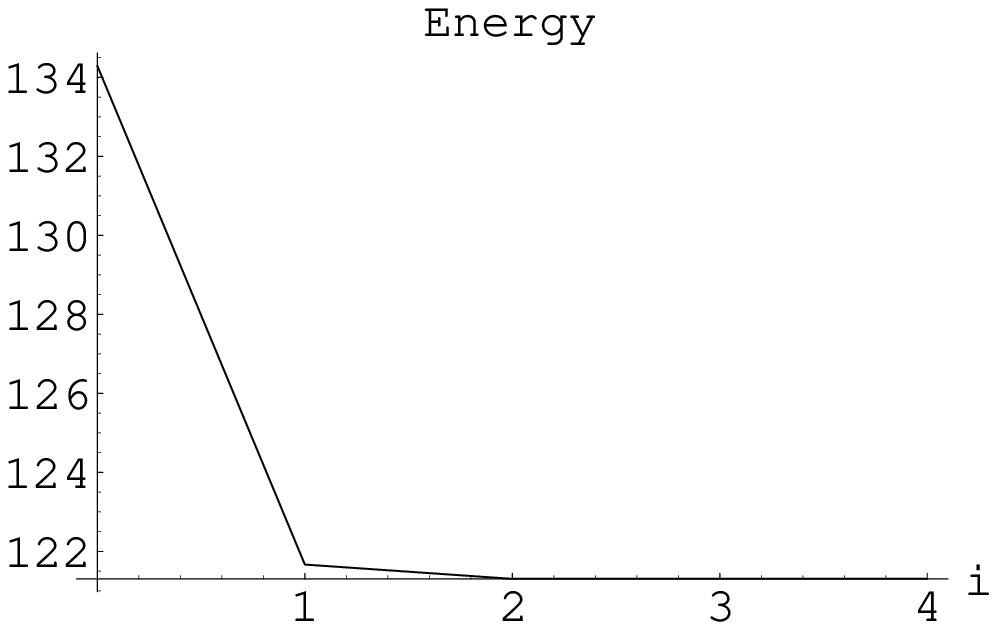, width= 50mm}
\epsfig{file=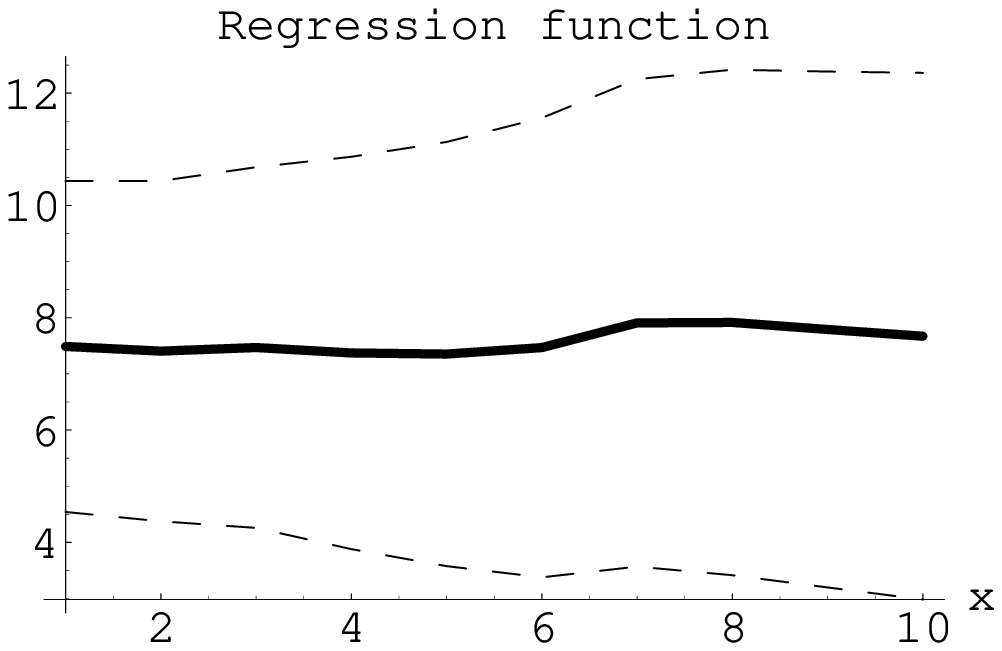, width= 50mm}\\
\epsfig{file=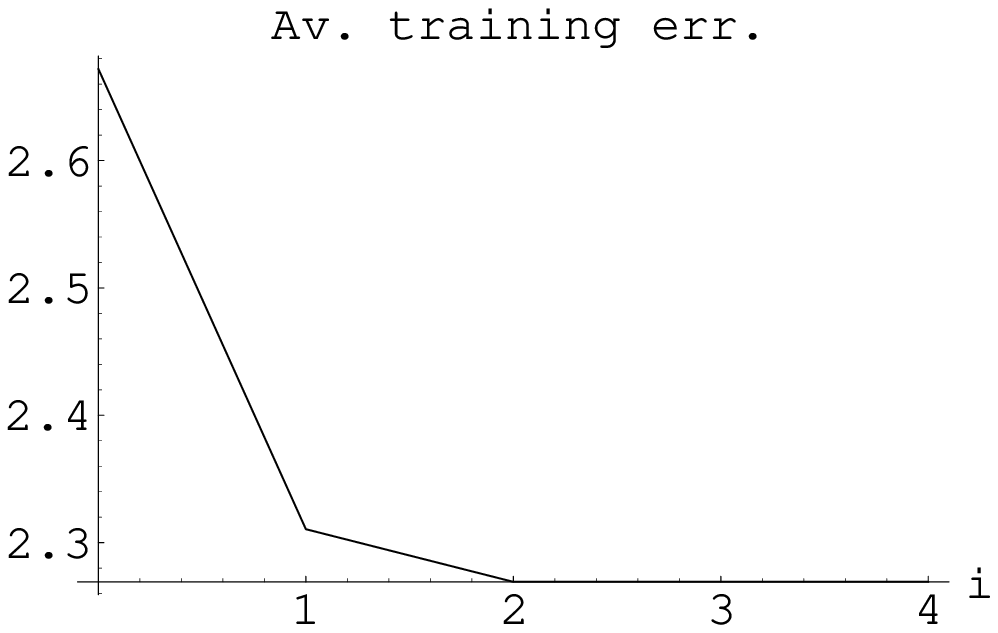, width= 50mm}
\epsfig{file=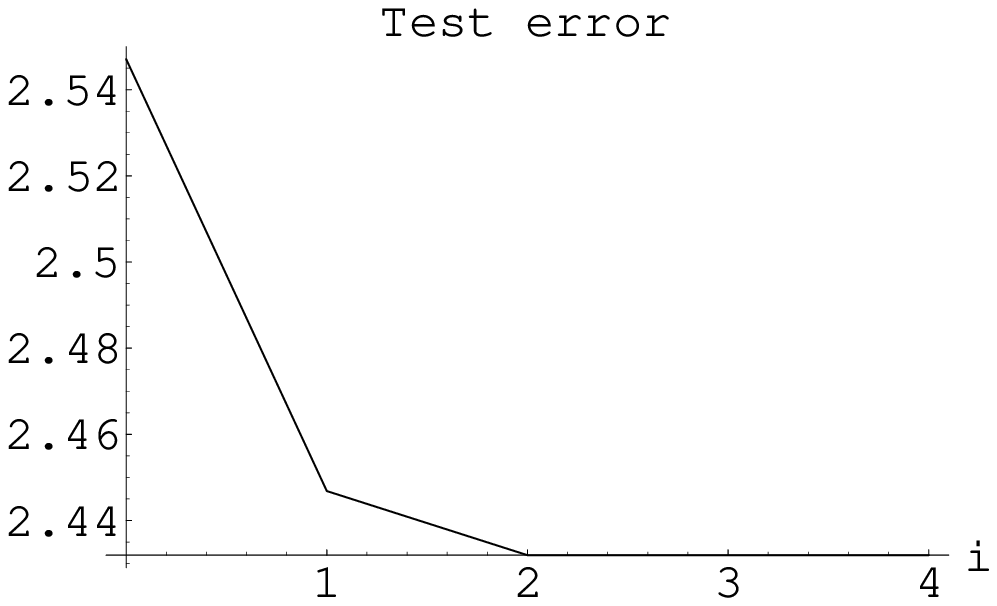, width= 50mm}
\end{center}
\vspace{-0.5cm}
\caption{
Density estimation with a Gaussian mixture prior 
for log--probability $L$ 
with 50 data points, Laplacian prior and the two template functions
shown in Fig.\  \ref{mixTrue}.
Top row: Final solution $P(x,y)$ = $p(y|x,h)$ and $L=\log P$.
Second row: Energy Energy $E_L$ (\ref{L-mixture-functional1})
during iteration and final regression function.
Bottom row: 
Average training error
-$(1/n)\sum_{i=1}^n\log p(y_i|x_i,h)$ 
during iteration
and
average test error 
$-\int \!dy\,dx\,  p(x)p(y|x,h_{\rm true}) \ln p(y|x,h)$
for uniform $p(x)$. 
(Two mixture components with 
$\lambda$ = 0.5 and
smoothness prior with ${\bf K}_1$  = ${\bf K}_2$ 
of the form (\ref{smoothness-prior-for-examples})
with
$\lambda_x$ = 2,
$\lambda_y$ = 1,
$\lambda_0$ = 0,
$\lambda_2$ = 1,
$\lambda_4$ = 
$\lambda_6$ = 0,
massive prior iteration with
${\bf A}$ = ${\bf K}+m^2 {\bf I}$ and squared mass $m^2$ = $0.01$,
initialized with $L$ = $t_1$.
At each iteration step the factor $\eta$ 
has been adapted by a line search algorithm.
Mesh with $l_x$ = $10$ points in $x$-direction and 
$l_y$ = $15$ points in $y$--direction, 
$n$ = $2$ data points at $(3,3)$, $(7,12)$,
periodic boundary conditions in $y$.
Except for the inclusion of two mixture components 
parameters are equal to those for Fig.\ \ref{mix50g}. )
}
\label{mix50a}
\end{figure}

\begin{figure}[ht]
\begin{center}
\epsfig{file=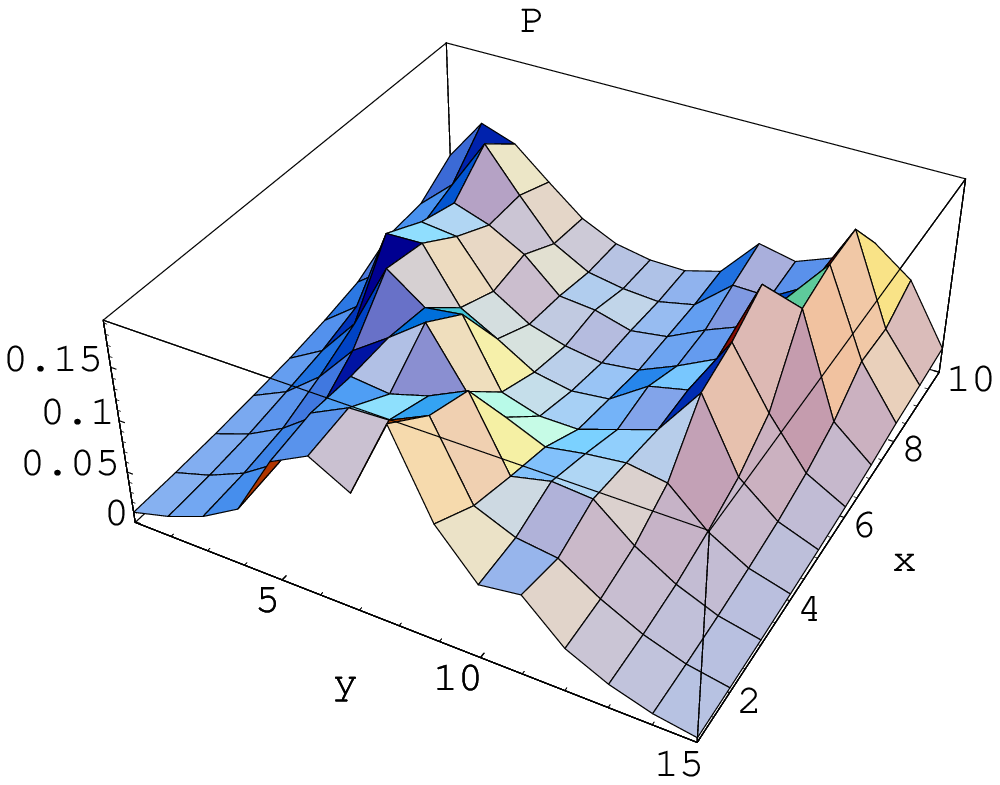, width= 65mm}
\epsfig{file=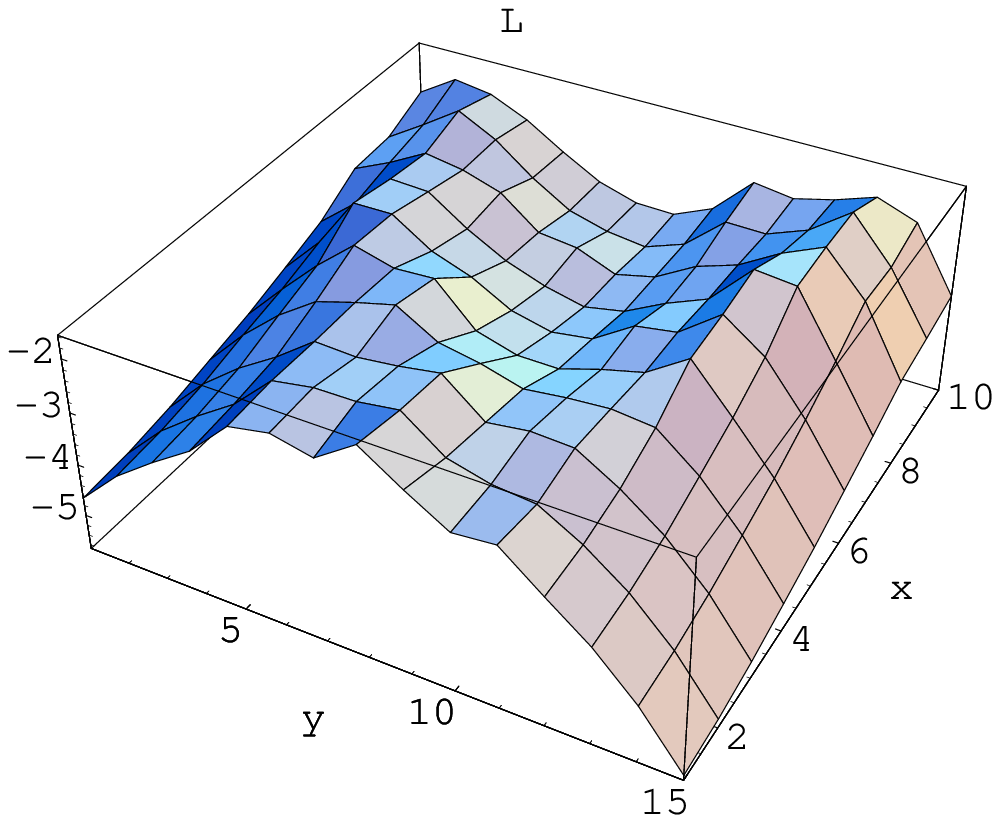, width= 65mm}\\
\vspace{0.5cm}
\epsfig{file=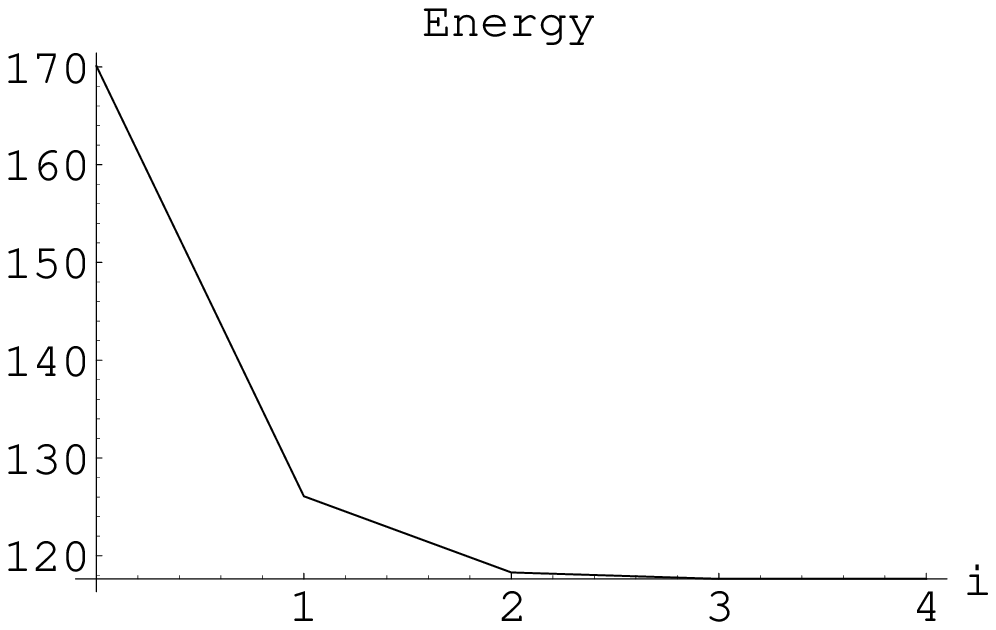, width= 50mm}
\epsfig{file=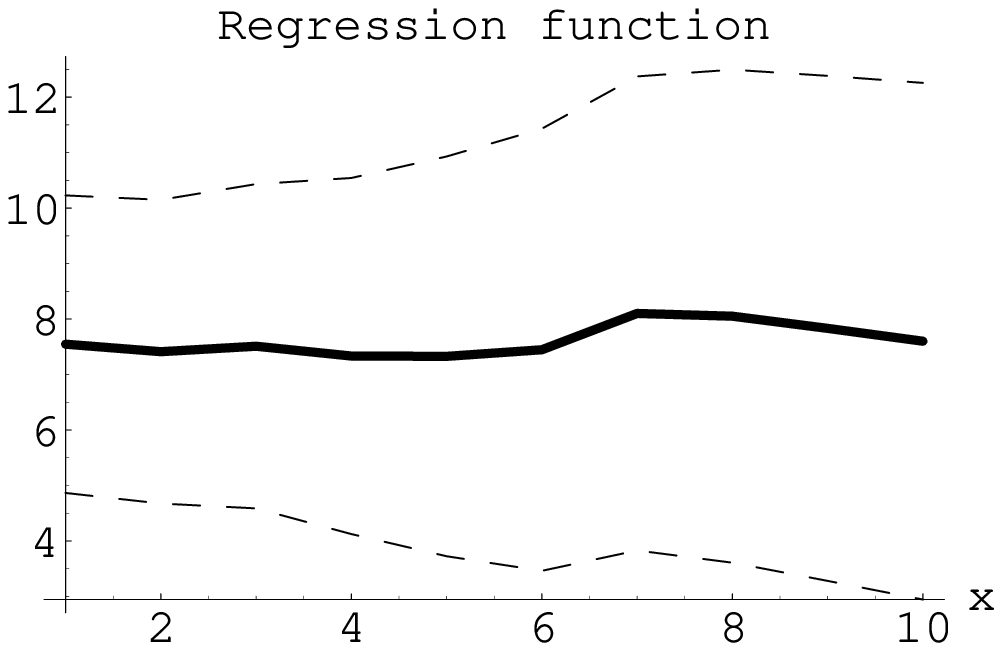, width= 50mm}\\
\epsfig{file=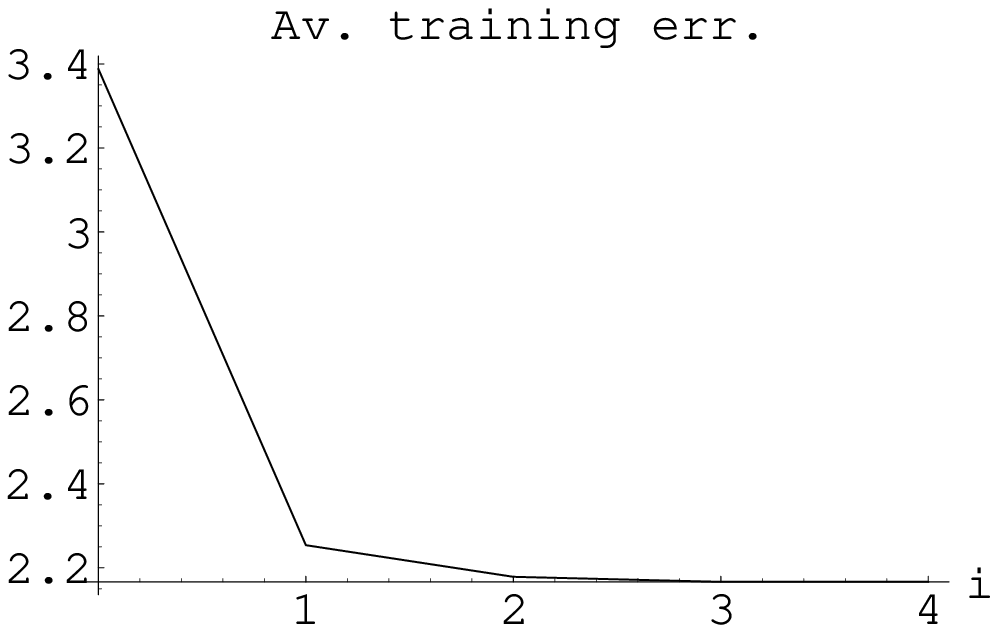, width= 50mm}
\epsfig{file=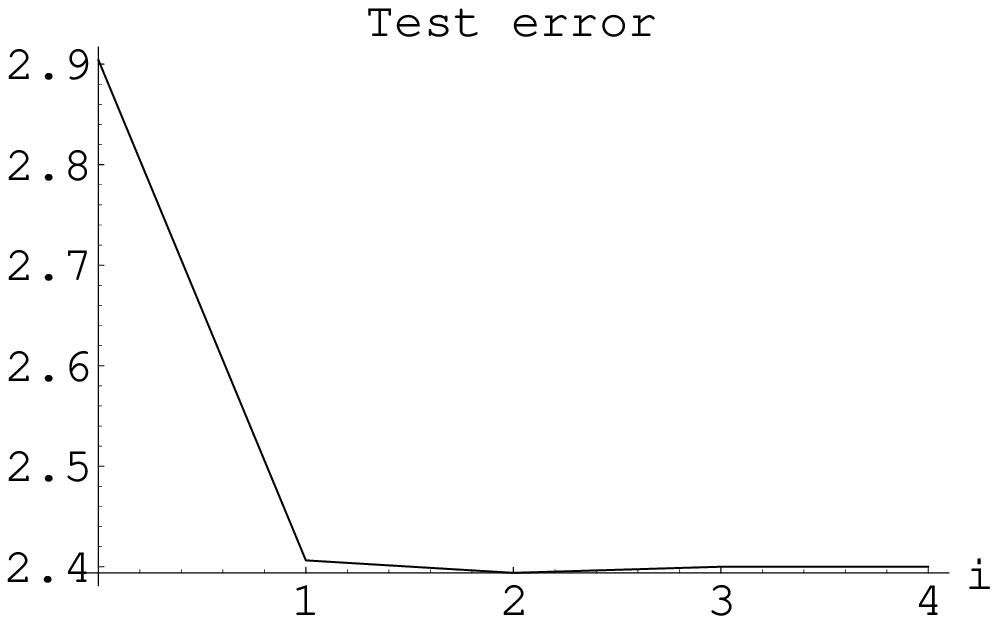, width= 50mm}
\end{center}
\caption{
Using a different starting point.
(Same parameters as for Fig.\ \ref{mix50a}, 
but initialized with $L$ = $t_2$.)
While the initial guess is worse 
then that of Fig.\ \ref{mix50a},
the final solution is even slightly better.
}
\label{mix50b}
\end{figure}

\begin{figure}[ht]
\begin{center}
\epsfig{file=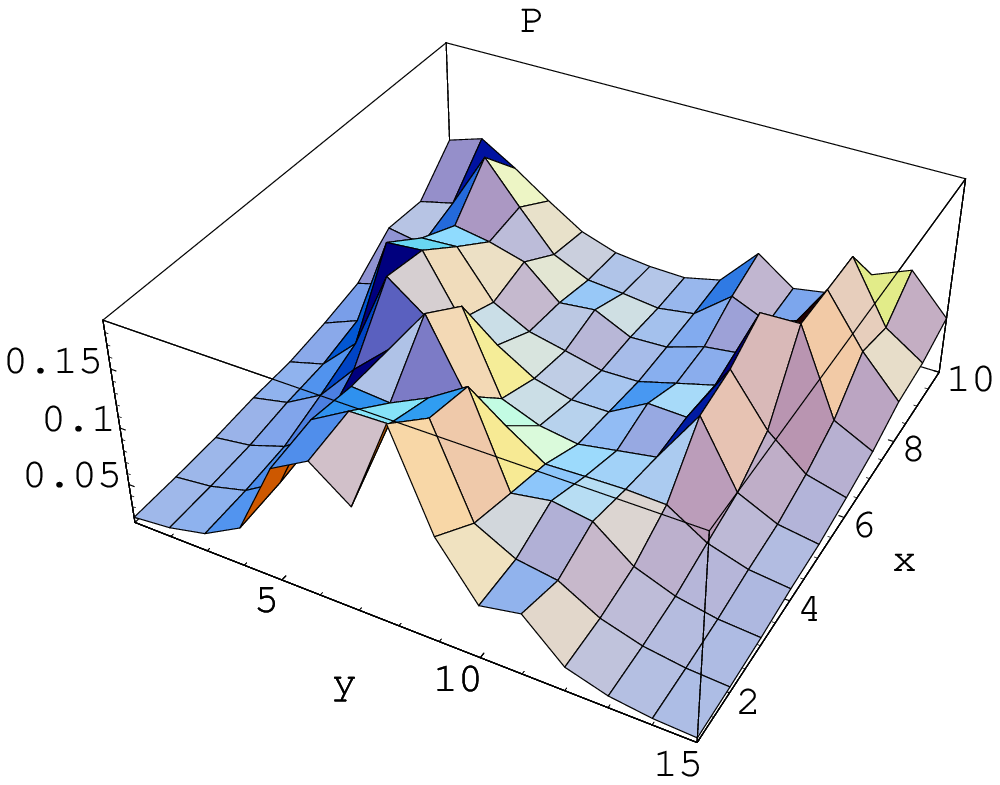, width= 65mm}
\epsfig{file=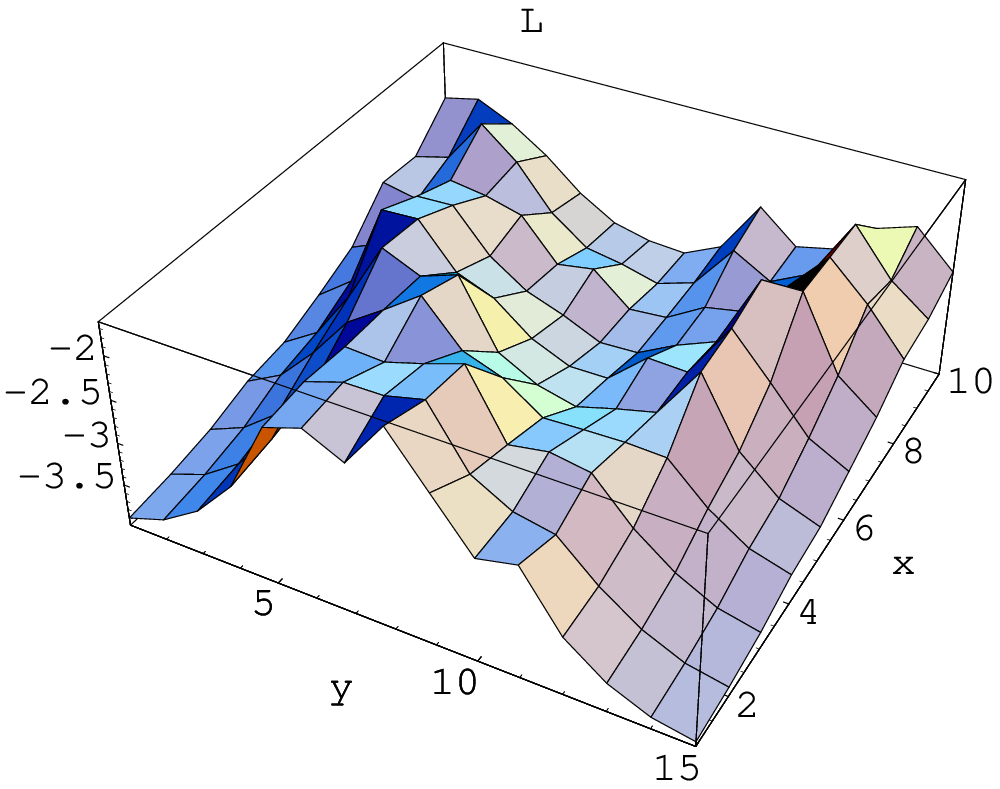, width= 65mm}\\
\vspace{0.5cm}
\epsfig{file=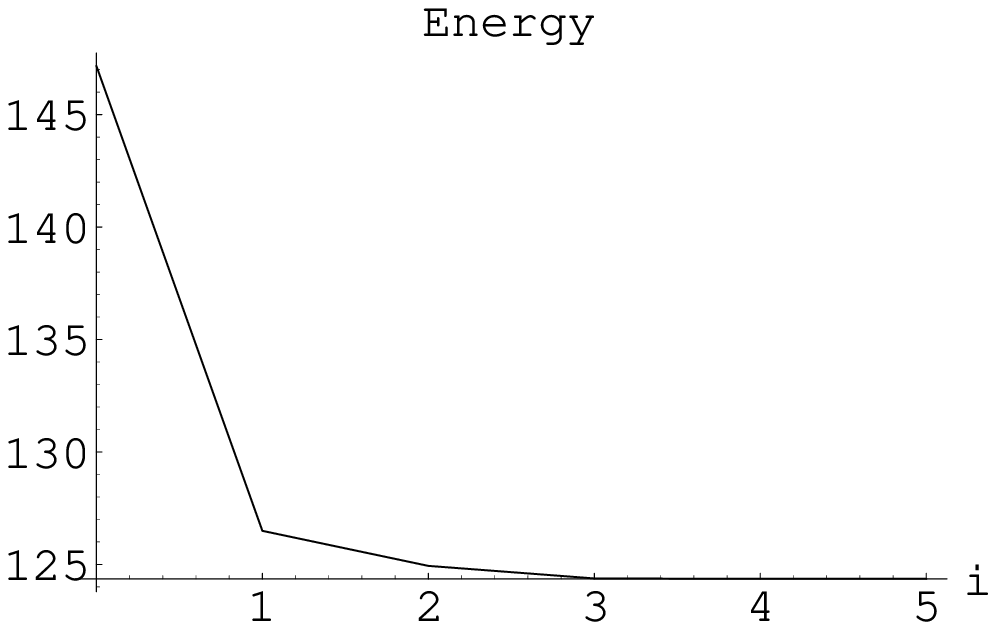, width= 50mm}
\epsfig{file=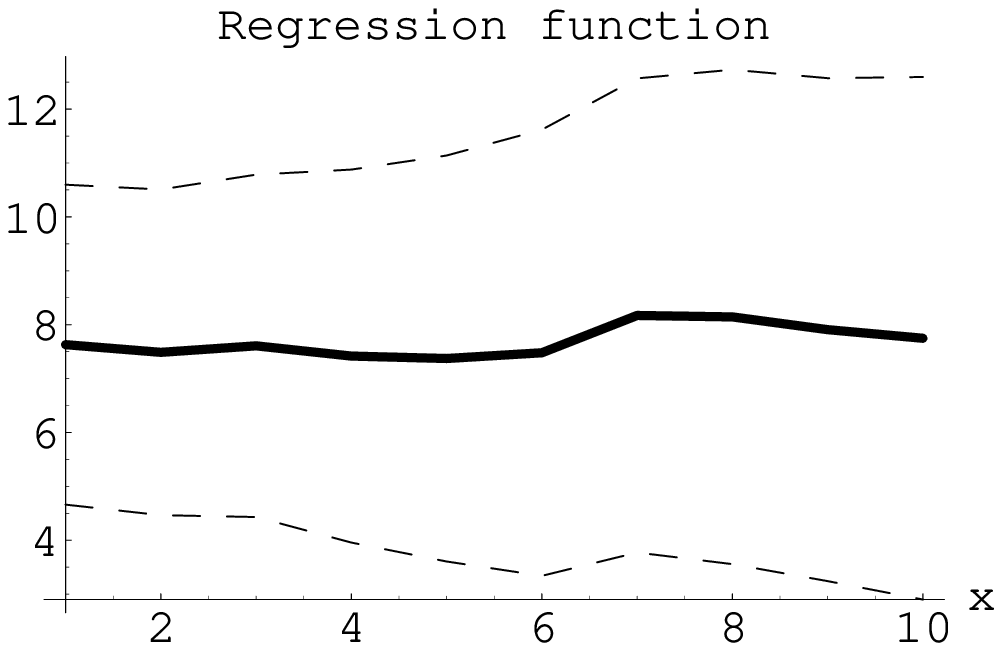, width= 50mm}\\
\epsfig{file=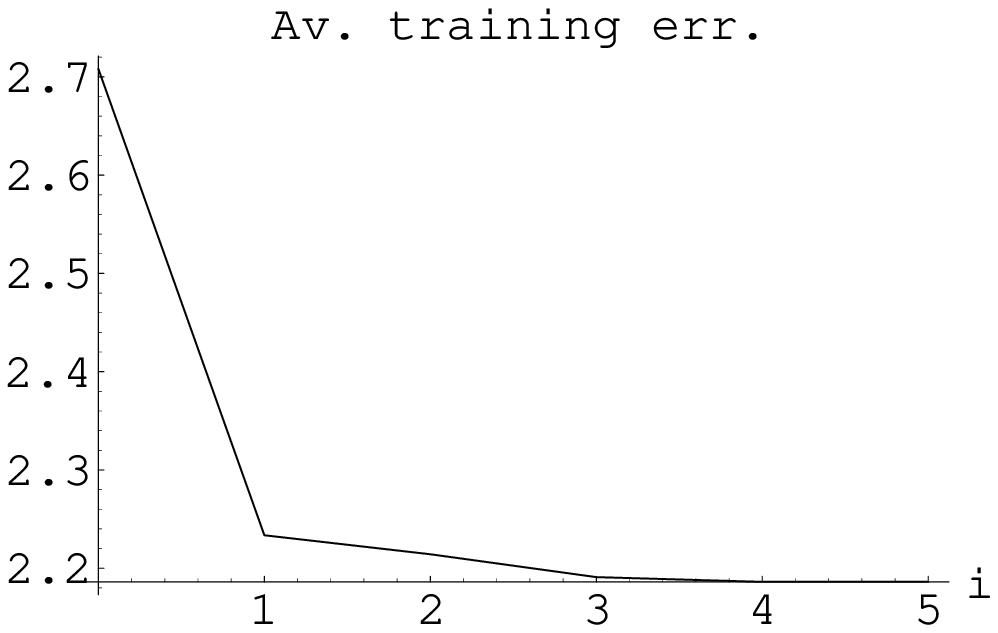, width= 50mm}
\epsfig{file=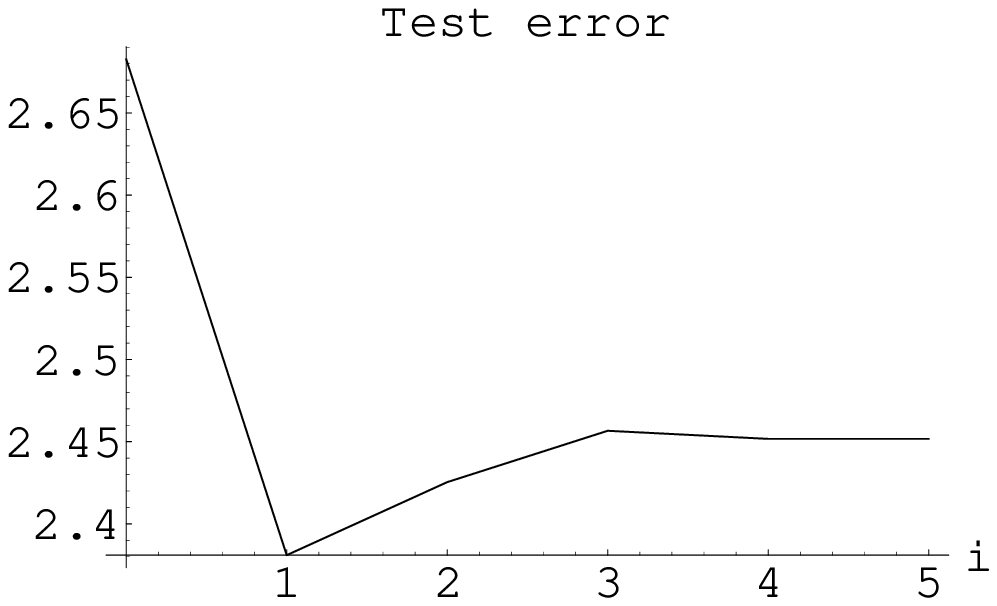, width= 50mm}
\end{center}
\caption{
Starting from a uniform initial guess. 
(Same as Fig.\ \ref{mix50a}, 
but initialized with uniform $L$.)
The resulting solution is, compared to 
Figs.\ \ref{mix50a} and \ref{mix50b},  
a bit more wiggly, i.e., more data oriented.
One recognizes a slight ``overfitting'',
meaning that the test error increases
while the training error is decreasing.
(Despite the increasing of the test error during iteration
at this value of $\lambda$,
a better solution cannot necessarily be found
by just changing $\lambda$--value.
This situation can for example occur, if
the initial guess is better then the implemented prior.)
}
\label{mix50c}
\end{figure}

\begin{figure}[ht]
\begin{center}
\epsfig{file=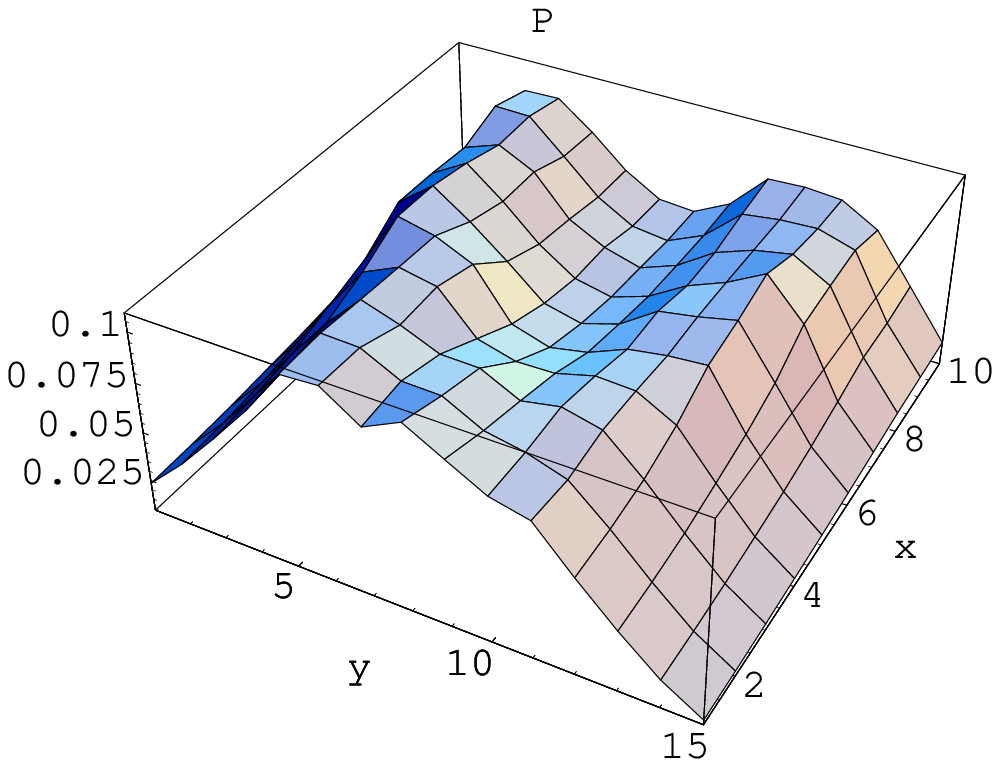, width= 65mm}
\epsfig{file=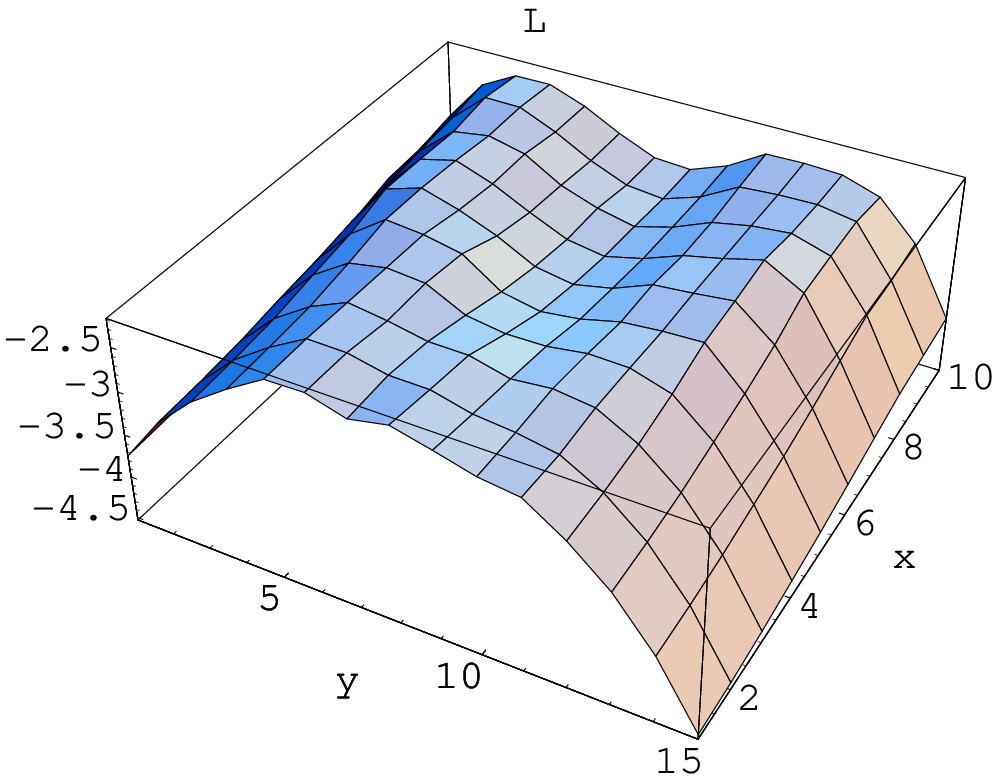, width= 65mm}\\
\vspace{0.5cm}
\epsfig{file=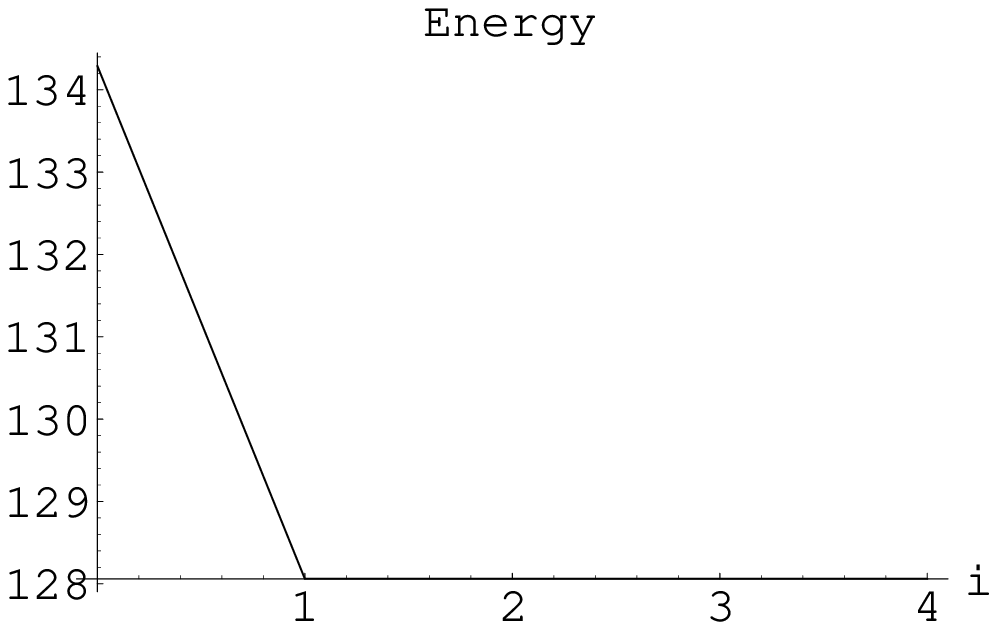, width= 50mm}
\epsfig{file=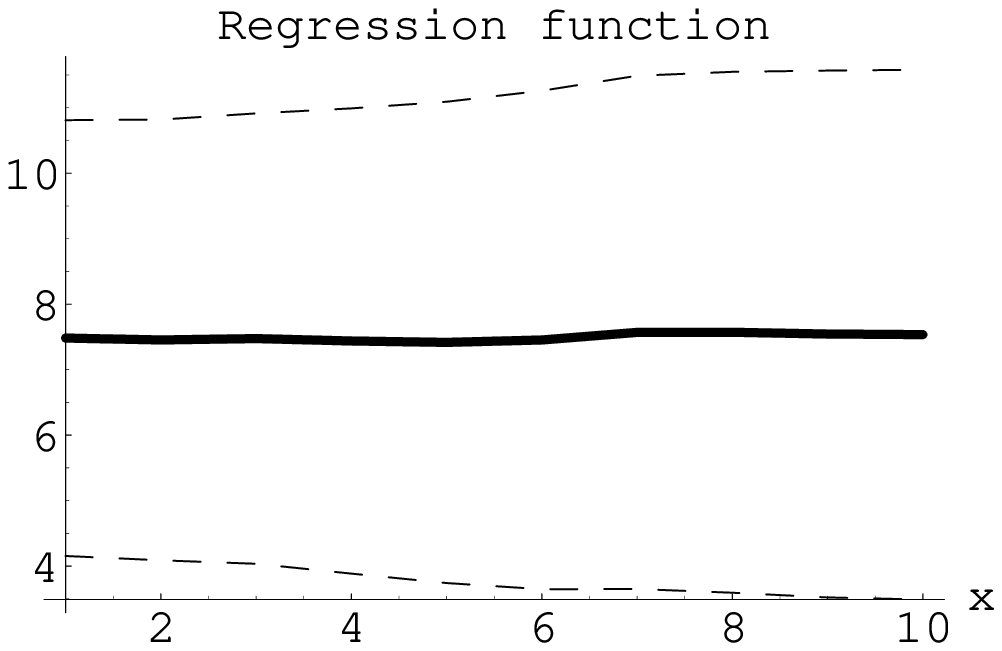, width= 50mm}\\
\epsfig{file=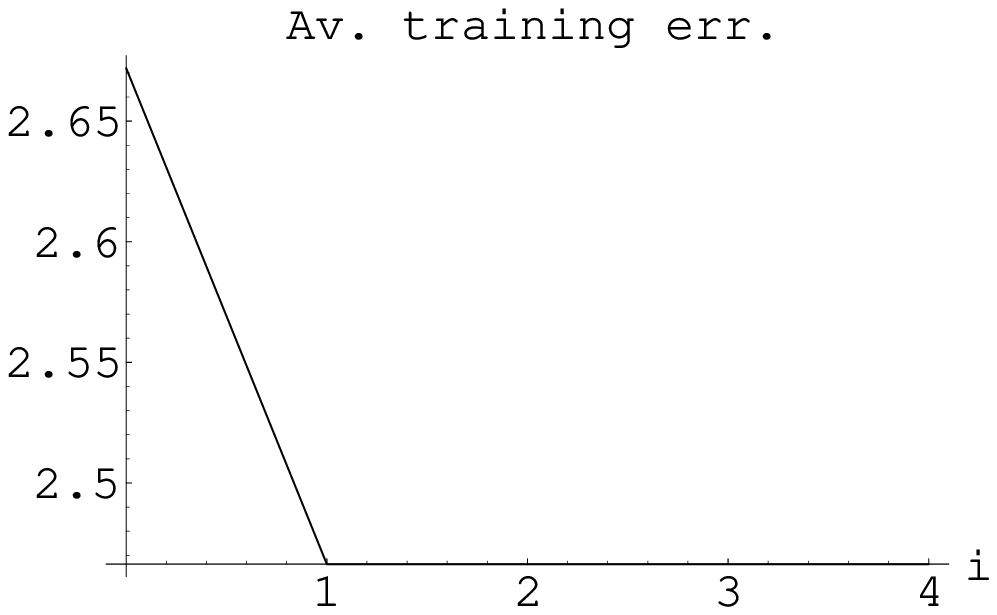, width= 50mm}
\epsfig{file=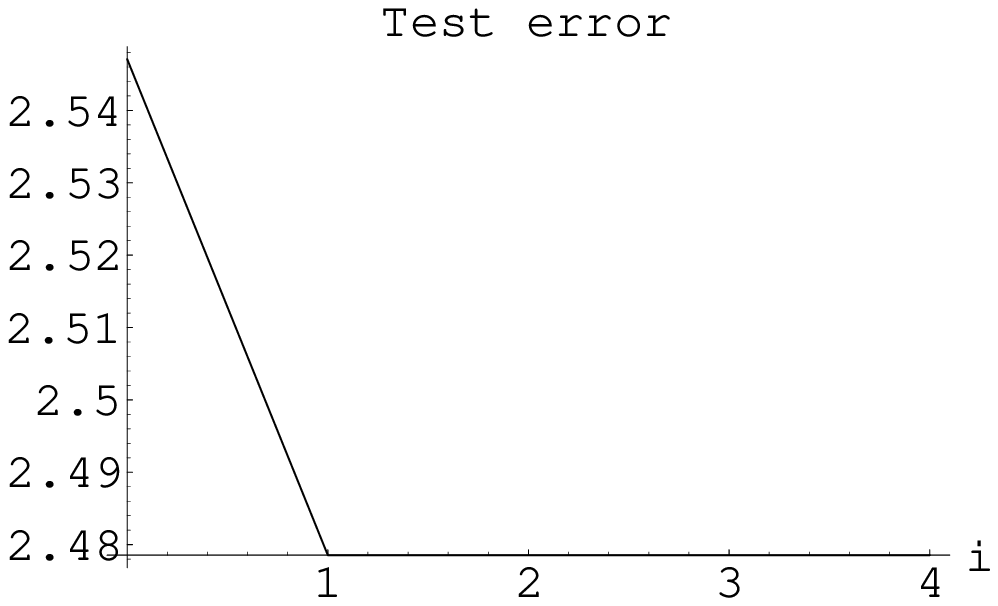, width= 50mm}
\end{center}
\caption{
Large $\lambda$.
(Same parameters as for Fig.\ \ref{mix50a}, 
except for $\lambda$ = 1.0.)
Due to the larger smoothness constraint
the averaged training error is larger than in Fig.\ \ref{mix50a}.
The fact that also the test error is larger than in Fig.\ \ref{mix50a}
indicates that the value of $\lambda$ is too large.
Convergence, however, is very fast.
}
\label{mix50d}
\end{figure}

\begin{figure}[ht]
\begin{center}
\epsfig{file=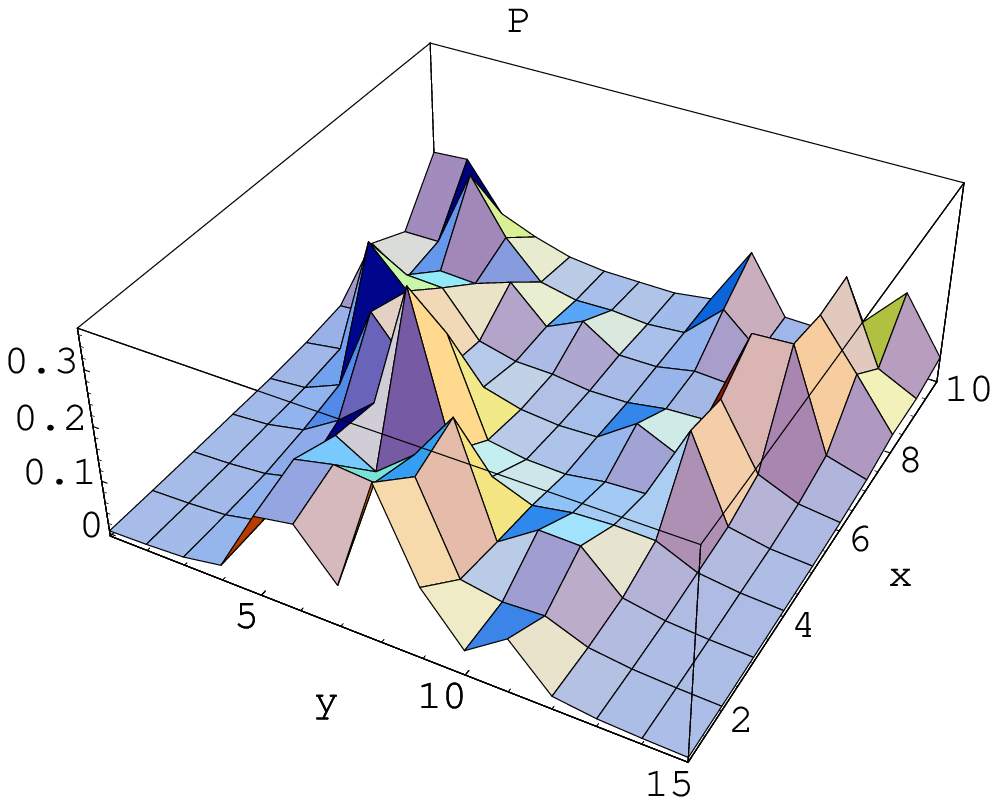, width= 65mm}
\epsfig{file=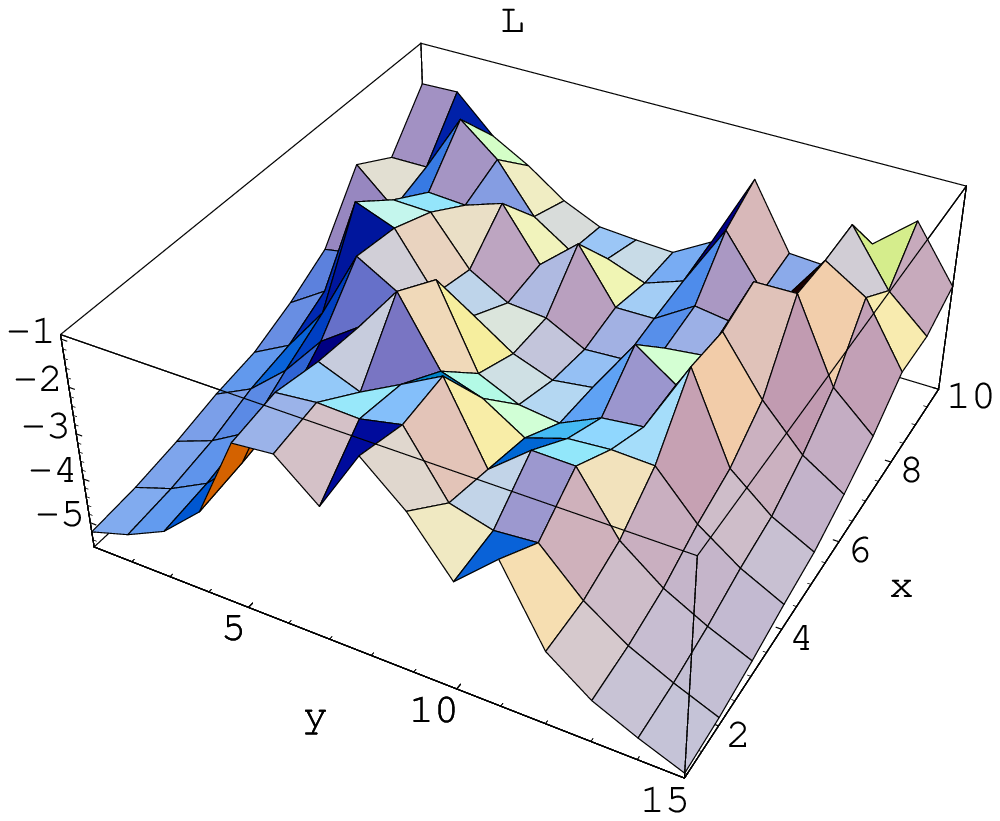, width= 65mm}\\
\vspace{0.5cm}
\epsfig{file=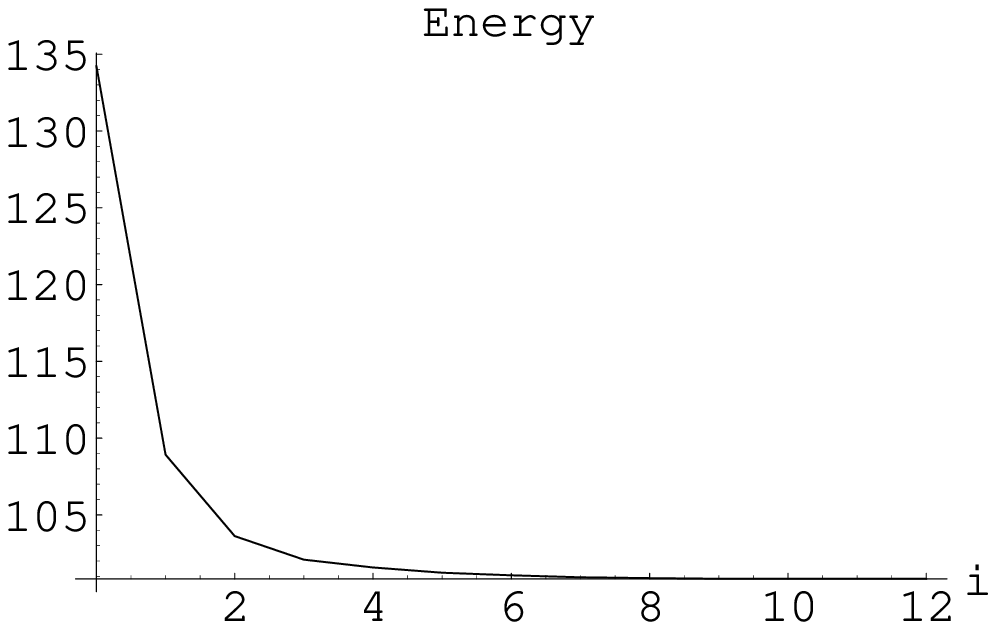, width= 50mm}
\epsfig{file=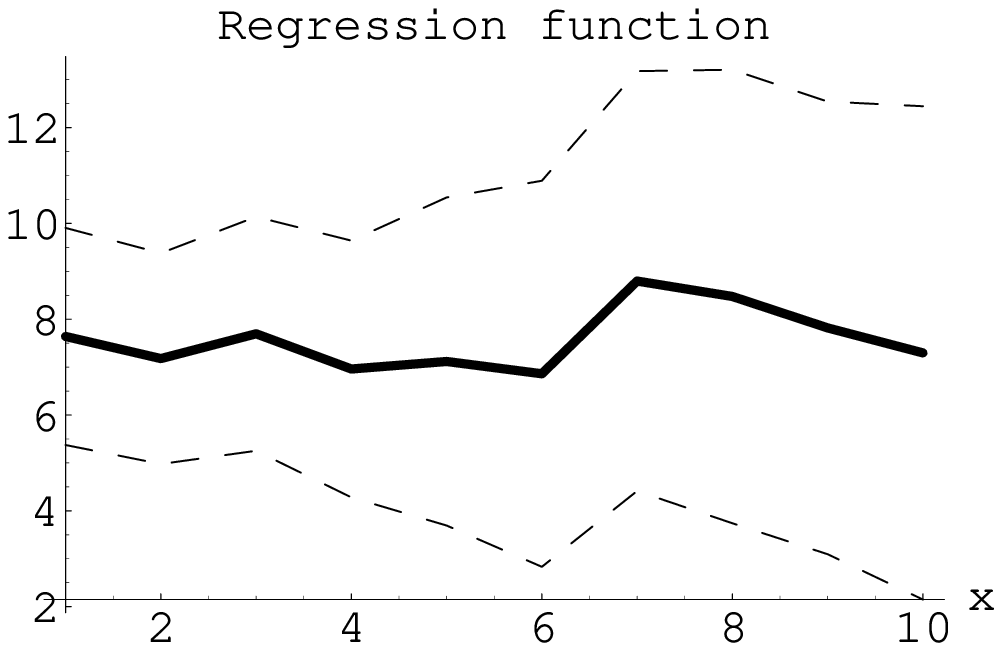, width= 50mm}\\
\epsfig{file=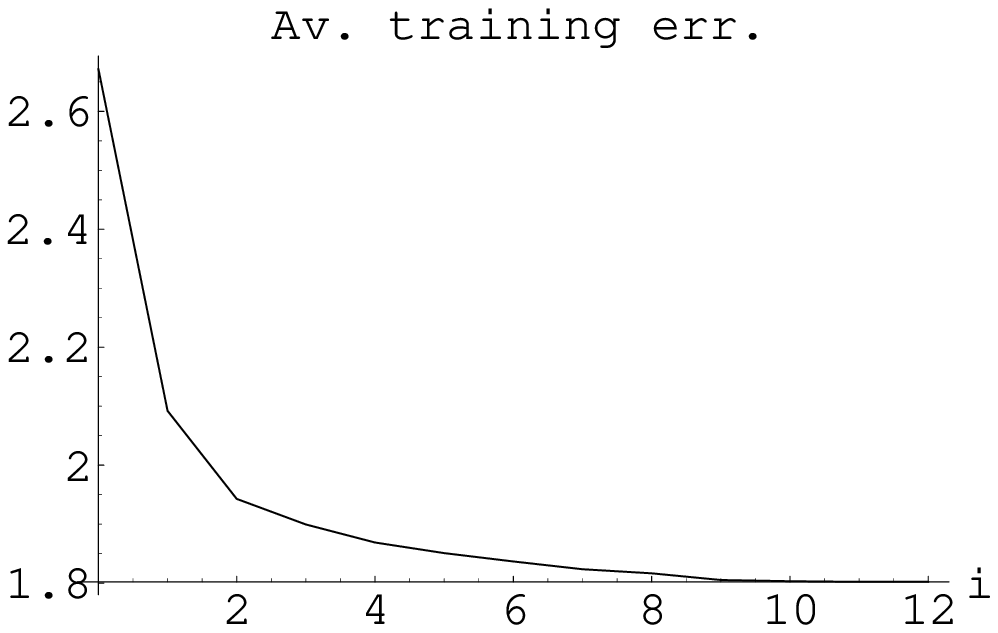, width= 50mm}
\epsfig{file=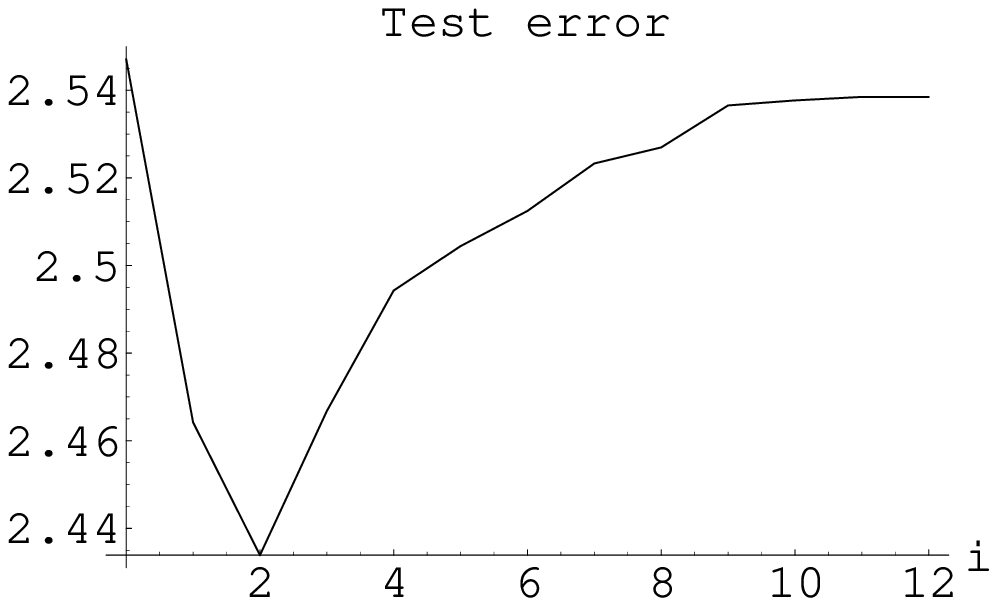, width= 50mm}
\end{center}
\caption{
Overfitting due to too small $\lambda$.
(Same parameters as for Fig.\ \ref{mix50a}, 
except for $\lambda$ = 0.1.)
A small $\lambda$ allows the average training error 
to become quite small.
However, the average test error grows
already after two iterations.
(Having found at some $\lambda$--value
during iteration an increasing test error, 
it is often but not necessarily the case that
a better solution can be found
by changing  $\lambda$.)
}
\label{mix50e}
\end{figure}

\begin{figure}[ht]
\begin{center}
\epsfig{file=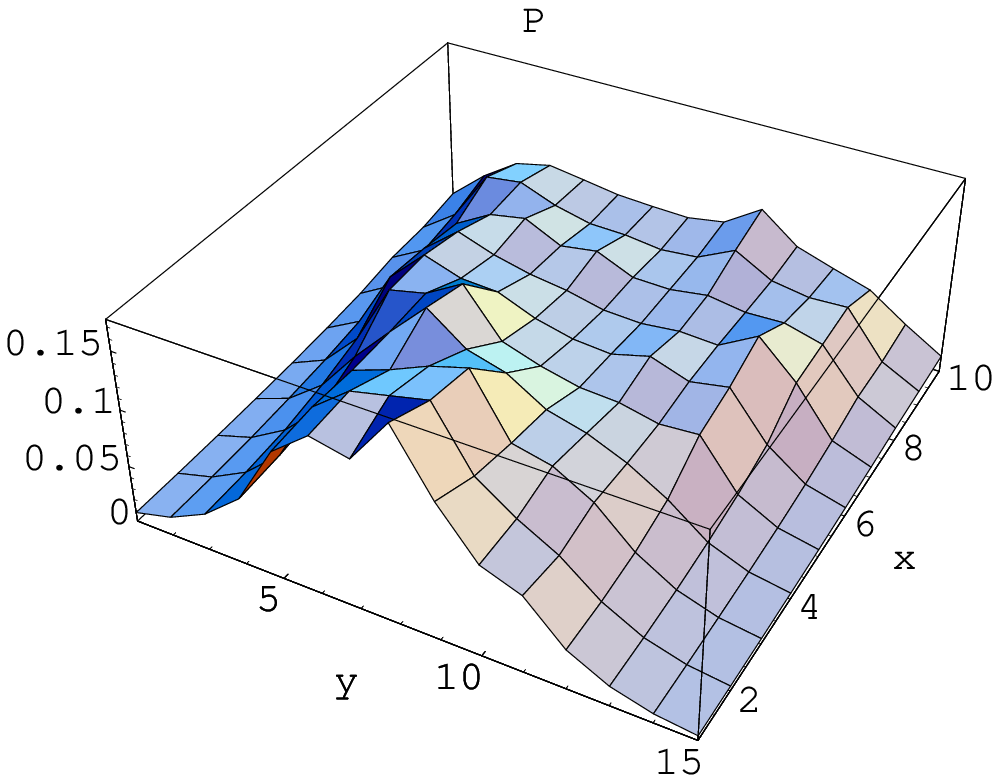, width= 65mm}
\epsfig{file=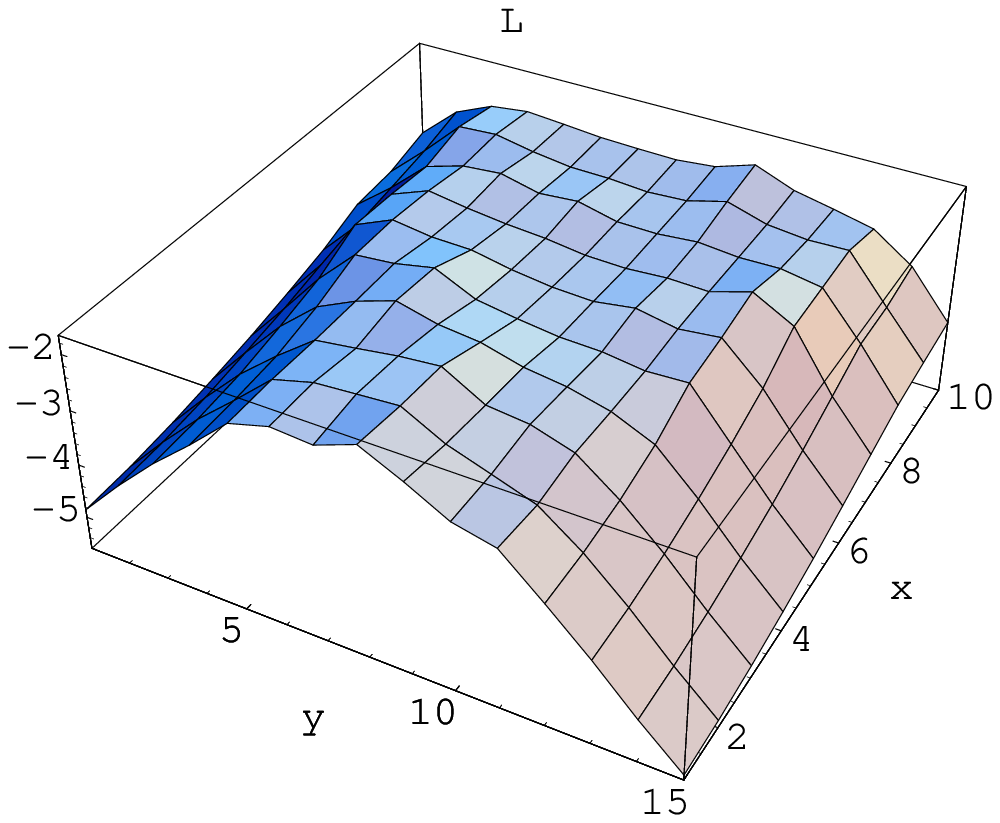, width= 65mm}\\
\vspace{1cm}
\epsfig{file=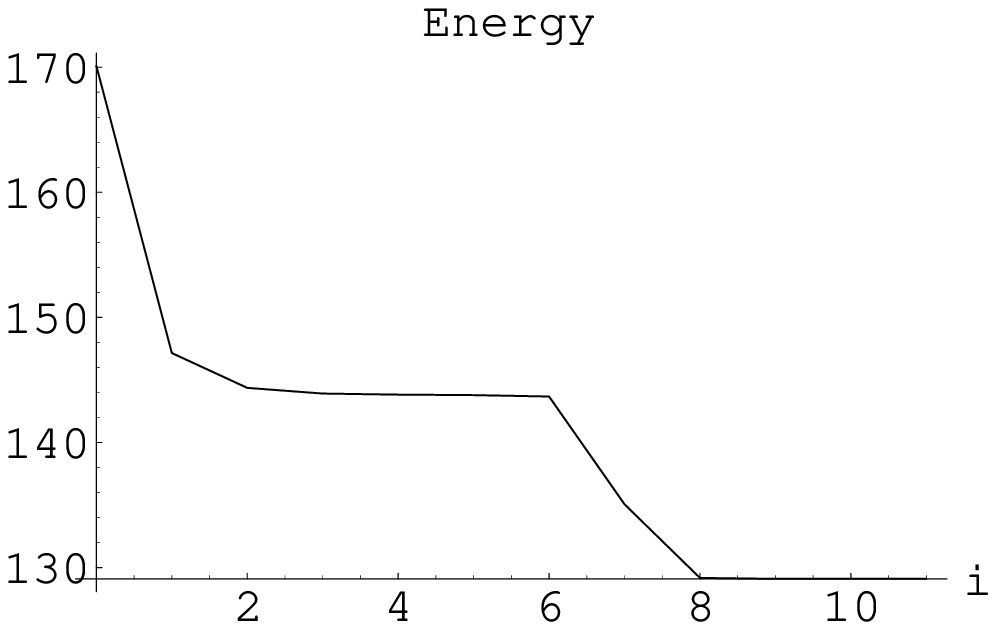, width= 50mm}
\epsfig{file=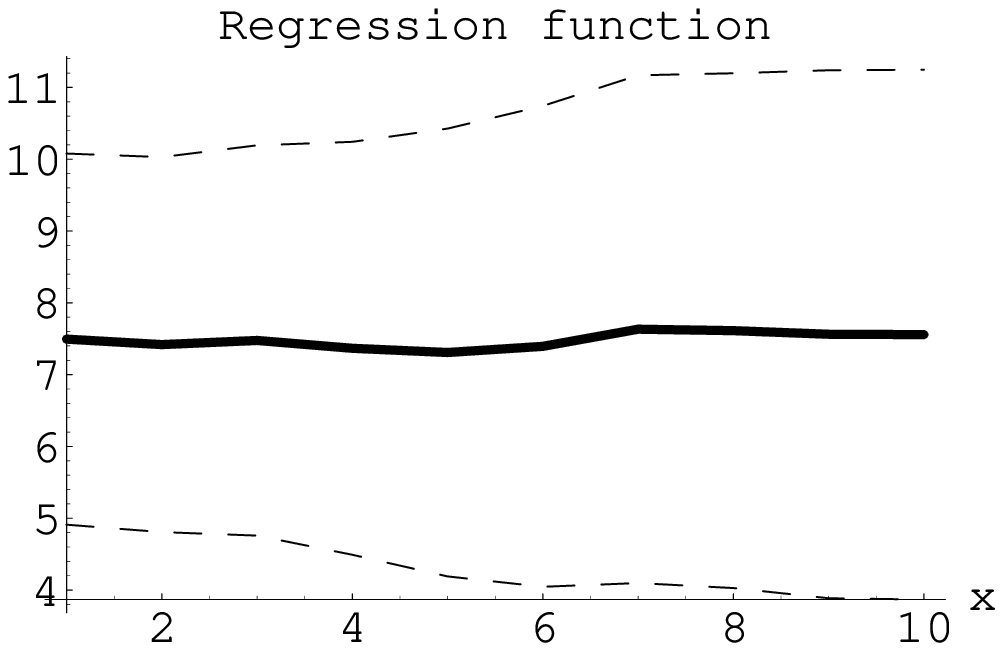, width= 50mm}\\
\epsfig{file=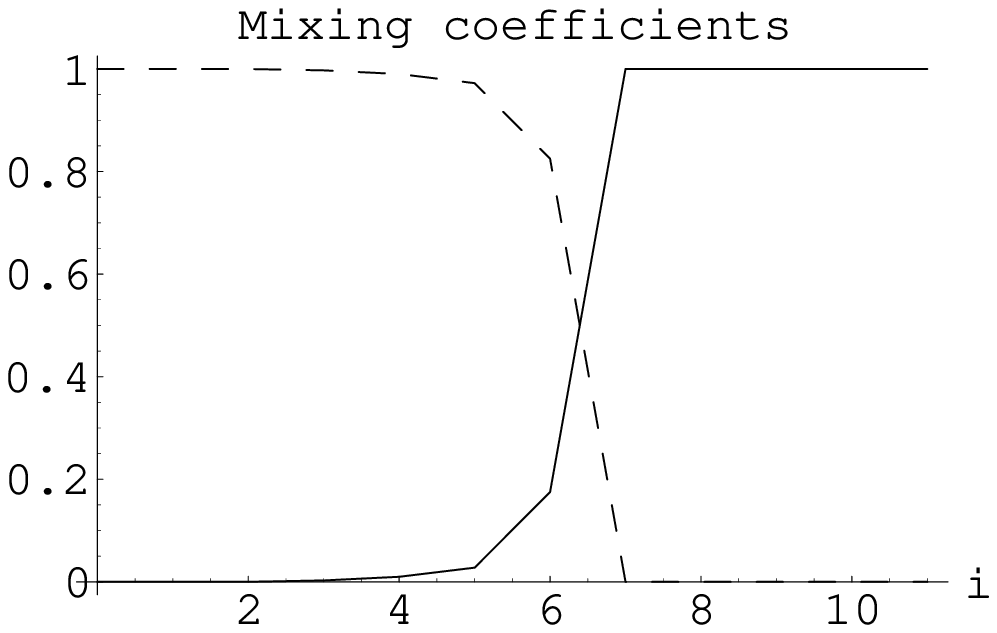, width= 50mm}
\epsfig{file=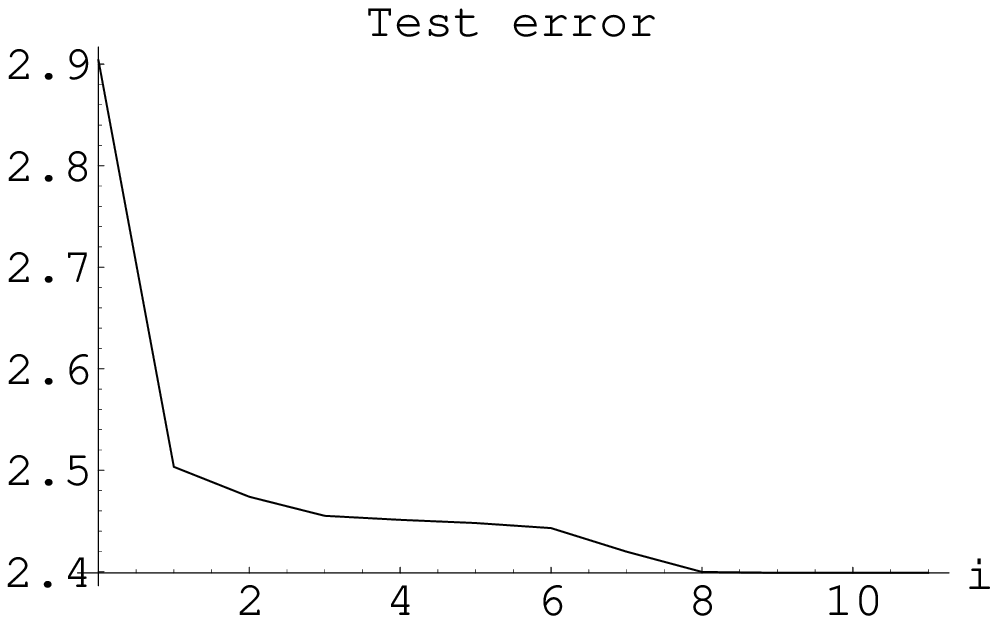, width= 50mm}
\end{center}
\caption{
Example of an approximately stable solution.
(Same parameters as for Fig.\ \ref{mix50a}, 
except for $\lambda$ = $1.2$, $m^2$ = $0.5$,
and initialized with $L$ = $t_2$.)
A nearly stable solution is obtained
after two iterations,
followed by a plateau between iterations 2 and 6.
A better solution is finally found
with smaller distance to template $t_1$.
(The plateau gets elongated with growing mass $m$.)
The figure on the l.h.s.\ in the bottom row
shows the mixing coefficients $a_j$ 
of the components of the prior mixture model
for the solution during iteration
($a_1$, line and $a_2$, dashed).
}
\label{mix50f}
\end{figure}

Typical numerical results of a prior mixture model
with two mixture components 
are presented in Figs.\ \ref{mix50a} -- \ref{mix50f}.
Like for Fig.\ \ref{mix50g}, the true likelihood 
used for these calculations is given by Eq.\ (\ref{trueP})
and shown in Fig.\ \ref{mixTrue}. 
The corresponding true regression function 
is thus that of Fig.\ \ref{mixTrueR}.
Also, the same training data 
have been used as for the model of Fig.\ \ref{mix50g}
(Fig.\ \ref{mixData}).
The two templates $t_1$ and $t_2$ 
which have been selected for the two prior mixture components
are (Fig.\ \ref{mixTrue})
\bea
t_1(x,y) 
& = &
\frac{1}{2\sqrt{2\pi}\sigma_t}
 \left(
                 e^{-\frac{(y-\mu_a)^2}{2\sigma_t^2}}
 +
                 e^{-\frac{(y-\mu_b)^2}{2\sigma_t^2}}
 \right)
,
\\
t_2(x,y) 
&= &
\frac{1}{\sqrt{2\pi}\sigma_t}e^{-\frac{(y-\mu_2)^2}{2\sigma_t^2}}
,
\eea
with
$\sigma_t$ = $2$,
$\mu_a$ = $\mu_2+25/9$ 
= $10.27$,
$\mu_b$ = $\mu_2-25/9$  
= $4.72$,
and $\mu_2$ = $15/2$.
Both templates capture a bit of the structure of the true likelihood,
but not too much, so learning remains interesting.
The average test error of $t_1$ is equal to 2.56
and is thus lower than that of $t_2$ being equal to 2.90.
The minimal possible average test error 2.23 
is given by  that of the true solution $P_{\rm true}$.
A uniform $P$, being the effective template in
the zero mean case of Fig.\ \ref{mix50g},
has with 2.68 an average test error
between the two templates $t_1$ and $t_2$.

Fig.\ \ref{mix50a} proves 
that convergence is fast for massive prior relaxation
when starting from $t_1$ as initial guess $L^{(0)}$.
Compared to  Fig.\ \ref{mix50g} the solution is a bit smoother,
and as template $t_1$ is a better reference than the uniform likelihood
the final test error is slightly 
lower than for the zero--mean Gaussian prior on $L$.
Starting from $L^{(0)}$ = $t_2$ 
convergence is not much slower
and the final solution is similar,
the test error being in that particular case even lower
(Fig.\ \ref{mix50b}).
Starting from a uniform $L^{(0)}$ 
the mixture model produces a solution very similar
to that of Fig.\ \ref{mix50g} (Fig.\ \ref{mix50b}).

The effect of changing the 
$\lambda$ parameter of the prior mixture
can be seen in 
Fig.\ \ref{mix50d}  
and Fig.\ \ref{mix50e}.
Larger $\lambda$
means a smoother solution
and faster convergence
when starting from a template likelihood
(Fig.\ \ref{mix50d}).
Smaller $\lambda$
results in a more rugged solution
combined with a slower convergence.
The test error in Fig.\ \ref{mix50e}
already indicates overfitting.

Prior mixture models tend to produce metastable
and approximately stable solutions.
Fig.\ \ref{mix50f}  presents an example
where starting with $L^{(0)}$ = $t_2$
the learning algorithm seems to have produced a stable solution
after a few iterations.
However, iterating long enough
this decays into a solution 
with smaller distance to $t_1$ and with lower test error.
Notice that this effect can be prevented by starting
with another initialization,
like for example with $L^{(0)}$ = $t_1$
or a similar initial guess.

We have seen now that, and also how, 
learning algorithms for Bayesian field theoretical models
can be implemented.
In this paper, the discussion of numerical aspects 
was focussed on general density estimation problems.
Other Bayesian field theoretical models,
e.g., for regression and inverse quantum problems,
have also been proved to be numerically feasible.
Specifically, 
prior mixture models for Gaussian regression 
are compared
with so--called Landau--Ginzburg models
in \cite{Lemm-1996}.
An application of prior mixture models
to image completion,
formulated as a Gaussian regression model,
can be found in \cite{Lemm-1999a}.
Furthermore, hyperparameter have been included in numerical calculations
in \cite{Lemm-1998} and also in \cite{Lemm-1999a}.
Finally, learning algorithms for 
inverse quantum problems are treated
in \cite{Lemm-1999b} for inverse quantum statistics, 
and, in combination with a mean field approach,
in \cite{Lemm-1999c} for inverse quantum many--body theory.
Time--dependent inverse quantum problems
will be the topic of \cite{Lemm-1999d}.

In conclusion, we may say that 
many different Bayesian field theoretical models
have already been studied numerically
and proved to be computationally feasible.
This also shows that such nonparametric Bayesian approaches
are relatively easy to adapt
to a variety of quite different learning scenarios.
Applications of Bayesian field theory
requiring further studies
include, for example, the prediction of time--series
and the interactive implementation of unsharp 
{\it a priori} information.

\vspace{1cm}
\noindent{\bf Acknowledgements}
\noindent
The author wants to thank 
Federico Girosi, Tomaso Poggio,
J\"org Uhlig, and Achim Weiguny 
for discussions.


\end{document}